\documentclass[11pt]{report}         
\usepackage{cite}
\usepackage{graphicx}
\usepackage{psfrag}
\usepackage{ubcthesis}               
\usepackage{url}

\makeatletter
  \newif\if@openbib
  \@openbibfalse 
\makeatother

\newcommand{\be}{\begin{equation}}
\newcommand{\ee}{\end{equation}}

\newcommand{\bea}{\begin{equation}\begin{array}{rcl}\displaystyle}
\newcommand{\eea}{\end{array}\end{equation}}

\newcommand{\abs}[1]{\left|#1\right|}     

\newcommand{\expect}[1]{\left<#1\right>}  
\newcommand{\floor}[1]{\left\lfloor#1\right\rfloor}
\newcommand{\erf}{\rm erf}
\newcommand{\half}{\frac{1}{2}}
\newcommand{\var}[1]{{\rm Var}\left[#1\right]}	
\newcommand{\cov}[2]{{\rm Cov}\left[#1, #2\right]}	
\newcommand{\kurt}{{\rm Kurt}} 

\newcommand{\ap}[1]{Appendix \ref{ap:#1}}
\newcommand{\ch}[1]{Chapter \ref{ch:#1}}
\newcommand{\enum}[1]{\ref{enum:#1}}
\newcommand{\eq}[1]{Eq.\ \ref{eq:#1}}
\newcommand{\eqs}[2]{Eqs.\ \ref{eq:#1}--\ref{eq:#2}}
\newcommand{\fig}[1]{Fig.\ \ref{fig:#1}}
\newcommand{\sect}[1]{Section \ref{sect:#1}}
\newcommand{\tbl}[1]{Table \ref{tbl:#1}}

\newcommand{\widebox}[1]{\\ \framebox{\parbox{\columnwidth}{#1}}}

\leftchapter

\singlespace

\renewcommand{\thesisauthor}{Hendrik J. Blok} 
\renewcommand{\thesismonth}{November}              
\renewcommand{\thesisyear}{2000}              

\renewcommand{\thesistitle}{ON THE NATURE OF THE STOCK MARKET: SIMULATIONS AND EXPERIMENTS}     

\renewcommand{\thesisauthorpreviousdegrees}   
{ B.Sc., University of British Columbia, 1993 \\
  M.Sc., University of British Columbia, 1995 }
\renewcommand{\thesissupervisor}              
{Birger Bergersen}
\renewcommand{\thesisauthoraddress}           
{ \#301 - 988 W.\ 16th Ave. \\                 
  Vancouver, B.C., V5Z 1T2 }
  
\renewcommand{\thesiscommitteesize}{5}
\renewcommand{\thesisdepartment}{Department of Physics and Astronomy}

\begin{document}

\thesistitlepage

\begin{thesisabstract}

Over the last few years there has been a surge of activity within the physics community in the emerging field of {\em Econophysics}---the study of economic systems from a physicist's perspective.  Physicists tend to take a different view than economists and other social scientists, being interested in such topics as phase transitions and fluctuations.

In this dissertation two simple models of stock exchange are developed and simulated numerically.  The first is characterized by centralized trading with a market maker.  Fluctuations are driven by a stochastic component in the agents' forecasts.  As the scale of the fluctuations is varied a critical phase transition is discovered.  Unfortunately, this model is unable to generate realistic market dynamics.

The second model discards the requirement of centralized trading.  In this case the stochastic driving force is Gaussian-distributed ``news events'' which are public knowledge.  Under variation of the control parameter the model exhibits two phase transitions: both a first- and a second-order (critical).  

The decentralized model is able to capture many of the interesting properties observed in empirical markets such as fat tails in the distribution of returns, a brief memory in the return series, and long-range correlations in volatility.  Significantly, these properties only emerge when the parameters are tuned such that the model spans the critical point.  This suggests that real markets may operate at or near a critical point, but is unable to explain why this should be.  This remains an interesting open question worth further investigation.

One of the main points of the thesis is that these empirical phenomena are not present in the stochastic driving force, but emerge endogenously from interactions between agents.  Further, they emerge despite the simplicity of the modeled agents; suggesting complex market dynamics do not arise from the complexity of individual investors but simply from interactions between (even simple) investors.

Although the emphasis of this thesis is on the extent to which multi-agent models can produce complex dynamics, some attempt is also made to 
relate this work with empirical data.  Firstly, the trading strategy applied by the agents in the second model is demonstrated to be adequate, if not optimal, and to have some surprising consequences.  

Secondly, the claim put forth by Sornette {\em et al.} \cite{sornette96} that large financial crashes may be heralded by accelerating precursory oscillations is also tested.  It is shown that there is weak evidence for the existence of {\em log-periodic precursors} but the signal is probably too indistinct to allow for reliable predictions.

\end{thesisabstract}

\tableofcontents
\listoftables
\listoffigures

\begin{thesisacknowledgments}
{

I would first like to thank my supervisor, Dr.\ Birger Bergersen, for his help and financial support over the last many years.  His remarkable ability to always keep both the ``forest'' {\em and} the ``trees'' in focus has been an inspiration to me.  I shall always try to do the same.

I am also particularly appreciative of the many helpful discussions with Dr.\ Alan Kraus of the Faculty of Commerce at UBC, whose input has been invaluable.  Thanks also to Casey Clements whose curiosity provided the inspiration for one of the models.

I gratefully acknowledge the financial support, through graduate fellowships, of the Crisis Points Group, Peter Wall Institute of Advanced Studies.  This unique interdisciplinary group has also afforded me the opportunity to learn about complex systems in many diverse fields.  I would especially like to thank the group's coordinator, Dr.\ Cindy Greenwood, for her helpful ideas and comments.

I would also like to thank a number of software developers whose free code segments were incorporated into my simulations: Grahame Grieve, Lucian Wischik, Fedor Koshevnikov, Igor Pavluk, Serge Korolev, and Troels Skovmand Eriksen...thanks!

Finally, I am greatly indebted to my wife and friend, Kathy, for always supporting me in my studies (even when it wasn't clear where they would lead) and for taking an interest in my work.  You always let me follow my dreams; now let's find out where {\em your} dreams lead.

}
\end{thesisacknowledgments}


\newpage
\thispagestyle{empty}
\begin{center}
(This page is intentionally blank.)
\end{center}

\chapter{Introduction}

\section{Financial markets}

Financial markets include {\em stock markets}, such as the New York Stock Exchange (NYSE), which deal in ownership {\em shares} of publicly-owned companies.  Companies owned privately can raise equity capital through an {\em initial public offering} (IPO) which releases part-ownership to the public.  When the public {\em stockholders} wish to sell some of their shares they do so on a financial market.  The market typically charges the company a fee to list it and may require the company to meet certain standards in order to protect investors.  

On the NYSE, trades are handled by a restricted number of {\em brokers} who are governed by the market's rules and regulations.  Brokers receive trading orders---consisting of a quantity of shares to trade and (optionally) a price---from the public and bring them to {\em specialists} who deal only with specific stocks.  

The specialist's role is to compare the highest {\em bid} (buy order) price with the lowest {\em offer} (sell order) price and if they meet, execute the trade.  At the beginning of trading each day the specialist also finds a fair market price for a stock by balancing the outstanding {\em supply} (total offers) with {\em demand} (total bids).  Although actually more complicated, for our purpose this is a sufficient explanation of the specialist's role.

\section{Motivation for research}

Neglecting {\em dividends} a company may pay to its shareholders, investors make money on the market by buying stocks at low prices and selling them at higher prices.  But given identical (publicly available) information one would expect (similar) investors to have the same prediction for how the price would move and they should all place similar orders.  

For example, if Betty hears that company XYZ has discovered oil, she may well expect the company's stock price to climb and so she places a bid order.  The trade will not be exercised, however, until another investor, say Sam, offers to sell his shares in XYZ.  But the question is then raised in Betty's mind, ``Why does Sam want to sell?''  Is he being irrational?  Did he miss the good news?  Does he know something Betty doesn't? 

Conventional wisdom assumes Betty and Sam have slightly different expectations about the future price of XYZ, perhaps due to imperfect information.  Thus the market dynamics are driven by random fluctuations.  But this assumption leads to predictions that price fluctuations should be normally distributed (or log-normally) and that trading volume should be low and steady.

What is actually observed on all markets is bursts of activity with very high volume and/or extremely large fluctuations in price which occur much more frequently than the conventional wisdom can account for.  The reason for these bursts has not been established and is an interesting topic of research.

\subsection{Motivation for the physicist}

Statistical physicists and condensed matter theorists have developed a significant arsenal of tools for analyzing many-particle systems with strong, localized interactions.  Methods such as mean-field theory, the renormalization group, and finite-scaling analysis allow physicists to explore complex, irreducible systems such as spin glasses (highly disordered magnetic systems) where the important details are in the interactions between the particles, rather than the individual particles themselves.

Recently, physicists have realized that the methods developed above may also be useful for non-physical systems such as ecological and social systems.  The leap of faith required is the assumption that it is not necessary to fully understand the individuals in the system themselves (their motivations, for instance), but only to the point that one can construct reasonable rules for the interactions between individuals.

Whether this leap of faith is justified remains an open question but interest is mounting within the physics community in complex, socio-economic systems like the stock market.  In 1995 the Los Alamos National Laboratory (LANL) condensed-matter preprint archive (\url{http://arXiv.org}) accepted three papers containing the word ``market'' in their abstracts (a check was done confirming they were all finance-related) representing 0.16\% of the submissions that year.  Since then it has doubled every year through 1999 when fifty of the 5,490 (0.91\%) submissions were market-related.  (The foray of physicists into economics has come to be known as {\em Econophysics}.)

\subsubsection{Phase transitions}

Together with the analytic tools physicists bring to the subject, they also bring a fresh perspective and new questions.  For instance, it has been empirically observed that price fluctuations exhibit {\em scaling} \cite{cont97b, mandelbrot97, gopikrishnan98, gopikrishnan99}, meaning the fluctuations appear invariant under a change of scale, over orders of magnitude from a few minutes to a few days.  Scaling is characterized by power-law distributions which are very familiar to physicists because they occur near second-order or {\em critical} phase transitions.  

Phase transitions, in the context of thermodynamics, are well understood phenomena.  The terminology of {\em order parameter}, a dependent variable which undergoes a ``sharp'' change, and {\em control parameter}, the variable which is smoothly adjusted to produce the change, is used to quantify the transition.  In the case of {\em first-order transitions} (such as melting) the order parameter undergoes a discontinuity---it jumps to a new value.  The jump is accompanied by an absorption or liberation of energy (latent heat).  Usually, fluctuations within the substance can be ignored for first-order transitions.

To demystify the above definitions, consider a pot of water boiling at 1 atmosphere of pressure and $100 ^\circ$C.  If we choose temperature as the control parameter then density could play the role of the order parameter.  Below $100 ^\circ$C water is a liquid with a relatively high density.  Above this point, all the water is in the form of steam which has a significantly lower density.  At the transition we observe fluctuations in the form of small, uniform steam bubbles.  This is a first-order transition.

\begin{figure}\centering
	\psfrag{order}[r][r]{order}
	\psfrag{control}[r][r]{control}
	\psfrag{First-order}[c][c]{First-order}
	\psfrag{Second-order}[c][c]{Second-order}
\includegraphics[width=\columnwidth]{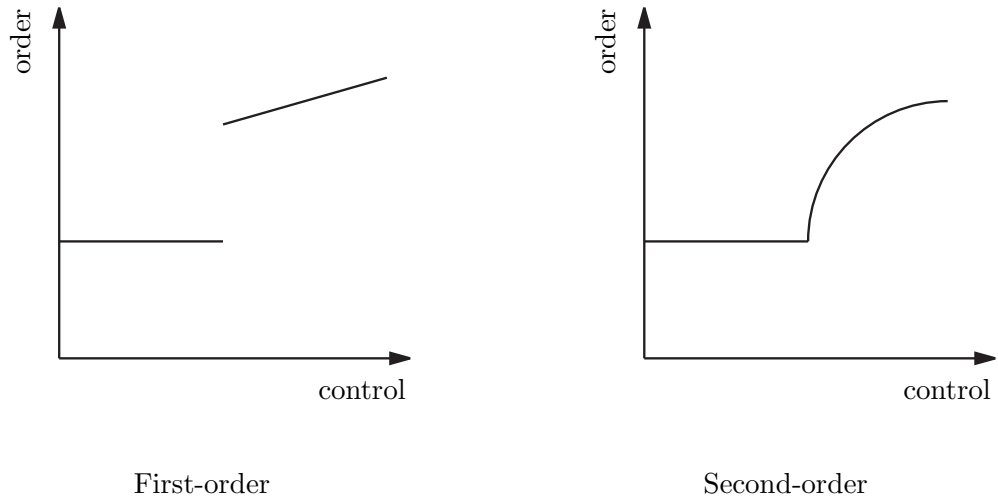}
	\caption{Sample phase transitions.  A first-order transition (left) is characterized by a discontinuity in the order parameter, while a second-order (critical, right) is discontinuous in the first derivative.}
\label{fig:introPhaseTrans}
\end{figure}

In contrast, second-order, or {\em critical}, transitions are characterized by a discontinuity in the derivative of the order parameter (see \fig{introPhaseTrans}).  In fact, the derivative diverges at the critical point.  Further, near the transition, properties are dominated by internal fluctuations on all scales.  For example, let us revisit our pot of boiling water.  We raise the pressure to 218 atm and the temperature to $374 ^\circ$C (the critical point of water).  Again, we observe steam bubbles but in this case the bubbles exist on all scales---from microscopic to the size of the pot itself \cite{mouritsen84}.  Also, the density (but not its derivative) is continuous across the transition.

Near a critical point, many thermodynamic properties obey diverging power laws.  Early studies of critical phenomena revealed that the characteristic exponents for the power laws clustered around distinct values for a variety of systems.  This suggested that some of the features of separate systems were irrelevant---they belonged to the same {\em universality class}.  Some of the irrelevant variables in a universality class are usually the type of local interactions, the number of nearest neighbours, et cetera.  On the other hand, dimensionality and symmetry, for example, appear to be relevant variables---that is, change the number of dimensions or the symmetry laws and the system changes its class (or may even cease being critical).  Critical systems with the same relevant variables but different irrelevant variables have the same critical exponents and are said to belong to the same universality class.  
	
Returning to our discussion of markets, the estimated power law exponent in the financial data seems to exhibit universality.  That is, the exponents seem to be similar for a number of different markets and stocks and they also seem not to change over time \cite{mandelbrot63, mantegna95, gopikrishnan98, gopikrishnan99}.  This evidence suggests that markets operate at or near a dynamical critical point as studied by physicists.

\subsubsection{Self-organized criticality}

To a physicist, the question of whether the market operates at a critical point is especially interesting.  The traditional theory of critical phenomena states that a system will approach a critical point via deliberate tuning of the control parameter.  In the above example, by adjusting both the temperature and pressure, water was brought to its critical point.

This description does not seem to apply to markets, however.  The rules governing market dynamics were not chosen in order to put the market in a critical state.  In fact, there does not appear to be any analog for temperature, which could be used to explain why the market might be at a critical point.  If it is critical, it appears to have arrived there spontaneously, without any tuning of a control parameter.  This phenomenon has come to be known as {\em self-organized criticality} (SOC) and was originally proposed as a possible explanation for scaling in many natural phenomena \cite{bak87, bak88}.  

The canonical example of SOC is a pile of sand to which grains are added very slowly.  As each grain is dropped it may cause the local slope of the pile to exceed a threshold and collapse, dispersing grains within a local neighbourhood.  These grains may cause further instabilities producing a cascade reaction.  Measuring the total effect of dropping each grain yields a power-law distribution of avalanche sizes, indicating the presence of a critical point.  The criticality is said to be self-organizing because it emerges spontaneously from the simple process of dropping grains periodically.

In some cases, SOC can be mapped back onto traditional criticality by a separation of timescales: systems which responds quickly to very slow driving forces are candidates for SOC.  In particular, the sand pile model described above qualifies for this mapping because the driving force (dropping of grains) is much slower than the duration of the avalanches \cite{sornette95}.  

More generally, the appearance of SOC can be an artifact of how the system is constructed.  Some natural choices of parameter values (such as an infinitesimal driving rate, as discussed above) automatically lead to dynamics which can be critical or very nearly so.  Traditional criticality is only revealed when the parameter is manually varied \cite{blok99} (for example, by increasing the rate at which sand is added to the pile).

Whether the markets operate at a critical point and, if so, how they develop towards and maintain this state is of interest to physicists.

\section{Anticipated challenges}

Although synthetic constructs, the markets are difficult to study scientifically for many of the same reasons as natural phenomena.  Firstly, stocks are strongly coupled to each other and to other systems, both natural and man-made.  For example, an earthquake in Taiwan on September 20, 1999 which cut off electrical power at Taiwan Semiconductor Manufacturing (TSM) and significantly disrupted production, had only a minimal impact on the company's stock price.  However, their South Korean competitors' stock prices soared in anticipation of increased demand.

This highlights the second challenge in studying the market: investors' responses (and hence stock price fluctuations) to incoming news can be strongly non-linear.  A company's quarterly forecast of a loss of ten cents per share could conceivably have much more than double the impact of five cents.  The precise response function (if one exists) is unknown.

Thirdly, the impact of an exogenous event may be practically, or even theoretically, unquantifiable.  Investors may receive imperfect information and/or the necessary calculations to assess the impact of the news on a stock's price may be too complex, beyond the rational abilities of the investor.  

Lastly, only some of the information which drives investors' actions is broadcast to all.  The rest (a rumour, for instance) is transmitted through a complex network of friends, families, and co-workers.  It is not clear if this information can be neglected and, if not, how the network is to be represented, structurally.

\section{Modeling}

The natural sciences are well acquainted with these types of challenges and their reaction is to study the system in two ways: first empirically, then with an idealized representation.

Empirical analysis is the first and best way to understand the world around us.  By collecting data and studying statistical properties thereof we can learn about the underlying distributions governing many phenomena.  Then, once sufficient empirical data have been collected idealized models may be constructed to try and account for the data.   

A vast store of financial market data is available.  For instance, precious metal price data are on record all the way back to the 1200s \cite{global99}.  A large number of these data sets have been analyzed and the results indicate that large market fluctuations (outliers) occur much more frequently than would be expected (the frequency distributions exhibit {\em fat tails}) and, unexpectedly, fluctuations occur in clustered bursts of volatility rather than uniformly.  This thesis will not focus on empirical analysis of financial data, rather relying (mainly) on these published results.

Instead, this thesis will focus on idealized representation or modeling of financial markets.  
Social scientists have developed simple analytic models of the stock market.  For tractability they assume a small number of investors who have {\em perfect rationality} (unlimited computational power) and complete information \cite{grossman76, youssefmir94, hogg95}.  These models are interesting to economists because they can explain equilibrium stock prices \cite{brown97}.  However, they are uninteresting to the physicists for precisely the same reason---since they are {\em equilibrium} models they fail to exhibit fat-tailed fluctuation distributions or clustered volatility.

\subsection{Computer simulations}

My hypothesis is that the complex dynamical behaviour of the stock market is an emergent property arising from the interactions of many agents and is largely independent of the complexity of the agents themselves.  In order to test my hypothesis I will construct some simple models meant to capture the essence of the stock market and study them experimentally via computer simulation.

Computer simulation is necessary because many-agent models are impractical or even impossible to analyze by hand---the number of interactions which need to be accounted for typically grows as the square of the number of agents.  (Even the simplest models tend to be too complex for an analytic treatment.)  So we turn to computers, which are capable of performing millions of calculations rapidly, with no (significant) errors.  

There are many objections to working with computer simulations but some of these apparent shortcomings are actually advantages.  For instance, it is impossible to construct a many-agent simulation with perfect rationality and complete information: each agent's expectations are formed on the basis of every other agent's expectations, which are formed on the basis of every other agent's expectations, ad infinitum.  (In some cases this infinite regress can be collapsed and solved.)  Besides being impossible to incorporate into a simulation, I hope the reader will agree this is an unrealistic account of investor behaviour.

Simulation may also seem inappropriate because, to develop a {\em stochastic} model, random events must be incorporated but computers are incapable of generating truly random numbers.  As an alternative, a number of algorithms have been constructed to produce {\em pseudo}-random numbers which pass all known statistical tests for randomness \cite{press92, gammel98, matsumoto98}.  However, these generators still require a {\em seed} from the user---a random, initial number to begin the sequence.  This flaw can often be a blessing because it offers {\em replicability} in one's experiments---by seeding the simulation with the exact same number as a previous iteration, the entire time series can be reproduced.  (To generate independent time series different seeds are used.)

Finally, market micro-simulation may be objected to because the events (for example, news releases) which drive the dynamics must be explicitly coded into the simulation.  As discussed above, these events are often not even quantifiable and, hence, can not be accurately coded.  However, turning this argument around, this is yet another advantage of the simulation methodology.  Any number of alternate hypotheses of the structure of the driving events can be encoded and their impact on the dynamics tested experimentally.  One of the interesting questions this thesis will address is ``How complex does the input (news) need to be to produce realistic output (price fluctuations)?''

\subsection{An appeal for simplicity}

A common temptation when constructing computer simulations is to try to capture as much detail as possible in order to make the simulation realistic.  But there are a number of reasons the model should be kept simple: Firstly, model complexity must be balanced against the constraints of current computational speed.  Simple models require less computational power and produce larger datasets.  Since large quantities of data are required to test the frequency distributions of rare events (such as price crashes), simpler is better for our purposes.

Secondly, as a model's complexity grows its capacity for being understood diminishes.  Some global climate models (GCMs), for example, have reached sufficient complexity that the modelers specialize in only a particular subroutine of the model, such as cloud formation.  Very few (if any) of the researchers have a full grasp of every detail of these simulations.  A problem with this approach is that the model becomes as difficult to understand as the system it was meant to idealize---a problem known as Bonini's Paradox \cite{dutton71}.  (Of course, GCMs are extremely useful for predictive purposes, but perhaps not for furthering scientific understanding.)

Thirdly, by starting with a trivial model and gradually adding layers of complexity, it is possible to determine the minimum requirements for a model which captures the essence of the system under investigation.  In the case of the market model this could mean building on a simple model until fat-tailed distributions (for example) are observed in the price fluctuations.  Then we can say with some confidence, ``These ingredients are the minimum requirements to explain market fluctuations.''

Finally, there is the issue of {\em Occam's razor}.  In the 1300s the Franciscan monk, William of Occam stated, ``Causes are not to be multiplied beyond necessity'' \cite{cover91, jaynes96} or, to paraphrase, ``The simplest explanation is best,'' guiding the course of science for centuries.  Notice this claim is aesthetic, not epistemological---it does not claim that the simplest explanation is {\em true}, but simply to be preferred, at least until evidence comes to light which requires us to reject it.  In Bayesian probability theory, Occam's razor has an even more precise role: given two theories which explain a phenomenon equally well, the one with fewer adjustable parameters is assigned a greater numerical likelihood \cite[Ch.\ 24]{jaynes96}.  Similarly, we should construct models which contain as few parameters, or assumptions, as possible.

\section{Organization of the thesis}

In this thesis I will develop and implement via simulation two hypothetical models of stock exchange.  An early model, which introduces the idea of a centralized market, will be described in \ch{csem}, and a later model, which discards the centralized trading restriction, in \ch{dsem}.  In \ch{results} the phase space of these models will be explored revealing some interesting phase transitions, including a critical point in either model.  Then, in \ch{results2} experiments will be performed and the results of the two models will be compared with each other and empirical data.  It will be discovered that the centralized model is incapable of generating the desired dynamics but the decentralized model can exhibit both fat tails and clustered volatility.  

Some interesting results of an experiment in investing, using a hypothetical portfolio, will be discussed in \ch{portfolio}.  The thesis will close with a discussion of some conclusions which can be drawn from the research and some ideas for future research.

\chapter{Centralized Stock Exchange Model}

\label{ch:csem}		

\section{Inspiration}

In this chapter we will explore the {\em Centralized Stock Exchange Model} (CSEM), a microscopic model which is built upon the premise of {\em centralization}; each agent on the market is restricted to trading with a single, monopolistic {\em market maker} who has complete control over the execution price.  No direct trades between agents are allowed.  This situation approximates some actual, thinly traded stocks on the New York Stock Exchange (NYSE) and other markets \cite{garman76}.  

There are two reasons this approach was chosen: firstly, there exists a significant collection of literature following this methodology \cite{palmer94, levy95, arthur97, caldarelli97, cont97, chen98, busshaus99, chowdhury99, iori99, lux99}.  I hoped to familiarize myself with this literature by constructing a model along the same vein.

Secondly, it allows for the construction of very simple agents.  By having the trading price set exogenously the agents need only react rather than formulate their own trading schedules.  In particular, the standard game theoretic approach is applicable only to reactive agents, as will be seen.  

Hence, the development of CSEM was a natural starting point for my research.

\section{Theory}

In this section the model's structure will be explained.

\subsection{Assumptions}

I will begin, for the sake of clarity, by laying out some common assumptions used in CSEM.

\subsubsection{Heterogenous agents}

The market consists of many agents interested in trading.  If all the agents had identical beliefs then we might expect their actions to be identical.  Hence, we would effectively have a market of just one meta-agent unable to execute any orders.  Similarly, if any subset of the population is homogeneous then that subset can be equally well represented by a single agent.

Therefore, it is natural to require that all the agents be unique.  Notice that heterogeneity can arise from imperfect rationality or incomplete knowledge, qualities which seems reasonable for the simple agents which will be constructed.  In most cases the agents will differ in fundamental parameters describing their preferences but transient differences alone (such as cash or shares held) may be allowed too, provided these factors influence the agents' actions.

\subsubsection{Single risky asset and single riskless asset}

For simplicity a market consisting of just one risky asset (public company stock) and one riskless asset (cash, for instance) will be used.

The total number of shares available on the market will be conserved.  Since the company pays no dividend the stock has no fundamental value and stock price is maintained solely by expectations of satisfactory returns on the sale of shares.  (The stock must at least have a chance of paying a dividend eventually or the stock price will be identically zero for all time, but the payout date is assumed to be far in the future.)  

For simplicity, the stock price will be assumed to be a continuous variable.  (In contrast, real stock prices are discretized, but on a sliding scale---dollar stocks usually have increments of one sixteenth of a dollar but penny stocks may be incremented by one tenth of a penny.)

The riskless asset (which we will call {\em cash}, though it could as easily represent some other stable equity such as gold) is defined to have a fixed intrinsic value in terms of which the value of the stock is measured.   (By measuring all value in terms of cash some of the difficulties of utility theory in comparing utilities of disparate objects \cite{vonneumann44} are sidestepped.)

The total cash in the market will also be conserved.  To achieve this, cash will pay no interest and no commissions will be charged on trades.  This restriction may be unrealistic but it has a significant advantage: the ratio of cash-to-shares is conserved.  This means that the market can avoid moving into a regime dominated by one or the other and instead establish a balance between the two.  For instance, if transaction costs were implemented cash would flow out of the system and, eventually, most of each agent's wealth would be held in stock.  Conversely, if interest was paid on cash, eventually the market might be cash-dominated.  In either case it is conceivable that the dynamics of the market would change, adding a complicating factor.  Fixing the amount of cash in the system simplifies the model and allows for the collection of large datasets. 

\subsubsection{No intraday trading}

CSEM uses a trading model which assumes all trades are executed only once daily (simultaneously).  This approach is common in the literature \cite{palmer94, levy95, arthur97, caldarelli97, busshaus99, iori99} and mimics trading which occurs in real markets on unprocessed orders before opening each day.

\subsubsection{Centralized trading}

As mentioned above, the agents in this model are restricted to trading only with a single, monopolistic market maker or specialist.  They are not allowed to trade directly with each other.  This has a empirical basis but is also a simplifying factor.  A discussion of how trading is implemented follows.

\subsection{Utility theory}

Each agent can adjust a portfolio consisting of $s$ shares of a single risky stock and riskless cash $c$.  If the share price is $p$ then the agent's total capital at time $t$ is $w_t = c_t + p_t s_t$.  With interest and fluctuations in the stock price the agent's capital after one day (defined as one time unit) becomes
\begin{eqnarray}
	w_{t+1} & = & c_t + p_{t+1} s_t \\
\label{eq:csemFutureWealth} & = & w_t + \left[ p_{t+1} - p_t \right] s_t
\end{eqnarray}
and the trading behaviour reduces to an optimization problem with respect to the holdings $s_t$.

If the agent could know what tomorrow's price of the stock $p_{t+1}$ will be in advance, finding the optimum strategy would be trivial: if $p_{t+1} > p_t$ then move all one's capital into the stock, otherwise move it all into cash.  But of course the future price is unknown.  Nevertheless each agent assumes it is a stochastic variable and has some expectations of the underlying probability distribution, based on historical prices.

A naive goal, then, might be to maximize one's expected future wealth $\expect{w_{t+1}}$ with respect to one's current holdings $s_t$.  Unfortunately, \eq{csemFutureWealth} simply tells us to invest all our capital into stock if $\expect{p_{t+1}} > p_t$ and otherwise into cash, almost exactly as before.  The problem with this approach is that it doesn't factor in {\em risk}.  What if there was a non-zero probability that the stock price would crash $\Pr(p_{t+1}=0)>0$?  Then, under repeatedly application of this strategy the agent would eventually lose all its wealth with certainty.  Even if the price can't drop to zero (which it can't if there is any expectation of a non-zero price in the future) this strategy can perform poorly, particularly if the price is a multiplicative stochastic process \cite{marsili98} because it assigns disproportionate weights to extremely unlikely events which would have exorbitant payoffs.  This strategy is said to be {\em risk neutral}.

We define our agents as simple expected utility maximizers where the {\em utility function} is monotonically increasing with wealth but has a negative second derivative (concave)
\be
	\frac{dU}{dw}>0,\, \frac{d^2U}{dw^2}<0.
\ee
These requirements for a utility function are well established within financial economics \cite{vonneumann44, pratt64} and basically mean that an agent is unwilling to make a ``double-or-nothing'' wager of any amount if the odds are even.  (Notice that the risk neutral agent $U=w$ would be ambivalent towards this wager and a {\em risk preferring} agent $\frac{d^2U}{dw^2}>0$ would willingly take the wager.)

\subsubsection{Exponential utility function}

\begin{figure}
	\begin{center}\input{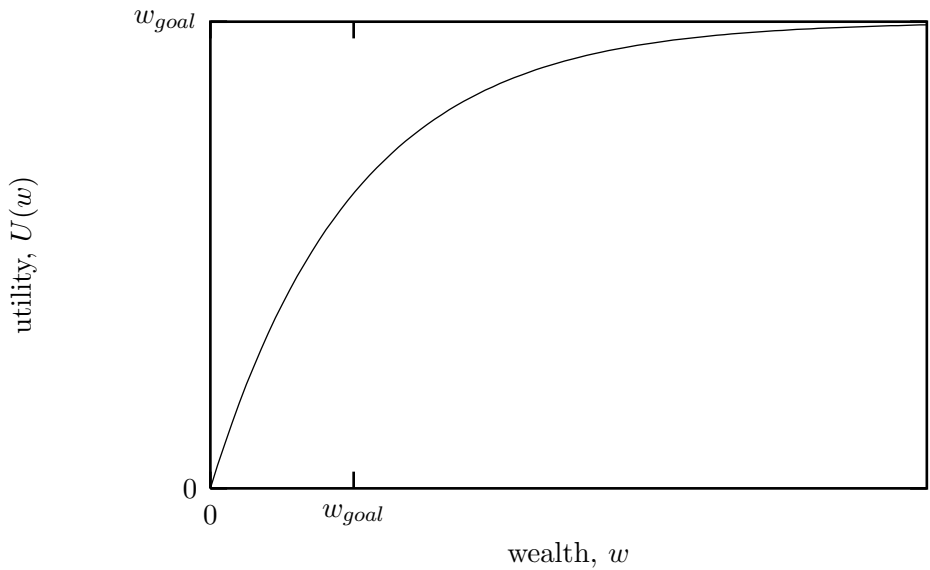}\end{center}
	\caption{The exponential utility function defined in \eq{csemExpUtility} is often applied in finance.  The goal wealth parameter $w_{goal}$ implicitly sets the risk aversion.}
\label{fig:csemExpUtility}
\end{figure}

An often-chosen form \cite{brown97} is the exponential utility $U(x)=-e^{-\alpha x}$ or equivalently (because utilities are defined only up to a linear transformation \cite{vonneumann44})
\begin{equation}
\label{eq:csemExpUtility}
	U(w) = w_{goal}\left(1-e^{-w/w_{goal}} \right)
\end{equation}
where $w_{goal}$ is called the goal wealth and sets a natural scale for the utility.  As shown in \fig{csemExpUtility}, the utility crosses over from a linear dependence on $w$ at small wealth $U(w\ll w_{goal}) \approx w$ to an asymptote at large wealth $U(w\gg w_{goal})\rightarrow w_{goal}$.  The interpretation of $w_{goal}$ as a ``goal wealth'' is justified because below $w_{goal}$ the agent is willing to take risks for the chance of high payoffs but above $w_{goal}$ it sees little reward in amassing greater wealth, being more concerned with maintaining its current level.

\subsection{Optimal holdings}

The exponential utility function is useful because it provides an analytic solution to the maximization problem \cite{grossman76} if we assume tomorrow's wealth $w_{t+1}$ is Gaussian distributed (a reasonable assumption by the Central Limit Theorem, if it is a cumulation of many additive stochastic components).  Then the expectation of the future utility is
\begin{eqnarray}
	\expect{U(w_{t+1})} & = & \int dw_{t+1} U(w_{t+1}) \Pr(w_{t+1}) \\
\label{eq:csemExpectedWealth}
	                    & = & w_{goal} \left[ 1 -\exp \left( \frac{\var{w_{t+1}}}{2 w_{goal}^2} - \frac{\expect{w_{t+1}}}{w_{goal}} \right) \right]
\end{eqnarray}
which is maximized by simply minimizing the argument of the exponential.

The future wealth depends on the price movement through \eq{csemFutureWealth} so the mean and variance become
\begin{eqnarray}
	\expect{w_{t+1}} & = & w_t + s_t \left\{ \expect{p_{t+1}} - p_t \right\} \\
	\var{w_{t+1}} & = & s_t^2 \var{p_{t+1}}.
\end{eqnarray}

\eq{csemExpectedWealth} can be maximized with respect to the free variable $s_t$ yielding the optimum quantity of shares to hold,
\be
\label{eq:csemOptimalHoldings1}
	s_t^*(p_t) = \frac{ w_{goal} (\expect{p_{t+1}} - p_t) }{ \var{p_{t+1}} }
\ee
with the additional constraints $s_t^* \geq 0$ (no short selling) and $w_t \geq p_t s_t^*$ (no borrowing cash).  The agent's strategy is to sell shares if $s_t>s_t^*$ or buy if $s_t<s_t^*$.  The above equation is intuitively appealing: only hold shares if the expected return on your investment is positive and decrease your investment when the uncertainty (variance) is large (indicating an aversion to risk).

\subsection{Risk aversion}

\label{sect:csemRiskAversion}


The goal wealth $w_{goal}$ in the utility function sets an undesirable, arbitrary scale for the agents behaviour: they will be become increasingly risk neutral as their wealth falls far below this scale, and conversely, increasingly risk averse far above it.  The arbitrary scale can be removed by setting the goal wealth proportional to the current wealth
\be
	w_{goal} = \frac{w_t}{a}
\ee
where $a$ is a dimensionless constant which describes risk aversion (which increases monotonically with $a$).  

Notice that introducing the dependence on the current wealth does not interfere with the optimization problem because $w_t$ is a constant at any time $t$, independent of any changes in the portfolio $s_t$ (assuming no trading costs).  Therefore the optimal portfolio simply becomes
\begin{equation}
\label{eq:csemOptimalHoldings}
	s_t^*(p_t) = \frac{ w_t (\expect{p_{t+1}} - p_t) }{a\, \var{p_{t+1}} }.
\end{equation}

From \fig{csemExpUtility} it is clear that the extremes of intense risk aversion and risk neutrality can be avoided by choosing $a$ on the order of unity.  A rough estimate provides an even more precise scale: empirically, the market appears to prefer to divide wealth equally between cash and stock when the annual expected return is 8\% better than cash with an uncertainty on the order of 25\%:
\begin{eqnarray}
	\expect{p_{t+1}} & \approx & (1+8\%) p_t \\
	\var{p_{t+1}} & \approx & (25\% p_t)^2 \\
	\Rightarrow s_t^* & \approx & \half \frac{w_t}{p_t}
\end{eqnarray}
where $t$ is scaled by years instead of days (but this does not interfere with the argument).  The $a$-value to satisfy these conditions is $a\approx 2.5$.  

Thus, the first agent-specific parameter introduced in CSEM is the risk aversion $a$ which is constrained to lie in $a\in [1,3]$.

\subsection{Optimal investment fraction}

For ease of comparison with the Decentralized model (to be presented in \ch{dsem}) the above discussion will be presented in terms of the fraction of one's wealth invested in stock.  The investment fraction $i_t$ at time $t$ is given by
\begin{equation}
\label{eq:csemInvestFraction}
	i_t = \frac{s_t p_t}{w_t}
\end{equation}
and the {\em optimal} investment fraction is denoted by $i_t^*$.

Let us also define the return on investment from time $t$ to $t+1$:
\begin{equation}
\label{eq:csemReturn}
	r_{t+1} = \frac{p_{t+1} - p_t}{p_t}
\end{equation}
which has a mean and variance (given a known current price $p_t$)
\begin{eqnarray}
	\expect{r_{t+1}} & = & \frac{\expect{p_{t+1}} - p_t}{p_t} \\ 
	\var{r_{t+1}}    & = & \frac{ \var{p_{t+1}} }{ p_t^2 }.
\end{eqnarray}

Then, substituting \eq{csemOptimalHoldings} into \eq{csemInvestFraction}, we find that the optimal investment fraction is
\widebox{\begin{equation}
\label{eq:csemOptimalFraction}
	i_t^* = \frac{ \expect{r_{t+1}} }{ a\, \var{r_{t+1}} }
\end{equation}}
with the constraints $0\leq i_t^* \leq 1$.

This relation has some intuitively attractive properties:
\begin{enumerate}
	\item{All else being equal, given two agents with different risk aversions, the one with the higher aversion will invest less.}
	\item{Only invest if the expected return is strictly positive, and invest in proportion to it.}
	\item{As your certainty of a good return increases (variance decreases), increase your investment.}
\end{enumerate} 

However, it also has one glaring fault: when the expected return exceeds some limit,
\begin{equation}
	\expect{r_{t+1}} \geq a\, \var{r_{t+1}} 
\end{equation}
the recommendation is to invest {\em all} capital in the stock, despite risk.  This arises because the agents assume the returns are Gaussian-distributed, with no higher moments than the variance, but as we will see, higher moments do exist, increasing the risk.

\subsubsection{Investment limit}

To avoid complications of this kind a limit of $\delta$ is imposed: the investment fraction is constrained to lie within $i\in[\delta,1-\delta]$.  Hence agents never take an absolute stance of investing all their money or withdrawing it all from the market. The motivation for this restriction is purely mathematical: it prevents the occurrence of all the agents simultaneously selling all their stock and driving the price down to zero (or conversely, selling all their cash and driving the stock price up to infinity).  

As the population size increases, the probabilities of these events diminish simply as a result of fluctuations so the parameter $\delta$ becomes less important.  To minimize its impact on the dynamics it should be assigned a small, positive value. Mathematically, however, it is allowed to be as large as 1/2 in which case the investment fraction would be a constant 1/2, never responding to \eq{csemOptimalFraction}.

Thus, the second agent-specific parameter is $\delta$ which is constrained to lie within $\delta\in(0,0.5)$.

\subsection{Forecasting}

With \eq{csemOptimalFraction} the optimization problem becomes one of forecasting one's future return $r_{t+1}$.  In order to solve the optimization problem estimates of the expectation and variance of one's future return are required.  The only information available to the agents is the history of returns so a reasonable choice is to try and extrapolate the series forward in time.

Although more complicated forecasting algorithms involving nonlinearity and chaos exist \cite{fang95, farmer87, sugihara90, casdagli91, palus95, palus95b}, I chose to extrapolate a simple curve-fitting algorithm to produce forecasts.  The goal of this model is not to test complicated forecasting models but to understand the effect of interactions between many simple investors, so the forecasting algorithm need only be adequate, not optimal.  Linear least-squares curve fitting is well understood so we don't have to worry about it generating unexpected side-effects in the dynamics.

The time series could be represented by a few parameters, one being the raw prices.  However, a natural choice is the returns (as defined by \eq{csemReturn}) because a Gaussian-distributed future wealth $w_{t+1}$ was assumed.  This assumption can be validated by assuming a Gaussian distribution for returns as well, because \eq{csemFutureWealth} can be written as
\be
	w_{t+1} = (1-i_t)w_t + i_t w_t r_{t+1}
\ee
where the stochastic variable is the return $r_{t+1}$.  Since least-squares fitting assumes Gaussian errors, the returns are a convenient choice.  Note that assuming a Gaussian distribution of returns is equivalent to a log-Brownian price series as is observed empirically on long timescales \cite{osborne59} (with interesting deviations on short timescales).

\begin{figure}
	\begin{center}\input{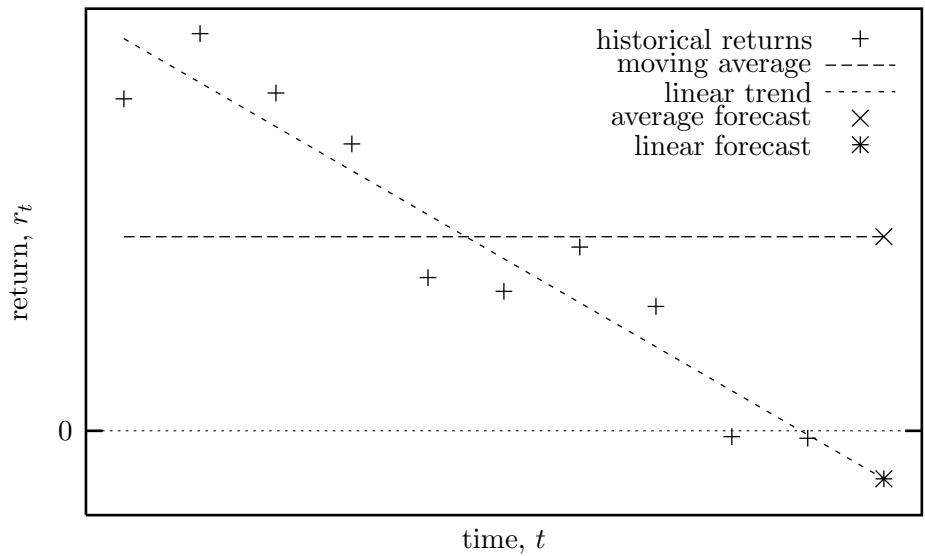}\end{center}
	\caption{Demonstration of forecasting via polynomial curve extrapolation.  Shown are forecasts produced by a simple moving average and a linear trend.  The linear trend is able to anticipate reversals in returns.}
\label{fig:csemSamplePoly}
\end{figure}

For simplicity, only low-degree polynomials will be  used as fitting functions.  The degree zero polynomial, a simple moving-window average, is already robust enough to project exponential growth in the stock's price.  Increasing to degree one (linear) also gives the agents the ability to forecast trend reversals (such as an imminent crash, as shown in \fig{csemSamplePoly}) {\em assuming} the return history has meaningful trends.

By choosing higher degree polynomials we can effectively make the agents {\em smarter} (better able to detect trends in the return series) but, in practice, it is unreasonable to go beyond a degree two, quadratic fit.  If too high a degree is chosen agents begin to ``see'' trends where none exist by fitting curves to noise.

Thus, the third agent-specific parameter, degree of fit $d$, is constrained to the integer values $d=0,\ldots,2$.

\subsubsection{Memory}

Obviously, as time progresses and the latest returns are acquired, the older data in the time series become irrelevant.  The standard methodology for handling this is to set a finite moving-window which only keeps the $M$ most recent data points, discarding the rest.  Then the curve fitting is performed only with respect to the remaining data.  However, this technique has a drawback: it suffers from shocks as outliers (strongly atypical data) get dropped from memory.  

To minimize this effect I constructed a method which uses an exponentially decaying window rather than the square window described above.  The contribution of each point to the curve fit is weighted exponentially by how old it is.  The technique is described in detail in \ap{dls} but a few points will be mentioned here:

The exponential weighting is characterized by a single parameter, the memory $M$ (denoted by $N^*$ in \ap{dls}) indicating the effective number of data points stored, which is approximately the decay constant of the exponential.

Using the exponential window allows compression of the data into just a few numbers regardless of the memory $M$ and, as such, is computationally efficient in terms of storage and speed.

An agent's memory also says something about its expectations.  A short memory produces fast responses to changes in returns and hence, more active trading.  Conversely, a long memory results in slow variations in expectations and, therefore, slow changes in investment strategy.  Hence, the memory implicitly also sets the (future) timescale, or horizon, over which the agent expects to collect.

As with standard curve-fitting the parameter $M$ is required to be greater than the number of parameters to be fit ($=d+1$ where $d\leq2$ so $M\geq 10$ (two trading weeks) is satisfactory) but there is no maximum value.  But to draw parallels with real markets it is reasonable to choose scales on the order of real market investors.  Many online stock-tracking sites allow one to compare a stock's current value to its moving average over windows up to 200 trading days (almost one year).

Thus, the fourth agent parameter in CSEM is the memory $M$ which is allowed to take on values in the range $M\in[10,200]$ (between two weeks and roughly one year).

\subsection{Fluctuations}

\label{sect:csemFluctuations}

To this point we have not explicitly identified the source of stochasticity.  (Thus, since the simulation begins with no memory of any fluctuations no trading will occur whatsoever.)  To mimic the noisy speculation which drives movements in real markets, stochastic fluctuations are introduced into CSEM.  The fluctuations are meant to represent the agents' imperfect information which can produce errors in their expectations of tomorrow's price.  Given that the return-on-investment time series is already assumed to have Gaussian distributed errors a natural extension is to introduce normally-distributed fluctuations into the agents' forecasts
\widebox{\begin{equation}
\label{eq:csemExpReps}
	\expect{r_{t+1}}_\epsilon \equiv \expect{r_{t+1}} + \epsilon_t 
\end{equation}}
where $\epsilon_t$ is a Gaussian-distributed stochastic variable with mean zero and variance $\sigma_\epsilon^2$.

It is assumed that agents are aware that their forecasts contain uncertainties so the variance of their forecasts is increased by
\widebox{\begin{equation}
\label{eq:csemVarReps}
	\var{r_{t+1}}_\epsilon \equiv \var{r_{t+1}} + \sigma_\epsilon^2
\end{equation}}
since the forecasted return $r_{t+1}$ is also assumed to be Gaussian-distributed (and the variance of the sum of two normally-distributed numbers is the sum of their variances).

Fluctuations are handled by determining a random deviate for each agent at each time step and adding it to the expected return, as discussed above.  Once the deviate is chosen, it is a constant (but unknown by the agent) for that time interval, so the expected return is also constant.  This is necessary for technical reasons (it keeps the agents' demand curves consistent for the auctioning process which will be discussed in \sect{csemAuction}) but it also seems intuitively reasonable---one would not expect an investor to forecast a different return every time she was asked (in the absence of new information).

The dynamics are driven solely by the presence of noise (as will be discussed below) so we require strictly non-zero standard deviations.  On the other hand, the standard deviation also sets the typical scale of errors in the forecasted return.  From personal experience, on a daily basis one would expect this error to be on the order of two percent.  However, to fully explore the effect of the noise parameter CSEM will allow errors as large as 1/2 (which represents daily price movements up to $\pm 50$\%).

Thus, the fifth agent parameter introduced into CSEM is the scale of the uncertainty $\sigma_\epsilon$ which is chosen to lie within $\sigma_\epsilon\in(0,0.5)$.

\subsection{Initialization}

The discussion so far has focused on how the agents evolve from day to day.  But we must also consider in what state they will be started.  It is important to choose starting conditions which have a minimal impact on the dynamics or a long initial transient will be required before the long-run behaviour emerges.

The simulations will be initialized with $N$ agents; each agent will have a fraction of some total cash $C$ and total shares $S$ available.  The effect of different initial distributions of cash and shares will be explored, but---unless otherwise specified---the cash and shares will usually be distributed uniformly amongst the agents.  This allows the simulations to test the performance of other parameters; that is, to see if there is a correlation between parameter values and income.

As mentioned above, agents will also be initialized to have zero expectation $\expect{r_1}=0$ and zero variance $\var{r_1}=0$ of tomorrow's return-on-investment.  However, this is subject to \eqs{csemExpReps}{csemVarReps} so the actual initial expectation is a Gaussian deviate with mean zero and variance $\sigma_\epsilon^2$.

The first trading day is unique in that there exists no prior price from which to calculate a return-on-investment (for future forecasts).  So the first day is not included in the agents' histories.  Thus, the dynamics for the first two days of trading are due solely to fluctuations.

In this section three market parameters were introduced: the number of agents $N$ on the market, as well as the total cash $C$ and total shares $S$ which are initially divided equally among the agents (unless otherwise stated).

\subsection{Market clearing}

\label{sect:csemAuction}

Having discussed how the agents respond to prices and choose orders we now turn our attention to how the trading price is set.  As mentioned before, this model is {\em centralized} in the sense that the agents are not allowed to trade directly with each other but all transactions must be processed through a {\em specialist} or {\em market maker} \cite{palmer94, levy95, caldarelli97, cont97, chen98, chowdhury99, busshaus99, iori99, lux99}.

In real markets, the role of the market maker is more complex than in this simulation: here the market maker simply negotiates a price such that the market {\em clears}; that is, all buyers find sellers and no orders are left open.  (All mechanisms by which the market maker may make a profit have been removed from the simulation for the sake of simplicity.)

A simple way for the market maker to establish a trading price is via an auction process: repeatedly call out prices and receive orders until buy and sell orders are balanced.  If buy orders dominate, raise the price in order to encourage sellers, and vice versa.

However, CSEM provides a simpler (and faster) method for arriving at the trading price.  Assuming the market maker knows each agent is using a {\em fixed investment strategy} as given by \eq{csemOptimalFraction}, it can be deduced that the optimal holdings for agent $j$ (with cash $c_j$ and shares $s_j$) at price $p$ is
\begin{equation}
	s_j^* = \frac{c_j+s_j p}{p} i_j^*.
\end{equation}
Effectively, by reporting their ideal investment fractions $i_j^*$ (and current porfolios ($c_j$, $s_j$)), the agents submit an entire demand curve (demand versus price) for all prices instead of just replying to a single price called out by the auctioneer.

The market maker's goal of balancing supply with demand can be achieved by choosing a price which preserves the total number of shares held by the investors:
\begin{eqnarray}
	0 & = & \sum_j (s_j^* - s_j) \\
	  & = & \frac{1}{p} \sum_j c_j i_j^* + \sum_j (i_j^* - 1) s_j
\end{eqnarray}
which has a solution
\widebox{\begin{equation}
\label{eq:csemTradingPrice}
	p = \frac{ \sum_j i_j^* c_j }{ \sum_j (1-i_j^*) s_j }
\end{equation}}
where, the values $i_j^*$, $c_j$, and $s_j$ are all from before any trading occurs on the current day.

So, instead of requiring an auction, the trading price is arrived at with a single analytic calculation.  Note that this method is possible because the optimal investment fraction $i_j^*$ does not depend on the current day's price but only on the history of prior returns.  (Once the trading price is established, the latest price is included in the history and contributes to the determination of tomorrow's optimal investment fraction.)

\subsubsection{Initial trading price}

\label{sect:csemInitialPrice}

In general, the calculation of the trading price is complicated and depends intricately on the history of the run but there is a special case where it is possible to determine explicitly the expected trading price---the first day.  Let us assume that the initial distribution of cash and shares is such that each agent has equal numbers of both so that \eq{csemTradingPrice} reduces to
\begin{eqnarray}
	p_0 & = & \frac{\sum_j i_j^*}{\sum_j 1-i_j^*} \\
\label{eq:csemInitialPrice}
	    & = & \frac{\expect{i^*}}{1-\expect{i^*}}.
\end{eqnarray}

To calculate the expected investment fraction recall that initially the return history is empty so the expected returns are simply Gaussian-distributed with mean zero and variance $\sigma_\epsilon^2$ so, from \eq{csemOptimalFraction},
\begin{equation}
	i_j^* = \frac{\epsilon_j}{a \sigma_{\epsilon}^2} \equiv \frac{x_j}{k},
\end{equation}
defining $x=\epsilon / \sigma_\epsilon$, $k=a \sigma_\epsilon$ and assuming the risk aversion $a$ and forecast uncertainty $\sigma_\epsilon$ are identical for all agents.

Neglecting the limits $i\in[\delta,1-\delta]$ on the investment fraction ($\delta=0$) simplifies the calculation of the expected investment fraction:
\begin{eqnarray}
	\expect{i^*} & = & \int_0^{i(x)=1} i(x) \Pr(x) dx + \int_{i(x)=1}^\infty \Pr(x) dx \\
	             & = & \frac{1}{\sqrt{2\pi}} \left[ \frac{1}{k} \int_0^k x e^{-x^2/2} dx + \int_k^\infty e^{-x^2/2} dx \right] \\
\label{eq:csemAvgInvestFrac}
	             & = & \frac{1}{\sqrt{2\pi}k} \left( 1 - e^{-k^2/2} \right) + \frac{1}{2}\left( 1 - \erf(k/\sqrt{2}) \right)
\end{eqnarray}
where $\erf(\cdot)$ is the error function.

\begin{figure}
	\begin{center}
		\input{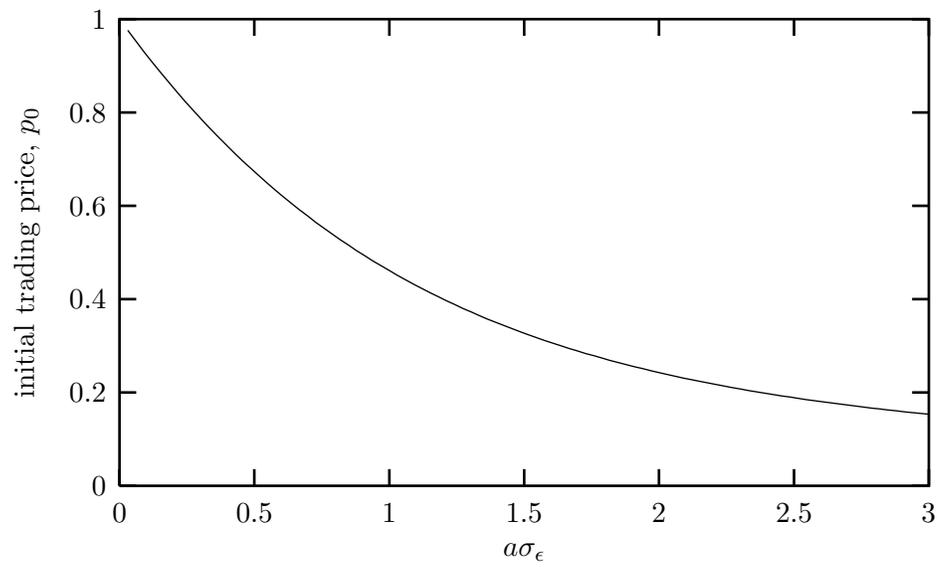}
	\end{center}
	\caption{The expected initial trading price depends only on the risk aversion multiplied by the uncertainty of returns, $a \sigma_\epsilon$.  As the aversion or uncertainty increases the initial value of the stock drops.}
\label{fig:csemInitialPrice}
\end{figure}

Substituting this equation into \eq{csemInitialPrice} gives the trading price as a function of the single parameter $k=a \sigma_\epsilon$, as shown in \fig{csemInitialPrice}.  Notice the value of the stock drops with increased risk aversion or uncertainty of return, properly capturing the essence of risk aversion.

It is interesting to note that the price drops to zero as $p_0\propto k^{-1}$ for large $k$.  To see how this occurs, notice that as the parameter $k$ approaches infinity the second term in \eq{csemAvgInvestFrac} drops out (falling off faster than $1/k$), as does the exponential in the first term, leaving only
\begin{equation}
	\expect{i^*}(k\rightarrow\infty) \approx \frac{1}{\sqrt{2\pi} k},
\end{equation} 
which diminishes to zero rapidly.  The power law tail in the price emerges from simply substituting this relation into \eq{csemInitialPrice}.

Now we briefly review the structure of the model.

\subsection{Review}

The Centralized Stock Exchange Model (CSEM) consists of a number $N$ of agents which trade once daily (simultaneously) with a single market maker, whose goal is to set the stock price such that the market clears (no orders are pending).  In this section, the structure of the model will be reviewed.

The agents are simple utility maximizers which extrapolate a fitted polynomial to the return history to predict future returns and, therefrom, optimal transactions.  Each agent has a portfolio of cash $c$ and shares $s$ and is characterized by the parameters listed in \tbl{csemParameters}.

\subsubsection{Algorithm}

Events are separated into days.  After the model has been initialized the agents place orders and have them filled once each day.  The basic algorithm follows:
\begin{enumerate}
	\item{Initialization.  Cash and shares distributed amongst agents.  Agents clear histories.}
	\item{Start of new day.  Agents forecast return-on-investment from history (and noise).
\label{enum:csemAlgStartDay}}
	\item{Agents calculate optimal investment fraction and submit trading schedules (optimal holdings as a function of stock price).}
	\item{Market maker finds market clearing price (supply balances demand).}
	\item{Trades are executed.}
	\item{Agents calculate stock's daily return-on-investment and append to history.}
	\item{End of day.  Return to step \enum{csemAlgStartDay}.}
\end{enumerate}

\subsubsection{Parameters}

\begin{table}
	\begin{center}\begin{tabular}{r|l|c}
		\hline \hline
		Symbol & Interpretation & Range \\
		\hline
		\multicolumn{3}{c}{Market parameters} \\
		\hline
		$N$ & number of agents & $2+$\\
		$C$ & total cash available \\
		$S$ & total shares available \\
		\hline
		\multicolumn{3}{c}{Market state variables} \\
		\hline
		$p_t$ & stock price at time $t$ \\ 
		$v_t$ & trade volume (number of shares traded) at time $t$ \\
		\hline
		\multicolumn{3}{c}{Agent parameters} \\
		\hline
		$a_j$ & risk aversion of agent $j$ & $[1,3]$ \\
		$\delta_j$ & investment fraction limit of agent $j$ & $(0,0.5)$ \\
		$d_j$ & degree of agent $j$'s fitting polynomial & $0,1,2$ \\
		$M_j$ & memory of agent $j$'s fit & $[10,200]$ \\
		$\sigma_{\epsilon,j}$ & scale of uncertainty of agent $j$'s forecast & $(0,0.5)$ \\
		\hline
		\multicolumn{3}{c}{Agent state variables} \\
		\hline
		$c_j$ & cash held by agent $j$ \\
		$s_j$ & actual shares held by agent $j$ \\
		$s_j^*$ & optimum shares held by agent $j$ \\
		$w_j(p)$ & wealth of agent $j$ at stock price $p$ \\
		$i_j$ & actual investment fraction of agent $j$ \\
		$i_j^*$ & optimum investment fraction of agent $j$ \\
		\hline \hline
	\end{tabular}\end{center}
	\caption{All parameters and variables used in the Centralized Stock Exchange Model (CSEM).}
\label{tbl:csemParameters}
\end{table}

For convenience all the variables used in CSEM are listed in \tbl{csemParameters}.  The parameters are inputs for the simulation and the state variables characterize the state of the simulation at any time completely.  For each run, the agent-specific parameters are set randomly; they are uniformly distributed within some range (a subset of the ranges shown in the table).  Each dataset analyzed herein will be characterized by listing the market parameters and the ranges of agent parameters used.  

\section{Implementation}

The above theory completely characterizes CSEM.  The model is too complex for complete analysis so it is simulated via computer.  The model was encoded using Borland C++Builder 1.0 on an Intel Pentium II computer running Microsoft Windows 98.  The source code and a pre-compiled executable are available from \url{http://rikblok.cjb.net/phd/csem/}.

\subsection{Pseudo-random numbers}

\label{sect:csemRandomNumbers}

Coding the model as it has been described is fairly straight-forward.  The only complication is that modern computers are unable to produce truly random numbers (required for the fluctuations in the forecasts) because computers are inherently deterministic.  

Many algorithms for generating numbers which {\em appear} random have emerged.  A good pseudo-random number generator must have three qualities: it must be fast, it must pass statistical tests for randomness and it must have a long period.  The period exists because there are only a finite number of states (typically $2^{32}$) a random number may take on.  Hence, it must eventually return to its original seed and once it does, since the series is deterministic, it is doomed to cycle endlessly.  If the period is less than the number of times the generator is called within a single run, the periodicity will contaminate the dataset.

One of the earliest and simplest pseudo-random number generators is the {\em linear congruential generator} \cite[Section 7.1]{press92} which is defined recursively for an integer $I_j$:
\begin{equation}
	I_{j+1} = a I_j + c\; ({\rm mod}\; m).
\end{equation}
While this algorithm is fast it is not a good choice because it exhibits correlations between successive values.  

More complicated generators have been developed which pass all known statistical tests for randomness \cite{press92, gammel98}.  One of these, the {\em Mersenne Twister} \cite{matsumoto98} is also fairly fast and has a remarkable period of $2^{19937}\approx 10^{6000}$.  Unless otherwise specified, the Mersenne Twister will be the generator of choice for CSEM.

\subsubsection{Seed}

All pseudo-random number generators require an initial seed: a first number ($I_0$ in the linear congruential generator, for example) chosen by the user which uniquely specifies the entire set of pseudo-random numbers which will be generated.  This seed should be chosen with care: using the same seed as a previous run will generate the exact same time series (all other parameters being equal).  

CSEM is coded to optionally accept user-specified seeds or it defaults to using the current time (measured in seconds since midnight, January 1, 1970, GMT).  Since no two simulations will be run simultaneously, this provides unique seeds for every run.  Unless explicitly specified, the default (time) seed will be used in the simulations.

\section{Parameter space exploration}

With CSEM coded into the computer, time series data can be generated for numerical analysis.  As presented CSEM requires at least eight parameters to fully describe it.  To fully explore the space of all parameters, then, means exploring an eight dimensional manifold\ldots a daunting task.  Before starting any experiments, then, it would be a good idea to check if any of these dimensions can be eliminated.

\subsection{Number of agents $N$}

The effect of changing the number of traders will be explored in detail in \ch{results} and is left until then.

\subsection{Total cash $C$ and total shares $S$}

\label{sect:csemCSscaling}

In this section the effect of rescaling the total cash $C$ and total shares $S$ will be explored.  Let us denote rescaled properties with a prime.  Then rescaling cash by a factor $A$ and shares by $B$ is written
\begin{eqnarray}
	C' & = & A C \\
	S' & = & B S.
\end{eqnarray}
Cash and shares are rescaled equally for each agent so the distribution remains constant.

To see how these rescalings affect the dynamics let us begin by assuming that each agent's ideal investment fraction $i_t^*$ is unchanged (this will be justified below).  Then from \eq{csemTradingPrice} the price is rescaled by
\begin{equation}
	p_t' = \frac{A}{B}p_t
\end{equation}
and each agent's total wealth is rescaled by
\begin{equation}
	w_t' = A w_t.
\end{equation}
(The rescaling of price can be interpreted as the ``Law of supply and demand'' because when either cash or stock exists in overabundance, it is devalued relative to the rarer commodity.)

Thus, the optimal holdings become
\begin{equation}
	s_t^{*\prime} = \frac{w_t'}{p_t'}i_t^* = B s_t^*
\end{equation}
and the volume an agent trades becomes
\begin{equation}
	\Delta s_t' = \abs{s_t^{*\prime} - s_t'} = B \Delta s_t.
\end{equation}

To justify that the optimal investment fraction remains unchanged, recall that it depends only on the return series through \eq{csemOptimalFraction}.  The return series, under rescaling, becomes
\begin{equation}
	r_t' = \frac{p_t'-p_{t-1}'}{p_{t-1}'} = r_t
\end{equation}
assuming the price series is rescaled by $A/B$.  Thus, if the investment fraction remains unscaled then the price series is scaled by $A/B$, so the investment fraction remains unscaled\ldots

This would be a circular argument except for the fact that the investment fraction is initialized by a Gaussian fluctuation, which depends only on the parameters $a$ and $\sigma_\epsilon$.  Thus the investment fraction begins unchanged (under rescaling of $C$ and $S$) and there exists no mechanism for changing it, so it remains unchanged throughout time.

So, when cash is rescaled by some factor $A$ and shares by $B$, the only effects are:
\begin{enumerate}
	\item{Trading price is rescaled by $A/B$.}
	\item{Trading volume is rescaled by $B$.}
\end{enumerate}

\begin{table}
	\begin{center}\begin{tabular}{c|c|c|c}
		\hline \hline
		Parameter & Run 1 & Run 2 & Run 3 \\
		\hline
		$N$ & 100 & 100 & 100 \\
		$C$ & \$1,000,000 & \$10,000,000 & \$1,000,000 \\
		$S$ & 1,000,000 & 10,000,000 & 10,000,000 \\
		$\sigma_\epsilon$ & 0.5 & 0.5 & 0.5 \\
		$M$ & $40\pm 20$ & $40\pm 20$ & $40\pm 20$ \\
		$a$ & $2\pm 1$ & $2\pm 1$ & $2\pm 1$ \\
		$\delta$ & 0.001 & 0.001 & 0.001 \\
		$d$ & $1\pm 1$ & $1\pm 1$ & $1\pm 1$ \\
		seed & -2 & -2 & -2 \\
		\hline \hline
	\end{tabular}\end{center}
	\caption{Parameter values for CSEM Runs 1, 2 and 3.}
\label{tbl:csemRuns123}
\end{table}

\begin{figure}
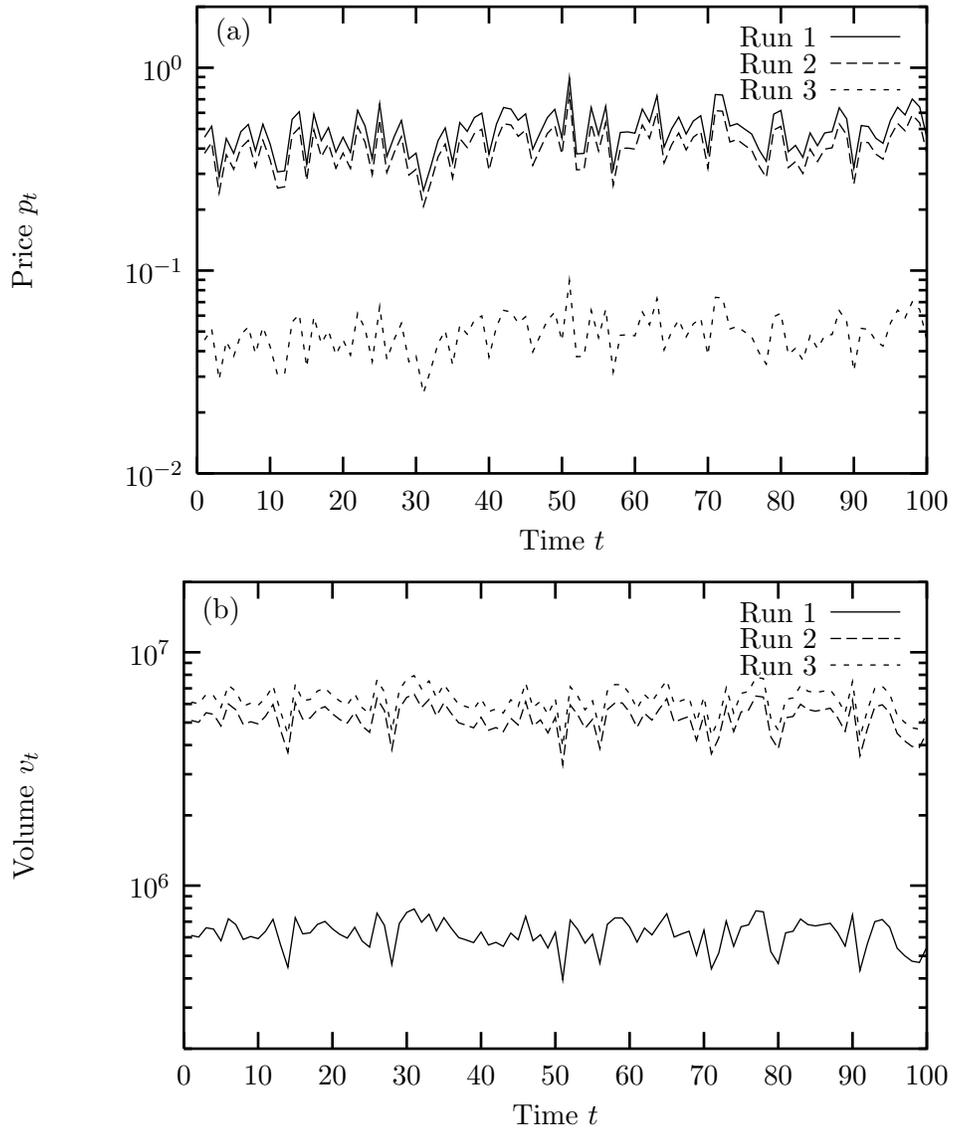

	\begin{center}
		\input{chModelCsem/runs123price.tex} \\
		\input{chModelCsem/runs123volume.tex}
	\end{center}
	\caption{Comparison of time evolutions of (a) price and (b) volume for Runs 1, 2 and 3 as defined in \tbl{csemRuns123}.  The price scales as the ratio of cash to shares and the volume scales as the number of shares.  (In both plots Run 2 is offset to improve readability.)}
\label{fig:csemCashSharesScaling}
\end{figure}

To clarify this point in the mind of the reader, three identical runs were performed, with the parameter values shown in \tbl{csemRuns123}.  Notice that Run 2 is Run 1 repeated with the scaling factors $A=B=10$, and Run 3 is Run 1 with $A=1, B=10$.  The resulting time series, shown in \fig{csemCashSharesScaling}, confirm the claim that price scales as $A/B$ and volume scales as $B$.  

Neither the absolute value of the price nor the volume are items of interest in this dissertation.  Instead we are interested in fluctuations, in the form of price returns and relative change of volume.  Neither of these properties are affected by rescaling the total cash or total shares so we are free to choose a convenient scale.  I have arbitrarily chosen a market with $C=\$1,000,000$ total cash and $S=1,000,000$ total shares, thereby reducing the degrees of freedom by two.

\subsection{Investment fraction limit $\delta$}

\label{sect:csemInvestLimit}

The investment fraction limit parameter $\delta$ sets a bound on the minimum and maximum allowed investment fractions $\delta\leq i\leq 1-\delta$.  This is purely a mathematical kludge to prevent singularities which could otherwise occur in \eq{csemTradingPrice}.  

Effectively, $\delta$ sets an upper and lower bound on the price itself: assume the total cash and shares are equal ($C=S$).  Then, the minimum price is realized when all agents want to discard their stocks, $i_j^* = \delta$ for all $j$, giving
\begin{equation}
\label{eq:csemMinPrice}
	p_{min} = \frac{\delta}{1-\delta}.
\end{equation}
Conversely, given maximal demand, $i_j^* = 1-\delta$, the price will climb to a maximum of
\begin{equation}
\label{eq:csemMaxPrice}
	p_{max} = \frac{1-\delta}{\delta}.
\end{equation}

So the choice of $\delta$ sets the price range for the stock.  Obviously, to allow reasonable freedom of price movements the limit should be significantly less than one half, $\delta \ll 1/2$.  To mimic the observed variability in some recent technology-sector stocks, a limit of $\delta=0.001$ will generally be used, allowing up to a thousand-fold increase in stock value---except in \ch{results} where we explore the effect of varying this parameter.

\subsection{Risk aversion $a$ and forecast uncertainty $\sigma_\epsilon$}

One's intuition may lead one to suspect that the risk aversion factor $a$ and the forecast uncertainty $\sigma_\epsilon$ are over-specified, and should be replaced by a single parameter $k=a\sigma_\epsilon$ as was done to calculate the initial trading price in \sect{csemInitialPrice}.  However, a closer inspection of \eq{csemOptimalFraction} demonstrates this is not quite true.  The optimal investment fraction is
\begin{equation}
	i_t^* = \frac{ \expect{r_{t+1}} + \epsilon_t }{ a (\var{r_{t+1}} + \sigma_\epsilon^2) }.
\end{equation}
Since $\sigma_\epsilon$ is the only parameter to set a scale for the returns, in \eqs{csemExpReps}{csemVarReps}, it is reasonable to expect the returns to scale linearly with $\sigma_\epsilon$ so
renormalizing gives 
\begin{equation}
\label{eq:csemScaleAversionSigma}
	i_t^* = \frac{1}{a \sigma_\epsilon} \left[ \frac{ \expect{r_{t+1}}/\sigma_\epsilon + \epsilon_t/\sigma_\epsilon }{ \var{r_{t+1}}/\sigma_\epsilon^2 + 1} \right]
\end{equation}
where the second factor is invariant under rescaling of $\sigma_\epsilon$.

Then, since $a$ and $\sigma_\epsilon$ occur nowhere else in the model, one may expect that the simultaneous rescaling
\begin{eqnarray}
	a' & = & C a \\
	\sigma_\epsilon' & = & \sigma_\epsilon/C
\end{eqnarray}
would preserve the dynamics.  

\begin{table}
	\begin{center}\begin{tabular}{c|c|c}
		\hline \hline
		Parameter & Run 4 & Run 5 \\
		\hline
		$N$ & 100 & 100 \\
		$C$ & \$1,000,000 & \$1,000,000 \\
		$S$ & 1,000,000 & 1,000,000 \\
		$\sigma_\epsilon$ & 0.25 & 0.5 \\
		$M$ & $40\pm 20$ & $40\pm 20$ \\
		$a$ & 2 & 1 \\
		$\delta$ & 0.001 & 0.001 \\
		$d$ & $1\pm 1$ & $1\pm 1$ \\
		seed & -2 & -2 \\
		\hline \hline
	\end{tabular}\end{center}
	\caption{Parameter values for CSEM Runs 4 and 5.}
\label{tbl:csemRuns45}
\end{table}

\begin{figure}
	\begin{center}
		\input{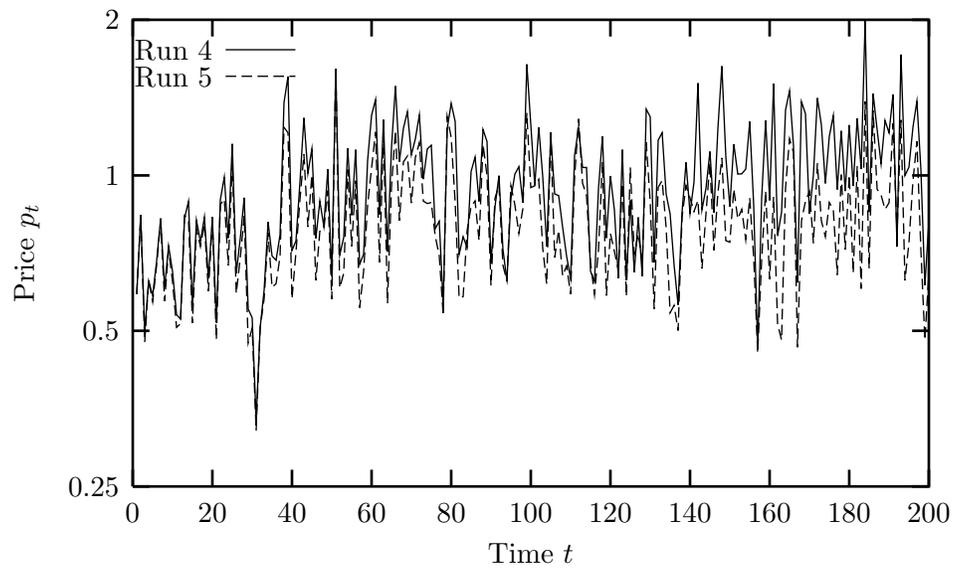}
	\end{center}
	\caption{Comparison of time evolutions of price for Runs 4 and 5 as defined in \tbl{csemRuns45}.  The price is not perfectly invariant under rescalings which preserve the constant $a \sigma_\epsilon$.}
\label{fig:csemRuns45}
\end{figure}

However, as the price series of Runs 4 and 5 (see \tbl{csemRuns45}) show in \fig{csemRuns45}, there are small deviations which grow with time until eventually the time series are markedly different.

To see why this occurs, let us consider a simple thought experiment: Consider a run with equal amounts of cash and shares ($C=S$) where the last trading price---for the sake of convenience---is $p_{t-1} = 1$.  Now assume that on the next day all the agents have negative fluctuations in their forecasts which drive their optimal investment fractions to their lower limits $i_t^* = \delta$.  Then, from \eq{csemTradingPrice}, the day's stock price will be given by \eq{csemMinPrice} and the return will be
\begin{equation}
	r_t = \frac{p_t-1}{1} = -\frac{1-2\delta}{1-\delta}
\end{equation}
which does not scale with $\sigma_\epsilon$ as was hypothesized in the derivation of \eq{csemScaleAversionSigma}.  

Occasional events like the one described in the above thought experiment are responsible for the deviations seen in \fig{csemRuns45}.  However, apart from these rare deviations (which, neglecting trends in the return history, should occur with decreasing frequency $1/2^N$ as the number of investors increases) the risk aversion parameter $a$ and uncertainty $\sigma_\epsilon$ appear to be over-specified.  Therefore, the risk aversions will always be chosen from a uniform deviate in the range $a\in[1,3]$ and only the forecast error $\sigma_\epsilon$ will be manipulated---excepting the following section in which the relative performance of different values of $a$ and $\sigma_\epsilon$ will be evaluated.

\section{Parameter tuning}

Thus far we have isolated three parameters ($C$, $S$, and $\delta$) which can be fixed at particular values without loss of generality.  We now want to choose reasonable ranges for the remaining parameters ($\sigma_\epsilon$, $M$, $a$, and $d$).  {\em Reasonable}, in this context, refers to agents with parameter combinations that tend to perform well (accumulate wealth) against dissimilar agents.  These parameter combinations are of interest because one would expect that, in real markets, poorly performing investors who consistently lose money will not remain in the market for long.

Note that, as discussed in the Introduction, parameter tuning generally diminishes an explanatory model's validity.  This, however, does not quite apply in this case because we are not tuning the parameters in order to produce a model which better fits the empirical data (i.e.. exhibits known market phenomena, such as fat tails and clustered volatility)---rather, we are simply trying to select ``better'' investors.  However, it must be acknowledged that this {\em may} concurrently tune the simulation towards realism. 

Further, the point of this exercise is not to completely specify the model but merely to avoid wasteful parameter combinations which should be driven out of the system by selective (financial) pressures.  In the model, ``dumb'' agents (with parameter combinations which tend to underperform) will lose capital and may eventually hold a negligible portion of $C$ and $S$.  Hence, these agents won't contribute to the market dynamics and will simply be ``dead weight'', consuming computer time and resources.  Hopefully, at this point the reader agrees that it would be helpful to cull ``dumb'' agents by finding the more successful parameter ranges.

It may be discovered, in the course of this investigation, that some parameters are irrelevant; they may be take on a wide variety of values with little or no impact on the dynamics.  In this case, these parameters may be assigned arbitrary ranges without loss of generality.

\begin{table}
	\begin{center}\begin{tabular}{c|c}
		\hline \hline
		Parameter & Run 6 \\
		\hline
		$N$ & 400 \\
		$C$ & \$1,000,000 \\
		$S$ & 1,000,000 \\
		$\sigma_\epsilon$ & $0.25 \pm 0.25$ \\
		$M$ & $105 \pm 95$ \\
		$a$ & $1.5 \pm 1.5$ \\
		$\delta$ & 0.001 \\
		$d$ & $1\pm 1$ \\
		seed & random \\
		\hline \hline
	\end{tabular}\end{center}
	\caption{Parameter values for CSEM Run 6.}
\label{tbl:csemRun6}
\end{table}

\begin{figure}
	\begin{center}
		\input{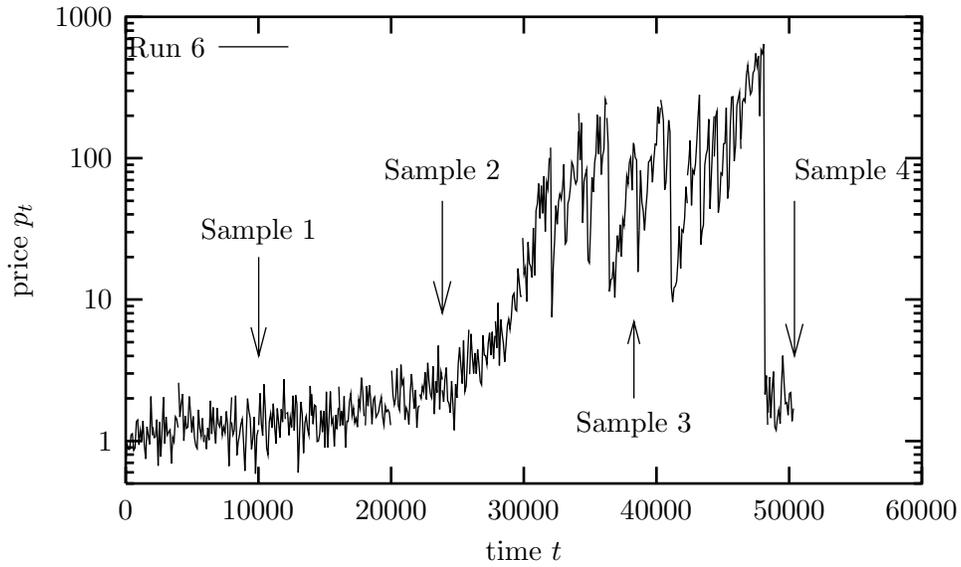} \\
	\end{center}
	\caption{Price history generated by CSEM with parameters listed in \tbl{csemRun6} (Run 6).  The price almost reaches its theoretical maximum of \$999 (see \eq{csemMaxPrice}) before collapsing.  The agent state variables were sampled at the times indicated.}
\label{fig:csemRun6Price}
\end{figure}

To determine successful values, a large parameter space should be explored.  To this end a long data set was collected with more agents and with broader parameter ranges, as indicated in \tbl{csemRun6}.  The price history for the run is shown in \fig{csemRun6Price}.  

The results were analyzed by looking for correlations between an agent's wealth and the following parameters: forecast error $\sigma_\epsilon$, memory $M$, risk aversion $a$, and degree of curve-fit $d$.  

Note that the point of this work is not to determine an optimal investment strategy (set of optimal parameter values), but simply to establish reasonable ranges for these parameters such that the agents perform reasonably well.  Thus, a complete correlation analysis is unnecessary.  Instead, a simple graphical description of the results should be sufficient, with a simple regression analysis for emphasis.

\begin{table}
	\begin{center}\begin{tabular}{c|c|c|c|c}
		\hline \hline
		& \multicolumn{4}{c}{Correlation with $\log w$} \\
		\hline
		Parameter & Sample 1 & Sample 2 & Sample 3 & Sample 4 \\
		\hline
		$\sigma_\epsilon$ & + & 0 & 0 & + \\
		$M$ & 0 & 0 & 0 & + \\
		$a$ & + & + & + & + \\
		$d$ & 0 & 0 & 0 & 0 \\
		\hline \hline
	\end{tabular}\end{center}
	\caption{Regression analysis of $\log w$ versus agent parameters for different samples of Run 6 (\tbl{csemRun6}).  The symbols indicate the sign of the regression-line slope, or zero if it is insignificant (relative to its standard error).  The results indicate that $a$ is positively correlated with wealth but $\sigma_\epsilon$, $M$ and $d$ are largely irrelevant.}
\label{tbl:csemRun6Corr}
\end{table}

\tbl{csemRun6Corr} shows the results of linear regression analyses of $\log w$ versus agent parameters for different samples of Run 6.  The logarithm of wealth is fitted to a straight line with respect to the parameter of interest and the sign of the slope is recorded.  If the slope $m$ has a standard error larger than 100\% then the parameter is interpreted as being uncorrelated with performance.  This method was constructed only because it lent itself to the computational tools available to the author.  However, it is reasonable: recall that the linear correlation coefficient (which is typically used to test for correlations) is related to the slope $r \propto m$.  Also, the standard error estimates the significance of the slope; a value greater than 100\% suggests that the sign of the slope is uncertain.

The results of the analysis indicate that the risk aversion parameter $a$ is positively correlated with performance (wealth).  However, the forecasting parameters $\sigma_\epsilon$ (forecast error), $M$ (memory) and $d$ (degree of polynomial fit) appear to be uncorrelated with performance.  Hence, these parameters can be set arbitrarily.  The memory from Run 6 ($M=105\pm 95$) will be used in all further simulations to maintain diversity.  However, the degree of the fitting polynomial will be constrained to $d=0$ (a moving average) because it boosts simulation speed.  The forecast error $\sigma_\epsilon$ requires further inspection.

\begin{figure}
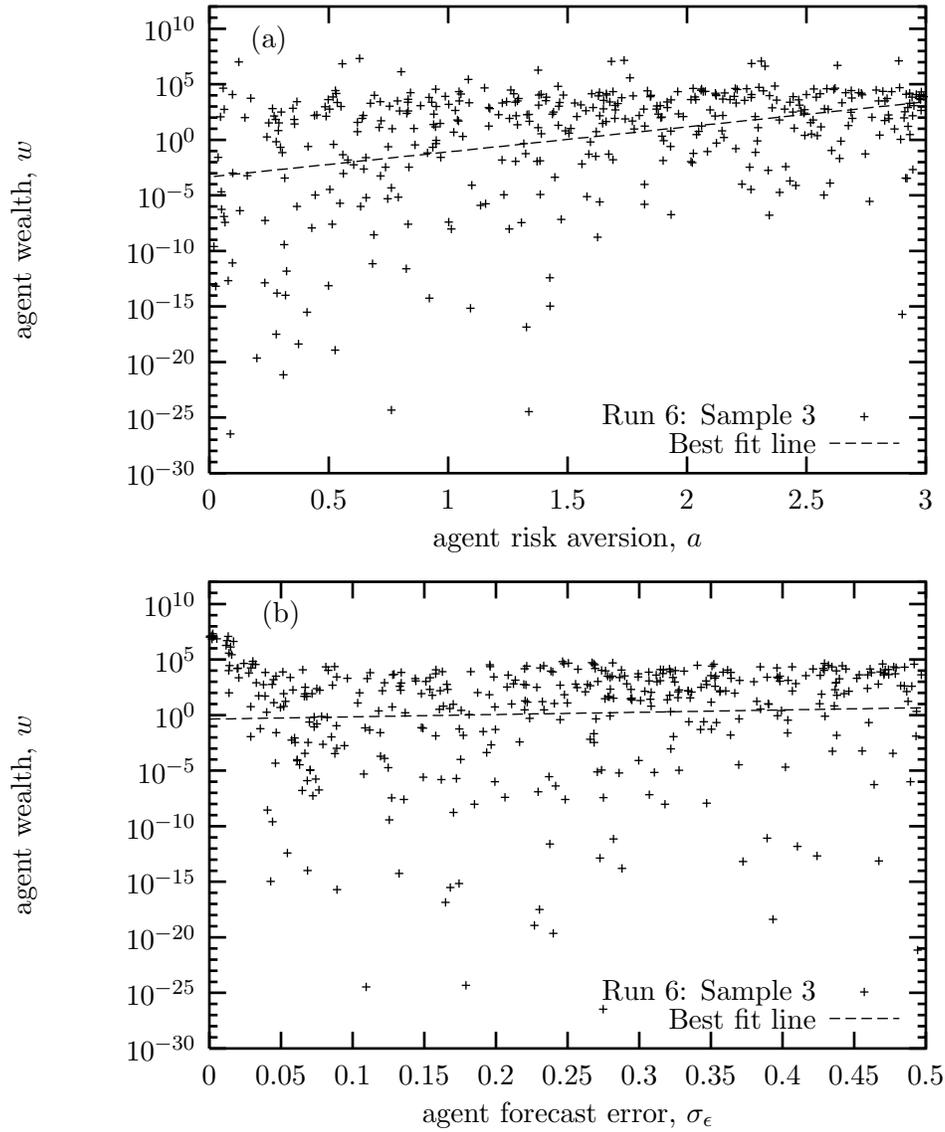

	\begin{center}
		\input{chModelCsem/run6risk} \\
		\input{chModelCsem/run6sigma}
	\end{center}
	\caption{Plot of agent wealth versus (a) risk aversion and (b) forecast error.  The best fit lines have slopes $5.2 \pm 1.4$ (positive correlation) and $4.7 \pm 8.8$ (no correlation), respectively.}
\label{fig:csemRun6Corr}
\end{figure}

Representative graphs of $\log w$ versus the parameters $a$ and $\sigma_\epsilon$ (using Run 6: Sample 3) are shown in \fig{csemRun6Corr}.  The slopes are used to estimate correlations, as discussed above.  The evidence suggests that risk aversion is positively correlated with performance.  Hence, small values of $a$ (high-risk behaviours) tend to underperform.  Thus, the range of $a$ is restricted to $a\in[1,3]$ instead of $a\in[0,3]$ as set in Run 6.

\subsection{Forecast error}

\label{sect:csemForecastError}

The only free parameter left is the forecast uncertainty $\sigma_\epsilon$.  Although \fig{csemRun6Corr} indicates no correlation between wealth and forecast error, a closer inspection reveals a small peak for the smallest errors $\sigma_\epsilon<0.05$.  

\begin{table}
	\begin{center}\begin{tabular}{c|c}
		\hline \hline
		Parameter & Run 7 \\
		\hline
		$N$ & 400 \\
		$C$ & \$1,000,000 \\
		$S$ & 1,000,000 \\
		$\sigma_\epsilon$ & $0.025 \pm 0.025$ \\
		$M$ & $105 \pm 95$ \\
		$a$ & $1.5 \pm 1.5$ \\
		$\delta$ & 0.001 \\
		$d$ & $1\pm 1$ \\
		seed & random \\
		\hline \hline
	\end{tabular}\end{center}
	\caption{Parameter values for CSEM Run 7.}
\label{tbl:csemRun7}
\end{table}

\begin{figure}
	\begin{center}
		\input{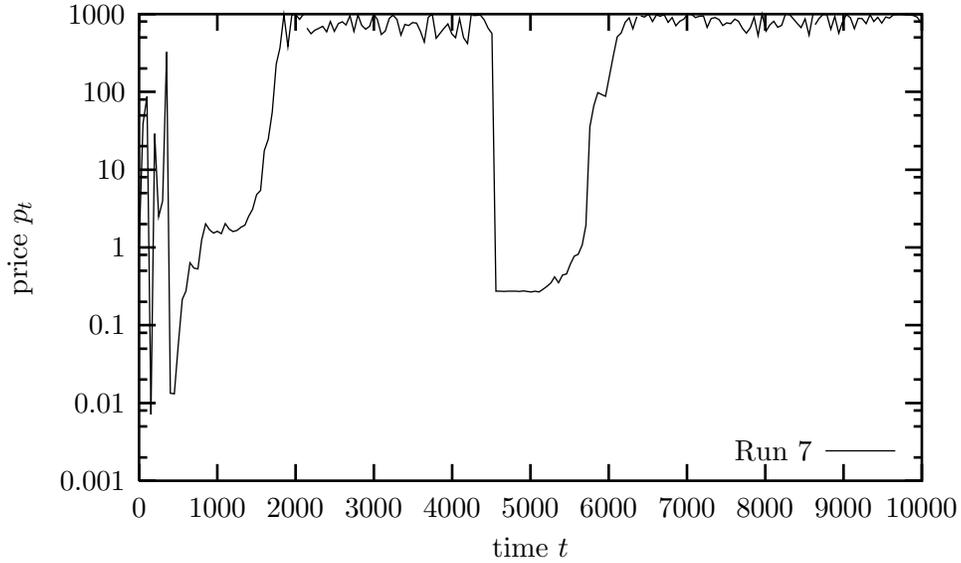} \\
	\end{center}
	\caption{Price history generated by CSEM with parameters listed in \tbl{csemRun7} (Run 7).  The series has the undesirable property that the price spends much of its history at or nearing its ceiling (\$999).}
\label{fig:csemRun7Price}
\end{figure}

To test this range, a new dataset was collected with all the parameters as in Run 6 except the forecast error scaled down by a factor of ten, as indicated in \tbl{csemRun7}.  The time series, shown in \fig{csemRun7Price}, exhibits wildly chaotic fluctuations which regularly test the price limits (\eq{csemMaxPrice}) imposed by $\delta$.  Since $\delta$ was an arbitrarily chosen parameter, we do not want it to significantly affect the dynamics as it does in Run 7.

Thus, the choice of $\sigma_\epsilon = 0.025 \pm 0.025$ causes problems.  This issue will be revisited in \ch{results}.

\subsection{Finalized parameter ranges}

\begin{table}
	\begin{center}\begin{tabular}{r|l|c}
		\hline \hline
		Symbol & Interpretation & Range \\
		\hline
		\multicolumn{3}{c}{Market parameters} \\
		\hline
		$N$ & number of agents & $2+$\\
		$C$ & total cash available & \$1,000,000 \\
		$S$ & total shares available & 1,000,000 \\
		\hline
		\multicolumn{3}{c}{Agent parameters} \\
		\hline
		$a_j$ & risk aversion of agent $j$ & $2\pm 1$ \\
		$\delta_j$ & investment fraction limit of agent $j$ & 0.001 \\
		$d_j$ & degree of agent $j$'s fitting polynomial & $0$ \\
		$M_j$ & memory of agent $j$'s fit & $105\pm 95$ \\
		$\sigma_{\epsilon,j}$ & scale of uncertainty of agent $j$'s forecast & $[0,0.5]$ \\
		\hline \hline
	\end{tabular}\end{center}
	\caption{As \tbl{csemParameters} except with updated parameter ranges.  These ranges will be used in subsequent simulations.}
\label{tbl:csemParameters2}
\end{table}

The finalized ranges of the model parameters are shown in \tbl{csemParameters2}.  The risk aversion and memory will always be assigned the shown ranges but the effect of varying $N$ and $\sigma_\epsilon$ will be explored further in \ch{results}.

\section{Discussion}

In this section some observed properties of the model (both theoretical and empirical) will be discussed.

\subsection{Fundamentalists versus noise traders}

This model borrows heavily from other work in the area \cite{palmer94, levy95, arthur97, caldarelli97, cont97, chen98, busshaus99, chowdhury99, iori99, lux99}.  However, it differs from most of these papers in that it does not divide the traders into types.  Many other models assign the agents one of two roles: either fundamentalists or noise traders \cite{youssefmir94, bak97, chen98, eliezer98, lux99}.  Fundamentalists believe the stock has a real value (for instance, if it pays a dividend) and trade when they believe the stock is over- or under-valued.  Noise traders (or {\em chartists}), on the other hand, have no interest in the stock's fundamental value, but simply try to anticipate price fluctuations from the historical data, and trade accordingly.

CSEM deliberately eliminates the fundamental value of the stock so the agents are necessarily what would be called noise traders.  Hence, the dynamics which emerge from the simulations are of a completely different nature than those mentioned above.

\subsection{Forecasting}

\tbl{csemRun6Corr} indicates that the forecasting parameters are largely irrelevant to performance.  This suggests that there are no serial correlations in the stock price and, therefore, no reward for increased effort to forecast (by increasing $M$ and/or $d$).  Whether the time series actually does have auto-correlations will be explored in \ch{results2}.   But the ineffectiveness of forecasting raises the question of whether a model based on forecasting is even relevant.  Perhaps the agents would do better to ignore the return history and just rely on fluctuations to make their estimates of future returns.  This may indeed be a valid argument but forecasting has another purpose---it adds a degree of heterogeneity to the agents through systematic differences in their investment fractions.

On the other hand, forecasting may provide a mechanism for herding.  As the history develops, the returns may be correlated for short periods.  If so, then the agents may converge in opinion regarding future returns and act in unison, with significant consequences in the price history.  For this reason, the forecasting algorithm will be retained.

\subsection{Portfolios}

\begin{figure}
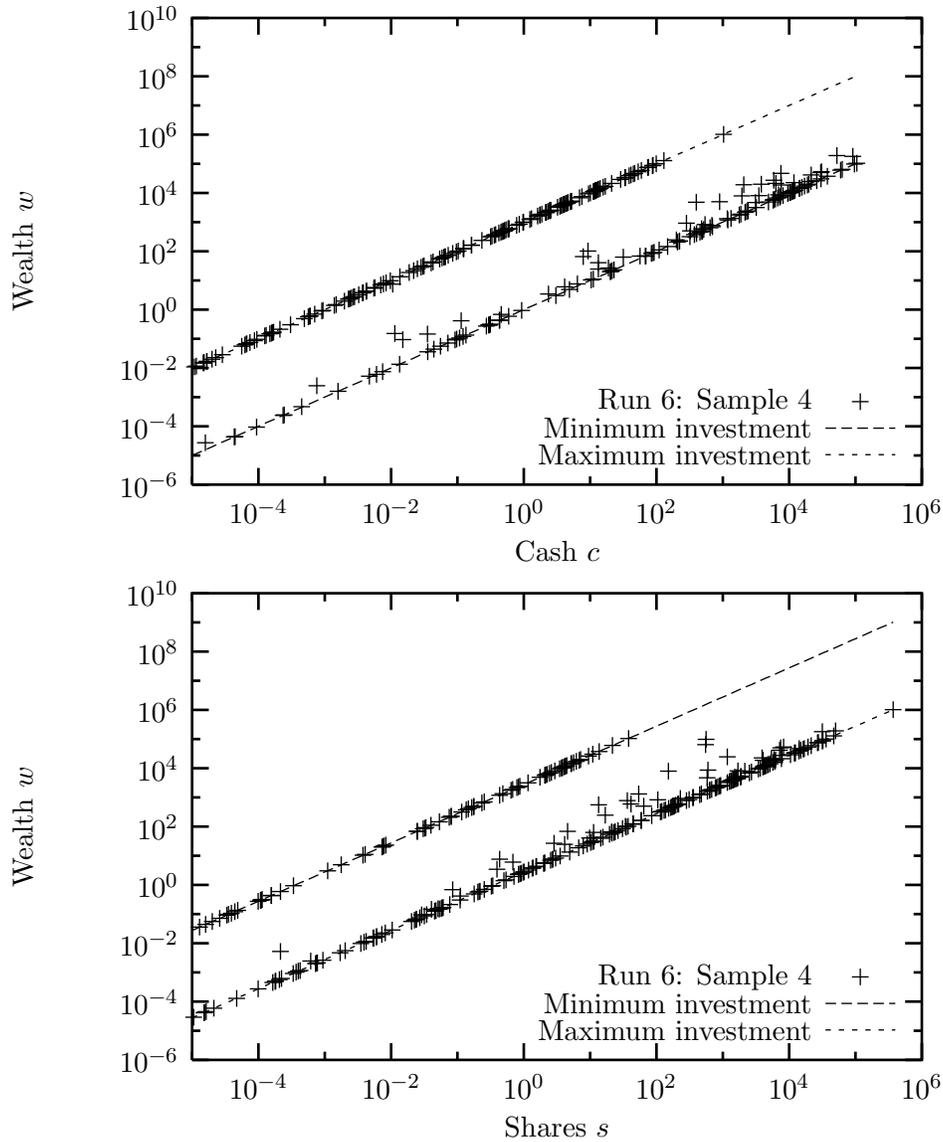

	\begin{center}
		\input{chModelCsem/run6cash} \\
		\input{chModelCsem/run6shares}
	\end{center}
	\caption{Plot of agent wealth versus (a) cash and (b) shares held showing that most agents hold extreme portfolios of maximum cash and minimum shares, or vice versa.  It appears that the method of calculating the investment fraction in CSEM (\eq{csemOptimalFraction}) is too sensitive to fluctuations.  (It should be acknowledged the plots are truncated since the lowest wealth actually extends down to $10^{-25}$, an unrealistic quantity since real money is really discretized with a minimum resolution of one penny.)}
\label{fig:csemRun6Portfolio}
\end{figure}

Given an investment fraction $i$, wealth $w$, and stock price $p$, an agent's distribution of cash $c$ and shares $s$ is given by the relations
\begin{eqnarray}
	iw & = & sp \\
	(1-i)w & = & c.
\end{eqnarray}

Given the investment limits $\delta\leq i \leq 1-\delta$, a linear relation between wealth and the maximum or minimum holdings of both cash and shares can be found.  (Recall, agents are not allowed to sell all their shares or cash.)  \fig{csemRun6Portfolio} shows the distribution of cash versus shares for the agents of Run 6: Sample 4.  Notice that the agents almost exclusively hold extremal portfolios dominated by either cash or stock.  Very few actually hold mixed portfolios.  This indicates that \eq{csemOptimalFraction}, which gives an agent's optimal investment fraction, may be too sensitive.  But the only freedom one has in reducing the sensitivity is through the parameters $a$ and $\sigma_\epsilon$, which have other consequences, as has been discussed.

\subsection{Difficulties}

Although this model showed promise, I had some technical and ideological problems which encouraged me to abandon it in favour of a different approach.  On the technical side, as the reader can see, the number of parameters is somewhat unwieldy.  Although some of the parameters could be determined, those remaining were difficult to manage.  The investment fraction limit $\delta$, for instance, is a necessary but unappealing result of the derivation, which imposes arbitrary limits on the stock price's range.  Another difficult parameter is the forecast error $\sigma_\epsilon$: if too large a value is chosen then the dynamics are dull and dominated by noise (see \fig{csemRuns45}), but too small a value produces wildly chaotic behaviour completely unlike empirical market data (\fig{csemRun7Price}).  This seems to be the critical parameter for determining the character of the dynamics, and the effect of varying it will be explored in more detail in \ch{results}.

One of the ideological problems was the use of parallel updating (all agents trading at a single moment each day).  Evidence is mounting that employing a parallel updating scheme (without strong justification) introduces chaotic artifacts into the dynamics which are generally not observed in the actual, continuous-time system being modeled \cite{huberman93, bersini94, rajewsky97, rolf98, blok99}.

Further, in this model the price of a share is artificially fixed by the market maker.  In most markets the price is an emergent phenomena: auction-type orders are placed at hypothetical prices (eg. limit prices) and the price is realized when a trade occurs.  Forcing the price to balance supply and demand destroys its emergent character.

For these reasons, this line of research was replaced with the model presented in the next chapter.  Nevertheless, the Centralized Stock Exchange Model is included here because it follows a prevalent line of reasoning in stock market simulations and falls into many of the same pitfalls encountered by others \cite{palmer94, youssefmir94, levy95, caldarelli97, chen98, chowdhury99, eliezer98, lux99, solomon99}.  Wherever possible, CSEM data will be analyzed alongside the output of the next model, the {\em Decentralized} Stock Exchange Model.


\chapter{Decentralized Stock Exchange Model}

\label{ch:dsem}			

\section{Inspiration}

I was growing disillusioned with CSEM when a collaboration with---then undergraduate student---Casey Clements inspired me to consider a radically different approach.  Casey also expressed dissatisfaction with the price being set by a centralized control (market maker) and wondered how real markets worked.  I explained to him that the stock (ticker) price was simply the last price at which a trade had occurred.  Casey expressed an interest in modeling this approach but I explained to him the hurdles: namely that the agents would have to be a great deal more complicated than in current models because they would have to make complicated decisions involving two or more parameters.  

Previously, agents were simple utility maximizers, applying \eq{csemOptimalHoldings} in order to calculate the quantity of shares they wanted to trade in response to a given price (set by the centralized control) but these new agents would have to decide on both the volume to trade and the price they wanted to trade at.

Nevertheless, Casey was enthusiastic so I obliged him and we constructed a simple event-driven model of stock exchange (neglecting the difficulties with the agents' decision processes) with the following properties:
\begin{enumerate} 
\item{agents trade by calling out and replying to orders,}
\item{trades can be called at any time (continuous time), and}
\item{agents can choose both the volume of the trade and the price.}
\end{enumerate}

In this model prices would be decentralized, arising directly from the agents decisions rather than being governed by a market maker.  Thus was born the {\em Decentralized Stock Exchange Model} (DSEM).

\section{Basic theory}

This model is of a more original nature than CSEM was and, therefore, requires more explanation.  For this reason the theory is divided into two sections, one which explains the basic structure of the model and another which describes how fluctuations are incorporated.  First, the structure of the model will be developed.

\subsection{Assumptions}

The decentralized model discussed here contains many of the same assumptions as CSEM (heterogenous agents; the market is composed of a single risky asset and a single riskless asset; cash and shares are both conserved; et cetera).  For the sake of brevity, only the differences in assumptions between the two models will be discussed.

\subsubsection{Decentralization}

The primary difference between the two models is, obviously, the move to a decentralized market.  This means that the agents are allowed to trade directly with one another without any interference from a market maker or specialist.  The market maker may be interpreted as having been relegated to the mechanical role of matching buyers with sellers, with no influence on the price.  This interpretation resembles intra-day trades of a fairly active stock \cite{garman76}.  CSEM, in contrast, was meant to mimic low-frequency trading, on timescales no shorter than a single day.

\subsubsection{Continuous time with discrete events}

By moving to intra-day trading, the natural periodicity of the market opening and closing daily, which gave rise to discretized time in CSEM, is eliminated.  In fact, in DSEM it is assumed that the market never closes; it is open around the clock, 24 hours per day.  Trades may be executed at any instant and time is a continuous variable.  (Another interpretation is that the market does close but when it re-opens it continues from where it left off without any effect from the close.)

To implement continuous-time (at least to some fine resolution) on fundamentally discrete devices (digital computers) a shift of paradigms is required.  Traditionally, time evolution is simulated by simply incrementing ``time'' by a fixed amount (as in CSEM).  This approach is cumbersome and inefficient when the model consists of discrete events (eg. trades) occurring at non-uniform time intervals.  Instead, an event-driven approach is preferred \cite{lawson00} in which waiting times (delays) for all possible events are calculated and ``time'' is immediately advanced to the earliest one.  (As each event time is calculated the event is placed in an ordered queue so the earliest event is simply the first event in the queue.)  The basic algorithm follows:
\begin{enumerate}
	\item{Calculate waiting times for all events.}
	\item{\label{enum:dsemEventsStartLoop} Find event with shortest waiting time.}
	\item{Advance time to this event and process it.}
	\item{Recalculate waiting times as necessary.}
	\item{Return to Step \enum{dsemEventsStartLoop}.}
\end{enumerate}

\subsubsection{Separation of time scales}

The model allows two types of events: {\em call} orders and {\em reply} orders.  By distinguishing between the two a simplifying assumption can be made: no more than one call order is active at any time.  

In many real markets orders are good (active) until filled or until they expire.  When new orders are placed they are first treated as reply orders by checking if they can satisfy any outstanding orders, and then, if they haven't been filled, they are placed on a {\em auction book}, and become call orders until they are removed.

In DSEM, however, it is assumed that reply orders occur on a {\em much} faster timescale than calls.  As soon as a call order is placed, all reply orders are submitted and executed (almost) instantaneously and the call (if not filled) can immediately be expired (because no more replies are expected).  Hence, the probability that two (call) orders are active simultaneously becomes negligible and it is assumed that at most one is ever active at any given time.  

This assumption improves performance at the cost of realism.  Allowing multiple orders to be simultaneously active would require more complicated book-keeping and would degrade simulation performance.  (It was observed that DSEM exhibited rich enough behaviour that the assumption did not need to be discarded.)

\subsection{Utility theory}

As with CSEM, we again begin with a game theoretic approach.  However, instead of an exponential utility function, a power-law is used,
\begin{equation}
	U(w) = \left\{\begin{array}{l|l}
	       	\frac{1}{1-a}w^{1-a} & a\neq 1 \\
	       	\ln w                & a = 1.
	       \end{array}\right.
\end{equation}
Notice these forms are effectively identical as $a\rightarrow 1$ because $\lim_{a\rightarrow 1} U' = w^{-1}$ for both, and utility is only defined up to a (positive) linear transformation.

The advantage of the power-law utility (sometimes known as the Kelly utility \cite{kelly56}) is that it has a constant {\em relative} risk aversion
\begin{equation}
	R(w) \equiv - \frac{w U''}{U'} = a,
\end{equation}
unlike the exponential utility which, from \eq{csemExpUtility}, has constant {\em absolute} risk aversion $A(w) \equiv -U''/U' = 1/w_{goal}$.  The constant relative risk aversion eliminates the scaling problems which had to be worked around in \sect{csemRiskAversion} and keeps an agent equally cautious regardless of its absolute wealth.  (See Ref.\ \cite{merton92} for a more detailed description of absolute and relative risk aversion).

\subsection{Optimal investment fraction}

Let us assume an agent is interested in holding a fixed fraction $i$ of its capital in the risky asset.  If the price moves from $p(0)$ to $p(t)$ the return-on-investment is
\begin{equation}
	r(t) = \frac{p(t)-p(0)}{p(0)}
\end{equation}
and the final wealth can be written as
\begin{eqnarray}
	w(t) & = & w(0)(1-i)+w(0)i(r+1) \\
	     & = & w(0)(1+ir).
\end{eqnarray}

The goal is to find the value of $i$ which maximizes the expected utility at some future time $t$.  First, the expected utility must be calculated:
\begin{equation}
\expect{U} = \frac{w(0)^{1-a}}{1-a} \expect{(1+ir)^{1-a}}.
\end{equation}
Unfortunately, finding a closed-form solution for $\expect{U}$ is difficult, even with very simple probability distributions.  

However, if we assume the timescale is relatively short then we expect the returns to be small and the above equation can be expanded around $r=0$, as
\begin{equation}
	\expect{U} \approx w(0)^{1-a} \expect{ \frac{1}{1-a} + ir - \half a i^2 r^2 + \cdots}.
\end{equation}

Thus, the optimization condition becomes
\begin{eqnarray}
	0 & = & \frac{d\expect{U}}{di} \\
	                       & = & w(0)^{1-a} \left[ \expect{r} - a i^* \expect{r^2} \right] \\
	                       & = & w(0)^{1-a} \left[ \expect{r} - a i^* (\var{r}+\expect{r}^2) \right],
\end{eqnarray}
giving an optimal investment fraction
\begin{equation}
\label{eq:dsemOptimalFraction}
	i^* = \frac{\expect{r}}{a(\var{r}+\expect{r}^2)}
\end{equation}
with the constraints $0\leq i^*\leq 1$.  (This derivation implicitly assumed that the higher moments in the expansion are negligible---an assumption which may not be valid even for short timescales, as will be seen in \ch{results2}).

Note how closely this corresponds with \eq{csemOptimalFraction} in CSEM, differing only by an extra term in the denominator.  However, the use of a power-law utility allowed us to drop many of the assumptions originally required, such as explicitly hypothesizing that the returns were Gaussian-distributed.

\subsection{Fixed investment strategy}

\eq{dsemOptimalFraction} states that, given some expectation and variance of the future return, one should hold a constant fraction of one's capital in the risky asset.  The same result was derived by Merton \cite[Ch.\ 4]{merton92} assuming a constant consumption rate (which can be taken to be zero, as in DSEM).  Further, Maslov and Zhang \cite{maslov98} demonstrated that keeping a fixed fraction of one's wealth in the risky asset maximizes the ``typical'' long-term growth rate (defined as the {\em median} growth rate).

These references suggest that the optimal strategy is to keep a fixed fraction of one's wealth in stock---a Fixed Investment Strategy (FIS).  The FIS is empirically tested with a hypothetical portfolio in \ch{portfolio} and developed, here, for use with DSEM.  The strategy discards the attempt to forecast returns (which proved problematic in CSEM) in favour of the basic principle of maintaining some fixed fraction invested in stock.

\subsubsection{Ideal price}

The total value of a portfolio consisting of $c$ cash and $s$ shares at price $p$ is
\begin{equation}
	w(p) = c+ps.
\end{equation}

The goal of FIS is to maintain a balance 
\begin{equation}
\label{eq:dsemBalance}
	ps = iw(p)
\end{equation}
by adjusting one's holdings $s$.  (The ideal investment fraction is simply denoted by $i$ henceforth, without the cumbersome asterisk.)  If an agent currently holds $c$ cash and $s$ shares then the agent would achieve the optimal investment fraction $i$ if the stock price was
\widebox{\begin{equation}
\label{eq:dsemIdealPrice}
	p^* = \frac{ic}{(1-i)s},
\end{equation}}
which will be called the agent's {\em ideal price}.

If the price is higher than the ideal price then the agent holds too much capital in the form of stock and will want to sell and, conversely, at a lower price the agent will buy in order to buoy up its portfolio.

\subsubsection{Optimal holdings}

The fixed investment strategy specifies not only what type of order to place at a given price, but also precisely how many shares to trade.  At a price $p$ an agent would ideally prefer to hold stock
\widebox{\begin{equation}
\label{eq:dsemOptimalHoldings}
	s^* = \frac{iw(p)}{p}.
\end{equation}}

Given the ideal price above, the agent's wealth can be written
\begin{equation}
	w(p)=\frac{1-i}{i}p^*s + ps.
\end{equation}

\begin{figure}
	\begin{center}
		\input{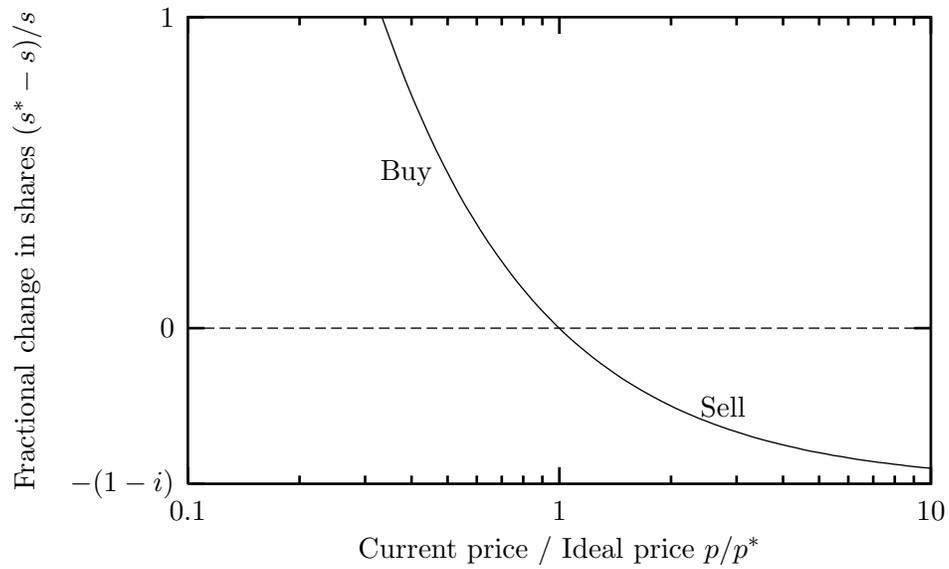}
	\end{center}
	\caption{The fixed investment strategy specifies how many shares to trade at a price $p$ given an investment fraction $i$ and an ideal price $p^*$ (from \eq{dsemIdealPrice}).  As the current price drops toward zero the fractional change in shares diverges.}
\label{fig:dsemOptimalTrade}
\end{figure}

Hence, an agent's optimal trade at price $p$ is
\begin{eqnarray}
	s^*-s & = & \frac{iw}{p}-s \\
	      & = & (1-i)s\frac{p^*}{p} + is - s \\
\label{eq:dsemOptimalTrade}
	      & = & (1-i)\left( \frac{p^*}{p}-1 \right) s.
\end{eqnarray}
The fractional change of stock $(s^*-s)/s$ is plotted as a function of price in \fig{dsemOptimalTrade}.

So, given a portfolio ($c$,$s$) and an optimal investment fraction $i$, FIS prescribes when to buy and sell and precisely how much.

\subsection{Friction}

The derivations of the fixed investment strategy here and elsewhere \cite{merton92, maslov98} assume no transaction costs, which makes it impractical in real markets.  Every minuscule fluctuation of a stock's price would require a trade in order to rebalance the portfolio but if the fluctuation was too small the trade would cost more than the value of the shares traded, resulting in a net loss.  

To circumvent this problem in a simulated portfolio with a commission on every trade (see \ch{portfolio}) I imposed limits on the buying and selling prices:
\widebox{\begin{eqnarray}
\label{eq:dsemLimitPriceBuy}
	p_B & = & p^*/(1+f) \\
\label{eq:dsemLimitPriceSell}
	p_S & = & p^*(1+f)
\end{eqnarray}}
where $f$ is defined as the trading {\em friction}.

The same approach can be used here: don't buy until the price drops below the limit $p_B$ and don't sell until it rises above the limit $p_S$.  This allows the simulation to mimic the effect of transaction costs while conserving the total cash.  It is also required to make the simulation a discrete-event model.  Without it, agents would trade on a continuous-time basis and the model would be difficult to simulate (and less realistic).

Obviously, the larger the friction, the larger a price fluctuation will be required before an agent decides to trade.  So increasing friction decreases the trade frequency and hinders market activity---which explains why the term ``friction'' is used.

Each agent's friction must be strictly positive because a zero value would allow an agent to place an order to trade zero shares at its ideal price---a null order.  The simulation does not forbid this but, as will be seen, such an order would never be accepted.

There is no fixed upper limit on the friction but it seems reasonable to impose $f<1$.  This would mean an agent places buy and sell limits at one half and double the ideal price, respectively.  (It would be peculiar for an investor not to sell when a stock's price doubles!)

Thus, the friction $f$ is the first agent-specific parameter in DSEM and it is chosen such that $f\in(0,1)$.

\subsection{Call orders}

As mentioned above, DSEM is a discrete-event simulation.  The events are orders which are called out.  An order consists of a price $p_o$, type (``Buy'' or ``Sell'') and quantity of shares to trade.

\begin{figure}
	\begin{center}
		\input{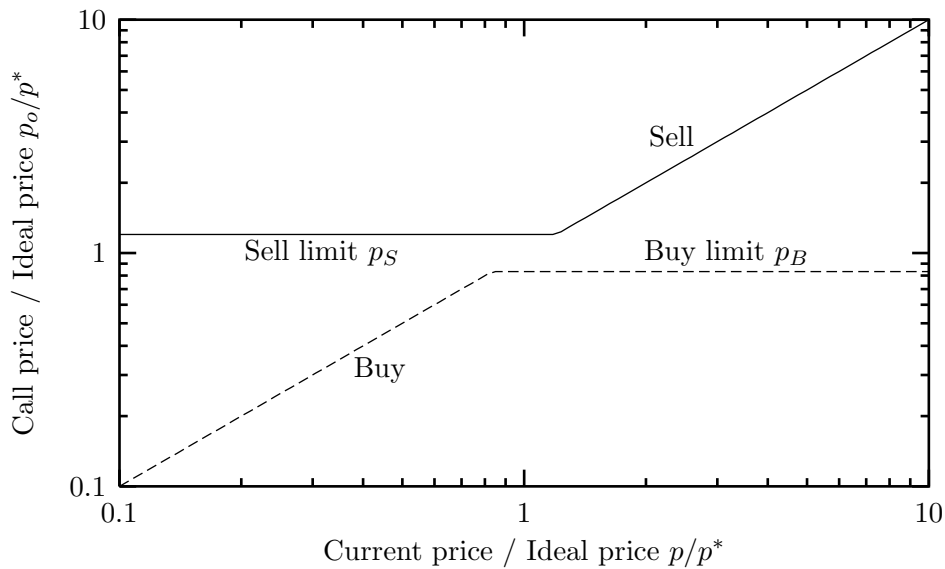}
	\end{center}
	\caption{Call orders $p_o$ are placed at either the current price $p$ or the limit prices $p_B$ or $p_S$, whichever is better.  The spread between the limits increases with friction $f$.}
\label{fig:dsemOrderPrice}
\end{figure}

The agents use the limit prices discussed in the last section to set their trading prices, thereby solving the problem of how to design agents which can choose both a trade volume {\em and} price.  (The volume is set by \eq{dsemOptimalTrade}.)  Of course, if the current market (last traded) price $p$ is ``better'' than the limit price ($p>p_S$ or $p<p_B$) then the agents rationally choose to trade at that price:
\widebox{\begin{equation}
	p_o = \left\{ \begin{array}{ll}
	      	\min(p,p_B) & \mbox{for ``Buy'' orders} \\
	      	\max(p,p_S) & \mbox{for ``Sell'' orders}.
	      \end{array} \right.
\end{equation}}

These order prices, shown in \fig{dsemOrderPrice}, are substituted into \eq{dsemOptimalTrade} to compute the volume of shares to trade.  Notice that there are two options (buy or sell) for every price.  Before discussing how the agent decides which action to take, we must understand the process which governs most discrete-event simulations.

\subsection{Poisson processes}

A Poisson process is a stochastic counting process defined by the probability of an event occurring in an infinitesimal interval $dt$,
\begin{equation}
	\pi(dt) = dt/\tau
\end{equation}
where $\tau$ is the average event interval \cite{vankampen81}.  All events are assumed to be independent.  The cumulative probability of {\em no} events having fired within a finite interval $t$ is then given by
\begin{equation}
	\Pr(0,t) = e^{-t/\tau}
\end{equation}
which accurately describes many natural processes, such as radioactive decay of nuclei.

The advantage of the Poisson process from a simulation perspective is that it may be interrupted.  Consider a time interval $t$ which is divided into two intervals $t_1$ and $t_2$ ($t=t_1+t_2$).  The probability of no events within $t$ can be written
\begin{eqnarray}
	\Pr(0,t) & = & e^{-(t_1+t_2)/\tau} \\
	       & = & \Pr(0,t_1) \Pr(0,t_2)
\end{eqnarray}
which simply means that if the event did not occur within the interval $t$ then it must not have occurred within either of the sub-intervals $t_1$ or $t_2$.

This equivalence relation may be interpreted to mean that a clock which measures the stochastic time to a Poisson event may be reset at any time before the event fires, without changing the probability distribution.  This property is very useful for discrete-event simulations because it allows one to proceed to the first event time, update the system, and recalculate all the event times from this point without disturbing the process.  (Event times would need to be recalculated, for instance, if their average rate $\tau$ was affected by the event which transpired.)

We now have the background necessary to discuss how an agent chooses which action to take.

\subsection{Call interval}

Agents always have two options when deciding on a call order: place a ``Buy'' or a ``Sell'' order.  Since the simulation is event-driven, the easiest way to handle this is to allow both possibilities.  Each is an event which will occur at some instant in time.  The events of calling orders is modeled as a Poisson process, as discussed above.  

More precisely, ``Buy'' and ``Sell'' orders are modeled as independent Poisson processes, each with its own characteristic rate.  Intuitively, if the current stock price $p$ is very high then we expect agents will try to sell at or near the current price, rather than trying to buy at a much lower price, and vice versa if the price is low.  This intuition can be captured by making the Poisson rates for the ``Buy'' and ``Sell'' calls price dependent \cite{garman76}, as
\widebox{\begin{eqnarray}
\label{eq:dsemCallIntervalBuy}
	\tau_B(p) & = & \frac{p}{p_B} \tau \\
\label{eq:dsemCallIntervalSell}
	\tau_S(p) & = & \frac{p_S}{p} \tau
\end{eqnarray}}
where $\tau$ is an unspecified (constant) timescale and $\tau_B$ and $\tau_S$ are the average times between each type of call order.

The above linear price dependence is justified only on the basis of its simplicity, but it also has some reasonable consequences.  An agent makes the fewest calls (of either type) when the call rate $1/\tau_{B+S}$ is minimized,
\begin{eqnarray}
	0 & = & \frac{d}{dp} \frac{1}{\tau_{B+S}} \\
	  & = & \frac{d}{dp} \left( \frac{1}{\tau_B} + \frac{1}{\tau_S} \right) \\
	  & = & -\frac{p_B}{p^2} + \frac{1}{p_S}
\end{eqnarray}
which occurs at a price $p = \sqrt{p_B p_S} = p^*$, the agent's ideal price.  Hence, when an agent is satisfied with its current portfolio, it will place the fewest (speculative) orders.  

As the price moves away from the agent's ideal price the call rate increases reflecting an increased urgency.  With many agents independently placing call orders, the agent with the greatest urgency (shortest waiting time) will tend to place the first order, so urgency drives market fluctuations \cite{farmer98}.  This is in contrast with CSEM and many other simulations in which fluctuations are driven by supply vs.\ demand \cite{palmer94, levy95, arthur97, caldarelli97, cont97, chen98, chowdhury99, busshaus99, iori99, lux99}. 

Incidentally, the minimum call rate is
\begin{eqnarray}
	\frac{1}{\tau_{B+S}(p^*)} & = & \frac{1}{\tau_B(p^*)} + \frac{1}{\tau_S(p^*)} \\
\label{eq:dsemMinCallRate}
	                         & = & \frac{2}{(1+f)\tau}
\end{eqnarray}
which decreases as the friction $f$ is increased, as discussed previously.

\begin{figure}
	\begin{center}
		\input{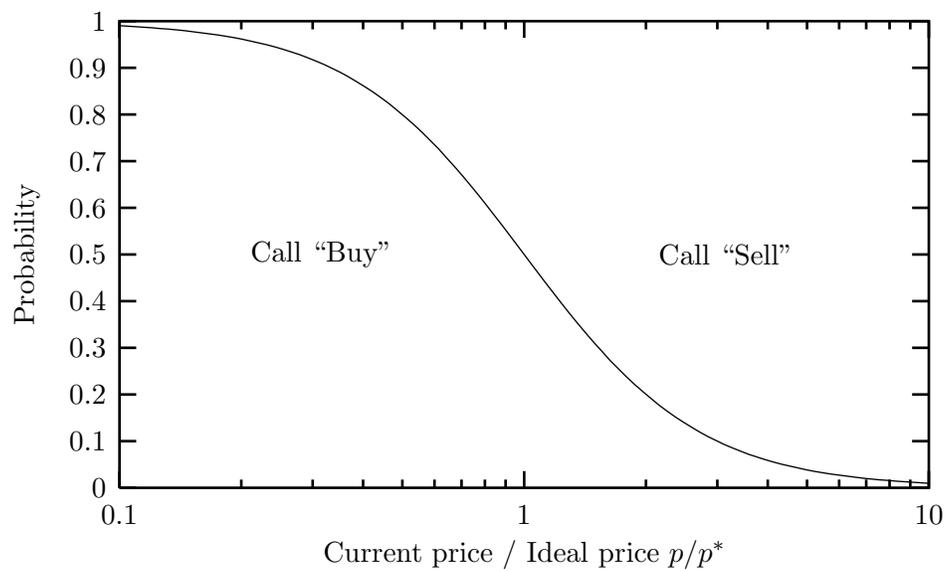}
	\end{center}
	\caption{``Buy'' and ``Sell'' call orders are modeled as independent Poisson processes with price-dependent rates.  As the last trading price increases, the probability of a ``Sell'' order being called becomes much more likely than a ``Buy.''}
\label{fig:dsemCallProb}
\end{figure}

\eqs{dsemCallIntervalBuy}{dsemCallIntervalSell} were introduced to increase the probability of buying if the last trading price was low and selling if the last trading price was high.  The probability of the next call being a ``Buy'' can be explicitly calculated: consider the probability of a Buy order being placed between times $t$ and $t+dt$ with neither a Buy nor a Sell having occurred yet.  Then, marginalizing over all $t$ gives the probability of a Buy occurring first,
\begin{eqnarray}
	\Pr({\rm Buy}) & = & \int_0^\infty \frac{dt}{\tau_B} e^{-t/\tau_B -t/\tau_S} \\
	               & = & \frac{1/\tau_B}{1/\tau_B + 1/\tau_S} \\
	               & = & \frac{p^*/p}{p^*/p + p/p^*}.
\end{eqnarray}
Of course, the probability of a Sell order being placed first is the complement, $\Pr({\rm Sell}) = 1 - \Pr({\rm Buy})$.  \fig{dsemCallProb} shows that Buy orders are much more likely at low prices and Sell orders more likely at high.

\subsection{Reply orders}

Orders are divided into two types: call orders and reply orders.  The call orders have been discussed above.  Reply orders are handled in much the same manner with two important changes:
\begin{enumerate}
	\item{When replying to a call order, the replier is not free to set a price but must accept the called price.}
	\item{Reply orders happen on a much faster timescale than call orders.}
\end{enumerate}

The first item requires that reply orders be handled slightly differently than call orders.  The agent still calculates price limits according to \eqs{dsemLimitPriceBuy}{dsemLimitPriceSell} but now this is used as the criterion for whether to place an order or not.  If the order price meets the Sell limit $p_o\geq p_S$ then a ``Sell'' reply is placed and if the price meets the Buy limit $p_o\leq p_B$ then a ``Buy'' reply is placed.  Otherwise the agent does not reply to the called order.

It is assumed that a called order is completely transparent; all potential repliers have full information about the order including the price, type of order (Buy or Sell) and quantity of shares to be traded.  This provides another criterion whether to reply or not: the quantity of shares in the called order must be sufficient to completely fill the replier's demand.  Although this requirement may seem too strict, it is useful because it prevents callers from swaying the price series with negligible orders (as could occur with zero friction or if the caller has negligible wealth compared to the replier).

Like call orders, replies receive stochastic execution times with average intervals from \eq{dsemCallIntervalBuy} or \eq{dsemCallIntervalSell} depending on whether the reply is a Buy or Sell order.  Replies compatible with the call order are then processed in a ``first come, first served'' queue until the call order is filled or all repliers are satisfied (partial orders are processed).

As mentioned above, replies occur on a much faster timescale than call orders, such that no more than one call order is ever active at a time.  The time for all replies to be processed is assumed to be infinitesimal compared to the calling time interval.

\subsection{Time scale}

\label{sect:dsemTimescale}

If each agent places Buy and Sell call orders at rates $1/\tau_B$ and $1/\tau_S$, respectively, then the net rate $\rho$ of call orders being placed (for all agents) is
\begin{equation}
	\rho = \sum_{i\in{\rm agents}} \left( \frac{1}{\tau_{B,i}} + \frac{1}{\tau_{S,i}} \right).
\end{equation} 
Let us assume that all agents are satisfied with their current portfolios; an assumption which, by \eq{dsemMinCallRate}, minimizes the net event rate,
\begin{equation}
	\rho_{min} = \frac{2}{\tau}\sum_i \frac{1}{1+f_i}.
\end{equation}

Now we can identify the minimum event rate with a real rate in order to specify the timescale $\tau$.  Note that fixing $\tau$ is only necessary in order to set a scale for comparing simulation data with empirical market data.  An arbitrary but convenient choice is to assume each agent trades once each day (on average).  Then, if the market contains $N$ agents, we expect $\rho_{min} = N/2$ (because each call order is a trade between at least two agents) so the timescale $\tau$ should be
\widebox{\begin{equation}
\label{eq:dsemTimescale}
	\tau = \frac{4}{N} \sum_i \frac{1}{1+f_i}.
\end{equation}}

Hence, one time unit is meant to represent one day.  This is a particularly useful choice because it allows us to draw some parallels between DSEM and the original model, CSEM, in which agents were constrained to exactly one trade per day.  Notice the parallel is not perfect for two reasons: firstly, there is no guarantee that a call order will have any repliers and secondly, if it is executed there may be multiple repliers.  Nevertheless, the scaling should be accurate within an order of magnitude.

\subsection{Initialization}

DSEM is initialized with $N$ agents amongst whom some total cash $C$ and total shares $S$ are distributed.  ($N$ sets the ``size'' of the market.  Heavily traded stocks would be represented by large $N$.)  As in CSEM, the cash and shares will usually be distributed uniformly between the agents.  

Each agent begins with an ideal investment fraction $i(0)=1/2$ which gives an ideal price
\begin{equation}
	p^*(0) = \frac{c}{s}.
\end{equation}
For simplicity, each agent begins satisfied with its current portfolio, believing that the current market price is actually its ideal price $p(0)=p^*(0)$.

Notice that \eq{dsemOptimalFraction} is not used to set the investment fraction.  Its purpose was only to establish that holding a fixed fraction $i$ of one's wealth in stock is rational.  How $i$ is updated is the subject of the next section.

If the cash and shares are initially distributed equally amongst the agents (as will be assumed for all runs, unless otherwise stated) then the simulation begins in stasis: any Sell order submitted must necessarily be set above all repliers ideal price ($C/S$) so it will not be filled, and vice versa for Buy orders.  Thus, no trading will occur and the price will never move away from $C/S$.  What is needed is some stochastic driving force to initiate the dynamics.  (Even starting with a non-uniform distribution produces only transient fluctuations before the price stabilizes.)

In this section three market parameters were introduced: the number of agents $N$, the total cash $C$, and the total shares $S$.

\section{Fluctuation theory}

The above theory completely specifies the basic model.  What remains is to incorporate fluctuations, as discussed in this section.  Recall that forecasting was problematic in CSEM (see \fig{csemRun6Portfolio}, for example) so in DSEM a different tack is taken.

\subsection{Bayes' theorem}

Before we can understand how fluctuations are incorporated it is necessary that we briefly review some results from Bayesian probability theory \cite[Ch.\ 4]{jaynes96}, which provides an inductive method for updating one's estimated probabilities of given hypotheses as new data arrive:

Let $H$ represent some hypothesis which one wishes to ascertain the truth value of.  If $X$ is our prior information then we begin with a probability of $H$ given $X$, denoted by $P(H|X)$.  Now notice that as new information $D$ (data) arrives the joint probability of both the hypothesis {\em and} the data becomes
\begin{eqnarray}
	P(HD|X) & = & P(H|X)P(D|HX) \\
	        & = & P(D|X)P(H|DX)
\end{eqnarray}
from the product rule of probabilities.

But these equations can be rewritten to give the probability of the hypothesis in light of the new information,
\begin{equation}
	P(H|DX) = P(H|X) \frac{P(D|HX)}{P(D|X)},
\end{equation}
which is known as Bayes' theorem.

Let us define {\em evidence} as the logarithm of the odds ratio $e\equiv \log\left[ P/(1-P) \right]$, which is just a mapping from probability space to the set of all real numbers $(0,1)\rightarrow (-\infty,\infty)$.

The advantage of the evidence notation over probabilities is that incorporating new information is an additive procedure
\begin{equation}
\label{eq:dsemBayesEvidence}
	e(H|DX) = e(H|X) + \log\left[ \frac{ P(D|HX) }{ P(D|\bar{H}X) } \right],
\end{equation}
where $\bar{H}$ is the negation of $H$.

With the understanding that assimilating new information is an additive process for evidence, we may proceed with extending DSEM to include fluctuations.

\subsection{News}

In real markets a stock's price is derived from expectations of its future earnings.  These expectations are formed from information about the company, which is released as {\em news}.  In other words, news drives market fluctuations.

In DSEM fluctuations are also driven by news.  How to represent news in this model, though, is problematic.  Inspiration comes from Cover and Thomas \cite[Ch.\ 6]{cover91} in which the optimal (defined as maximizing the expect growth rate of wealth) wagering strategy in a horse race (given fair odds) is to wager a fraction of one's capital on each horse equal to the probability of that horse winning---a fixed investment strategy, where the investment fraction is identified with a probability.  

In DSEM, this would have to be translated as the probability of the stock (or cash) ``winning,'' the interpretation of which is unclear.  (A loose interpretation might be that the stock wins if its value at some future horizon is greater than cash, otherwise cash wins.)

Regardless of what $i$ is a measure of, if we interpret $i$ as {\em some} probability measure then Bayesian probability theory offers an avenue for further development.  The evidence, as discussed in the last section, corresponding to the probability $i$ is
\widebox{\begin{equation}
	e = \log \frac{i}{1-i}.
\end{equation}}
As new information is acquired the evidence is updated via
\begin{equation}
	e' = e + \eta
\end{equation}
where $\eta$ represents the complicated second term of \eq{dsemBayesEvidence}.

This suggests modeling news as a stochastic process $\eta$ which affects an agent's confidence in the stock and, thereby, investment fraction.  (Modeling the news as $\eta$ directly, instead of through the information $D$ as presented in \eq{dsemBayesEvidence}, dramatically simplifies the calculations.)  A positive $\eta$ increases the evidence (and investment fraction), a negative value decreases it, while $\eta=0$ is neutral.  

Assuming news releases have a finite variance and are cumulative, the Central Limit Theorem indicates the appropriate choice is to model $\eta$ as Gaussian noise.  It should have a mean of zero $\expect{\eta}=0$ (unbiased) so that there is no long-term expected trend in investment (or price).  

The scale of the fluctuations $\sigma_\eta$, though, is difficult to decide; but this just opens up an opportunity to increase diversity amongst the agents---let each set its own scale.  We begin by setting an arbitrary scale $\sigma_\eta$ and then allowing agents to rescale it according to their own preferences.  Since each agent will apply its own scaling factor the universal scale $\sigma_\eta$ is arbitrary so it will be set to a convenient value later.

\subsubsection{News response}

First, we begin by defining individual scale factors.  Let us define a new agent-specific parameter $r_n$, which represents the agent's {\em responsiveness} to news.  Then evidence would be updated as
\widebox{\begin{equation}
\label{eq:dsemNewsResponse}
	e' = e + r_n \eta.
\end{equation}}

A responsiveness of zero would indicate the agent ignores news releases and maintains a constant investment ratio.  As the responsiveness increases the agent becomes increasingly sensitive to news and adjusts its ideal investment fraction more wildly, via
\begin{equation}
	i' = \frac{i \exp(r_n\eta) }{1 - i\left( 1-\exp(r_n\eta) \right) }.
\end{equation}

Notice there exists a symmetry between positive and negative news if the responsiveness also changes sign, $(-r_n)(-\eta) \equiv r_n \eta$, so we can impose the restriction $\expect{r_n}\geq 0$ (averaged over all agents) without loss of generality.

\subsubsection{News releases}

\label{sect:dsemNewsVariance}

Since DSEM is a discrete-event simulation the news must also be inserted as discrete events.  For reasons discussed in \ap{sampling} news events are modeled as a discrete Brownian process with some characteristic interval $\tau_n$ (unbiased Gaussian-distributed jumps occurring regularly at intervals of $\tau_n$).  Thus, on average, we expect $1/\tau_n$ news events each day.  If every news release has a variance of $\sigma_\eta^2$ then the variance of the cumulative news after one day is $\sigma_\eta^2/\tau_n$.  To minimize the impact of the news interval parameter the variance of the news over some fixed interval should be constant, independent of how often news is released.  Otherwise rescaling the news interval will rescale the evidences and hence impact upon the scale of the price series.  Let us take the variance to be one unit over one day, which is satisfied when
\begin{equation}
	\sigma_\eta^2 = \tau_n.
\end{equation}

To draw parallels with real markets the news release interval is chosen between $1/6.5\leq \tau_n \leq 5$ where the market is assumed to be open six and one half hours per day \cite{gopikrishnan99}.  Thus news releases occur at least once per week and at most once per hour.  A smaller interval is inappropriate because news is irrelevant until an investor is made aware of it---most investors (except professional traders) probably do not check for news more than once per hour.  (In fact, most people still get their news from the daily newspaper, suggesting $\tau_n=1$.)  On the other hand, a timescale longer than a week (5 days) is useless because we are particularly interested in fluctuations on the scale of hours to days, so the driving force should be on the same scale.

\subsubsection{Parameters}

In this section two new parameters related to news were presented: the agent-specific news response parameter $r_n$ which is allowed to be negative but is constrained such that $\expect{r_n}\geq 0$; and the global news interval parameter which is strictly positive and constrained to $\tau_n\in(1/6.5,5)$.

\subsection{Price response}

\label{sect:dsemPriceResponse}

DSEM with news-driven fluctuations is a complete model ready for experimentation.  However, it is lacking in that it neglects a significant source which real investors often construct their expectations from: the price history itself.  It is probable that feedback from the price series is integral to the clustered volatility and other complex phenomena found in empirical data.  
In CSEM this feedback was modeled by tracking the history of returns and forming future expectations therefrom, to set the investment fraction.  Since DSEM sets the investment fraction completely differently, this method is unavailable.  However, it is possible to construct a method which allows agents to extract information from the price series.

Consider how a single agent's ideal price is affected by news.  From \eq{dsemIdealPrice}, after a news release $\eta$ the ideal price becomes
\begin{eqnarray}
	{p^*}' & = & \frac{i'c}{(1-i')s} \\
	     & = & \exp(r_n \eta)p^*
\end{eqnarray}
which suggests that news is related to the logarithm of the price through
\begin{equation}
	\eta \propto \log \frac{p'}{p}.
\end{equation}

Therefore price movements imply news and if the price does change, an agent may infer that it missed some news which others are privy to.  Thus the price feedback may be inserted by extending the evidence dependence such that on a price move from $p$ to $p'$
\widebox{\begin{equation}
\label{eq:dsemPriceResponse}
	e' = e + r_p \log \frac{p'}{p}
\end{equation}}
where $r_p$, the {\em response to price}, is a new agent-specific parameter.  Setting $r_p=0$ eliminates the price feedback and reverts the model to being driven solely by news.  But with non-zero $r_p$ agents have a {\em chartist} nature: they presume the price series contains information (trends) and wager accordingly.  

Some market models separate agents into two groups: {\em fundamentalists} and {\em chartists} \cite{bak97, chen98, lux99}.  Fundamentalists (or ``rational'' traders) value the stock using fundamental properties such as dividends and reports of assets.  In the absence of dividends, this would correspond to responding strongly to news releases in DSEM.  Chartists (or ``noise'' traders) simply use the price history itself as an indicator of the stock's value.  This would correspond to a strong price response in DSEM.  However, in DSEM the two are not exclusive.  Instead of drawing a distinction between the two types of traders a continuum exists where $r_n$ and $r_p$ can take on a wide range of values so that agents may value the stock based on fundamentals {\em and} on its performance history.

Note that the price only changes when a trade occurs.  Only agents not involved in the trade (neither the caller nor a satisfied replier) should update their evidence since the traders can't be interpreted as having ``missed'' some information (because their trade {\em was} the information).  Further, allowing the caller and replier(s) to also update there evidence would cause complications because an agent could never reach equilibrium---a trade would bring them to their ideal investment fractions but then their evidences (and investment fractions) would be immediately changed.

The price response parameter is similar to an autocorrelation in returns.  Therefore we expect the dynamics to destabilize if we allow it to exceed unit magnitude.  As we will see in \ch{results}, imposing $-1<\expect{r_p}<1$ keeps the price from rapidly diverging.

In this section a new parameter, the price response $r_p$ which is constrained by $\abs{\expect{r_p}}<1$, was derived.

\subsection{Review}

The Decentralized Stock Exchange Model (DSEM) consists of a number $N$ of agents which trade with each other directly, without the intervention of a market maker.  In this section the structure of the model will be reviewed.

Game theory indicates the optimal strategy is to maintain a fixed fraction of one's capital in stock, the Fixed Investment Strategy (FIS).  Therefore the agents trade in order to rebalance their portfolios consisting of $c$ cash and $s$ shares.  News releases and price fluctuations cause agents to re-evaluate their investment fraction $i$ and up- or down-grade it as they see fit.  

\subsubsection{Algorithm}

Everything that happens in the model occurs as a discrete event in continuous time.  The basic algorithm follows:
\begin{enumerate}
	\item{Initialization.  Cash and shares distributed amongst agents.  Agents sample Poisson distribution to get waiting times until first call orders.  Waiting time for first news event is set to zero (first event).}
	\item{Next event is found (shortest waiting time).  Time is advanced to that of the event which is executed next.  If it is a news release proceed to Step \enum{dsemAlgNews}.  Otherwise it is a call order being placed, proceed to Step \enum{dsemAlgCall}.
\label{enum:dsemAlgNext}}
	\item{News event.  Sample Gaussian distribution to generate deviate for news.  Adjust all agents' ideal investment fractions.  Set news waiting time to $\tau_n$ and recalculate all agents' waiting times.  Return to Step \enum{dsemAlgNext}.
\label{enum:dsemAlgNews}}
	\item{Call order placed.
\label{enum:dsemAlgCall}
	\begin{enumerate}
		\item{Calculate all reply orders and corresponding waiting times.  Place compatible replies in queue, ordered by waiting times.}
		\item{If queue is empty, proceed to Step \enum{dsemAlgCallEnd}.  If queue is not empty, remove first reply order from queue and execute it.  
\label{enum:dsemAlgReply}}
		\item{Reduce outstanding call order by appropriate volume.  If not completely filled, return to Step \enum{dsemAlgReply}.}
		\item{Recalculate call-order waiting times for agents which traded.  If price did not change, return to Step \enum{dsemAlgNext}.
\label{enum:dsemAlgCallEnd}}
		\item{If price did change, recalculate ideal fractions for all other agents.  Recalculate call-order waiting times.  Return to Step \enum{dsemAlgNext}.}
	\end{enumerate}
}
\end{enumerate}

\subsubsection{Parameters}

\begin{table}
	\begin{center}\begin{tabular}{r|l|c}
		\hline \hline
		Symbol & Interpretation & Range \\
		\hline
		\multicolumn{3}{c}{Market parameters} \\
		\hline
		$N$ & number of agents & $2+$\\
		$C$ & total cash available \\
		$S$ & total shares available \\
		$\tau_n$ & average interval between news releases & $(1/6.5,5)$ \\
		\hline
		\multicolumn{3}{c}{Market state variables} \\
		\hline
		$p(t)$ & stock price at time $t$ \\ 
		$v(t)$ & trade volume (number of shares traded) at time $t$ \\
		\hline
		\multicolumn{3}{c}{Agent parameters} \\
		\hline
		$f_j$ & friction of agent $j$ & $(0,1)$ \\
		$r_{n,j}$ & news response of agent $j$ & $\expect{r_n}\geq 0$ \\
		$r_{p,j}$ & price response of agent $j$ & $\abs{\expect{r_p}}<1$ \\
		\hline
		\multicolumn{3}{c}{Agent state variables} \\
		\hline
		$c_j$ & cash held by agent $j$ \\
		$s_j$ & shares held by agent $j$ \\
		$w_j(p)$ & wealth of agent $j$ at stock price $p$ \\
		$i_j$ & optimum investment fraction of agent $j$ \\
		\hline \hline
	\end{tabular}\end{center}
	\caption{All parameters and variables used in the Decentralized Stock Exchange Model (DSEM).}
\label{tbl:dsemParameters}
\end{table}

For convenience all the variables used in DSEM are listed in \tbl{dsemParameters}.  The parameters are inputs for the simulation and the state variables characterize the state of the simulation at any time completely.  For each run, the agent-specific parameters are set randomly; they are uniformly distributed within some range (a subset of the ranges shown in the table).  Each dataset analyzed herein will be characterized by listing the market parameters and the ranges of agent parameters used.

\section{Implementation}

The Decentralized Stock Exchange Model (DSEM) is completely characterized by the above theory.  The model is beyond the scope of rigorous analysis in all but the most trivial of scenarios so it is simulated via computer.  The model was programmed in C++ using Borland C++Builder 1.0 on an Intel Pentium II computer running Microsoft Windows 98.  The source code and a pre-compiled executable are available for download from \url{http://rikblok.cjb.net/phd/dsem/}.

DSEM encounters some of the same issues as CSEM.  In particular, random numbers are handled using the same code as was discussed in \sect{csemRandomNumbers}.  Similarly, the random number seed will be specified with the other model parameters if the default (the current time) is not used.

\section{Parameter space exploration}

Having converted DSEM to computer code time series can be generated for numerical analysis.  Currently DSEM requires seven parameters (one fewer than CSEM started with) to fully describe it.  Again, it would be useful to try and reduce the parameter space before collecting any serious data.

\subsection{Number of agents $N$}

The effect of changing the number of traders will be explored in detail in \ch{results} and is left until then.

\subsection{Total cash $C$ and total shares $S$}

In this section the effect of rescaling the total cash $C$ and total shares $S$ will be explored.  Let us denote rescaled properties with a prime.  Then rescaling cash by a factor $A$ and shares by $B$ is written
\begin{eqnarray}
	C' & = & AC \\
	S' & = & BS.
\end{eqnarray}
Cash and shares are rescaled equally for each agent so the distribution remains constant.

We begin by noticing that the initial evidence is unchanged (being fixed at $e(0)=0$). The evidence depends on news releases (which are unaffected by rescaling) and price movements through \eq{dsemPriceResponse}.  Assuming price scales linearly ($p' \propto p$) the logarithm of the price ratios will be unchanged.  Thus, the evidence and the investment fraction will also be unchanged under rescaling.  

Assuming $i$ remains unscaled, an agent's ideal price scales as
\begin{equation}
	{p^*}' = \frac{ic'}{(1-i)s'} = \frac{A}{B}p^*
\end{equation}
and the buy and sell limits scale identically.  Since all agents ideal prices scale as $A/B$ it is reasonable to expect that the entire price series scales as
\begin{equation}
	p' = \frac{A}{B} p,
\end{equation}
as is required for the investment fraction to remain unchanged.  Thus, these two hypotheses are compatible and, through the initial conditions, are realized.

This argument took the same form as in \sect{csemCSscaling} for CSEM.  An identical argument for trade volume also applies giving the result that volume scales with shares as
\begin{equation}
	v' = B v.
\end{equation}

\begin{table}
	\begin{center}\begin{tabular}{c|c|c|c}
		\hline \hline
		Parameter & Run 1 & Run 2 & Run 3 \\
		\hline
		$N$ & 100 & 100 & 100 \\
		$C$ & \$1,000,000 & \$10,000,000 & \$1,000,000 \\
		$S$ & 1,000,000 & 10,000,000 & 10,000,000 \\
		$\tau_n$ & 1 & 1 & 1 \\
		$r_n$ & $0.025 \pm 0.025$ & $0.025 \pm 0.025$ & $0.025 \pm 0.025$ \\
		$r_p$ & $0.75 \pm 0.75$ & $0.75 \pm 0.75$ & $0.75 \pm 0.75$ \\
		$f$ & $0.05 \pm 0.05$ & $0.05 \pm 0.05$ & $0.05 \pm 0.05$ \\
		seed & -2 & -2 & -2 \\
		\hline \hline
	\end{tabular}\end{center}
	\caption{Parameter values for DSEM Runs 1, 2 and 3.}
\label{tbl:dsemRuns123}
\end{table}

\begin{figure}
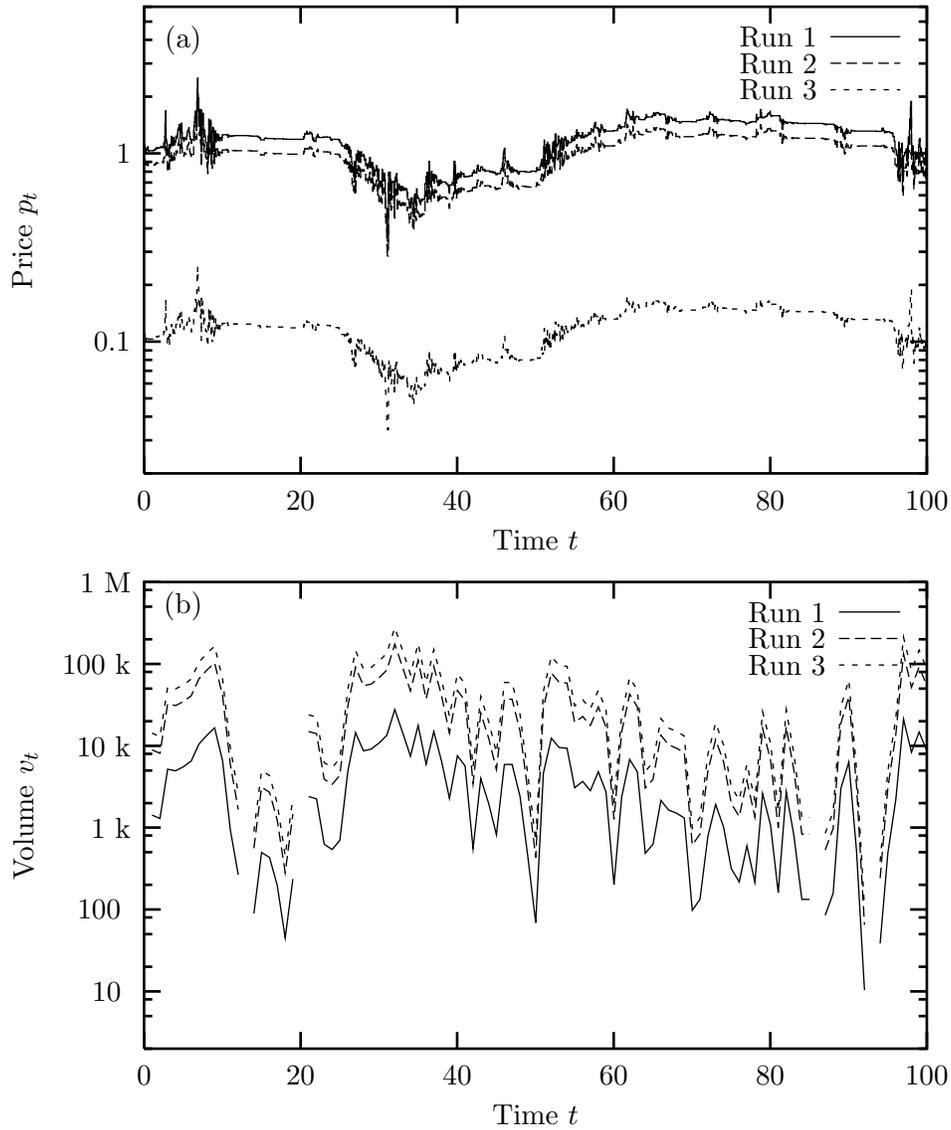

	\begin{center}
		\input{chModelDsem/runs123price.tex} \\
		\input{chModelDsem/runs123volume.tex}
	\end{center}
	\caption{Comparison of time evolutions of (a) price and (b) volume for Runs 1, 2 and 3 as defined in \tbl{dsemRuns123}.  The price scales as the ratio of cash to shares and the volume scales as the number of shares.  (Run 2 is offset to improve readability.  The gaps in (b) denote periods of zero volume.)}
\label{fig:dsemCashSharesScaling}
\end{figure}

To test these hypotheses three runs were performed, with the parameter values shown in \tbl{dsemRuns123}.  Contrasting Runs 2 and 3 with Run 1 give $A=10,B=10$ and $A=1,B=10$, respectively.  The results, shown in \fig{dsemCashSharesScaling}, confirm our hypothesis.

Thus, we again have a model where the roles of $C$ and $S$ are only to set the price and volume scales, which are irrelevant anyway.  Without loss of generality we can arbitrarily fix $C=\$1,000,000$ and $S=1,000,000$ thereby reducing the degrees of freedom by two.

\subsection{Further scaling}

Having fixed the total cash and shares it is possible to show that no further scaling arguments are possible---there is no way to transform the model parameters such that the dynamics are invariant.  We begin by noticing that, subject to $C$ and $S$ being fixed, each agent's cash $c$ and shares $s$ must also be fixed---constant under any transformation.

Under transformation a trade occurring between times $t_-$ and $t_+$
\begin{eqnarray}
	c(t_+) & = & c(t_-) - p\Delta s \\
	s(t_+) & = & c(t_-) + \Delta s
\end{eqnarray}
becomes
\begin{eqnarray}
	c(t_+)' & = & c(t_-) - p'\Delta s' \\
	s(t_+)' & = & c(t_-) + \Delta s'.
\end{eqnarray}
But requiring $c' = c$ and $s'=s$ for all times immediately imposes the restrictions $\Delta s' = \Delta s$ and
\begin{equation}
	p' = p.
\end{equation}
So price must also be a constant under the transformation.

In particular, an agent's ideal price, given by \eq{dsemIdealPrice}, and limit prices, given by \eqs{dsemLimitPriceBuy}{dsemLimitPriceSell}, must remain unscaled, which immediately implies $i'=i$ and 
\begin{equation}
	f'=f,
\end{equation}
so there is no way to rescale $f$ without impacting the dynamics.

The constraint that $i$ be invariant necessarily means that the evidence $e$ must also be.  After a number of news releases $\eta_j$ and price movements the evidence is
\begin{equation}
	e(t) = r_n \sum_j \eta_j + r_p \log \frac{p(t)}{p(0)}.
\end{equation}
Therefore $e$ is only invariant if the number of news releases is the same (requiring $\tau_n'=\tau_n$) and 
\begin{eqnarray}
	r_n' & = & r_n \\
	r_p' & = & r_p.
\end{eqnarray}

The conclusion which may be drawn is that there is no transformation of any model parameters which leaves the dynamics invariant.

\section{Parameter tuning}

In the absence of scaling arguments to reduce the parameter space further we must use tuning methods to choose appropriate parameter ranges.  (It is acknowledged this weakens the model's results somewhat, but it is necessary in order to establish a sufficiently small parameter space for experimentation.)

\subsection{News response}

\label{sect:dsemNewsResponse}

In this section we will explore the effect of varying the news response parameter $r_n$.  Let primes denote values after the arrival of a news event and unprimed quantities the same values before its arrival.  If we could neglect price response ($r_p=0$) then the price fluctuations would behave as
\begin{equation}
	\log \frac{{p^*}'}{p^*} = e'-e = r_n \eta
\end{equation}
where $\eta$ is the cumulative news in the interval.  So we would expect the price to have a log-Brownian motion with a standard deviation of $r_n\sqrt{t}$ (because $\eta$ has a variance $t$).

However, the agents' response to price movements clouds the picture somewhat.  Accounting for both news and price response, the evidence changes as
\begin{equation}
	e'-e = r_n \eta + r_p \log \frac{p'}{p}.
\end{equation}
If we assume, for simplicity, that the agent begins and ends at its ideal price ($p=p^*$ and $p'={p^*}'$) then the relationship becomes
\begin{equation}
\label{eq:dsemNewsPrice}
	\log \frac{{p^*}'}{p^*} = \frac{r_n}{1-r_p} \eta.
\end{equation}
Although the assumption is too restricting for the above equation to accurately describe price movements it at least sets a scale for the dependence of the price movements with respect to the model parameters.  

\begin{figure}
	\begin{center}
		\input{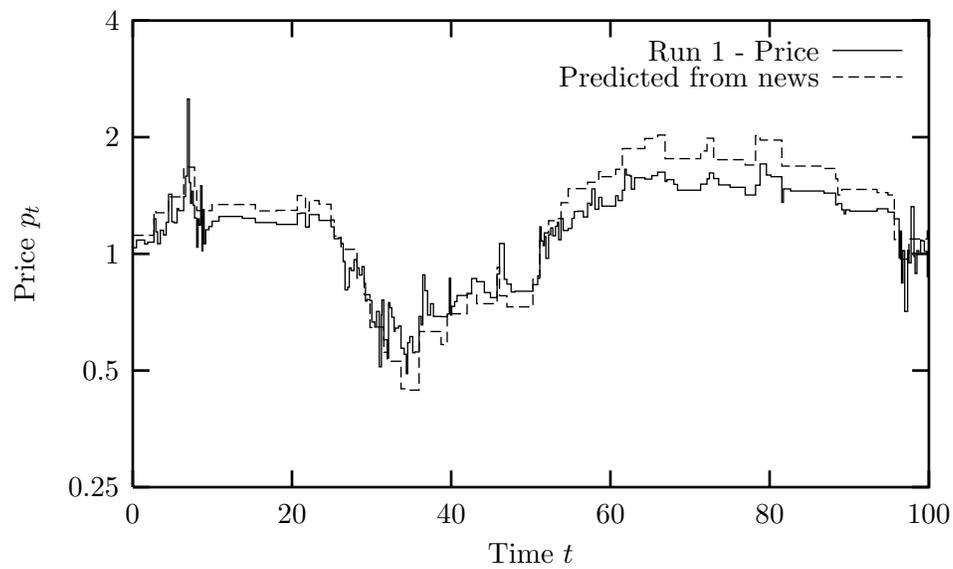}
	\end{center}
	\caption{The price series generated by Run 1 is compared with the expected price generated by \eq{dsemNewsPrice}, showing rough agreement (though with systematic deviations).}
\label{fig:dsemNewsPrice}
\end{figure}

To test \eq{dsemNewsPrice} the price series of Run 1 is reproduced in \fig{dsemNewsPrice} along with the expected price from the equation (initialized at $p(0)=1$).  The graph indicates a rough agreement between the series but with a systematic error for prices far from \$1, indicating the equation does not completely capture the dynamics.

Nevertheless, it is sufficient for the following purpose: it at least allows us to choose the scale of the news response $r_n$ such that the fluctuations are on roughly the same scale as observed in real markets.  If the log-return $r$ (not to be confused with responsiveness) over one day obeys
\begin{equation}
	r(t) \equiv \log \frac{p(t)}{p(t-1)} = \frac{r_n}{1-r_p} (\eta(t) - \eta(t-1))
\end{equation}
where $\eta(t)$ is the cumulative news up to time $t$, then the standard deviation $\sigma_r$ of the returns is
\begin{equation}
	\sigma_r = \frac{r_n}{1-r_p} \sqrt{\var{\eta}}
\end{equation}
where $\var{\eta}$ is the news variance over one day, which was set in \sect{dsemNewsVariance} to 1.  Rearranging the relationship, we can set a scale for the news response
\begin{equation}
	r_n = (1-r_p) \sigma_r.
\end{equation}

Daily returns for the New York Stock Exchange over 26 years covering the period 1962--1988 \cite{lebaron88} give $\sigma_r = 0.00959$ while the nine individual stocks studied in \ch{portfolio} are somewhat more variable with $\sigma_r=0.036\pm 0.014$ (see \tbl{portStats}).  A rough guideline, then, is $r_n = (1-r_p)\cdot 0.02$.  Arbitrarily taking $\expect{r_p}\approx 1/2$ suggests $\expect{r_n}=0.01$.  For all future simulations a range of $r_n=0.01 \pm 0.01$ will be used, unless otherwise stated.

\subsection{Friction}

In this section the effect of changing the friction parameter will be explored.  

One of the effects of changing the friction has already been presented in \sect{dsemTimescale} where it was found that, in order to preserve the number of trades per day, it was necessary to rescale time with \eq{dsemTimescale}.  This has an appealing interpretation: as friction (or cost per trade) increases the trade rate drops.

However, this effect is rather trivial and, by rescaling time via \eq{dsemTimescale}, it can be ignored.  Nevertheless, $f$ still influences the dynamics in subtle ways.

To choose the friction we again appeal to real market structure.  Most markets prefer trade quantities to be in {\em round lots} or multiples of one hundred shares.  So the minimum trade in DSEM should consist of one hundred shares.  Of course, the trade quantity depends on how many shares an agent holds and how the dynamics have unfolded.  But, at the very least, we can impose this condition initially (at $t=0$) because then we know each agent's portfolio and the market price precisely.  

Assuming the total cash $C$ and shares $S(=C)$ are distributed equally, each agent begins with an ideal investment fraction $i=1/2$, ideal price $p^*=1$, and limit prices $p_{BS}=(1+f)^{\pm 1}$.  A trade will be initiated at one of the limit prices so the quantity of shares to be traded, from \eq{dsemOptimalTrade}, is
\begin{eqnarray}
	s^* - s & = & \frac{S}{2N} \left( (1+f)^{\pm 1} - 1 \right) \\
	        & \approx & \pm \frac{Sf}{2N}.
\end{eqnarray}
Imposing the round lot restriction $\Delta s_{min} = 100$ sets the friction at
\begin{equation}
	f = 2\frac{\Delta s_{min} N}{S} = 200 \frac{N}{S}.
\end{equation}

From this argument it appears that the friction should increase with the number of agents $N$ in the model.  However, this is just an artifact of having fixed the total number of shares $S$.  As $N$ is increased each agent receives fewer shares since $S$ is constant so it must wait for a larger price move before trading, so that it can trade a full lot.  But fixing $S$ in this way is somewhat unnatural.  It is more natural to expect that if more agents are involved in a certain company then the company is probably larger and has more shares allocated.  So $S$ should probably have scaled with $N$.  

But the effect of scaling $S$ can be mimicked by scaling the round lot as $\Delta s_{min} \propto 1/N$.  Then the friction is a constant, regardless of $N$.  The most common system size to be used in this research will be $N=100$ so, given $S=1,000,000$, a good scale for the friction is $f=0.02$.  However, to incorporate heterogeneity future simulations will use $f=0.02 \pm 0.01$.

\subsection{News interval}

Previously it was argued that the average news release interval $\tau_n$ should be on the order of a day.  After more reflection the value of one day seems even more appropriate given the strong daily periodicity in real markets; daily market openings and closings, daily news sources (such as newspapers), and human behaviour patterns are just a few examples of daily cycles which influence market dynamics.  

Another advantage of using an interval of one day in DSEM is that it strengthens the connection with CSEM, in which all events occur simultaneously once per day.

For these reasons the news release interval will be fixed at $\tau_n=1$ (day).

\subsection{Finalized parameter ranges}

\begin{table}
	\begin{center}\begin{tabular}{r|l|c}
		\hline \hline
		Symbol & Interpretation & Range \\
		\hline
		\multicolumn{3}{c}{Market parameters} \\
		\hline
		$N$ & number of agents & $2+$\\
		$C$ & total cash available & \$1,000,000 \\
		$S$ & total shares available & 1,000,000 \\
		$\tau_n$ & average interval between news releases & $1$ \\
		\hline
		\multicolumn{3}{c}{Agent parameters} \\
		\hline
		$f_j$ & friction of agent $j$ & $0.02 \pm 0.01$ \\
		$r_{n,j}$ & news response of agent $j$ & $0.01 \pm 0.01$ \\
		$r_{p,j}$ & price response of agent $j$ & $\expect{r_p}\in(0,1)$ \\
		\hline \hline
	\end{tabular}\end{center}
	\caption{As \tbl{dsemParameters} except with updated parameter ranges.  These ranges will be used in subsequent simulations.  All parameters except $N$ and $r_p$ are firm.}
\label{tbl:dsemParameters2}
\end{table}

Via rescaling and tuning we have greatly narrowed the allowed ranges of the parameters.  The final ranges which will be used in all further simulations are shown in \tbl{dsemParameters2}.  All the parameters except the number of agents $N$ and the price response $r_p$ have well-defined values (or ranges, in the case of agent-specific parameters).  These two parameters will be the subject of further examination in the next chapter.







\chapter{Analysis and Results: Phase space}

\label{ch:results}

In the previous two chapters the Centralized and Decentralized Stock Exchange Models (CSEM and DSEM, respectively) were presented and in each case all but two parameters were fixed.  In this chapter the remaining parameter space will be investigated and it will be demonstrated that both models exhibit phase transitions for interesting values of these parameters.  We begin with CSEM.

\section{CSEM phase space}

\subsection{Review}

The Centralized Stock Exchange Model (CSEM), presented in \ch{csem}, consists of a number $N$ of agents which trade once daily with a centralized market maker.  The market maker chooses a trading price such that all orders are satisfied and the market clears (supply exactly balances demand).  The agents choose their orders based on a forecast of the daily return-on-investment which has a stochastic component modeled as a Gaussian deviate with standard deviation $\sigma_\epsilon$ (defined as the forecast error).  In \ch{csem} the model parameter space was reduced leaving only $N$ and $\sigma_\epsilon$ as free parameters.  In this section the remaining two-dimensional parameter space will be explored.

\subsection{Data collection}

\begin{table}
	\begin{center}\begin{tabular}{r|c|c|c|c|c}
		\hline \hline
		Parameters & \multicolumn{5}{c}{CSEM Dataset 1} \\
		\hline 
		\multicolumn{6}{c}{Particular values} \\
		\hline
		Number of agents $N$ & 50 & 100 & 200 & 500 & 1000 \\
		Investment limit $\delta$ & $10^{-3}$ & $10^{-3}$ & $10^{-3}$ & $10^{-3}$ & $10^{-3}$ \\
		Run length (time steps) & 10,000 & 10,000 & 20,000 & 20,000 & 30,000 \\
		Number of runs & 38 & 38 & 38 & 38 & 38 \\
		\hline 
		\multicolumn{6}{c}{Common values} \\
		\hline
		Forecast error $\sigma_\epsilon$ & \multicolumn{5}{c}{\begin{tabular}{c@{ to }c@{ by }c}
			0.01 & 0.25 & 0.01 \\ 
			0.26 & 0.50 & 0.02
		\end{tabular}} \\
		Total cash $C$ & \multicolumn{5}{c}{\$1,000,000} \\
		Total shares $S$ & \multicolumn{5}{c}{1,000,000} \\
		Memory $M$ & \multicolumn{5}{c}{$105 \pm 95$ (uniformly distributed)} \\
		Risk aversion $a$ & \multicolumn{5}{c}{$2 \pm 1$ (uniformly distributed)} \\
		Degree of fit $d$ & \multicolumn{5}{c}{0 (moving average)} \\
		seed & \multicolumn{5}{c}{random} \\
		\hline \hline
	\end{tabular}\end{center}
	\caption{Parameter values for CSEM Dataset 1.  Some of the parameters were established in \ch{csem} and are common to all the runs.  Dataset 1 explores two dimensions of phase space: $N$ and $\sigma_\epsilon$.}
\label{tbl:resultsCsemDataset1}
\end{table}

To explore the phase space thoroughly simulations were performed on systems of sizes $N$=50, 100, 200, 500, and 1000 with forecast errors in the range $\sigma_\epsilon\in[0.01,0.50]$ for each $N$, with increments of 0.01 up to $\sigma_\epsilon=0.25$ and increments of 0.02 thereafter, for a grand total of 190 experiments performed.  The complete list of parameter values used can be found in \tbl{resultsCsemDataset1}.

The choices of parameter values used (other than $N$ and $\sigma_\epsilon$) are justified in \ch{csem}.  To introduce heterogeneity amongst the agents some of the parameters, namely the memory $M$ and risk aversion $a$, were chosen randomly for each agent from the ranges indicated in the table (with the deviates uniformly distributed within the ranges).

Each run consisted of at least 10,000 time steps (days) and larger systems had longer runs to compensate for slower convergence to the steady state.  (With these run lengths the initial transient never accounted for more than one third of the total run).

\subsection{Phases}

\begin{figure}
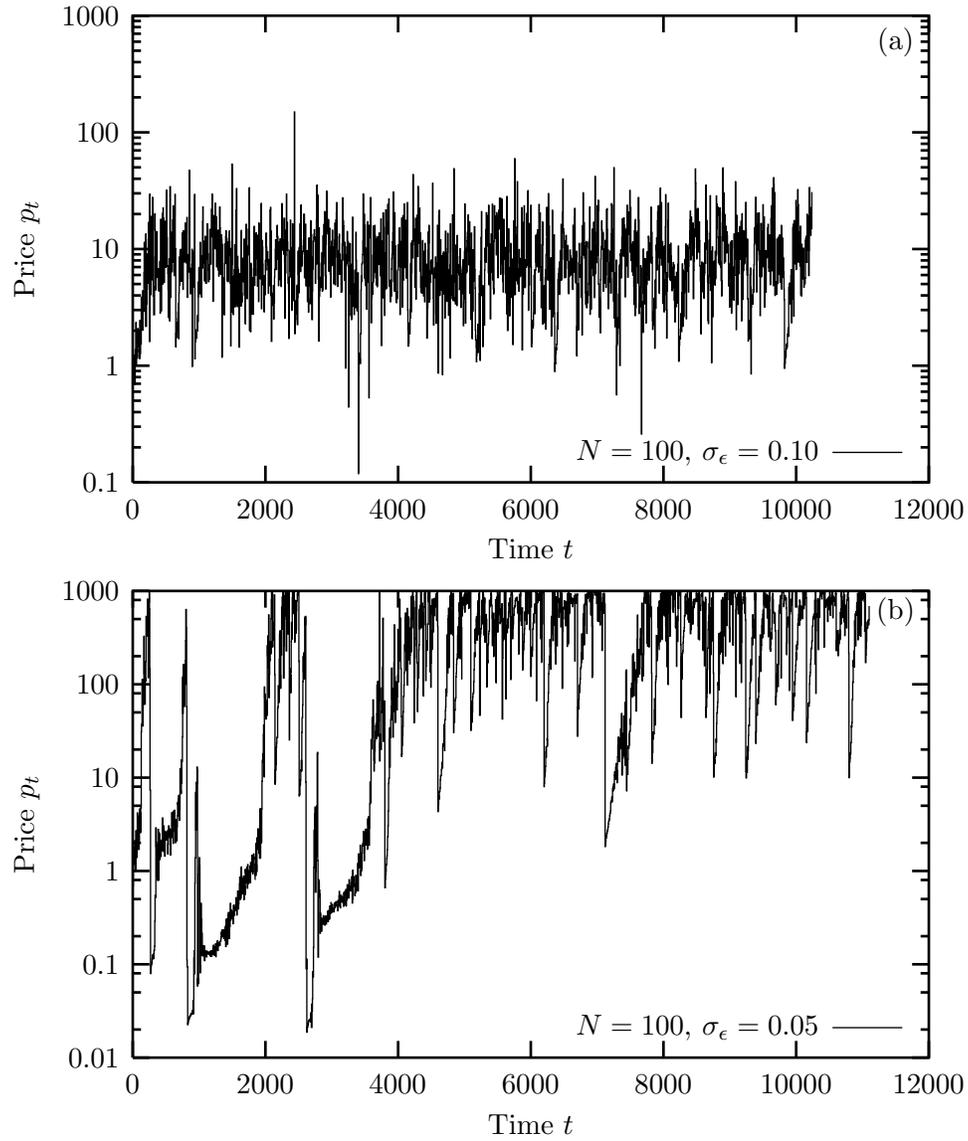
\centering
	\input{chResults/csemPriceN100s0.10.tex} \\
	\input{chResults/csemPriceN100s0.05.tex}
	\caption{The price series plots for CSEM with $N=100$ agents and $\sigma_\epsilon=0.10$ (a) and $\sigma_\epsilon=0.05$ (b) indicate a change of character of the dynamics.}
\label{fig:resultsCsemPhases}
\end{figure}

In most of the runs an initial transient period was observed before the price converged to a steady state value around which it fluctuated.  The only discrepancy was for small forecast errors where the price climbed quickly until it reached a maximum value which it often returned to.  Representative plots of these behaviours are shown in \fig{resultsCsemPhases}(a) and (b), respectively.

The transition between these two behaviours was observed for all system sizes near $\sigma_\epsilon \approx 0.08$ and is most dramatic when looking at the maximum price observed in a run.

\subsubsection{Maximum price}

\begin{figure}\centering
	\input{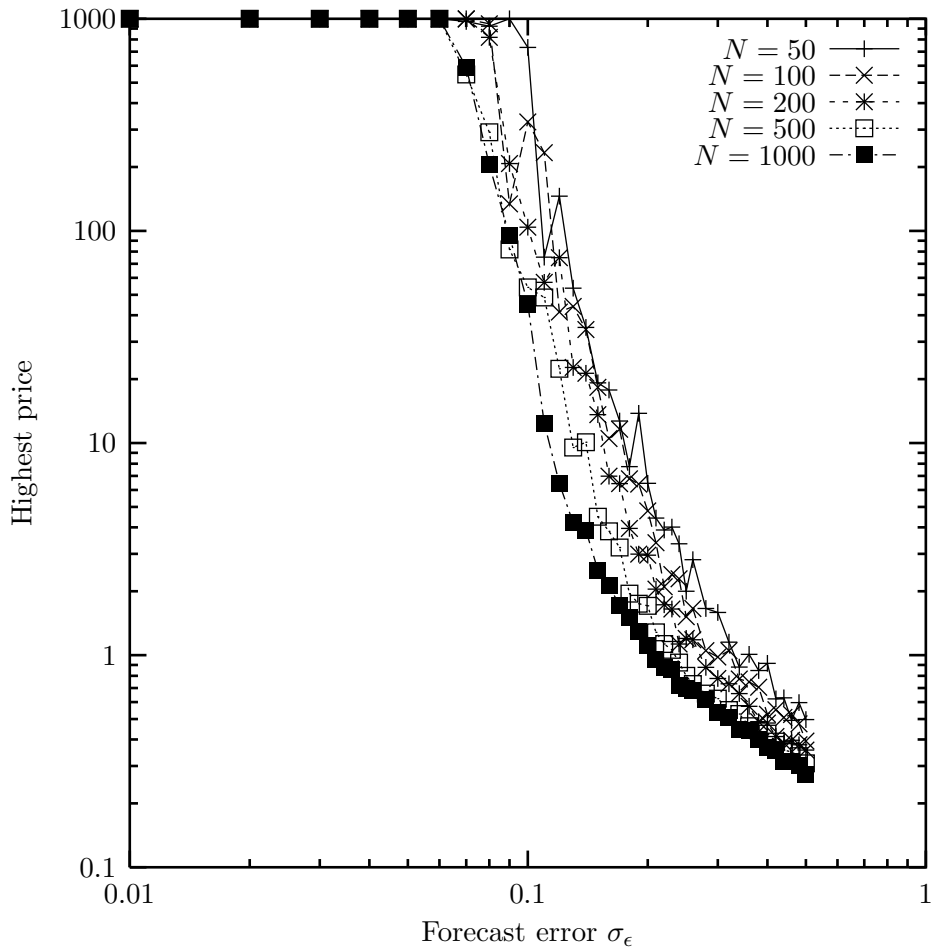}
	\caption{The highest price in any given simulation increases as the forecast error decreases until it reaches its theoretical limit, creating two separate phases for the dynamics.}
\label{fig:resultsCsemHighestPrice}
\end{figure}

As is demonstrated in \fig{resultsCsemPhases}(a) the price in each run converged to some steady-state value after some time and then appeared to randomly fluctuate around that value, never exceeding some maximum.  As mentioned above, the only exception was when the price reached a limit which interfered with its natural fluctuations.

The limit price is a consequence of the investment limit parameter $\delta$ introduced in \sect{csemInvestLimit} where it was noted that the price may not exceed the limit $p_{max}=(1-\delta)/\delta$ (\eq{csemMaxPrice}).

\fig{resultsCsemHighestPrice} clearly captures the distinct character of the dynamics on both sides of $\sigma_\epsilon \approx 0.08$.  For larger $\sigma_\epsilon$ the price fluctuates freely while for smaller values the limit has a strong influence on the dynamics.

In the limit $\delta\rightarrow 0$ (which is disallowed because it can occasionally generate singularities in the price series) it appears that the maximum price would diverge at $\sigma_\epsilon=0$ producing a phase transition.  Although this cannot be tested by directly setting $\delta=0$ the limit can be explored by studying smaller values of $\delta$.

\subsection{Investment limit}

\begin{table}
	\begin{center}\begin{tabular}{r|c|c|c}
		\hline \hline
		Parameters & \multicolumn{3}{c}{CSEM Dataset 2} \\
		\hline 
		Number of agents $N$ & 100 & 100 & 100 \\
		Investment limit $\delta$ & $10^{-2}$ & $10^{-4}$ & $10^{-5}$ \\
		Number of runs & 38 & 38 & 38 \\
		Run length (time steps) & 10,000 & 10,000 & 10,000 \\
		\hline \hline
	\end{tabular}\end{center}
	\caption{Parameter values for CSEM Dataset 2.  These runs are a variation of Dataset 1 (all unspecified parameters are duplicated from
 \tbl{resultsCsemDataset1}, $N=100$) exploring a range of investment limits $\delta$.}
\label{tbl:resultsCsemDataset2}
\end{table}

A subset of Dataset 1 with $N$=100 agents was simulated again, but this time with $\delta=10^{-2}$, $10^{-4}$ and $10^{-5}$ as shown in \tbl{resultsCsemDataset2}.  It was suspected that reducing $\delta$ would increase the limit price and thereby allow the price to fluctuate freely for smaller values of $\sigma_\epsilon$, reducing the domain of the second phase.

\begin{figure}\centering
	\input{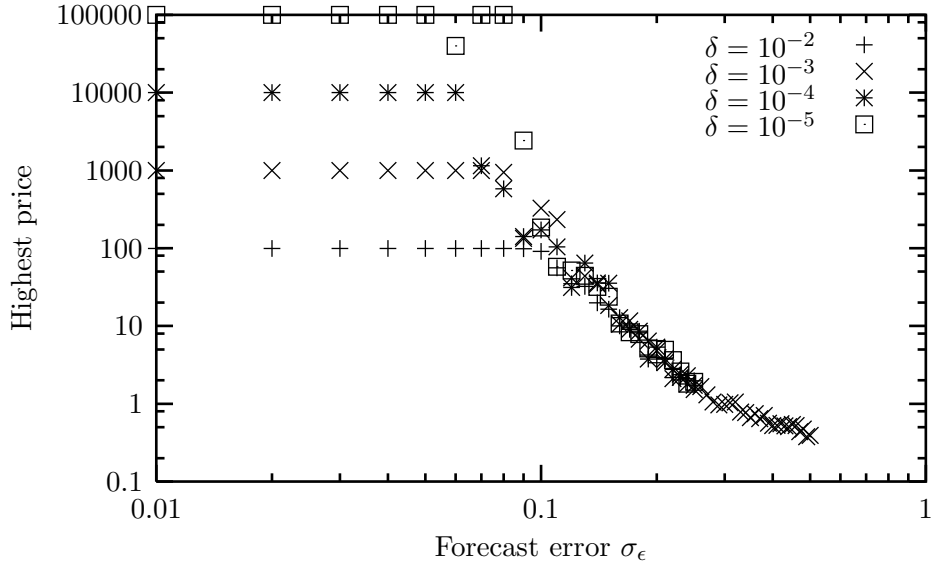}
	\caption{The maximum price in CSEM has a limit which depends on the investment limit $\delta$.  However, the threshold value of $\sigma_\epsilon$ for which the limit is first reached does not appear to depend on $\delta$.}
\label{fig:resultsCsemPriceDelta}
\end{figure}

\begin{table}
	\begin{center}\begin{tabular}{c|c}
		\hline \hline
		$\delta$ & Threshold $\sigma_\epsilon$ \\
		\hline 
		$10^{-2}$ & 0.08 \\
		$10^{-3}$ & 0.06 \\
		$10^{-4}$ & 0.06 \\
		$10^{-5}$ & 0.08 \\
		\hline \hline
	\end{tabular}\end{center}
	\caption{The threshold values of $\sigma_\epsilon$ separating the two phases of CSEM shown in \fig{resultsCsemPriceDelta} do not appear to depend on the investment limit $\delta$.}
\label{tbl:resultsCsemDeltaSigma}
\end{table}

However, as \fig{resultsCsemPriceDelta} demonstrates the threshold values of $\sigma_\epsilon$ (see \tbl{resultsCsemDeltaSigma}) for which the price first reaches its limit remains constant, even though the price limit increases.  This suggests that there exists a critical forecast error $\sigma_c>0$ at which the maximum price diverges.  Even though the critical point is only strictly defined in the limit of $\delta\rightarrow 0$  the term will also be used here to refer to systems with nonzero values of $\delta$.

\subsection{Critical regime}

Critical points are heralded by power law relationships of the form $f(x) \propto (x-x_c)^z$ where $x_c$ is the critical point and $z$ is known as the {\em critical exponent}.  (For a thorough explanation of critical phenomena see Ref.\ \cite{plischke94}.)  Many different quantities can play the role of the critical variable $f$. In a thermodynamic system it could be an order parameter such as the magnetization of a ferromagnet.  Alternatively, $f$ can be a response function such as
the susceptibility or the specific heat or it could be a correlation time or time for thermalization. The control parameter $x$ could be the temperature or an external field.  In the case of CSEM we will continue to use the maximum price as the order parameter and $\sigma_\epsilon$ as the control parameter.

To specify the transition in more detail it would be helpful to estimate the critical point $\sigma_c$ and the exponent from the data.  We begin by reconsidering the data from \fig{resultsCsemPriceDelta} and fitting it to a power law
\begin{equation}
\label{eq:resultsCsemCritical}
	p_{max} = C (\sigma_\epsilon - \sigma_c)^{-b}
\end{equation}
to estimate $\sigma_c$ and $b$ ($C$ is unimportant).  

The fitting algorithm used is a Levenberg-Marquardt nonlinear routine \cite[Section 15.5]{press92} and the fit is performed over the range $\sigma_\epsilon\geq 0.14$.  (Choosing a range which is too near the actual critical point tends to reduce the quality of the fit because critical points tend to be ``blurred'' on finite systems.)  The fitting algorithm attempts to minimize the sum-of-squares error between the curve and the data but it can get stuck in suboptimal solutions which depend on the initial parameter choices when performing the fit.  For these fits the parameters were initially set to $C=\exp(-3)$, $\sigma_c=0.08$, and $b=2$ because these values were observed to fit the data reasonably well.  (Though setting $\sigma_c=0.01$ initially, the fit still converged to the same solution.)

\begin{figure}
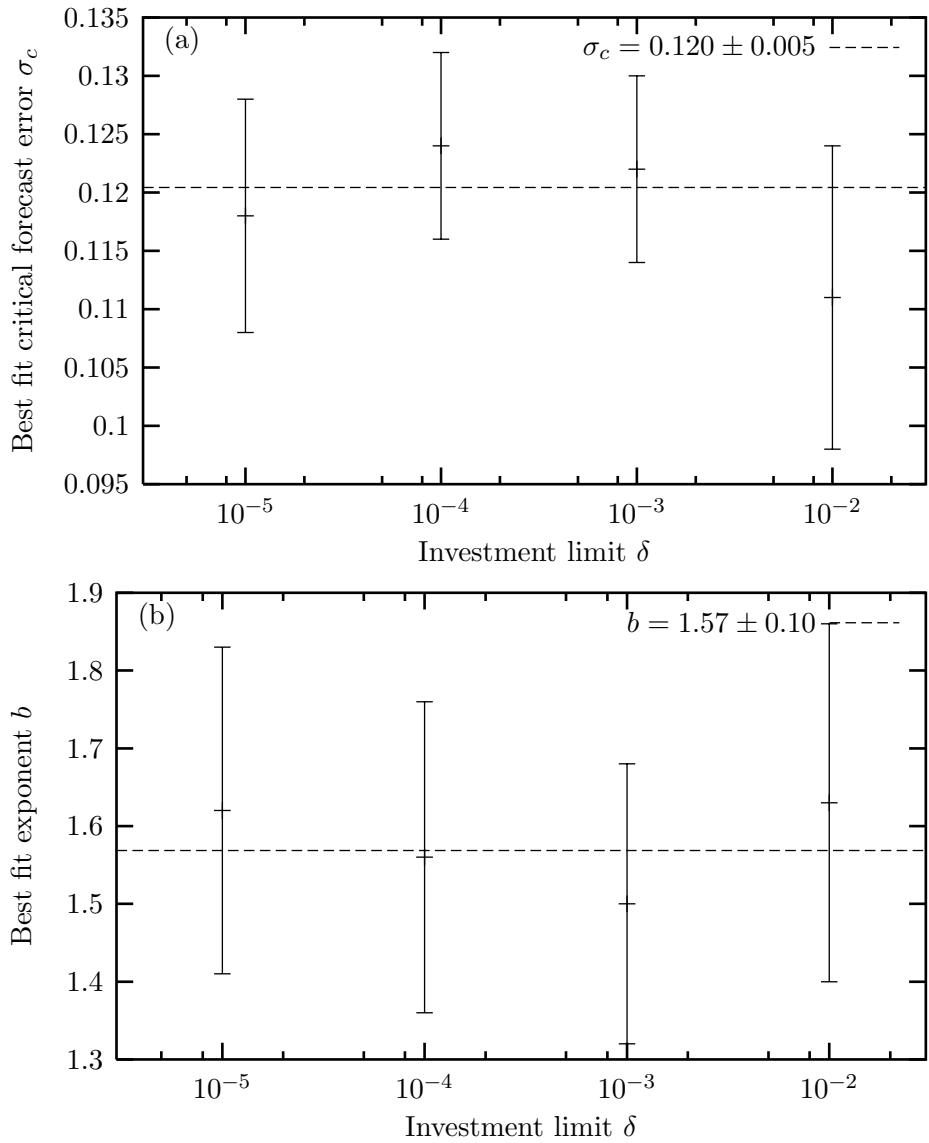
\centering
	\input{chResults/csemPowerDeltaSigma} \\
	\input{chResults/csemPowerDeltaExp}
	\caption{The best fits of power laws to CSEM Dataset 2 (and $N=100$ from Dataset 1) yield the critical points (a) and scaling exponents (b) shown.  The lines represent the weighted averages of the best fit values.}
\label{fig:resultsCsemPowerDelta}
\end{figure}

The resultant fits for each value of $\delta$ in Dataset 2 (all with $N=100$) and for $N=100$ in Dataset 1 give the critical points and exponents shown in \fig{resultsCsemPowerDelta}.  The weighted average of the exponents is $b=1.57\pm 0.10$ and the mean critical forecast error is $\sigma_c=0.120\pm 0.005$.  The fact that these values are similar for all values of $\delta$ tested strengthens the conclusion that $\sigma_c$ is a critical point.

Notice that the calculated value of $\sigma_c$ is significantly higher than the 0.08 originally hypothesized.  This is a common feature of experiments involving critical phenomena and is due to the finite size of the system under investigation.  A true critical or {\em second-order} phase transition is characterized by a discontinuity in the derivative of the order parameter.  In finite systems the discontinuity is smeared out and becomes more refined with larger systems.  In this case the smearing resulted in an inaccurate first guess of the critical point.  After exploring some alternative choices for the order parameter we will consider finite size effects in more detail.

\subsection{Alternative thermodynamic variables}

In the last section the maximum price over any run was chosen as the thermodynamic property whereby the phase transition was detected.  In this section we demonstrate that a number of alternative variables would be equally suitable.

In particular, we consider two alternatives: the (logarithmic) average of the price series
\begin{equation}
	\bar{p}  \equiv \exp \expect{\log p} 
\end{equation}
and the (logarithmic) variance around the average
\begin{equation}
	\sigma_p^2 \equiv \var{\log p}.
\end{equation}
The variance $\sigma_p^2$ measures the scale of the fluctuations and is analogous to magnetic susceptibility in non-equilibrium systems.

Notice a phase transition in the average price $\bar{p}$ is equivalent to one in the (time-averaged) wealth per agent $\bar{w}$ because they are related by
\begin{equation}
	N\bar{w} = C + \bar{p} S
\end{equation}
where $C$ and $S$ are the total amount of cash and shares, respectively.

\begin{figure}
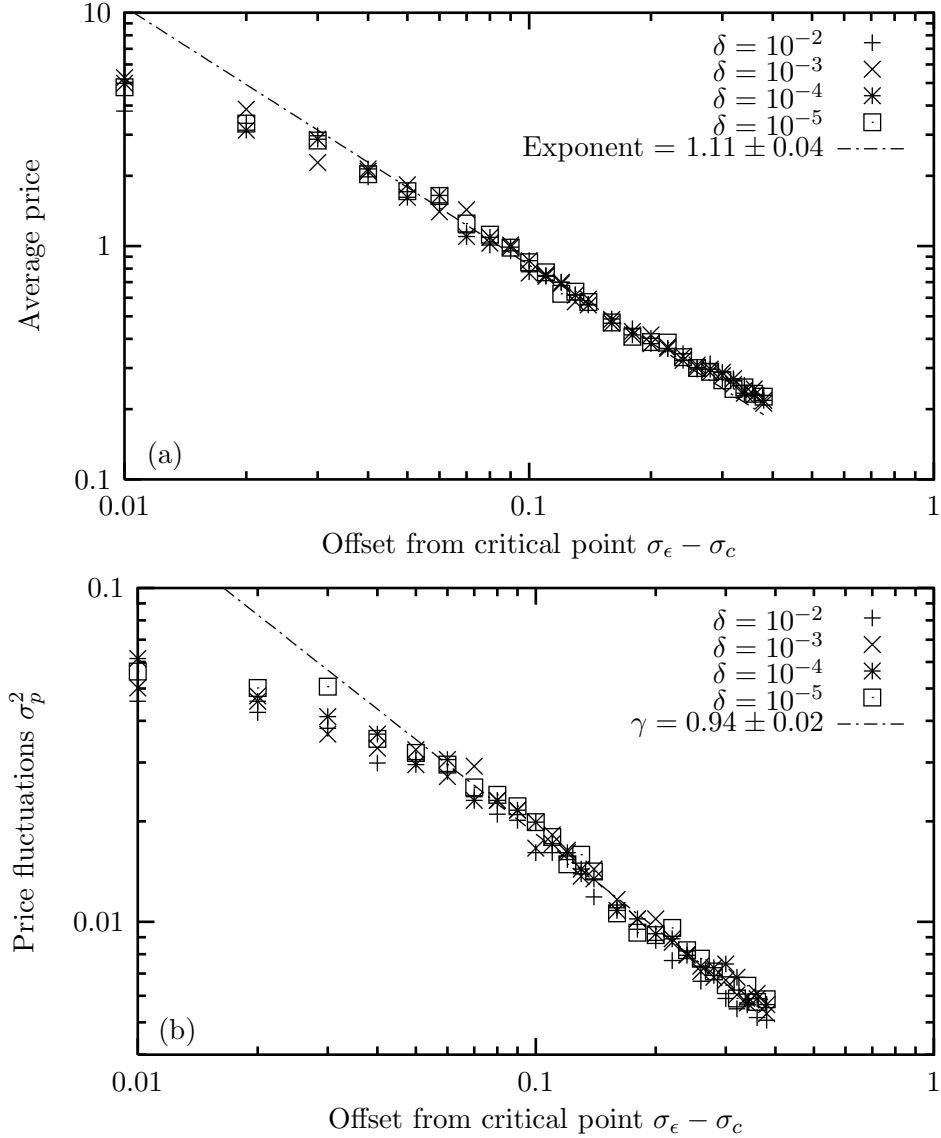
\centering
	\input{chResults/csemAvgPrice.tex}
	\input{chResults/csemPriceFluct.tex}
	\caption{The average price (a) and variance of fluctuations (b) also exhibit scaling near the critical point $\sigma_c=0.12$ for the data from CSEM Dataset 2.  The deviation from scaling observed near the critical point in (b) is due to the finite size of the system ($N=100$) as will be seen in \fig{resultsCsemPriceFluctN}.}
\label{fig:resultsCsemOrderParams}
\end{figure}

\fig{resultsCsemOrderParams} demonstrates that both these properties exhibit scaling, diverging at the critical point $\sigma_c=0.12$ (from \fig{resultsCsemPowerDelta}(a)).  The critical exponent for the average price power law is $1.11\pm 0.04$ and the exponent for the fluctuations is $0.94\pm 0.02$.  Clearly these properties would be equally suitable to determine the phase transition, but the maximum price has the advantage that the transition becomes very clear because it is a constant to the left of the critical point, being bounded by $\delta$.

As mentioned before, the deviation from scaling observed in the variance plot is due to finite size effects which we explore next.

\subsection{Finite size effects}

\label{sect:resultsCsemFiniteSize}

\begin{figure}
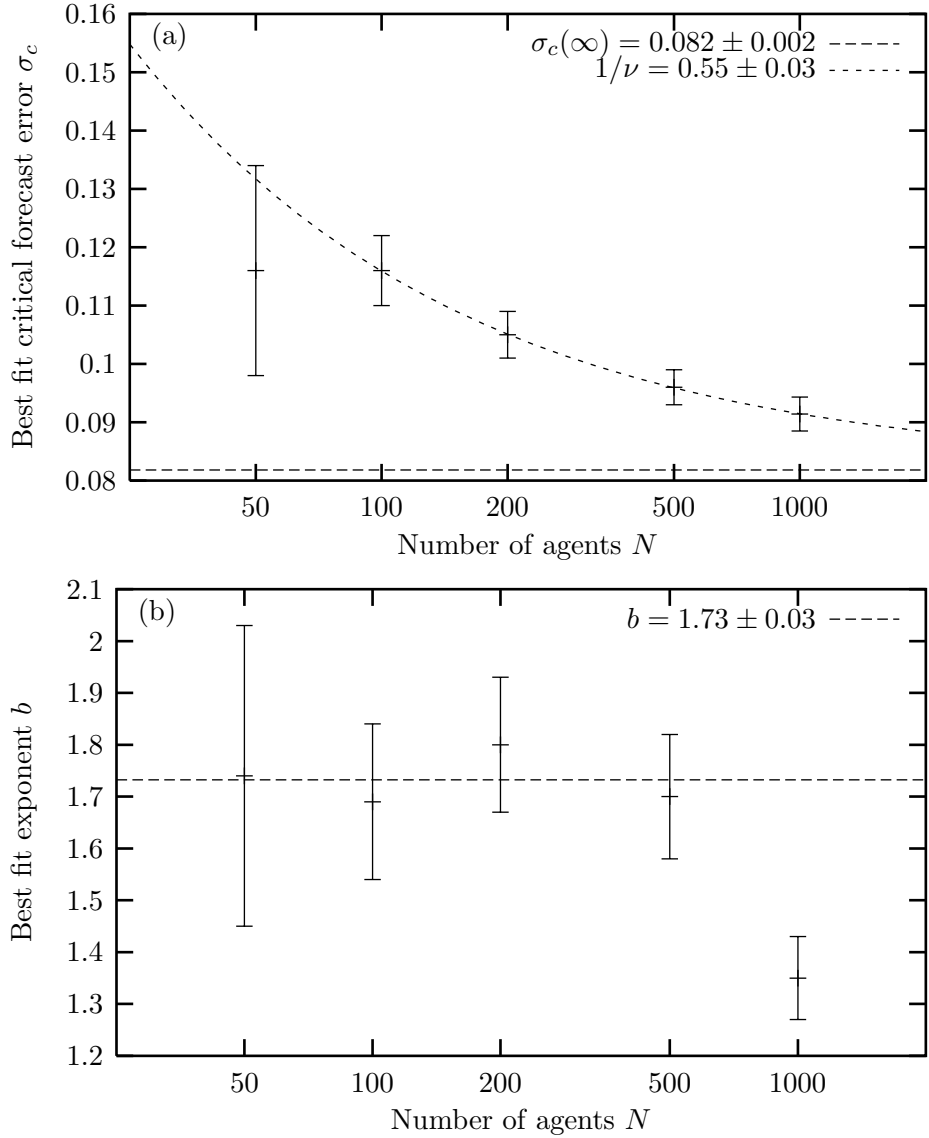
\centering
	\input{chResults/csemPowerNSigma} \\
	\input{chResults/csemPowerNExp}
	\caption{The best fits of power laws to CSEM Dataset 1 yield the critical points (a) and scaling exponents (b) shown.  A finite-size scaling analysis (neglecting $N=50$) reveals information on how the critical point changes with increasing investor numbers (a).  For reasons discussed in the text, the exponent for $N=1000$ is dropped from the estimate of the scaling exponent (b).}
\label{fig:resultsCsemPowerN}
\end{figure}

Now we return to the maximum price data in \fig{resultsCsemHighestPrice} and fit it to a power law using the technique described before.  The resultant estimates of the critical point are shown in \fig{resultsCsemPowerN}(a) which demonstrates that as the system size $N$ increases the critical point decreases systematically (neglecting the smallest system which is plagued by noise).  To derive the relationship between the system size and the associated critical point we need to understand the role of correlations.

\subsubsection{Correlations}

Near a critical point the dynamics are dominated by correlations between elements of the system (agents, in our case).  The degree to which the elements are correlated is measured by the correlation length $\xi$ which, in CSEM, counts the typical number of agents affected by any single agent's decision.  Far away from the critical point we don't expect one agent's decisions to affect (many) other agents so the correlation length is short.  But near the critical point the correlation length diverges as \cite{plischke94}
\begin{equation}
	\xi(\sigma_\epsilon - \sigma_c) \propto (\sigma_\epsilon - \sigma_c)^{-\nu}.
\end{equation}

When dealing with a finite system the correlation length is attenuated by the size of the system $N$.  It should reach a maximum at the critical point for that particular system size, denoted by $\sigma_c(N)$, and we expect the maximum to grow linearly with $N$,
\begin{equation}
	\xi(\sigma_c(N) - \sigma_c) \propto N.
\end{equation}
As a consequence of these two equations we expect the finite-size critical point to converge to the thermodynamic critical point as
\begin{equation}
	\sigma_c(N) - \sigma_c \propto N^{-1/\nu}.
\end{equation}

Applying this relationship to the data in \fig{resultsCsemPowerN}(a) gives a finite-size scaling exponent $1/\nu=0.55\pm 0.03$ which means that the correlation length grows as $\xi \sim (\sigma_\epsilon - \sigma_c)^{-\nu}$ with $\nu=1.82\pm 0.10$.  It also allows a more precise estimate of the (limit) critical point, giving $\sigma_c = 0.082\pm 0.002$.

\subsubsection{Scaling exponent}

The best estimates of the critical exponent $b$ from \eq{resultsCsemCritical} are shown in \fig{resultsCsemPowerN}(b), giving an average value of $b=1.73\pm 0.03$.  Notice the largest system size gives a markedly different result and is not used to compute the average $b$.  This is a consequence of the range of forecast errors over which scaling applies:

Notice that the tails of the highest prices ($p_{max}<1$) in \fig{resultsCsemHighestPrice} appear slightly ``flatter'' than the rest of the data.  In fact the tails fit quite well to the scaling relation $p_{max} \propto \sigma_\epsilon^{-1}$ which is probably directly related to the inverse relationship derived in \sect{csemInitialPrice}.  This inverse power law can obscure the critical scaling so the range over which the scaling was tested for was constricted to $p_{max}\geq 1$.  

Unfortunately, since the run for $N=1000$ had the lowest observed prices at any given value of $\sigma_\epsilon$ this restriction severely limited the available data and compromised the fit.  It appears that the tail is artificially drawing the estimate of $b$ down for this system size, so it was not included in the computation of $b\approx 1.73$.

\subsubsection{Fluctuations}

\begin{figure}\centering
	\input{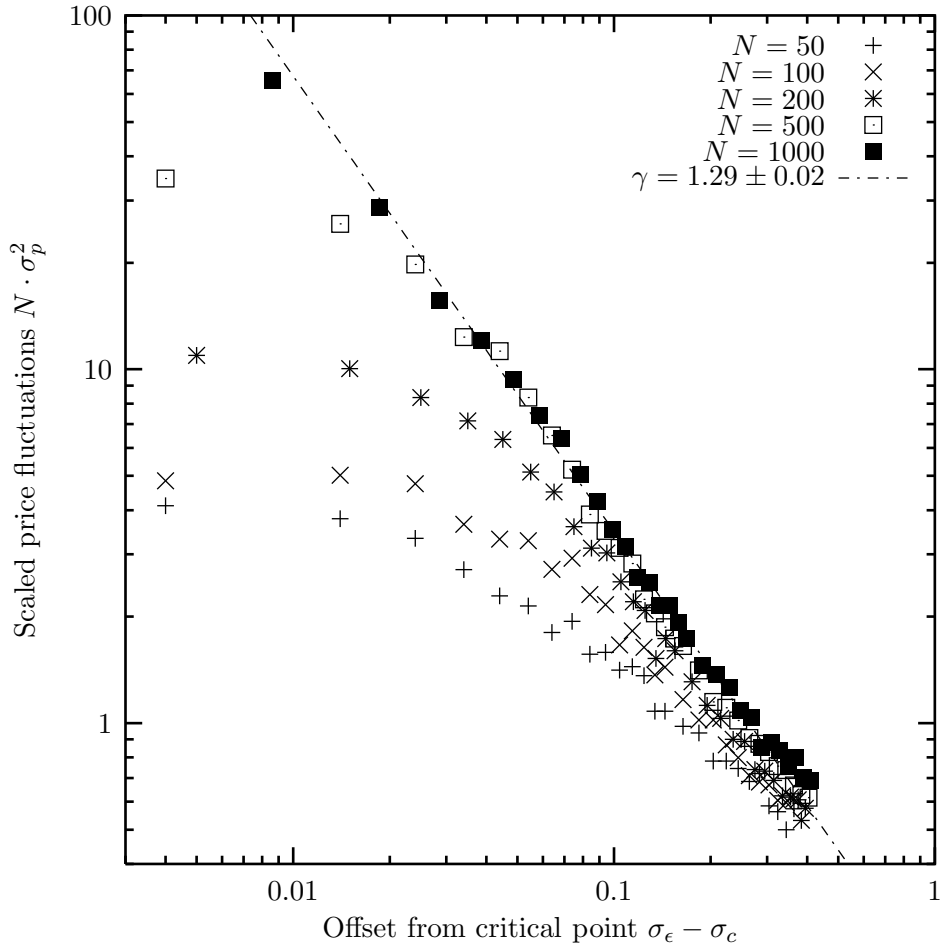}
	\caption{The variance of the log-price largely collapses to a single curve when multiplied by the system size $N$ for CSEM Dataset 1.  This curve diverges as the critical point is approached with an exponent $\gamma=1.29\pm 0.02$ calculated from the largest system $N=1000$.  (The critical points were taken from \fig{resultsCsemPowerN}(a).)}
\label{fig:resultsCsemPriceFluctN}
\end{figure}

In this section we briefly revisit the scaling seen in the fluctuation data (\fig{resultsCsemOrderParams}(b)) to demonstrate the round-off seen near the transition is a finite-size effect.

Since the fluctuations in the log-price series are due to stochasticity in each of the $N$ agent's trading decisions it is reasonable to expect the variance of the log-price to scale with system size as $\sigma_p^2(N) \propto 1/N$ so that $N \sigma_p^2$ should be independent of system size.  For the most part \fig{resultsCsemPriceFluctN} confirms this hypothesis with some deviations near the critical point.  Notice these deviations diminish with larger system sizes so these deviations are just finite-size effect---a ``blurring'' of the phase transition for small system sizes.

The best estimate of the scaling exponent, taken from the largest system is $\gamma=1.29\pm 0.02$.

\subsubsection{Universality class}

The main advantage of characterizing the critical point precisely is that the critical exponents $b$, $\gamma$, and $\nu$ may tell us to which {\em universality class} the critical point belongs.  At a critical point many of the particular details of a system become irrelevant and the scaling properties depend only on a few basic quantities, such as the dimensionality and symmetry of the system \cite{plischke94}.  As such, many disparate systems are observed to behave in the same way at a critical point and may be classified by their common exponents.  

Discovering which universality class a system belongs to leads to further understanding of the important features of the system.  For instance, the author has recently been involved in research into the ``game of Life'' (GL), a toy model of interactions between spatially-distributed individuals.  GL lies close to a critical point in the same universality class as directed percolation, a model of the spreading of a cluster through its nearest neighbours \cite{blok99}.  Making this connection teaches us that the important factor determining the dynamics at the critical point near GL is simply the probability of a disturbance spreading to its neighbours, not the particular details of GL.

I do not know which universality class CSEM belongs to, but by computing the exponents it is my hope that a reader will recognize them and classify the model.  The exponent $b$ probably is unrelated to physical systems but the variance of the fluctuations which gave $\gamma\approx 1.29$ is analogous to magnetic susceptibility.  The exponent for the correlation length $\nu\approx 1.8$ may also be relevant even though the model is {\em mean field} (each agent interacts with all other agents through the market maker).

\subsection{Transient}

In this section we explore the critical phase transition discovered above from a different perspective; namely, that of the transient period.  A close inspection of \fig{resultsCsemPhases} will reveal that the price series had an initial transitory phase before it settled down to its steady-state dynamics.  In this phase the memory of the initial conditions is slowly erased.

The transient can be systematically quantified by recognizing that the price series eventually converged to some steady-state value.  Then the transient is formally defined as the period until the price first crosses its (logarithmic) average for the entire series.

The above definition is satisfactory provided that the steady state price is far from the initial price (on the order of \$1, see \fig{csemInitialPrice}) but becomes irrelevant for large $\sigma_\epsilon$ when the steady-state price is on the same scale as the initial price (see \fig{resultsCsemHighestPrice}).

\begin{figure}\centering
	\input{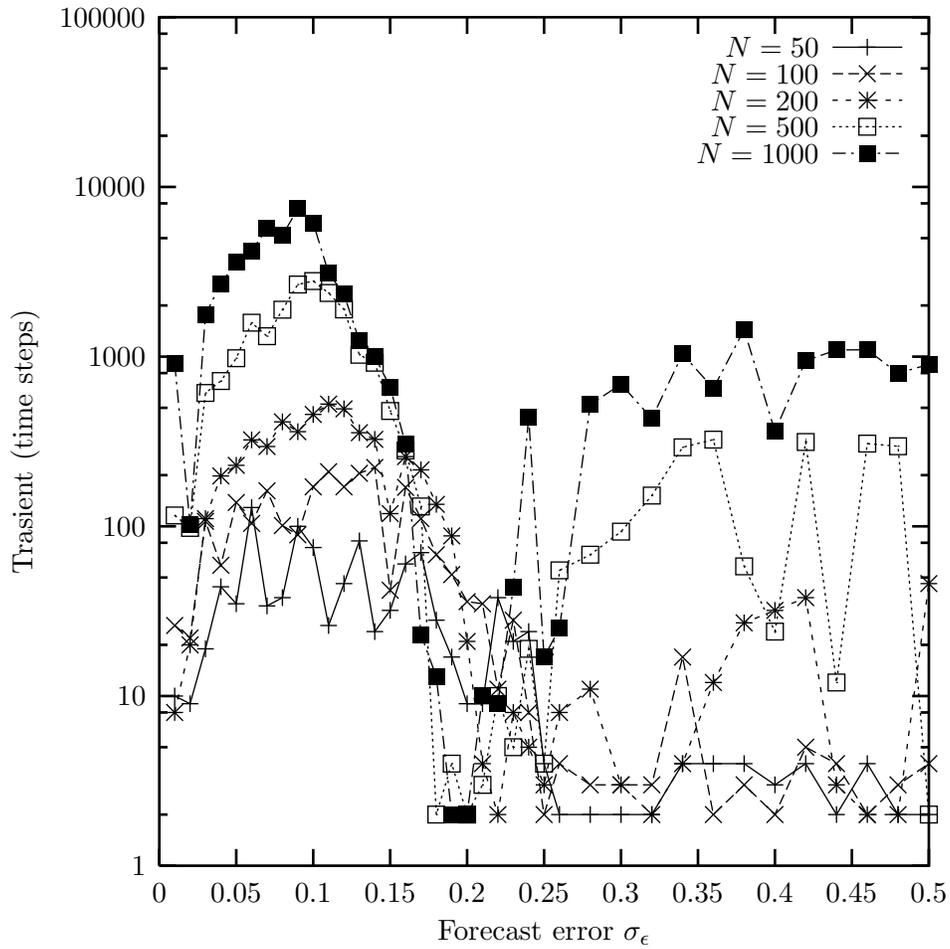}
	\caption{Duration of the transient period in CSEM (Dataset 1) before the price series settles down to some steady-state value.  The transient grows near the critical point $\sigma_c\approx 0.08$.}
\label{fig:resultsCsemTransients}
\end{figure}

A plot of the transient duration of each run in Dataset 1, shown in \fig{resultsCsemTransients}, exhibits some interesting properties.  For large forecast errors $\sigma_\epsilon\gg \sigma_c$ the transient measure is quite unstable---large for some runs and small for others---with some interesting system size dependence but these properties will not be analyzed further because the transient is a poor measure in this region.  

For small $\sigma_\epsilon$ another interesting pattern is observed: the transient appears to grow near the critical point declining away from it, on both sides.  This behaviour can be explained if the system does exhibit a second-order phase transition at $\sigma_c$.  Criticality arises from correlations between agents which extend further and further as the system approaches the critical point.  At the critical point the correlations span the entire system such that a perturbation in any element can have a cascade effect which may impact on any or all other elements.  However, the correlations are an emergent phenomena in critical systems and require time to set up---the initial transient period.  Away from the critical point the correlations do not span the entire system so they require less time to set up and, correspondingly, the transient is shorter.

\begin{figure}\centering
	\input{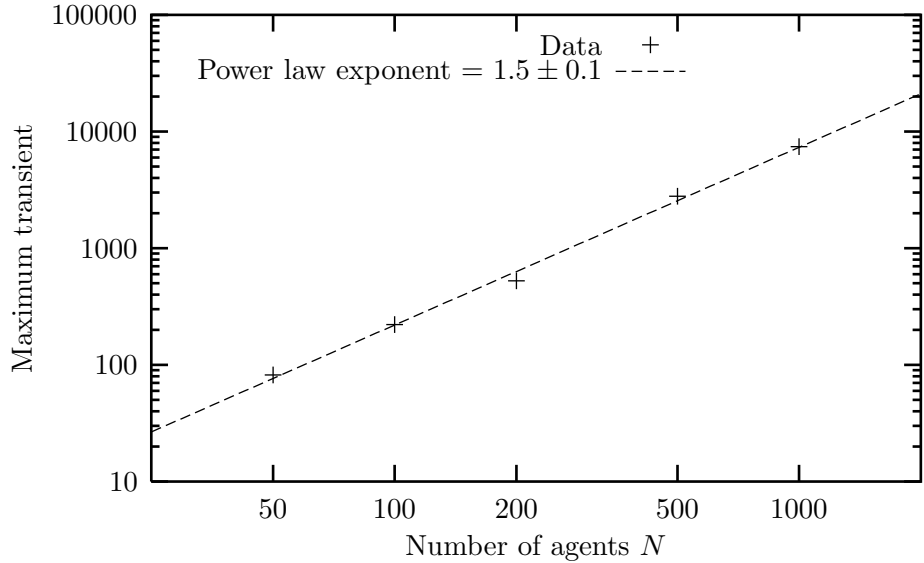}
	\caption{The maximum transient in CSEM appears to scale with the system size with an exponent $1.5\pm 0.1$.}
\label{fig:resultsCsemMaxTransN}
\end{figure}

Notice the maximum transient grows with the system size reflecting the longer time required for the correlations to span the system.  Critical theory predicts the maximum transient should scale with the system size and \fig{resultsCsemMaxTransN} confirms it.  The associated scaling exponent is estimated to be 
$1.5\pm 0.1$.  Since this exponent is greater than one the duration of the transient grows faster than $N$ as the system size is increased---a typical property of criticality, known as {\em critical slowing down}.

\subsection{Summary}

In this section the phase space of the Centralized Stock Exchange Model (CSEM) was explored in detail.  The main discovery was of a critical value of the forecast error $\sigma_c=0.082\pm 0.002$ above which the dynamics are relatively stable and below which the price fluctuates over many orders of magnitude, up to the maximum imposed by the investment limit $\delta$.  The transition exists for all values of $\delta$ explored ($10^{-5}\leq \delta \leq 10^{-2}$) and appears to be universal.  Naturally, the transition becomes more pronounced for larger systems.

The economic interpretation of this transition is unclear.  It is reasonable to expect the price to rise as uncertainty decreases, reflecting increasing confidence in the stock, but it is not obvious why the price would diverge for a non-zero uncertainty.

No other interesting phenomena were observed as the parameters were adjusted.  In the next section, a similar analysis is performed on DSEM.

\section{DSEM phase space}

\subsection{Review}

The Decentralized Stock Exchange Model (DSEM), presented in \ch{dsem}, was constructed as an alternative to CSEM, discarding the notion of a centralized control which sets the stock price.  In DSEM the price is an emergent property of agents placing and accepting orders with each other directly.  The agents use a fixed investment strategy and place orders when their portfolios become unbalanced.  The fraction of one's wealth each of the $N$ agents keeps invested in the stock is affected by (exogenous) news and price movements.  The degree of influence each of these factors has is parameterized by the news and price responsiveness, $r_n$ and $r_p$, respectively.  In \ch{dsem} arguments were presented which reduced the parameter space, leaving only $N$ and $r_p$ as free parameters.  In this section the role each of these parameters plays will be explored.

\subsection{Data collection}

\begin{table}
	\begin{center}\begin{tabular}{r|c|c|c|c|c}
		\hline \hline
		Parameters & \multicolumn{5}{c}{DSEM Dataset 1} \\
		\hline 
		\multicolumn{6}{c}{Particular values} \\
		\hline
		Number of agents $N$ & 50 & 100 & 200 & 500 & 1000 \\
		Number of runs & 39 & 39 & 39 & 39 & 39 \\
		\hline 
		\multicolumn{6}{c}{Common values} \\
		\hline
		Price response $r_p$ & \multicolumn{5}{c}{\begin{tabular}{c@{ to }c@{ by }c}
			--0.75 & 0.25   & 0.05 \\
			0.50   & 0.95   & 0.05 \\
			--0.34 & --0.31 & 0.01 \\
			0.91   & 0.94   & 0.01
		\end{tabular}} \\
		Total cash $C$ & \multicolumn{5}{c}{\$1,000,000} \\
		Total shares $S$ & \multicolumn{5}{c}{1,000,000} \\
		News interval $\tau_n$ & \multicolumn{5}{c}{1} \\
		News response $r_n$ & \multicolumn{5}{c}{$0.01 \pm 0.01$ (uniformly dist.)} \\
		Friction $f$ & \multicolumn{5}{c}{$0.02 \pm 0.01$ (uniformly dist.)} \\
		seed & \multicolumn{5}{c}{random} \\
		Run length (``days'') & \multicolumn{5}{c}{1,000} \\
		\hline \hline
	\end{tabular}\end{center}
	\caption{Parameter values for DSEM Dataset 1.  Some of the parameters were established in \ch{dsem} and are common to all the runs.  Dataset 1 explores two dimensions of phase space: $N$ and $r_p$.}
\label{tbl:resultsDsemDataset1}
\end{table}

The phase space was explored by varying the price response parameter $r_p$ and number of agents $N$.  The choices of system sizes were $N$=50, 100, 200, 500, and 1000 agents while the price response was initially explored at a coarse resolution with increments of 0.25 between $-0.75$ and $+0.75$ then again with a finer resolution of 0.05 in the ranges $-0.75$ to $+0.25$ and  0.50 to 0.95.  Finally, the regions $r_p\in[-0.34,-0.31]$ and $r_p\in[0.91,0.94]$ were explored at a higher resolution of 0.01 because they exhibited interesting properties (to be discussed).  

Although $r_p$ is an agent-specific parameter it was set to a single value for all the agents reducing diversity somewhat.  However, sufficient heterogeneity was maintained through the news response and friction parameters which were each spread over a range of values and each agent was randomly assigned a (uniformly distributed) deviate from within that range.  The complete list of parameters is listed in \tbl{resultsDsemDataset1}.

Each run lasted for 1,000 ``days'' as defined in \sect{dsemTimescale}.  Effectively, this means longer runs (more trades) as the number $N$ of agents increases because of how time is scaled.

\subsection{Phases}

\label{sect:resultsDsemPhases}

\begin{figure}
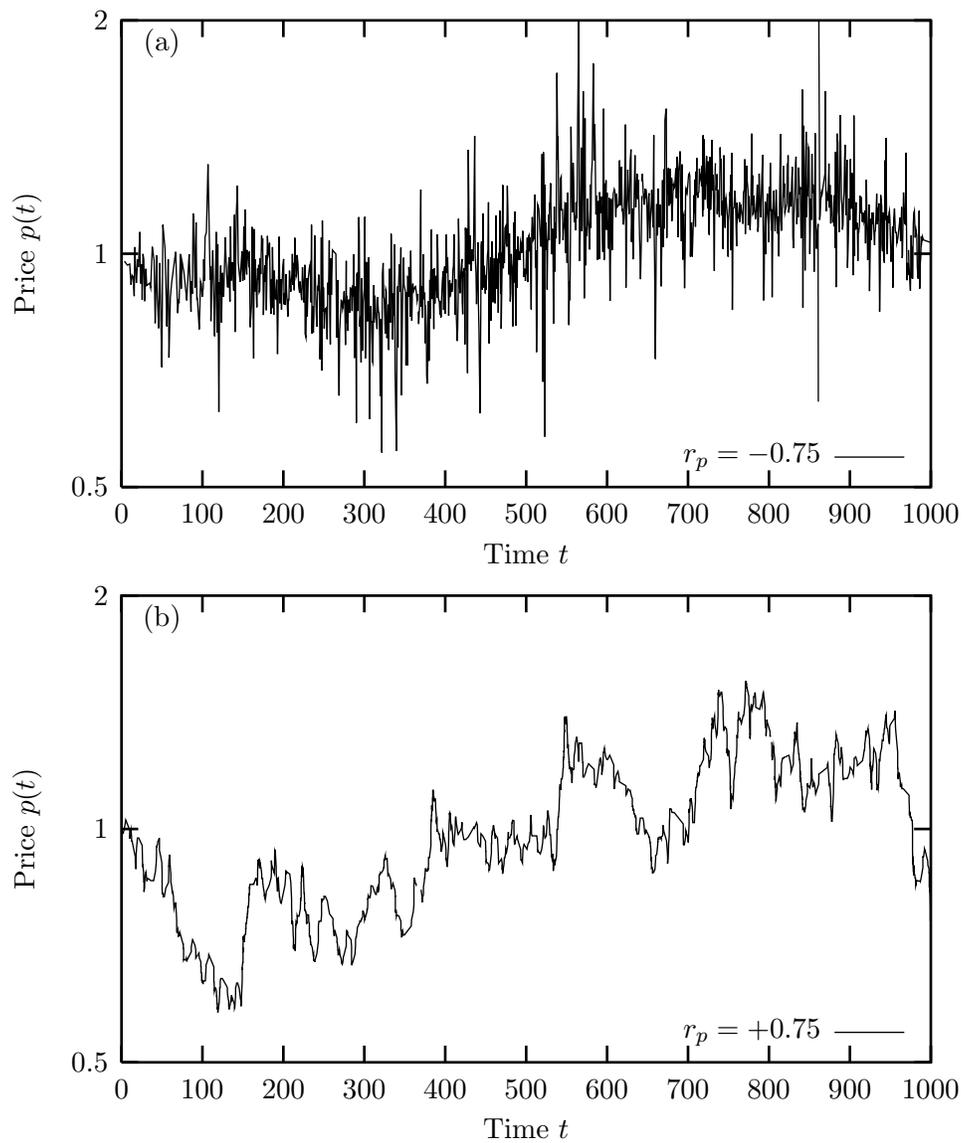
\centering
	\input{chResults/dsemPriceRp-0.75.tex} \\
	\input{chResults/dsemPriceRp+0.75.tex}
	\caption{Sample price series for DSEM with $N=100$.  Negative values of $r_p$ (a) produce an anticorrelated series while positive values (b) result in positive autocorrelations.}
\label{fig:resultsDsemPhases}
\end{figure}

\fig{resultsDsemPhases} shows sample price series for a strongly negative value of $r_p$ and a strongly positive value indicating a change of character as $r_p$ is varied.  For negative $r_p$ the dynamics are dominated by high frequency fluctuations overlaying a relatively small low-frequency component while the reverse seems to be true for positive $r_p$.  
This is not entirely surprising because the parameter $r_p$ acts as a kind of autocorrelation between successive price movements.  When $r_p$ is negative, an increase in the price will lower the agents' ideal investment fractions, decreasing demand which usually results in a price drop.  Conversely,  price increases tend to be followed by further increases when $r_p$ is positive because of increased demand.

\subsubsection{Hurst exponent}

To quantify the dynamics, then, an order parameter which characterizes the autocorrelations is called for.  The tickwise autocorrelation (between successive trades) was considered but rejected because it was found to be ``noisier'' than the alternative---the Hurst exponent.  The Hurst parameter $0\leq H \leq 1$, discussed in \ap{hurst}, quantifies the proportion of high-frequency to low-frequency fluctuations and measures the long-range {\em memory} of a process.  It is an alternative representation of temporal correlations in a time series with $H<1/2$ for anticorrelated series and $H>1/2$ for positively correlated data ($H=1/2$ indicates no correlations).  

The Hurst exponent was calculated by the application of dispersional analysis, a simple and accurate method described in \sect{hurstDispersion}, to the log-return series of the price sampled at discrete intervals of four ``minutes'' (interpreting a trading ``day'' to consist of 6.5 hours).  Sampling the data at regular intervals did not significantly affect the estimates of $H$ but it fixed the size of the dataset to be analyzed; for large systems many trades could be executed within a few minutes, producing exorbitantly large datasets which were cumbersome to analyze.  (Before discretization the largest dataset contained some 400,000 points.)  With a fixed sampling rate all system sizes generated the same volume of data, around 100,000 points.

\begin{figure}\centering
	\input{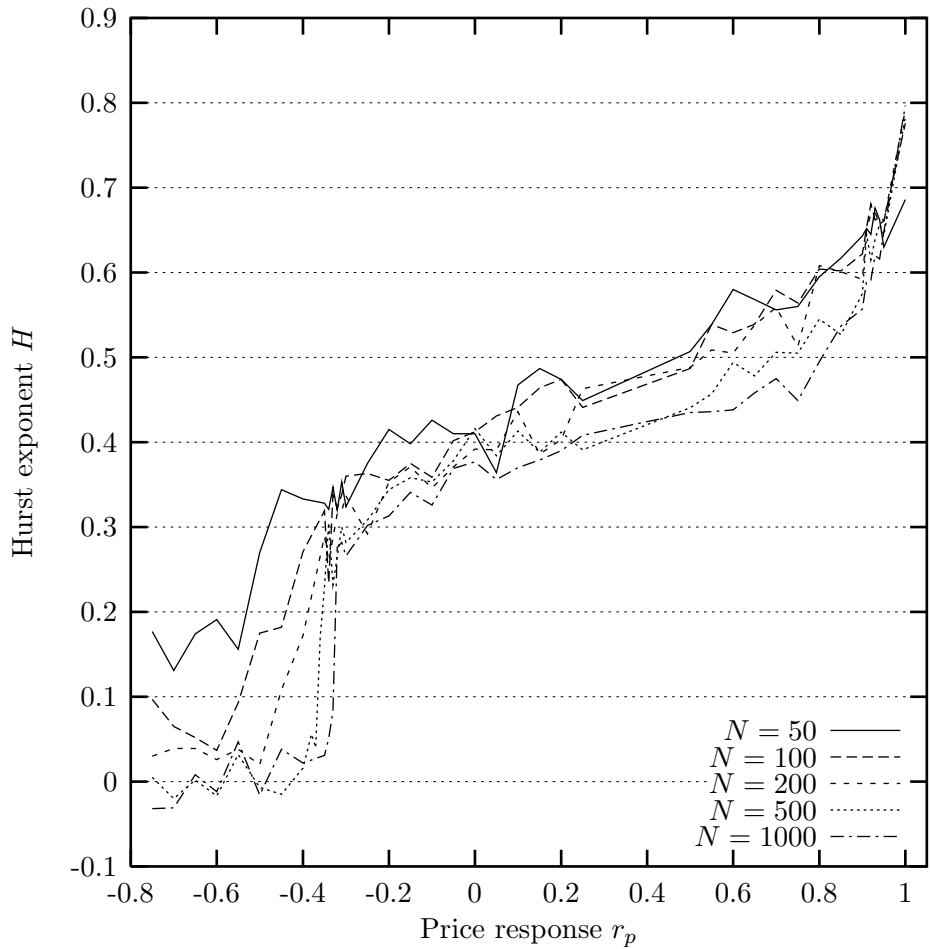}
	\caption{The Hurst exponent increases with $r_p$ in DSEM as expected but with two surprising phase transitions emerging at larger system sizes: one near $r_p\approx -0.4$ and the other near $r_p\approx 1$.}
\label{fig:resultsDsemPriceResponse}
\end{figure}

As the price response parameter $r_p$ increases from $-0.75$ up to 1.00 some interesting properties emerge: for large systems ($N\geq 200$) the Hurst exponent is effectively zero below $r_p\approx -0.4$, indicating very strong anticorrelations in the price series (independent of $r_p$).  Suddenly, near $-0.4$, the anticorrelations break down and the Hurst exponent climbs quickly (with increasing $r_p$) to roughly $H\approx0.4$ (suggesting weakly anticorrelated data).  It remains relatively constant until $r_p\approx 1$ where it climbs again, to $H\approx 0.8$.  

Interestingly, the Hurst exponent is less than one half at $r_p=0$ meaning that price movements in one direction tend to be followed by opposite movements even with no explicit price response coded into the agents' behaviour.  This arises from corrections to initial over-reaction to news---agents with extremist news responses react to news by placing orders with atypical prices which are corrected for when moderate agents have an opportunity to trade.

\subsubsection{Price response greater than unity}

Apparently, DSEM exhibits three distinct phases: the first two are demonstrated in \fig{resultsDsemPhases} but the third ($r_p>1$) is not shown so it is briefly discussed here.  It is characterized by very strong positive correlations in the price series, such that the price explodes or crashes exponentially, depending on the direction of the first price movement.  

Very rapidly (within a few ``days'') the price hits a boundary imposed by a mechanism identical to that in CSEM: the investment fraction in DSEM is actually constrained by $\delta$ such that $\delta\leq i \leq 1-\delta$.  The price is therefore bounded according to \eqs{csemMinPrice}{csemMaxPrice}.  This constraint was not mentioned in the development of the model because, in DSEM, the limit is $\delta=10^{-12}$, a value chosen only to avoid numerical round-off errors.  In practice $\delta$ was not observed to affect the dynamics whatsoever, except in the region $r_p>1$.

The exponential growth of the price in this region indicates that the Hurst exponent (which cannot be accurately measured because the price reaches the boundary too quickly) is identically one $H=1$, the same value obtained from a straight line (which this is, since we are analyzing the logarithm of the price).

\subsection{Phase transition to $H=1$ at $r_p=r_1$}

\label{sect:resultsDsemPhaseTrans1}

In this section the phase transition for positive values of $r_p$ will be explored.  This phase transition is easier to characterize than the one to be discussed in \sect{resultsDsemR2} because we have reason to expect the transition to occur at $r_p=r_1\equiv 1$ and may therefore eliminate one adjustable parameter from the fitting function.  (The fit was also performed with $r_1$ as an adjustable parameter (not shown) and the results corroborate the hypothesis that the transition is at $r_1=1$.)

From \fig{resultsDsemPriceResponse} it is clear that the phase transition at $r_1$ is not first-order (discontinuous) but appears to be second-order (critical).  In the last section it was argued that the transition is to $H=1$ for $r_p>r_1$ so the power-law to be fit takes the form
\begin{equation}
	1-H(r_p) = C (r_1-r_p)^b
\end{equation}
with fitting parameters $C$ and $b$.

\begin{figure}\centering
	\input{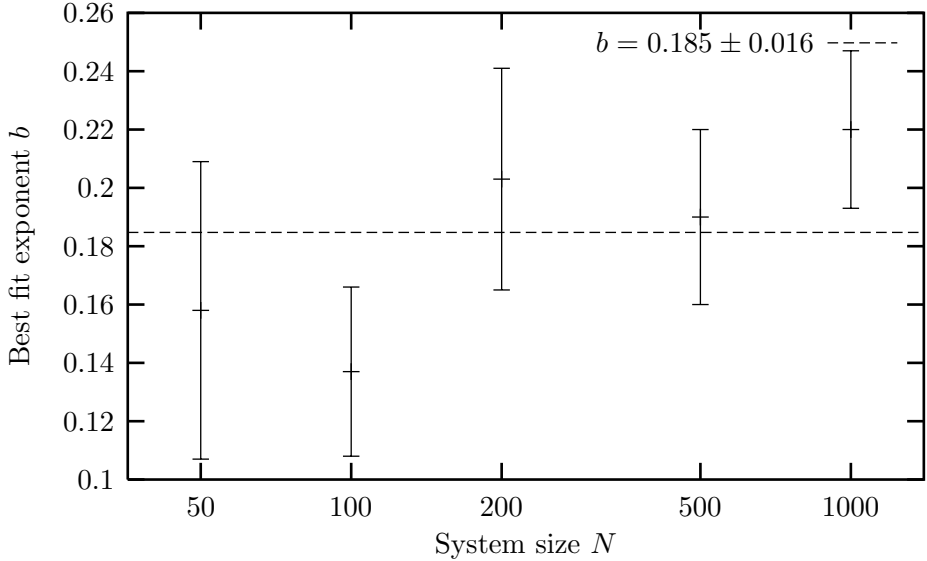}
	\caption{The best fits of power laws to DSEM Dataset 1 yield the scaling exponents shown.  The average exponent is $0.185\pm 0.016$.}
\label{fig:resultsDsemPowerExpH1}
\end{figure}

The fits were performed over the range $0.75\leq r_p \leq 0.95$ for each value of $N$ giving the 
exponents $b$ shown in \fig{resultsDsemPowerExpH1} with an average value of $b=0.185\pm 0.016$.  It is not known to what universality class (defined in \sect{resultsCsemFiniteSize}), if any, this transition belongs.  Next we explore the phase transition observed for negative values of $r_p$.

\subsection{Phase transition to $H=0$ at $r_p=r_2$}

\label{sect:resultsDsemR2}

Now we turn our attention to the other phase transition in the system, near $r_p\approx -0.4$.  For the smallest systems $N\leq 100$ the transition is not discernable in \fig{resultsDsemPriceResponse} but it comes into focus as the system size is increased.  For intermediate agent numbers, $N=200$ and 500, the transition looks very much second-order (continuous).  However, in the largest system $N=1000$ the transition is quite abrupt, so particular care must be taken to establish whether it is first- or second-order.

\subsubsection{Comments on phase transitions}

The distinction between first- and second-order transitions is not merely academic; it can greatly enhance our understanding of the underlying dynamics.  First-order phase transitions, characterized by a discontinuity in the order parameter, occur via nucleation: small pockets of the new phase emerge within the old phase and grow until the entire system is in the new phase.  On the other hand, second-order transitions, characterized by a continuous order parameter with a diverging derivative, exhibit system-spanning correlations such that the entire system undergoes the transition as a whole.

In the context of the models presented here, correlations would indicate correlated behaviour amongst investors and nucleation would refer to a small sub-group of investors acting differently from the larger population.

\subsubsection{Classification of $r_2$}

\begin{table}
	\begin{center}\begin{tabular}{r|c|c}
		\hline \hline
		Parameters & \multicolumn{2}{c}{DSEM Dataset 2} \\
		\hline 
		\multicolumn{3}{c}{Particular values} \\
		\hline
		Number of agents $N$ & 300 & 700 \\
		Number of runs & \makebox[0.75in][c]{16} & \makebox[0.75in][c]{16} \\
		\hline 
		\multicolumn{3}{c}{Common values} \\
		\hline
		Price response $r_p$ & \multicolumn{2}{c}{\begin{tabular}{c@{ to }c@{ by }c}
			$-0.35$ & $-0.30$ & 0.01 \\
			$-0.25$ & 0.25 & 0.05
		\end{tabular}} \\
		Run length (``days'') & \multicolumn{2}{c}{500} \\
		\hline \hline
	\end{tabular}\end{center}
	\caption{Parameter values for DSEM Dataset 2. These runs are a variation of Dataset 1 (all unspecified parameters are duplicated from
 \tbl{resultsDsemDataset1}) exploring a few other intermediate system sizes.}
\label{tbl:resultsDsemDataset2}
\end{table}

To explore the phase transition in detail some more data were collected at intermediate system sizes as shown in \tbl{resultsDsemDataset2} for a total of seven different values of $N$.  The datasets were analyzed by attempting to fit both first-order and second-order transitions.

Assuming a first-order transition the fit becomes trivial: we can just assume a linear dependence on $r_p$ in the neighborhood of $r_2$.  (In this case linearity was observed over $r_p\in(-0.3,0.1)$ for all runs.)  The transition point is simply read off the graph from the largest system $N=1000$ (the transition is resolved with greater accuracy as $N$ increases) giving $r_2=0.33\pm 0.01$.  The magnitude of the discontinuity in the Hurst parameter $H$ at $r_2$ is then $\Delta H(r_2) = 0.281 \pm 0.008$.

We now consider the possibility that the transition is second-order.  If so, then a power-law dependence
\begin{equation}
\label{eq:resultsDsemPowerRp2}
	H(r_p) = C (r_p - r_2)^b
\end{equation}
should characterize the behaviour near the transition $r_2$ with adjustable parameters $C$ and $b$.

\begin{figure}
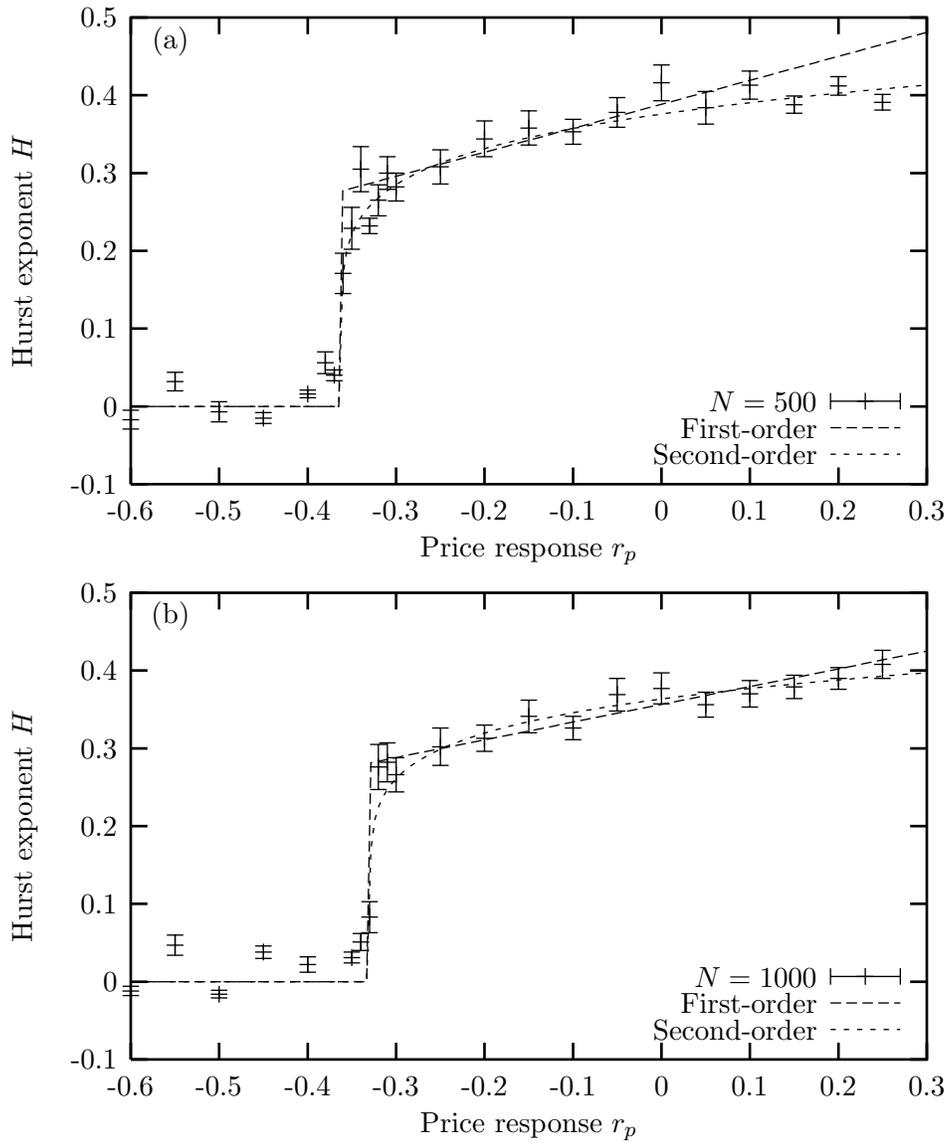
\centering
	\input{chResults/dsemPhaseFitN500} \\
	\input{chResults/dsemPhaseFitN1000}
	\caption{Sample fits of first- and second-order phase transitions to $N=500$ (a) and $N=1000$ (b) near $r_2$ in DSEM show that the power-law fits better for small $N$ but the first-order prevails for larger systems.}
\label{fig:resultsDsemPhaseFit}
\end{figure}

Sample fits for $N=500$ and 1000 are shown in \fig{resultsDsemPhaseFit} with a simple linear fit representing a first-order transition for comparison.  Notice that the $N=500$ system is better described by a critical transition but $N$ increases to 1000 the transition becomes sharper, more like a first-order transition suggesting that the scaling behaviour is only a finite-size effect.  

\subsubsection{Finite-size scaling}

\begin{figure}\centering
	\input{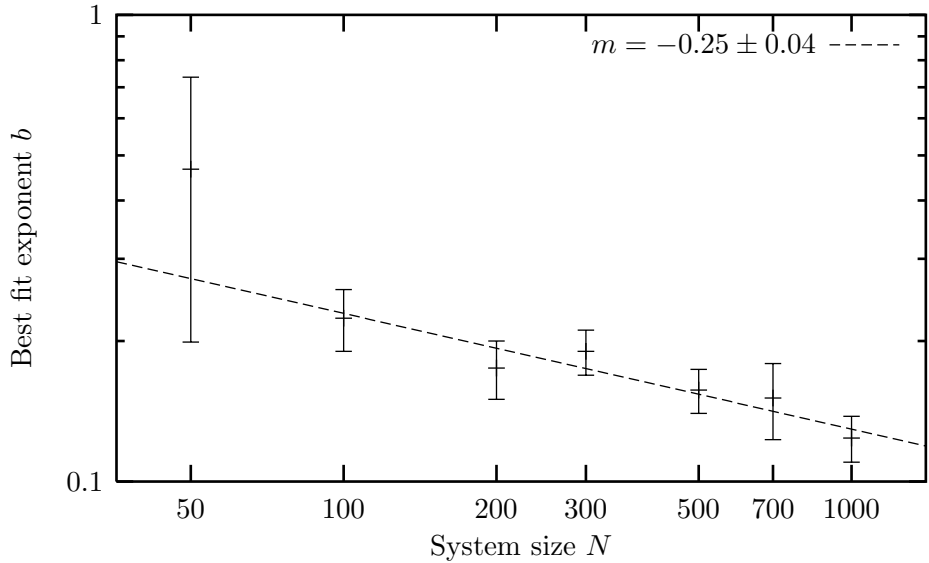}
	\caption{As the system size $N$ increases the critical exponent $b$ tends to zero.  The line represents a power-law fit $b\propto N^m$ giving an exponent $m=-0.25\pm 0.04$.}
\label{fig:resultsDsemFiniteScaling}
\end{figure}

Assuming criticality, each system in Datasets 1 and 2 were fit to \eq{resultsDsemPowerRp2} and the best-fit exponents $b$ are plotted in \fig{resultsDsemFiniteScaling}.  Clearly, the exponents exhibit a trend as the system size $N$ increases.  On a log-log graph the trend appears linear suggesting that the exponents scale with the system size as yet another power law $b\propto N^m$.  (Be warned, this is not a traditional---nor rigorous---finite-size scaling argument.)  The scaling exponent is found to be $m=-0.25\pm 0.04$ meaning that the exponent $b$ will be halved every time the system size is scaled up by a factor of 16 and in the thermodynamic limit ($N\rightarrow \infty$) $b$ drops to zero.

To understand what is going on here, consider the general scaling function
\begin{equation}
	y=c(x-x_c)^b
\end{equation}
which has a slope
\begin{equation}
	y' = bc(x-x_c)^{b-1}.
\end{equation}
If we demand that the scaling function mimic a first-order transition, requiring that both $y>0$ and $y'$ approach constants as $x\rightarrow x_c$ then the scaling exponent must vary such that
\begin{equation}
	b = \frac{y'}{y}(x-x_c).
\end{equation}
So a second-order transition ``mimics'' a first-order in the limit $b\rightarrow 0$.

Returning to the transition at $r_p=r_2$ in DSEM we see that the apparent criticality is an artifact of finite simulation size and in the limit $N\rightarrow \infty$ the transition is first-order.

\subsubsection{Intermittency}

As discussed above, there are important consequences of knowing a transition is first-order.  The foremost is that fluctuations are local, they do not spread throughout the entire system (as is found for critical points).  Another consequence is that the transition is a change of {\em quality} not simply of quantity.  That is, since the order parameter exhibits a discontinuity the nature of the system changes qualitatively, not just quantitatively.  A third important feature of first-order transitions is {\em nucleation}.  Near the transition stochastic fluctuations can often give rise to small pockets which exhibit one phase while the greater system is within the other phase.  A good analogy to keep in mind is a pot of boiling water: steam bubbles form on the bottom and sides of the pot where small variations of the surface exist.

\begin{figure}
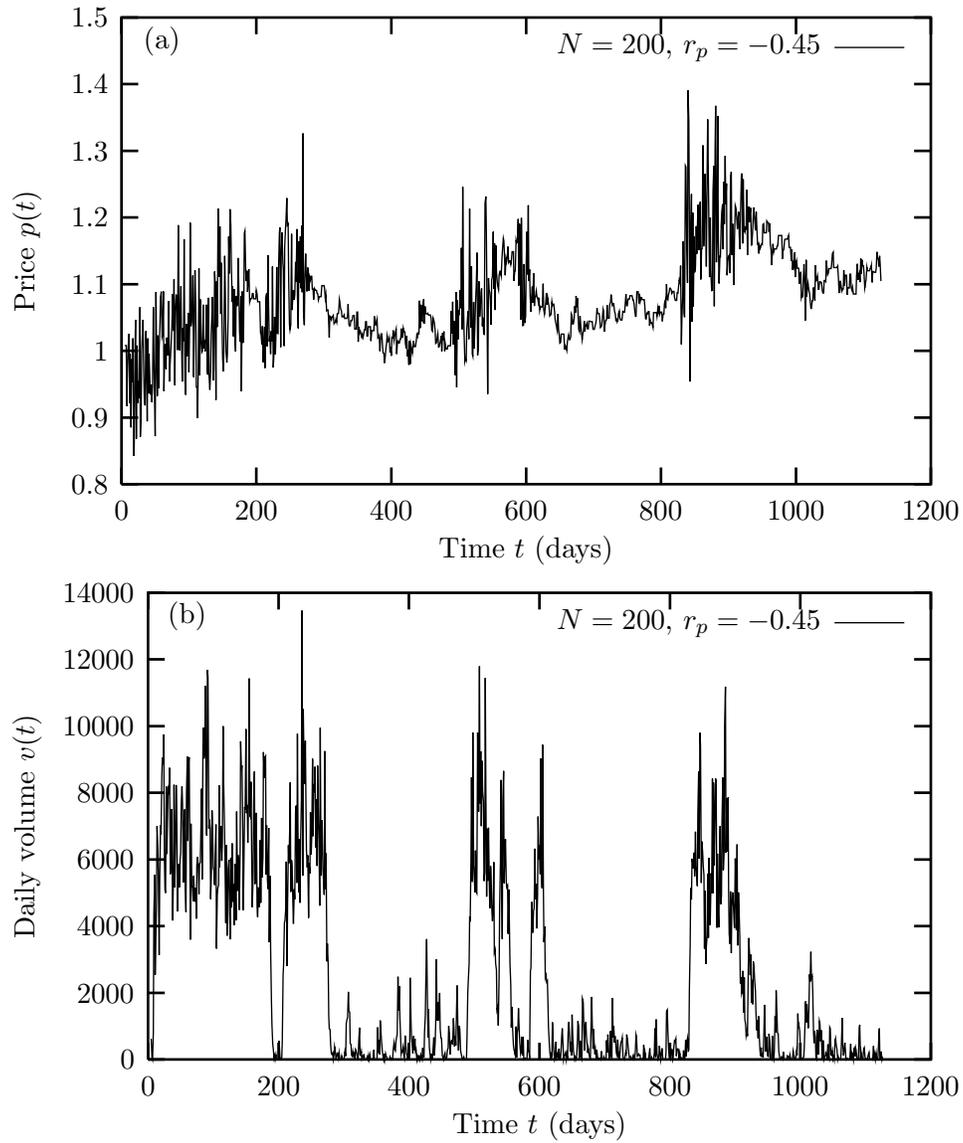
\centering
	\input{chResults/dsemN200rp-0.45price.tex} \\
	\input{chResults/dsemN200rp-0.45volume.tex}
	\caption{The price series (a) and daily volume (b) of DSEM with $N=200$ and $r_p=-0.45$ is a good example of intermittency.  The dynamics fluctuate between two phases.}
\label{fig:resultsDsemN200rp-0.45}
\end{figure}

The last interesting property of first-order transitions that will be mentioned here is intermittency.  As discussed above, near the phase transition bubbles of one phase form at nucleation points within the other phase.  In a spatially-extended system this causes intermittent periods of either phase at any particular point in the system.  Since DSEM is nonspatial the intermittent behaviour is captured in the price series which is observed to consist of periods of low activity separated by periods of high activity, as demonstrated in \fig{resultsDsemN200rp-0.45}.  

\subsection{Summary}

In this section we explored the phase space of the price response parameter $r_p$ and the number of agents $N$ in DSEM.  Two phase transitions were observed: at $r_p=r_1=1$ DSEM undergoes a critical transition to perfectly correlated price movements (the Hurst parameter $H$ goes to unity), and at $r_p=r_2\approx -0.33$ a first-order transition is observed.  Below the transition strong anticorrelations in the price series are observed but above it only weak anticorrelations exist.  Near the transition point the system spends time in both regimes giving rise to clusters of high volatility.

\section{Number of investors}

Before comparing the models with empirical data we should complete our exploration of the phase space.  In both CSEM and DSEM we explored a variety of system sizes (number of agents $N$) in order to enhance the resolution of the phase transitions.  In this section we will re-evaluate this data in light of the discovery that in many market simulations the dynamics reduce to being semi-regular in the limit of many investors \cite{kohl97, egenter99}.

Most past simulations were performed with investors numbering between 25 and 1,000 \cite{levy95, arthur97, bak97, caldarelli97, chen98, busshaus99} with the largest ranging between 5,000 and 40,000 \cite{youssefmir94, chowdhury99, iori99}.  Even the largest of these are minuscule when compared with natural systems exhibiting phase transitions, which have on the order of $10^{23}$ particles.  It is not clear that these market models (including CSEM and DSEM) are at all interesting in the limit of many investors.  In fact it has been discovered that the dynamics of many of these models become almost periodic as the number of investors grows \cite{kohl97, egenter99}.

Whether this behaviour detracts from the models is uncertain.  The models can only be tested by comparison with real markets but even the largest markets cater to an infinitesimal number of individuals when compared with natural systems.  After all, the natural world consists of only a few billion ``agents'' of which only a minute fraction are actively involved in stock trading.  Therefore, one may argue that these models do not need to exhibit rich behaviour in the limit of many investors in order to be realistic.  They need only exhibit realistic dynamics on the same scale as real markets.  

Nevertheless, it is interesting and useful to understand how the models predict the dynamics to evolve with increasing investor numbers.  One advantage is that testable predictions may be made with regard to how a market will scale as more investors come aboard.  In the last few years the number of investors in the markets have grown substantially, mainly due to the rise of the internet which allows traders to monitor their portfolios in (almost) real time and execute trades promptly.  The only research the author is aware of to explore the consequences of growing markets is a model which suggests that fluctuations increase with system size \cite{hogg95}.

Thus, it is useful to explore the effect of increasing the number of agents in both the Centralized and Decentralized models.

\subsection{Centralized Stock Exchange Model}

\begin{table}
	\begin{center}\begin{tabular}{r|c|c|c}
		\hline \hline
		Parameters & \multicolumn{3}{c}{CSEM Dataset 3} \\
		\hline 
		Number of agents $N$ & 10,000 & 10,000 & 10,000 \\
		Forecast error $\sigma_\epsilon$ & 0.01 & 0.08 & 0.15 \\
		Investment limit $\delta$ & $10^{-2}$ & $10^{-2}$ & $10^{-2}$ \\
		Number of runs & 1 & 1 & 1 \\
		Run length (time steps) & 10,000 & 10,000 & 10,000 \\
		\hline \hline
	\end{tabular}\end{center}
	\caption{Parameter values for CSEM Dataset 3.  These runs are a variation of Dataset 1 (all unspecified parameters are duplicated from
 \tbl{resultsCsemDataset1}) with many agents $N=10,000$.}
\label{tbl:resultsCsemDataset3}
\end{table}

\begin{figure}
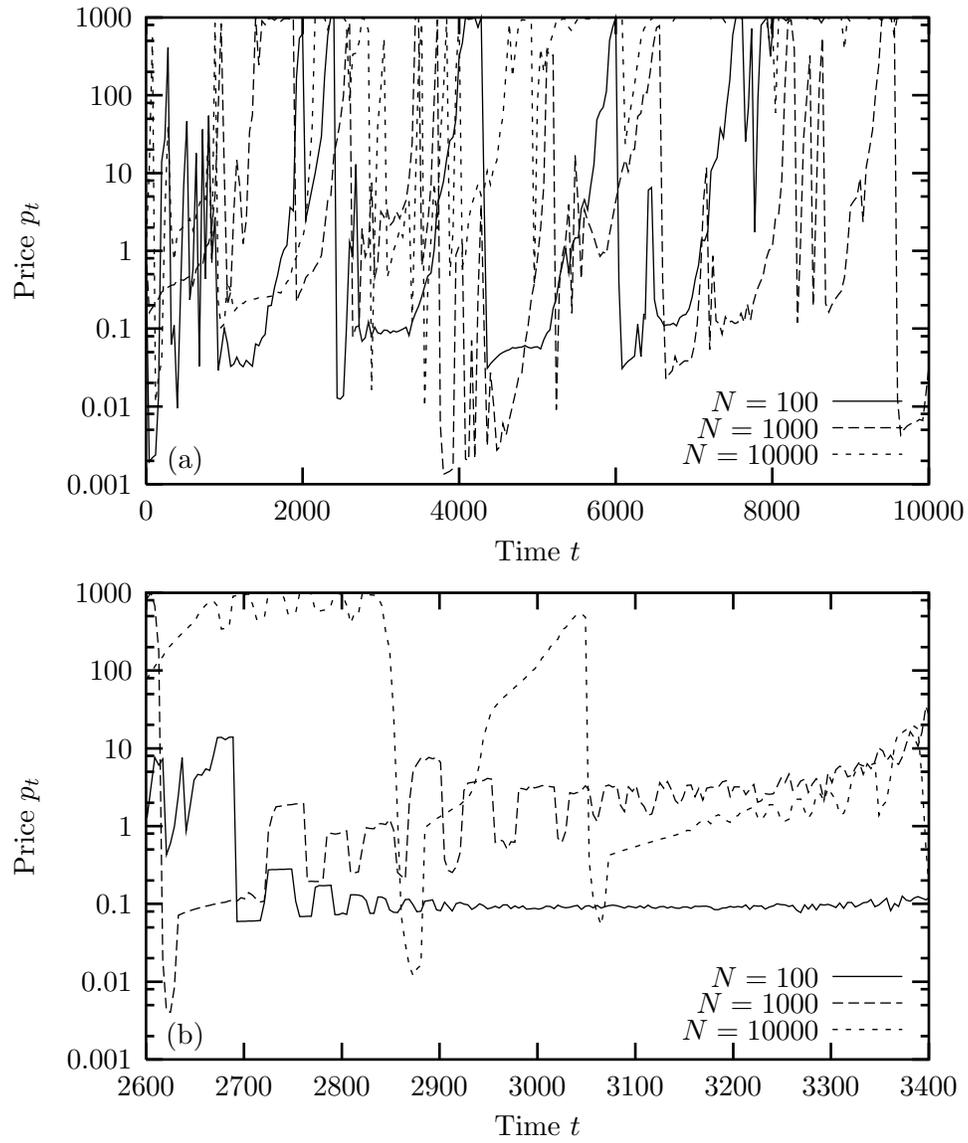
\centering
	\input{chResults/csemS0.01AgentsA.tex} \\
	\input{chResults/csemS0.01AgentsB.tex}
	\caption{The price series of CSEM for $\sigma_\epsilon=0.01$ (a) appears unaffected by changing the number of agents $N$.  In particular, occasional semi-periodic fluctuations (b) are observed for all system sizes.}
\label{fig:resultsCsemS0.01Agents}
\end{figure}

\begin{figure}\centering
	\input{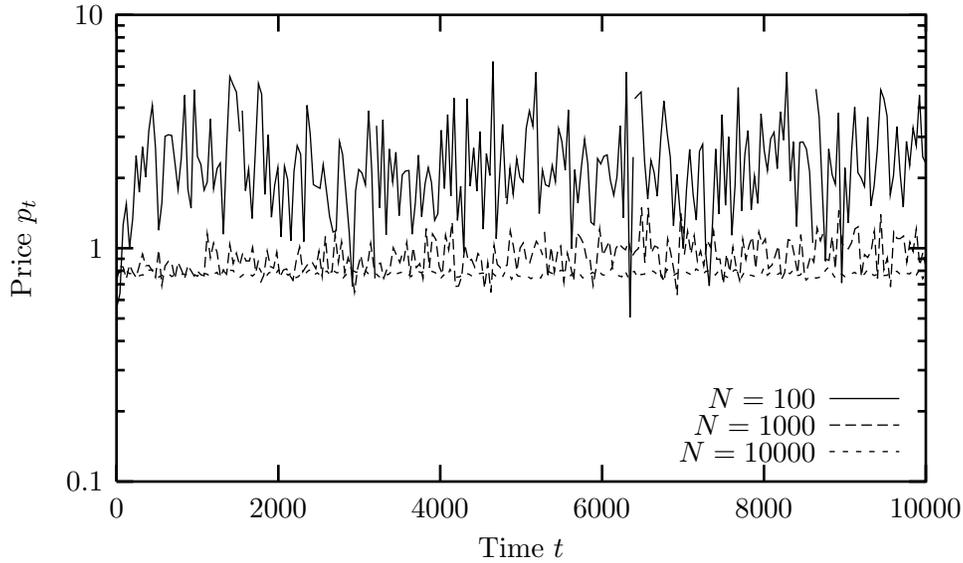}
	\caption{The price series of CSEM for $\sigma_\epsilon=0.15$ exhibits smaller fluctuations and a lower mean as the system size increases.  (The lower mean may simply be because the system has not reached a steady state yet.)}
\label{fig:resultsCsemS0.15Agents}
\end{figure}

To test the effect of changing the number of agents $N$ thoroughly a new dataset (see \tbl{resultsCsemDataset3}) was collected with markets containing $N=10,000$ agents in three regimes: far below the critical point $\sigma_\epsilon=0.01$, near the critical point $\sigma_\epsilon=0.08$, and above the critical point $\sigma_\epsilon=0.15$.  The price series for the first and last of these are shown in \fig{resultsCsemS0.01Agents} and \ref{fig:resultsCsemS0.15Agents}, respectively.

Far below the critical point the dynamics appear to be largely invariant under change of the number of investors.  Interestingly, the dynamics do display semi-periodic intervals, as demonstrated in \fig{resultsCsemS0.01Agents}(b), but this occurs for all system sizes and is not a stable phenomenon even for the largest system so it does not appear that the dynamics converge to being almost periodic in the limit $N\rightarrow\infty$ as found in other models \cite{kohl97, egenter99}.

Above the critical point the most obvious feature is that the fluctuations decline with the number of investors (see \fig{resultsCsemS0.15Agents}).  But this is to be expected since the trading price results from the interactions between $N$ ``noisy'' investors.  Therefore we expect the fluctuations (measured as the standard deviation of the log-price) to decrease as $1/\sqrt{N}$, a hypothesis which was found to hold fairly well for $\sigma_\epsilon=0.15$ between $N=50$ and $N=10,000$ but which held better further from the critical point (as can be seen in \fig{resultsCsemPriceFluctN} which demonstrates the variances multiplied by system sizes nearly collapse to a single curve).

The main conclusion to be drawn from this analysis is that the dynamics of CSEM do not appear to become trivial or converge to a semi-periodic pattern as the number of investors becomes infinite.  Above the critical point fluctuations do diminish but as one approaches the critical point they reappear and below the critical point the dynamics are largely independent of the number of agents.

\subsection{Decentralized Stock Exchange Model}

Collecting data for $N=10,000$ in DSEM proved prohibitive since each run would require a full week of computer run-time to collect sufficient data for analysis.  This occurs because DSEM requires on the order of $N^2$ operations per simulation ``day'' ($N$ calls per day with $N$ potential replies each) whereas CSEM grows linearly with $N$.  Therefore the data collected in \tbl{resultsDsemDataset1} will be used here.

\begin{figure}
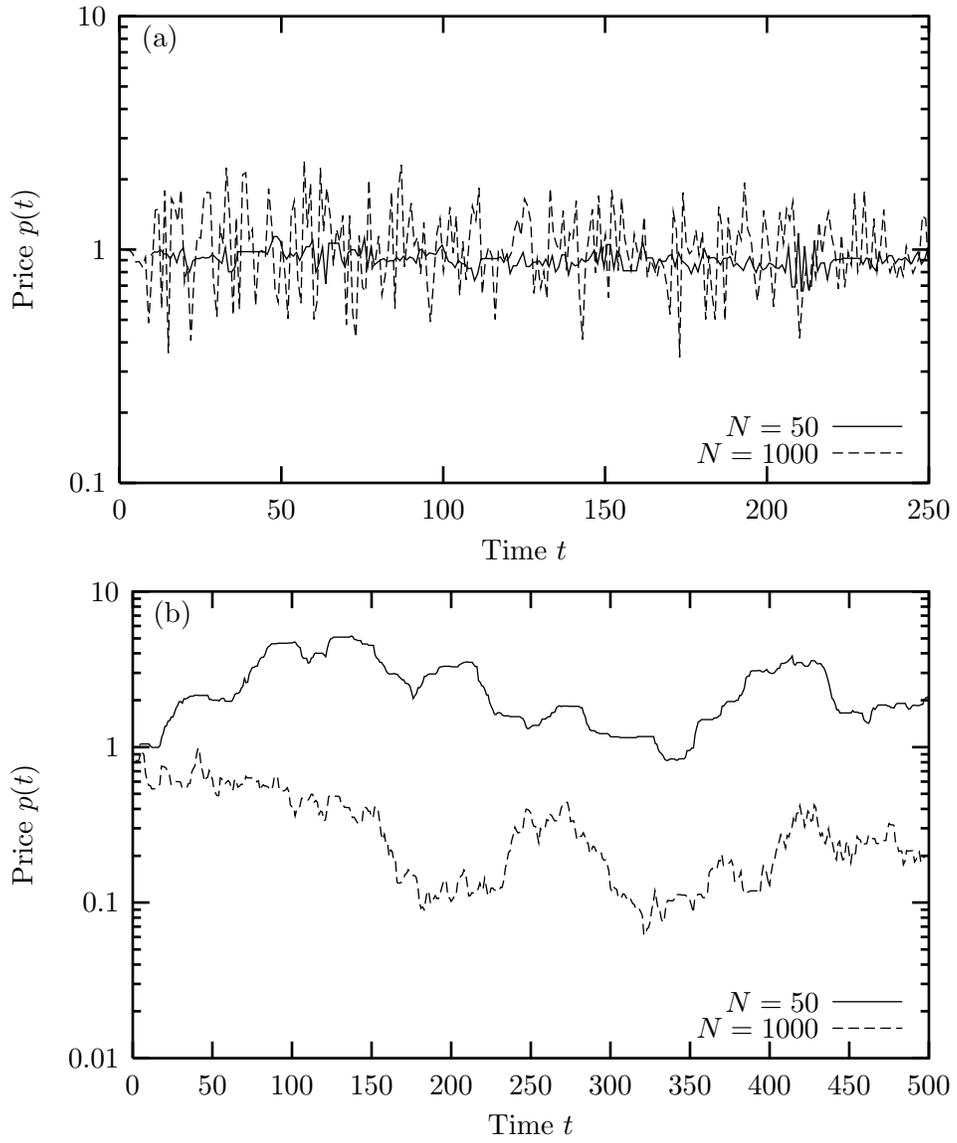
\centering
	\input{chResults/dsemRp-0.75Agents.tex} \\
	\input{chResults/dsemRp0.90Agents.tex}
	\caption{In DSEM the price series does not get more regular as the system size is increased---in fact the fluctuation grow.  This is especially true for $r_p=-0.75$ (a) but it is also indicated to a lesser degree at $r_p=0.90$ (b).}
\label{fig:resultsDsemAgents}
\end{figure}

Again, we observe the price series for a variety of system sizes---this time $N=$ 50, 200, and 1000 agents---and regions of phase space spanning the phase transitions---$r_p=-0.75$, 0.00, and 0.90---to estimate how the dynamics would change as the system size grew without limit.  In each region it was found that the fluctuations actually grew with system size, especially below the first-order transition $r_p<r_2\approx-0.33$.

The two plots in \fig{resultsDsemAgents} show the price series only for $N=50$ and $N=1000$ ($N=200$ exhibited predictably intermediate fluctuations and was not plotted to reduce clutter).  The plots show that the fluctuations are somewhat larger for the larger system when $r_p=0.90$ and significantly so for $r_p=-0.75$.  Thus the dynamics do not reduce to semi-regular in the limit of many investors.

\subsection{Summary}

In this section we tested the effect of increasing the number of agents in both CSEM and DSEM in light of recent research that indicates that some market models become quasi-periodic in the limit of many agents \cite{kohl97, egenter99}.  It was found that fluctuations did decline in CSEM when the forecast error $\sigma_\epsilon$ was significantly above its critical value but the dynamics were largely invariant under variation of the number of agents below the critical point.

In DSEM the fluctuations actually grew as more investors were introduced, especially below the first-order transition at a price response of $r_p\approx -0.33$.  This makes an interesting and testable prediction regarding empirical markets: it indicates that fluctuations in stock prices should be greater in larger markets.  (Some caveats are required: the size of the market is measured in terms of the number of independent investors (a fund group would be interpreted as a single investor) and not the total value of the outstanding stock.  Recall that in these simulations the total cash and shares were held fixed: with more investors each held a smaller portion of the total resources.)

Since the number of investors does not strongly affect the dynamics we are free to choose values which correspond well with observed market fluctuations.  Comparing the fluctuations with empirical data will be the subject of the next chapter.

\chapter{Analysis and Results: Empirical results}

\label{ch:results2}

In the last chapter we explored the phase space of the Centralized and Decentralized Stock Exchange Models (CSEM and DSEM, respectively).  This chapter is concerned with contrasting the data from these models with empirically known qualities of real markets.  Some of the properties we hope to uncover are leptokurtosis in the price returns and correlated volatilities.  As we will see, the emergence of these properties is closely related to the phase transitions discovered in the last chapter.

\section{Price fluctuations}

\label{sect:results2PriceFluct}

We begin by exploring the distribution of price fluctuations.

\subsection{Background}

It has long been known that stocks exhibit stochastic fluctuations in their price histories.  Originally it was hypothesized that the markets exhibited (discrete) Brownian motion and therefore had Gaussian-distributed price increments \cite{bachelier00}.  Later this was adapted to account for the strictly positive nature of stock prices via geometric Brownian motion \cite{osborne59} with the logarithm of the price following a random walk.  Much theoretical work on derivative pricing assumes geometrical Brownian motion including the famous Black-Scholes equation \cite{black73}.  

It was a startling discovery, then, when Mandelbrot pointed out that, empirically, the logarithm of price-returns did not have a Gaussian distribution \cite{mandelbrot63, mandelbrot97}.  In fact, on short timescales, large (exceeding a few standard deviations) fluctuations occurred much too frequently to be explained by the Gaussian hypothesis.  These large fluctuations contribute to the tails of the distribution resulting in ``fat tails''.  

Mandelbrot proposed the correct probability distribution function (on the logarithmic scale) was not the Gaussian but its generalization---the stable L\'evy distribution.  L\'evy distributions (see \sect{hurstLevy}) drop off as power laws
\begin{equation}
	p(x) \sim \frac{1}{x^{\alpha+1}}, \; \abs{x}\rightarrow \infty
\end{equation}
for $0<\alpha<2$, resulting in fatter tails than the Gaussian ($\alpha=2$).  An attractive feature of this hypothesis is that it scales: that is, the distribution (and the exponent $\alpha$) remains the same whether measured hourly, daily, or even monthly.  Mandelbrot measured the daily and monthly distribution of returns from cotton prices and found both fitted well to a L\'evy distribution with exponent $\alpha=1.7$ \cite{mandelbrot63}.  Studies of other markets have had similar results concluding $1.4\leq \alpha \leq 1.7$ \cite[and references therein]{mantegna95, cont97b}.  (\ap{sampling} demonstrates it is possible to simulate fat tails by regular sampling of a discrete Brownian process but Pal\'agyi and Mantegna \cite{palagyi99} demonstrate this is not responsible for the fat tails observed in return distributions.)

Since then the adequacy of the L\'evy distribution to describe price fluctuations has been called into question because it implies that the fluctuations have an infinite variance whereas experimental evidence indicates it is probably finite \cite{bouchaud99, gopikrishnan99}.  (Recent studies indicate the tails of the return distributions fall off fast enough that the variance is finite.)  Related to this is the observation that scaling is violated on long timescales (of more than a week \cite{cont97b}) where the distribution converges to a Gaussian (because the Central Limit Theorem applies if the variance is finite).  

This discrepancy was initially resolved by arbitrarily truncating the power law with an exponential weighting function for large events \cite{cont97b, gupta99}.  According to this theory, {\em gradually truncated L\'evy flight} (GTLF),  the tails of the cumulative distribution of (normalized) returns $r$ follow
\begin{equation}
	C(r) \sim \left\{ \begin{array}{ll}
		r^{-\alpha} & \abs{r}\leq l_C \\
		r^{-\alpha} \exp\left[ -\left( \frac{ \abs{r}-l_c }{ k } \right)^\beta \right] & \abs{r} > l_C.
		\end{array} \right.
\end{equation}
with an exponential decay beyond some cut-off $l_C$.  While this did improve the quality of the fit to observed returns it did so at the expense of three new fitting parameters---the cut-off $l_C$, the decay rate $k$, and the power of the exponential $\beta$---bringing the total to five adjustable parameters.

An appealing alternative is the idea that the tails of the return distribution do have a power law but with an exponent $\alpha\approx 3$.  (An interesting, and testable, consequence is that moments higher than three---such as the kurtosis---are divergent.)  Since the exponent is greater than two the distribution is not stable and converges to a Gaussian on long timescales.  It also implies that the variance is finite.  A very comprehensive analysis was performed across 1,000 companies yielding a (huge!) dataset of 40 million returns (Mandelbrot had only available 2,000 points) and the results strongly support the {\em inverse cubic} (IC) hypothesis \cite{gopikrishnan98, plerou99}.

To be precise, the theory is that a L\'evy law ($\alpha\approx1.4$) applies for small to intermediate returns (less than a few standard deviations) but then the distribution crosses over to the inverse cubic for larger returns.  Thus we have two fitting parameters in either scaling regime and a crossover point, for a total of five adjustable parameters, the same as GTLF. Although this research \cite{gopikrishnan98, gopikrishnan99, plerou99} is excellent the practical application of the theory is somewhat cumbersome and a simpler alternative exists.

\subsection{Alternative: Decaying power law}

In this section an alternative which appears to explain the data almost as well but with only three parameters is presented.  Koponen \cite{koponen95}---expanding on work done by Mantegna and Stanley \cite{mantegna94}---demonstrated that a power law with a smooth exponential cutoff has L\'evy increments on short timescales which converge to Gaussian after a long time.  This process is equivalent to GTLF with $l_C=0$ and $\beta=1$ so there exists no cut-off point, the exponential truncation applies for all returns.  The hypothesis is that the tails of the cumulative return distribution obey
\begin{equation}
\label{eq:results2DecayPowerLaw}
	C(\abs{r_i} > r) \sim r^{-\alpha} e^{-r/r_c}
\end{equation}
where $\alpha$ is the scaling exponent and $r_c$ is the decay constant, which sets a characteristic scale over which the power law dominates---in the limit $r_c\rightarrow\infty$ Mandelbrot's pure L\'evy flight hypothesis re-emerges.  (This functional form has been observed to accurately describe the distribution of fluctuations in the ``game of Life'' \cite{blok97}.)  

The use of this truncation hypothesis must be justified in light of the overwhelming empirical evidence supporting the inverse cubic hypothesis \cite{gopikrishnan98, gopikrishnan99, plerou99}.  To do so it is necessary to explicitly formulate the goals of this section: (1) to determine if the return distributions generated by the models scale with an exponent near $\alpha\approx 1.4$ over some range of returns, and, if so, (2) to estimate the range of returns over which the scaling holds.  

In doing so we can determine if the models reproduces truncated L\'evy flight observed empirically.  However, the amount of data collected will be insufficient to adequately determine how the truncation occurs, whether it is an inverse cubic or exponential decay.  Further, this detail is not of central importance to this work so it is left for future research.

Therefore we are free to choose the most convenient form for the truncation factor, that given by \eq{results2DecayPowerLaw}.  This form has a few technical advantages over the previously discussed methods.  The first is that it requires a fit of only three parameters and the fit is linear in each of them when performed on the logarithmic scale.  Hence, only one optimal solution exists and a number of algorithms exist for arriving there \cite[Ch.\ 15]{press92}.

Secondly, it is a single continuous function so it requires no manual searching for a crossover point between two regimes (which is an art in itself).  In fact, it automatically determines the crossover from power law behaviour to exponential decay with the fitted parameter $r_c$.  The larger $r_c$ is, the greater range the power law is valid over.

\begin{figure}
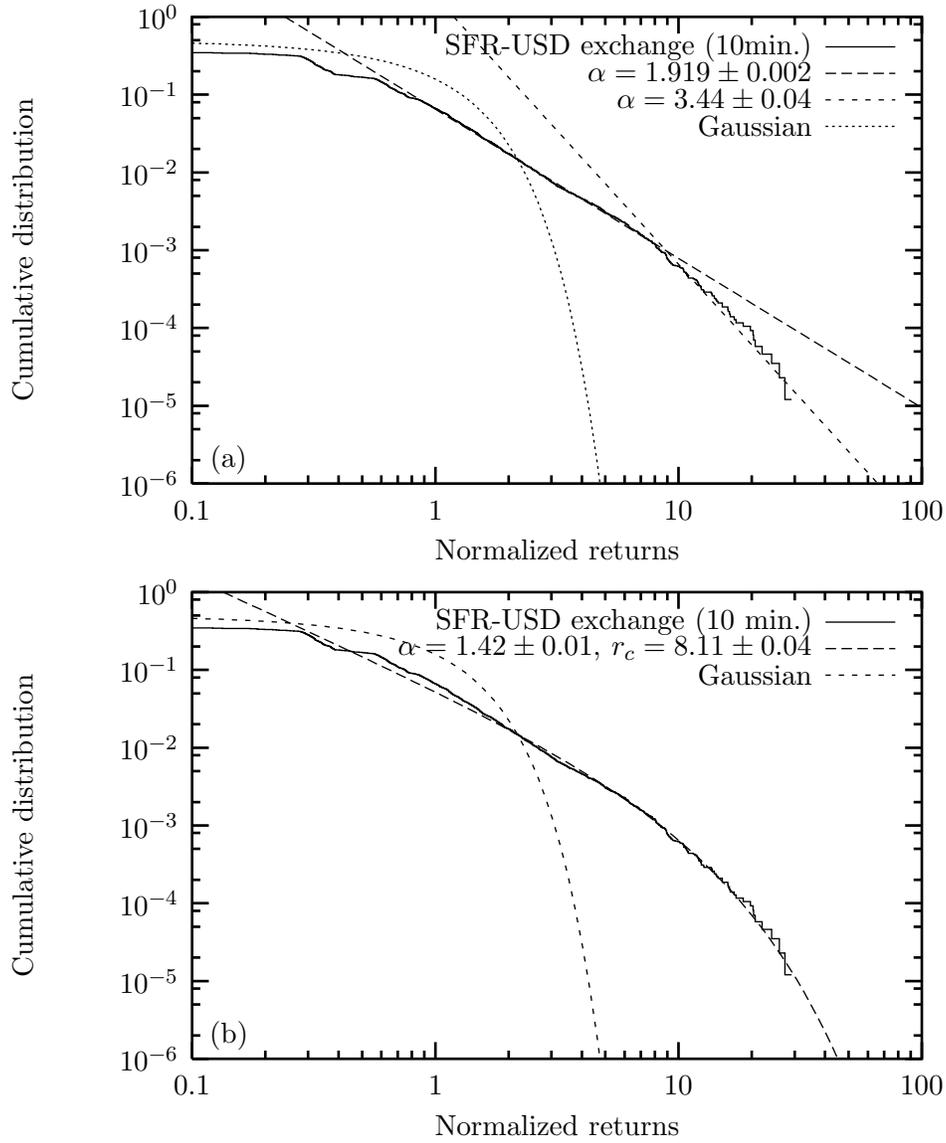
\centering
	\input{chResults2/sfr-usd10minPowNeg.tex} \\
	\input{chResults2/sfr-usd10minExpNeg.tex}
	\caption{Ten minute returns (86,000 data points) of the Swiss franc--U.S. dollar exchange rate \cite{weigend91} (negative tail) compared to power law with crossover to $\alpha\approx 3$ (a) and power law with exponential drop-off presented in this section (b).}
\label{fig:results2SfrUsdReturns}
\end{figure}

\fig{results2SfrUsdReturns} contrasts the fit of the inverse cubic hypothesis (a) with the decaying power law (b).  Both fit the high frequency exchange data quite well, but recall the inverse cubic requires two additional parameters to do so.  The decaying power law hypothesis indicates a power law with exponent $\alpha\approx 1.42$ applies for returns less than $r_c \approx 8.11$ (standard deviations), beyond which the power law is attenuated by an exponential decay.

\begin{figure}
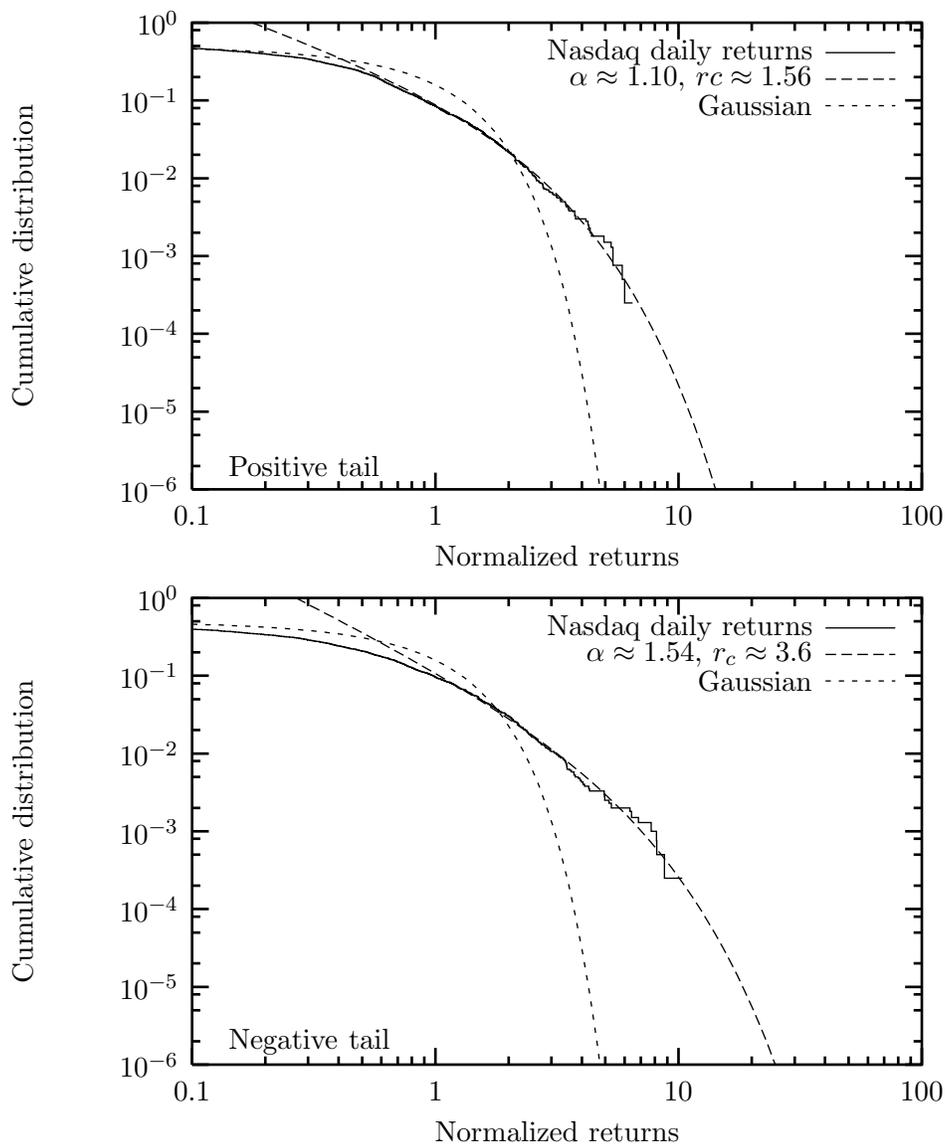
\centering
	\input{chResults2/nasdaqDailyPos.tex} \\
	\input{chResults2/nasdaqDailyNeg.tex}
	\caption{Both tails of the cumulative distribution of daily (normalized) returns for the Nasdaq Composite index between October 1984 and Jun 2000 (4,000 data points) fit well to a decaying power law.  The power law is truncated by two standard deviations in the positive tail but extends almost to four in the negative tail.}
\label{fig:results2NasdaqReturns}
\end{figure}

Also shown is the distribution of daily returns for the Nasdaq Composite index over almost 16 years \cite[ticker symbol=$^\wedge$IXIC]{yahoo99} in \fig{results2NasdaqReturns}.  This figure demonstrates an important point: when the crossover to the exponential occurs at a small value of $r_c$ (less than a few standard deviations, as in the positive tail) the estimate of the L\'evy exponent is unreliable.  The larger $r_c$ gets, the more meaningful the value of $\alpha$ becomes.

While the claim that real market fluctuations actually obey \eq{results2DecayPowerLaw} is largely unsubstantiated, a weaker claim that this functional form is an effective method to test for scaling in market data is also being made based on two observations: (1) it is systematic and does not require any intervention (tuning of parameters) on the part of the researcher, and (2) it characterizes both the range and exponent of the L\'evy region well with the parameters $r_c$ and $\alpha$, respectively.  For these reasons this method will be used in the following sections to test for scaling in CSEM and DSEM.

\subsection{Methodology}

To determine the distribution of returns over some time interval $\Delta t$ the price series will be regularly sampled producing a series of returns
\begin{equation}
	r_i \equiv \ln \left[ \frac{p((i+1)\Delta t)}{p(i\Delta t)} \right].
\end{equation}
The mean $\bar{r}$ and standard deviation $\sigma_r$ will be computed and the returns normalized
\begin{equation}
	\hat{r}_i \equiv \frac{r_i - \bar{r}}{\sigma_r}.
\end{equation}

The cumulative distribution of the normalized returns will be calculated and compared with the cumulative Gaussian (the error function).  Of particular interest are the tails of the distribution which are hypothesized to obey the scaling functions
\begin{eqnarray}
	C(\hat{r}_i \geq \hat{r}) \sim & \hat{r}^{-\alpha_+}\exp(-\hat{r}/r_{c,+}), & \; \hat{r}\rightarrow +\infty \\
	C(\hat{r}_i \leq \hat{r}) \sim & \abs{\hat{r}}^{-\alpha_-}\exp(-\abs{\hat{r}}/r_{c,-}), & \; \hat{r}\rightarrow -\infty
\end{eqnarray}
with exponents $\alpha_+$ and $\alpha_-$ for the positive and negative tails and crossover values $r_{c,+}$ and $r_{c,-}$.  (The cumulative distribution is preferred because cumulating effectively ``smooths'' the data, making it more amenable to analysis.)

The adjustable parameters $\alpha_\pm$ and $r_{c,\pm}$ will be acquired via a Levenburg-Marquardt nonlinear fit to $\log C$ for returns exceeding $\abs{\hat{r}}>1$ since we only want to fit the tails.  (A linear fit is also possible with a suitable choice of parameters.)

A small value of $r_c$ will be interpreted to mean that no scaling exists and the parameter $\alpha$ is irrelevant.

\subsection{Centralized stock exchange model}

For this experiment we return to Dataset 1 (\tbl{resultsCsemDataset1}) and apply the analysis to the largest system $N=1000$.  Fortunately, each run consists of over 30,000 days worth of data so the scaling can be tested on a wide range of timescales.  (In the analysis, the initial transient will be discarded.)

Since CSEM contains the free parameter $\sigma_\epsilon$ (the forecast error) we must sample a suitable spectrum of values in our search for scaling.  Obviously, the dynamics around the critical point $\sigma_c\approx 0.08$ (from \fig{resultsCsemPowerN}(a)) is of particular interest so samples are chosen which span the critical point.

The scaling regime is indicated by the characteristic return $r_c$: a small value indicates that the exponential drop-off occurs for small returns (before scaling becomes evident) and a large value indicates the power law applies over a broad range of returns.  In this experiment a threshold value of $r_c=3$ was observed to adequately distinguish between distributions which scaled and those which didn't.  This limit was also used in Ref.\ \cite{gopikrishnan99} to estimate the scaling exponent $\alpha$.

\begin{figure}\centering
	\input{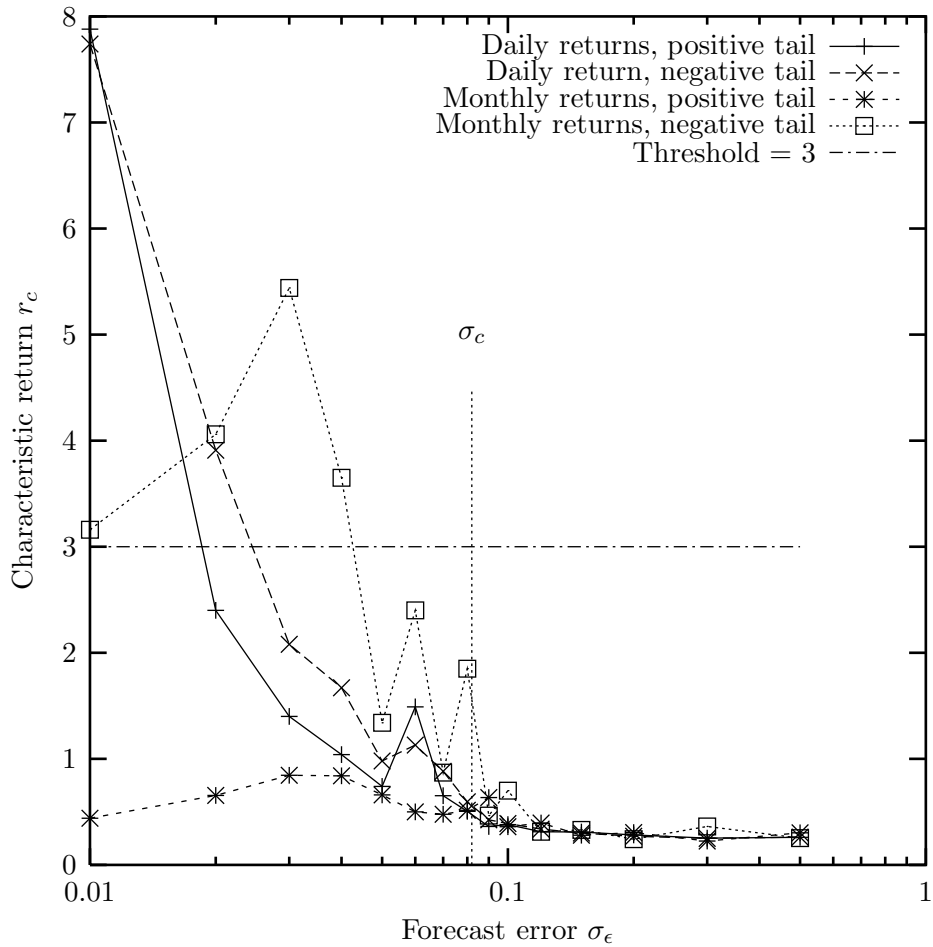}
	\caption{Scaling in the distribution of returns is only observed well below the critical point $\sigma_\epsilon\ll \sigma_c$ in CSEM as indicated by large values of the characteristic return $r_c$.  For small $\sigma_\epsilon$ scaling occurs in both tails for daily returns but only for negative returns in monthly returns.}
\label{fig:results2CsemCharReturns}
\end{figure}

Plotting the characteristic return $r_c$ for a variety of forecast errors $\sigma_\epsilon$ (\fig{results2CsemCharReturns}) shows that scaling is only observed well below the critical point $\sigma_c$.  The average scaling exponent for all distributions with $r_c>3$ is $\alpha=0.8\pm 0.4$, which compares poorly with the empirical value $1.4\leq \alpha \leq 1.7$.  

For positive returns scaling is found to disappear as the sampling interval is increased from daily to monthly (20 days), as expected.  Interestingly, the same is not true for negative returns: instead the returns scale for even more values of $\sigma_\epsilon$ on a monthly timescale than they do daily.  

\begin{figure}
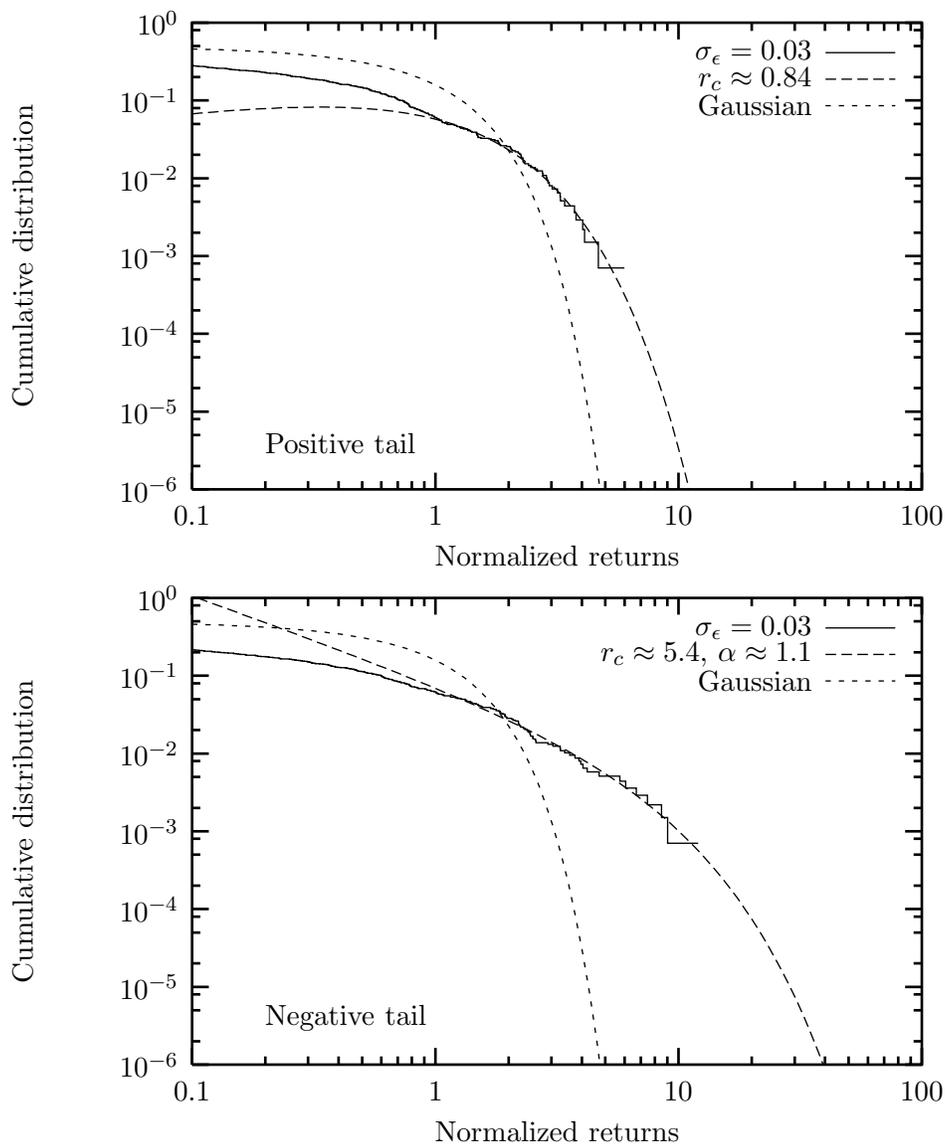
\centering
	\input{chResults2/csemN1000s0.03CumRetPos.tex}\\
	\input{chResults2/csemN1000s0.03CumRetNeg.tex}
	\caption{For $\sigma_\epsilon=0.03$ in CSEM ($N=1000$) the distribution of positive (monthly) returns (upper) almost converges to a Gaussian but still has a slightly heavy tail.  The negative returns (lower), however, exhibit scaling for $r<r_c\approx 5.4$ with an exponent $\alpha\approx 1.1$. }
\label{fig:results2CsemS0.03ReturnDist}
\end{figure}

The run $\sigma_\epsilon=0.03$ sampled monthly is an interesting case because it exhibits a strong asymmetry between up and down moves, having a characteristic return above the threshold for negative returns and below for positive returns, so its return distribution is plotted in \fig{results2CsemS0.03ReturnDist}.  This effect is due to an asymmetry between up- and down- movements which arises from the artificial price cap imposed by the parameter $\delta$.  See, for example, \fig{resultsCsemPhases}(b) which shows that 
occasional large crashes occur when the price approaches its upper limit while upwards movements are more normally distributed.

Unfortunately, the above only serves to further call in to question the validity of the CSEM model because scaling behaviour is only observed below the critical point, in the regime where we have already seen (\fig{resultsCsemPhases}, for instance) the dynamics are completely unrealistic.  So we turn to DSEM in the hope that it is a more realistic model of market dynamics.

\subsection{Decentralized stock exchange model}

In this section the distribution of returns in DSEM is analyzed.

\begin{table}
	\begin{center}\begin{tabular}{r|c|c|c}
		\hline \hline
		Parameters & \multicolumn{3}{c}{DSEM Dataset 3} \\
		\hline 
		\multicolumn{4}{c}{Particular values} \\
		\hline
		News response $r_n$ & 0.01 & 0.01 & 0.001 \\
		Price response $r_p$ & -0.75 to 0.75 by 0.25 & 0.95 & 0.99 \\
		Number of runs & 7 & 1 & 1 \\
		\hline 
		\multicolumn{4}{c}{Common values} \\
		\hline
		Number of agents $N$ & \multicolumn{3}{c}{100} \\
		Run length (``days'') & \multicolumn{3}{c}{20,000} \\
		\hline \hline
	\end{tabular}\end{center}
	\caption{Parameter values for DSEM Dataset 3. These runs are a variation of Dataset 1 (all unspecified parameters are duplicated from
 \tbl{resultsDsemDataset1}) run out for longer times (roughly 80 years).  Also notice that for $r_p=0.99$ the news response was reduced by an order of magnitude to keep the price within reason.}
\label{tbl:results2DsemDataset3}
\end{table}

To get the limit distribution a large quantity of data is required so DSEM was run with the parameter values from \tbl{results2DsemDataset3}, the most notable feature being that the run length was extended from 1,000 days ($\approx 4$ years) to 20,000 days ($\approx 80$ years).

\begin{figure}\centering
	\input{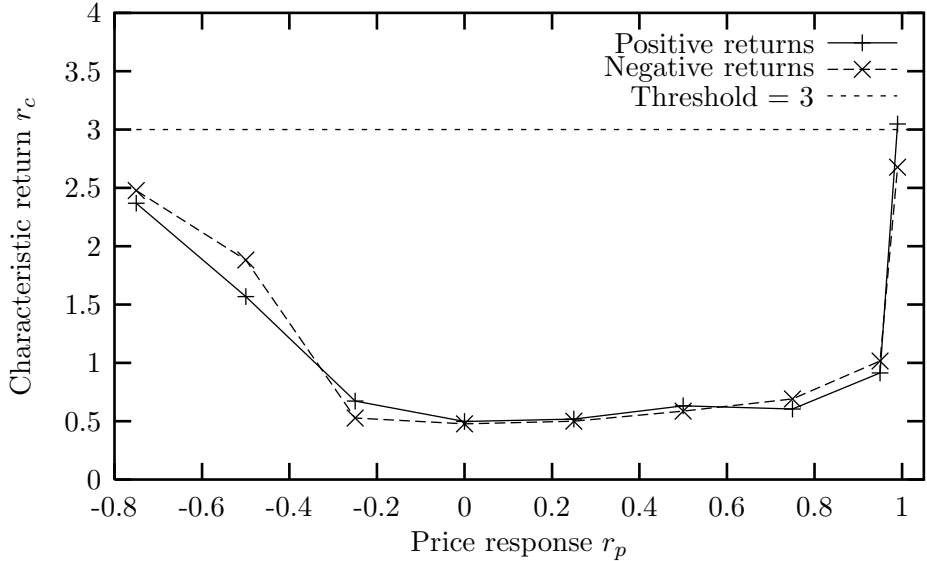}
	\caption{DSEM only begins to exhibit scaling, as measured by a characteristic return exceeding three standard deviations, for price responses well below the first-order transition $r_2=-0.33$ and as the price response approaches the critical point $r_1=1$.}
\label{fig:results2DsemCharReturns}
\end{figure}

The distribution of price returns (sampled ``daily'') was then cumulated and fitted with \eq{results2DecayPowerLaw}.  (Issues raised in \ap{sampling} regarding sampling are addressed on page \pageref{sect:results2DsemTickReturns}, below.)  The characteristic size of the returns for the different values of $r_p$ is shown in \fig{results2DsemCharReturns}.  Notice that they are almost exclusively below the threshold required to establish scaling indicating that the distributions do not exhibit scaling properties observed empirically.  The worst region appears to be intermediate values of $r_p$ with better performance near the endpoints.

Recall that DSEM exhibits three behaviours as $r_p$ is varied: (1) when $r_p>r_1=1$ the price is perfectly autocorrelated---every movement is followed by another (typically larger) movement in the same direction; (2) in the intermediate region $r_1>r_p>r_2$ the price series looks most realistic and has (at most) weak autocorrelations on long timescales; and (3) when $r_p<r_2\approx -0.33$ the price fluctuations have a strong negative autocorrelation extending over all timescales.  Thus, the price fluctuations appear only to obey (realistic) scaling distributions in the domains precisely where the dynamics were observed to be unrealistic!  To reconcile this dichotomy we need to expand our experimental parameter space.

\subsubsection{Two-point price response}

Thus far the price response had been fixed at a single value for all the agents.  But since realistic dynamics (characterized by both the lack of strong memory effects and scaling in the distribution of returns)  were not to be obtained by any single value of $r_p$ I was forced to allow multiple price responses.  Originally I explored allowing $r_p$ to span a broad range which covered all three phases but the range required to get scaling was so large that most of the agents were either in Phase 1 ($r_p>r_1$) or in Phase 3 ($r_p<r_2$) with only a few in Phase 2.  Therefore it seemed easier to just require that $r_p$ take on one of only two allowed values, $r_{lo}$ and $r_{hi}$.
\subsubsection{Data analysis}

\begin{table}
	\begin{center}\begin{tabular}{r|c}
		\hline \hline
		Parameters & DSEM Dataset 4 \\
		\hline
		Number of agents $N$ & 100 \\
		Lower price response $r_{lo}$ & --0.75 to 0.75 by 0.25 \\
		Higher price response $r_{hi}$ & 0.50 to 1.50 by 0.25 \\
		Number of runs & 35 \\
		Run length (``days'') & 20,000 \\
		\hline \hline
	\end{tabular}\end{center}
	\caption{Parameter values for DSEM Dataset 4. These runs are characterized by a two-point distribution of the price response.  Each agent chooses $r_p=r_{lo}$ or $r_{hi}$ with equal probability.  (All unspecified parameters are duplicated from
 \tbl{resultsDsemDataset1}.)}
\label{tbl:results2DsemDataset4}
\end{table}

Thirty five runs were executed spanning a two-dimensional region of parameter space with each agent choosing a price response of either $r_{lo}$ or $r_{hi}$ (with equal probability).  The lower price response was varied between $-0.75\leq r_{lo} \leq 0.75$, spanning the first order phase transition at $r_2\approx -0.33$, and the upper value was varied between $0.50 \leq r_{hi} \leq 1.50$, spanning the critical point at $r_1=1$, as indicated in \tbl{results2DsemDataset4}.

For each run the cumulative distributions of returns (both positive and negative) were calculated and the tails (returns exceeding one standard deviation) fitted to a decaying power law (\eq{results2DecayPowerLaw}).  As before, the tail was determined to scale if the decay constant $r_c$ exceeded three standard deviations, otherwise the region over which a power law is suitable is insubstantial.

\begin{figure}
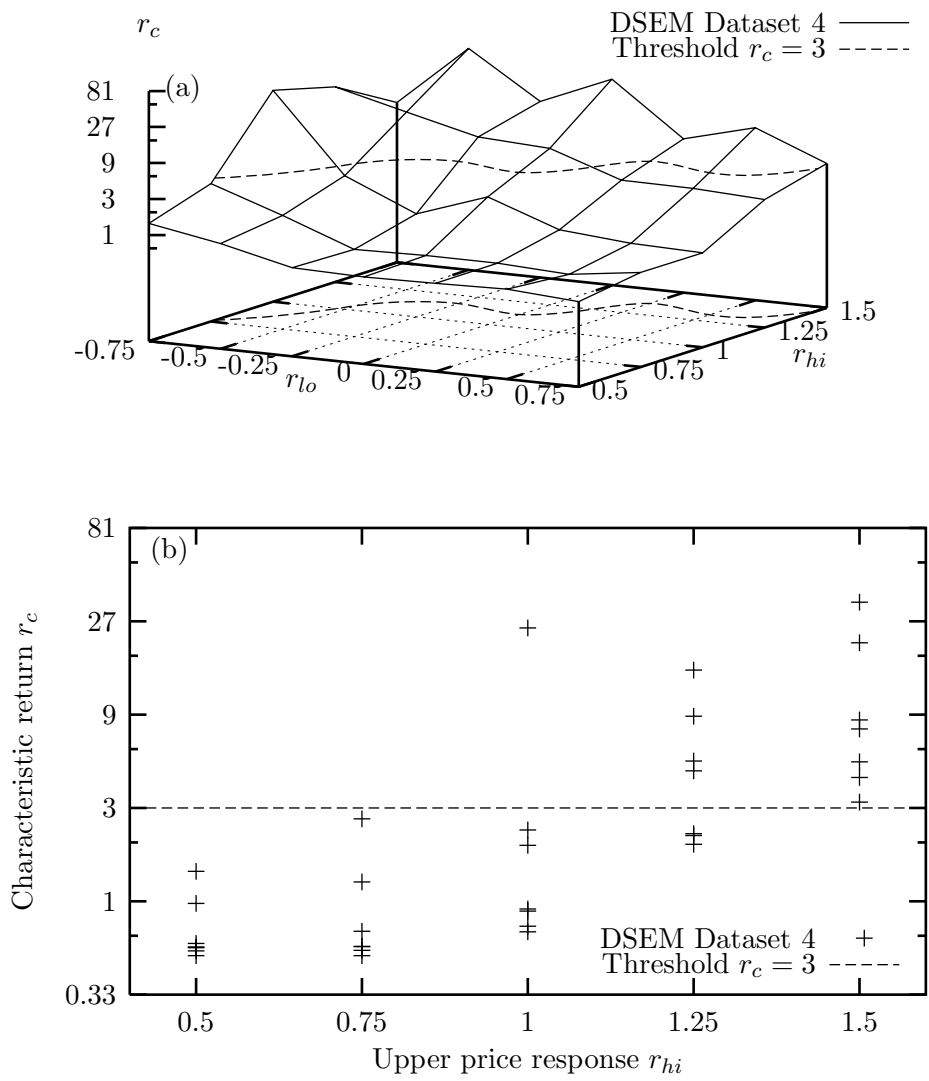
\centering
	\input{chResults2/dsemReturnTails3d} \\
	\input{chResults2/dsemReturnTailsSpread}
	\caption{The characteristic returns in DSEM with a two-point distribution of price responses ($r_{lo}$ and $r_{hi}$) exceeds the required threshold of $r_c=3$ when $r_{hi}$ is large (a).  Neglecting the dependence on $r_{lo}$ (b) it becomes clear that the characteristic return grows exponentially with the upper limit $r_{hi}$, crossing the threshold near $r_{hi}\approx 1$.}
\label{fig:results2DsemReturnTails}
\end{figure}

\begin{table}
	\begin{center}\begin{tabular}{c|c}
		\hline \hline
		Variable & Correlation with $\log r_c$ \\
		\hline 
		$r_{hi}$        & 78\% \\
		$r_{hi}-r_{lo}$ & 76\% \\
		$r_{lo}$        & --41\% \\
		\hline \hline
	\end{tabular}\end{center}
	\caption{Linear correlation analysis between said variable and the logarithm of the characteristic return from DSEM Dataset 4.  The correlation is strongest with the upper limit of the price response $r_{hi}$.}
\label{tbl:results2DsemScaling}
\end{table}

For returns smaller than the characteristic return $r_c$ the exponential in \eq{results2DecayPowerLaw} is almost flat so the power law dominates.  Larger values of $r_c$ indicate that scaling spans a greater range of returns.  As shown in \fig{results2DsemReturnTails}(a), scaling is observed for some parameter combinations.  To test which parameter produces scaling a linear correlation analysis was performed, as shown in \tbl{results2DsemScaling}, between the logarithm of the characteristic return and a few obvious possibilities: the upper price response $r_{hi}$, the lower limit $r_{lo}$, and the spread $r_{hi}-r_{lo}$.  The best predictor for scaling over a large range of returns was found to be $r_{hi}$ with a correlation of 78\%.  (The spread also correlated well but, as will be seen later in this chapter, is unable to account for other empirical qualities of the market.)  

\fig{results2DsemReturnTails}(b) shows the dependence of the characteristic return on the upper price response.  Notice that 85\% of the data points lie in the upper-right and lower-left quadrants if axes are drawn at $r_c=3$ (horizontal) and $1<r_{hi}<1.25$ (vertical).  Thus, the strongest condition for scaling appears to be that the upper price response $r_{hi}>1$, above the critical point $r_1=1$.

Of all the runs which exhibit scaling the average scaling exponent was calculated to be $\alpha=1.64\pm 0.25$, in line with the empirical value $\alpha\approx 1.40 \pm 0.05$ \cite{mantegna95}.

\subsubsection{Recanting continuous heterogeneity}

Some other market models characterize agents by types: either {\em fundamentalists} or {\em chartists}.  In the derivation of DSEM I claimed (\sect{dsemPriceResponse}) that allowing a continuous range of the parameter $r_p$ would be superior, reflecting a greater diversity of opinion as would be expected in the real world.  However, as we saw above, DSEM is only able to capture the essence of real market fluctuations (scaling) when $r_p$ is set to two discrete values, rather than a continuum.  (As mentioned before, a continuous range of $r_p$ can also produce scaling but only if the spread is set to a much greater value than required by the two point distribution.)  In other words, scaling appears to depend on the separation of agents into ``types''.

\subsubsection{Timescales}

An interesting empirical property of scaling in real markets is that the exponent appears to be invariant when measured on different timescales (except for timescales exceeding a few days when the distribution converges to a Gaussian).  To test if this also occurred in DSEM the run $r_{lo}=0.00$, $r_{hi}=1.25$ was chosen for further analysis.  This run was chosen because it was observed to exhibit scaling on timescales of one day, and because a value of $r_{lo}=0$ seems ``natural''---it divides the population into two types: one of which are pure fundamentalists, not responding to price fluctuations at all. 

\begin{figure}
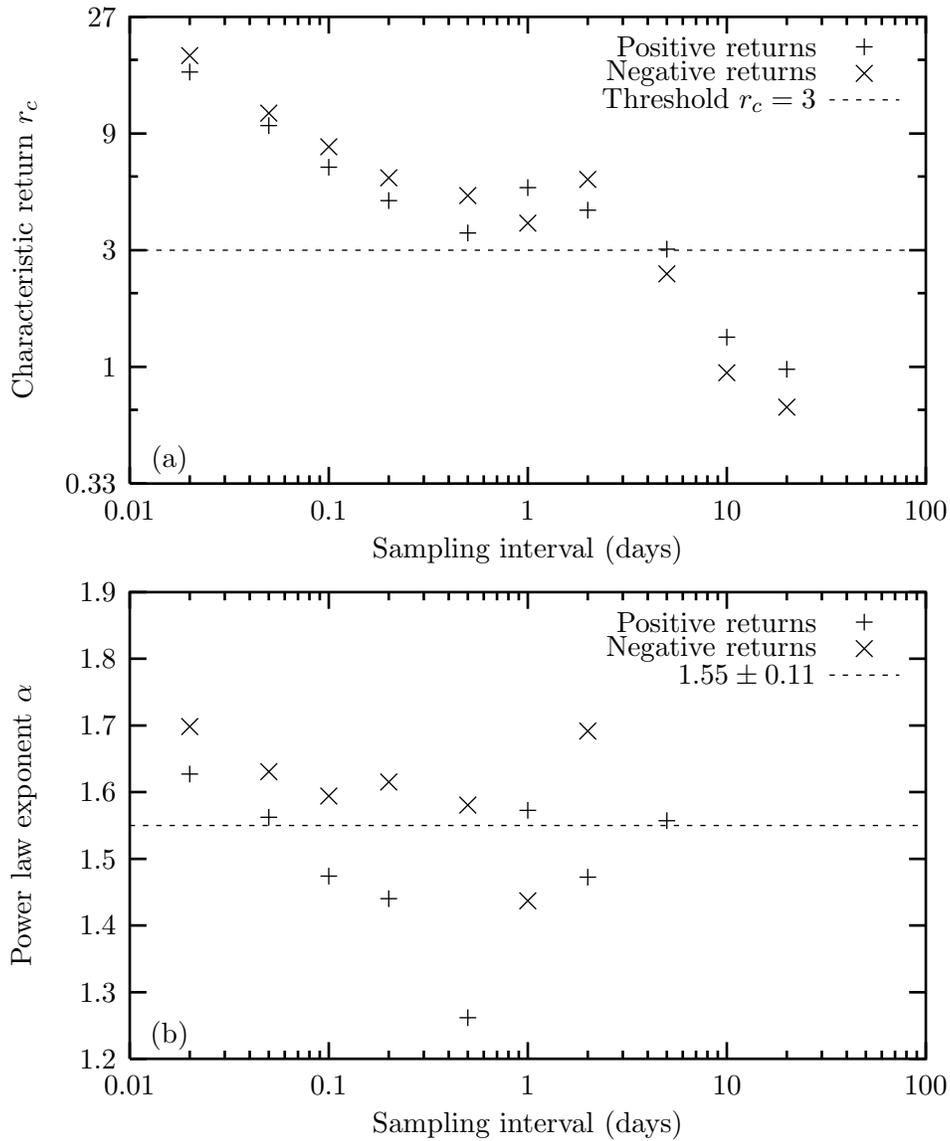
\centering
	\input{chResults2/dsemRp0.00and1.25Rc.tex} \\
	\input{chResults2/dsemRp0.00and1.25Alpha.tex}
	\caption{The characteristic return $r_c$ (a) and scaling exponent $\alpha$ (b) for DSEM with $r_{lo}=0.00$ and $r_{hi}=1.25$.  The characteristic return grows as the sampling interval is shortened, but the scaling exponent $\alpha$ is fairly constant ($1.55\pm0.11$).}
\label{fig:results2DsemRp0.00and1.25}
\end{figure}

This run was sampled at ten different intervals ranging from 0.02 days (roughly 8 minutes, assuming a 6.5 hour trading day), to 20 days (one month, neglecting weekends).  As can be seen in \fig{results2DsemRp0.00and1.25}(a) the characteristic return increases with smaller sampling intervals, and drops below the threshold for detecting scaling when the interval exceeds 5 days (one week).  On longer timescales the distribution indeed converges to a Gaussian (not shown).

As expected, (when scaling is detectable) the power law exponent does not appear to depend on the sampling interval, fluctuating around $\alpha=1.55\pm 0.11$.  

\subsubsection{Tickwise returns}

\label{sect:results2DsemTickReturns}

\ap{sampling} demonstrates that it is possible to generate the illusion of fat tails in a discrete Brownian process simply by sampling it at regular intervals, as was done for DSEM.  Therefore, it is important to establish that the fat tails discussed above are not an artifact of sampling, but are inherent to the fluctuations themselves.  This is easily tested by simply sampling the process in {\em trading time} rather than real time.  That is, a sample is taken directly after every trade (or {\em tick}).  

\begin{figure}\centering
	\input{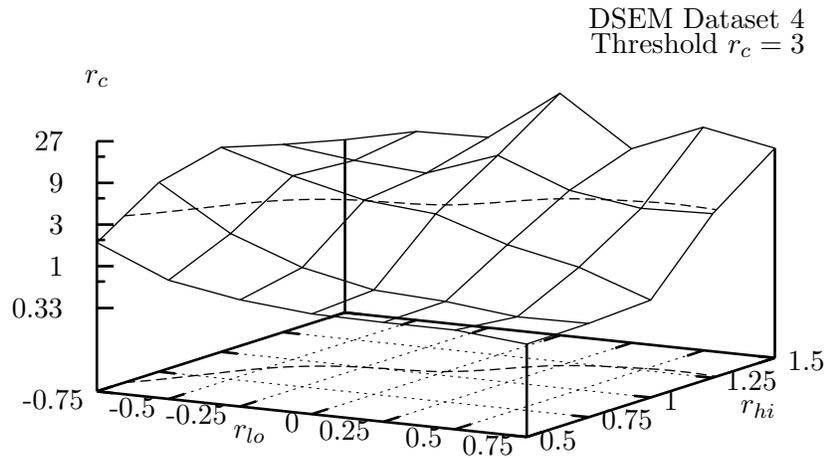}
	\caption{Fitting the decaying power law to DSEM with a two-point price response using returns on individual trades (rather than per unit time, as in \fig{results2DsemReturnTails}) shows scaling still occurs in the same region of parameter space.}
\label{fig:results2DsemReturnTailsDt0}
\end{figure}

Clark \cite{clark73} raised the issue of whether regular sampling may be producing the fat tails observed empirically but Pal\'agyi and Mantegna \cite{palagyi99} demonstrated fat tails are still observed when sampled in trading time.  To test if this was also the case for DSEM \fig{results2DsemReturnTails}(a) was reproduced, using trading time instead of daily samples, in \fig{results2DsemReturnTailsDt0}.  Clearly, scaling is still evident (in the same region of parameter space) when sampling in trading time so it is not an artifact of the sampling interval.

\subsection{Summary}

The distribution of price returns, measured as the logarithm of the ratio of successive prices, was the subject of investigation in this section.

Two theoretical curves meant to describe the tails of the distribution were presented.  An alternate form was also presented, whose main advantages are that it is linear in its parameters (of which there are two fewer than the competing models).  This curve appears to describe empirical data quite well, but even if it is found to be inaccurate, it is still useful because it provides a simple, mechanical method for estimating over what range the power law dependence applies ($\abs{r}<r_c$) and the scaling exponent itself.

Both CSEM and DSEM were tested for ``fat tails'' with this functional form with the requirement that the scaling extend for at least three standard deviations ($r_c\geq 3$) to be deemed significant.  CSEM was found only to exhibit scaling for $\sigma_\epsilon\ll \sigma_c$, in a region of parameter space where the dynamics are known to be unrealistic.

DSEM provided some surprises: if all the agents maintained an identical price response parameter $r_p$ then scaling did not occur except as $r_p$ moved into regions known to produce unrealistic dynamics.  However, if two values of $r_p$ were allowed, with each agent randomly picking one or the other, scaling was observed when the responses spanned the critical point $r_p=r_1=1$.  A test for a variety of values of $r_p$ established a scaling exponent $\alpha=1.64\pm 0.25$, with returns sampled daily, comparing favourably with the empirical quantity $\alpha=1.40\pm 0.05$ \cite{mantegna95}.  The scaling exponent was shown to be robust, independent of the sampling interval.

In short, DSEM was able to produce realistic return distributions while CSEM was not.  Furthermore, DSEM implied the mechanism which produces scaling is somehow related to having different ``types'' of agents interacting with each other: fundamentalists versus chartists, for instance.  The crucial determinant for scaling appeared to be that the range of parameter values spanned the critical point.

In the next section we explore a related phenomenon: autocorrelations in the price series.

\section{Price autocorrelation}

\subsection{Background: The efficient market hypothesis}

In this section we explore the possibility of serial correlations in the price series.  As discussed in the last section it has long been thought that the market behaves as a random walk \cite{bachelier00, osborne59}.  Related to this is the {\em efficient market hypothesis} (EMH) which, in its weakest form, states that new information received by investors is reflected in the stock's price almost instantly \cite{zhang99}.  Since new information cannot be predicted, neither can the future price of the stock so price movements should be independent of their histories.  If this were not the case then there would be a riskless way to exploit one's foreknowledge for profit (an {\em arbitrage} opportunity).  

The presence of transaction costs allows an even weaker form of the EMH: there may exist arbitrage opportunities (autocorrelations) but they are so small (brief) that any potential profit would be absorbed by commissions.  This form of the EMH is supported by evidence: a number of studies have concluded that autocorrelations in the price series decay exponentially over a scale of only a few minutes \cite{lo91, cizeau97, cont97b, liu97, rotyis97, arneodo98, busshaus99, lux99, mantegna99} (or a few trades \cite{palagyi99}).

In this section we will look for correlations in the price series generated by DSEM.  (Since it was established in the last section that CSEM is not a realistic market model an autocorrelation analysis of its price series will be dispensed with.)  Of interest are both short- and long-range correlations.

\subsection{News}

Before analyzing the correlations in the price series it should be reiterated that the price series is driven by news releases such that the logarithm of the price $p(t)$ roughly follows the cumulative news $\eta(t)$ as
\begin{equation}
	\log p(t) \propto \eta(t).
\end{equation}

\begin{figure}\centering
	\input{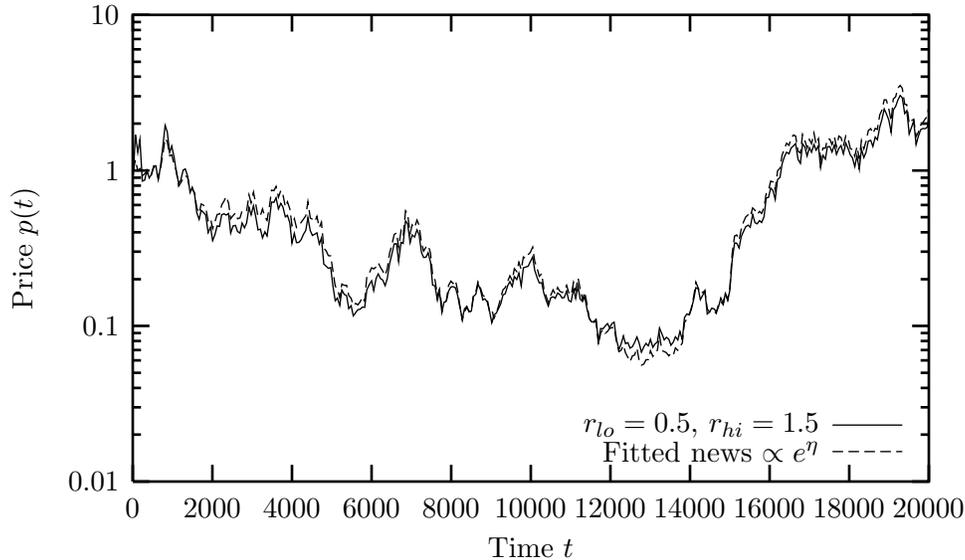}
	\caption{Sample price series for DSEM Dataset 4 ($r_{lo}=0.5$, $r_{hi}=1.5$) showing the price roughly tracks the exponential of the cumulative news $e^\eta$.  The proportionality constant is estimated from the data.}
\label{fig:results2DsemPriceNews}
\end{figure}

In \sect{dsemNewsResponse} it was argued that the proportionality constant should be $r_n/(1-r_p)$ but with a two-point price response distribution this is not applicable, especially given that the upper limit can be $r_{hi}\geq 1$.  Nevertheless the relation still holds but the proportionality constant is best estimated from the data as is done in \fig{results2DsemPriceNews}.

Since the news is a discrete Brownian motion we may naively expect the price series to be a simple geometric Brownian motion but as we have already seen the price series exhibits an abundance of outliers not observed in the news.  As we will see in the next section, the price series also contains a memory which the news does not.

\subsection{Short timescales}

\begin{figure}\centering
	\input{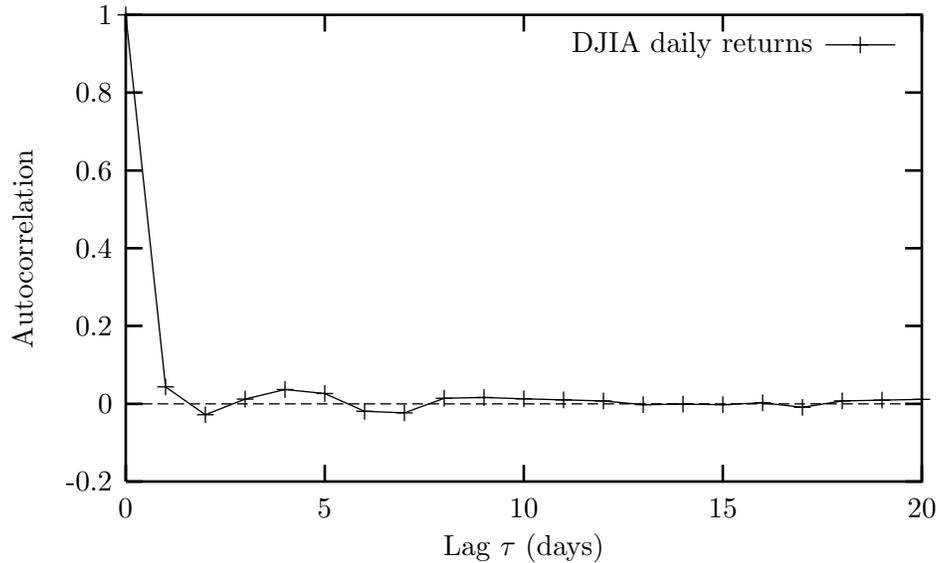}
	\caption{The autocorrelation between daily returns for the Dow Jones Industrial Average \cite{economagic99} decays rapidly to zero with an estimated characteristic timescale $\tau_c=0.4\pm 0.2$ days.  (Being less than the sampling interval, this estimate is not precise.)}
\label{fig:results2DjiacAutoCorr}
\end{figure}

The analysis for short timescales is fairly straightforward.  We need only compute the autocorrelation between returns for different lags.  For this analysis, {\em trading time} (or {\em ticks}), defined as the number of transactions executed, will be used as the time index since the quantity of interest is the correlation between successive trades.  For comparison, the autocorrelations between daily returns for the Dow Jones Industrial Average for the last hundred years \cite{economagic99} is shown in \fig{results2DjiacAutoCorr}, indicating no significant correlations in support of the efficient market hypothesis.  (Of course, this does not preclude correlations existing on timescales of less than one day but data were not available at this resolution.)

\begin{figure}\centering
	\input{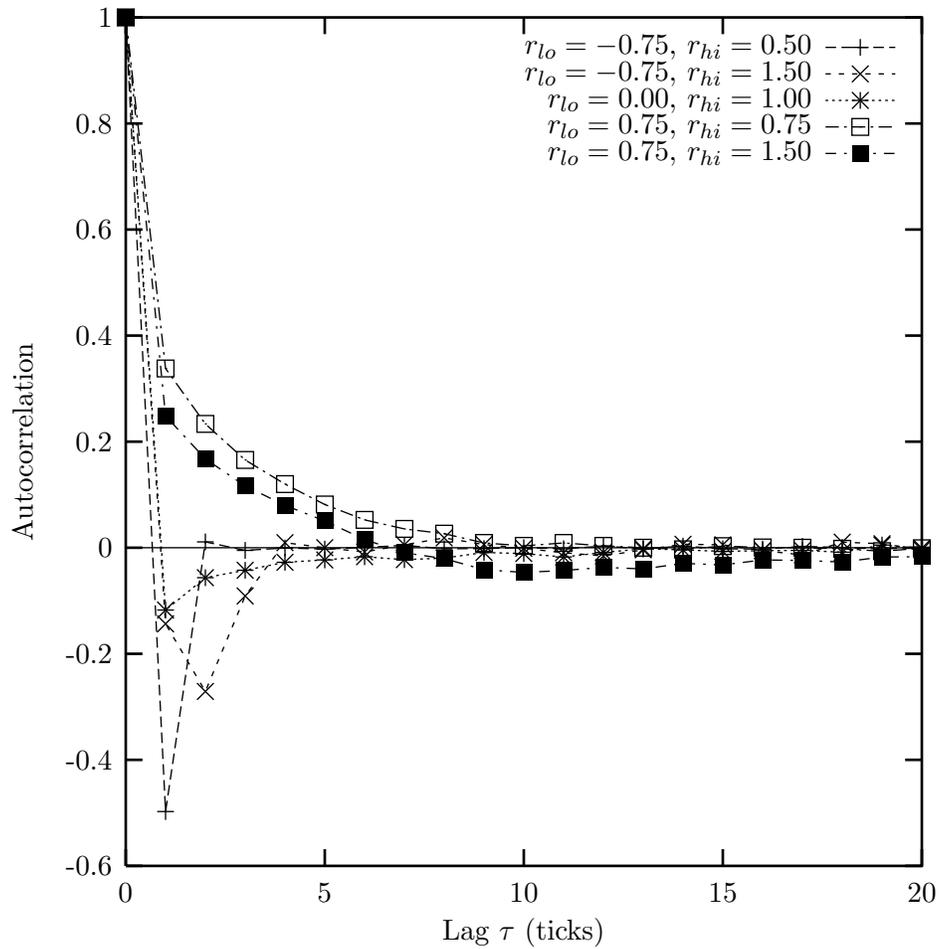}
	\caption{The autocorrelation between tickwise returns for DSEM (with a two-point price response distribution) decays rapidly to zero for all runs sampled.}
\label{fig:results2DsemAutoCorr}
\end{figure}

As demonstrated in \fig{results2DsemAutoCorr} correlations also decay quickly for DSEM regardless of the value(s) of the price response, with correlations only evident over a few successive trades.  This would seem to imply that the price series has no memory.  However, recall that in \sect{resultsDsemPhases} we observed two phase transitions in DSEM by directly measuring the memory of the price series, which challenges the results presented here.  

There are two possible reasons for the discrepancy: (1) the autocorrelations are measured tickwise whereas the Hurst exponent was originally measured from a daily sample, or (2) a plot of the autocorrelation does not fully describe temporal dependencies in the data.  To determine which is the case the long-range dependencies are again estimated from the Hurst exponent, this time calculated from the tickwise data.

\subsection{Long timescales}

To test for long-range temporal dependencies we again compute the Hurst exponents for the price returns in DSEM.  But first it should be mentioned that data from real markets have been found to have no memory, with Hurst exponents near $H\approx 0.5$.  Indeed, for the daily Dow Jones Industrial Average returns presented above, the Hurst exponent is estimated at $H=0.484\pm 0.013$ indicating no long-term memory effects.

\subsubsection{Crossover}

\begin{figure}\centering
	\input{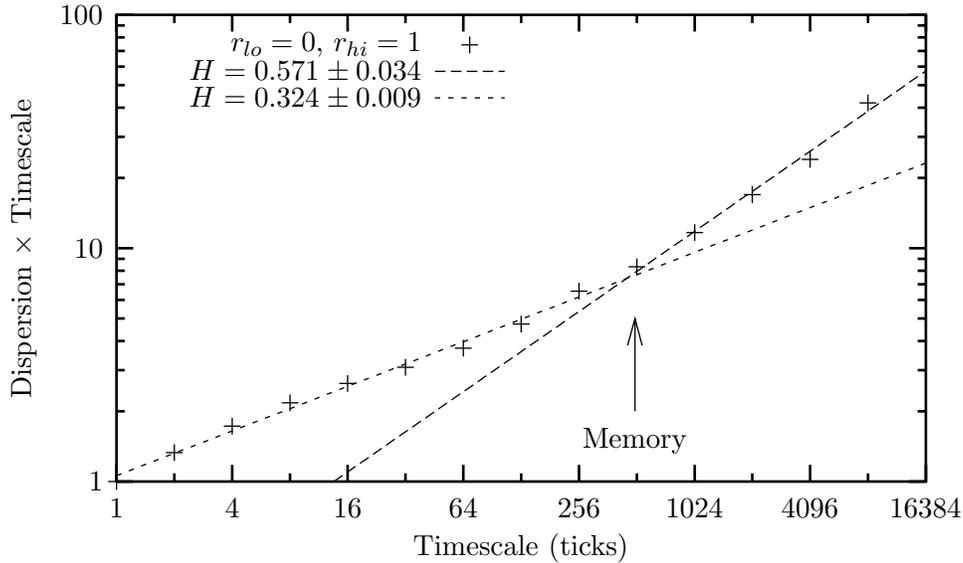}
	\caption{Sample dispersion plot (see \sect{hurstDispersion}) demonstrating the phenomenon of crossover in the Hurst exponent to $H\approx1/2$ on long timescales for DSEM with a two-point price response distribution.}
\label{fig:results2DsemHurstCrossover}
\end{figure}

On first glance the computed Hurst exponents appeared similar to the original results shown in \fig{resultsDsemPriceResponse} but on closer inspection some interesting qualities were revealed.  Namely, in almost all the runs there appeared to be two different scaling behaviours: for small timescales one Hurst exponent dominated but as the timescale grew there appeared a crossover to a different exponent.  The latter of these was invariably near $H=1/2$ indicating a lack of memory.  A sample graph demonstrating crossover is presented in \fig{results2DsemHurstCrossover}.

\begin{figure}\centering
	\input{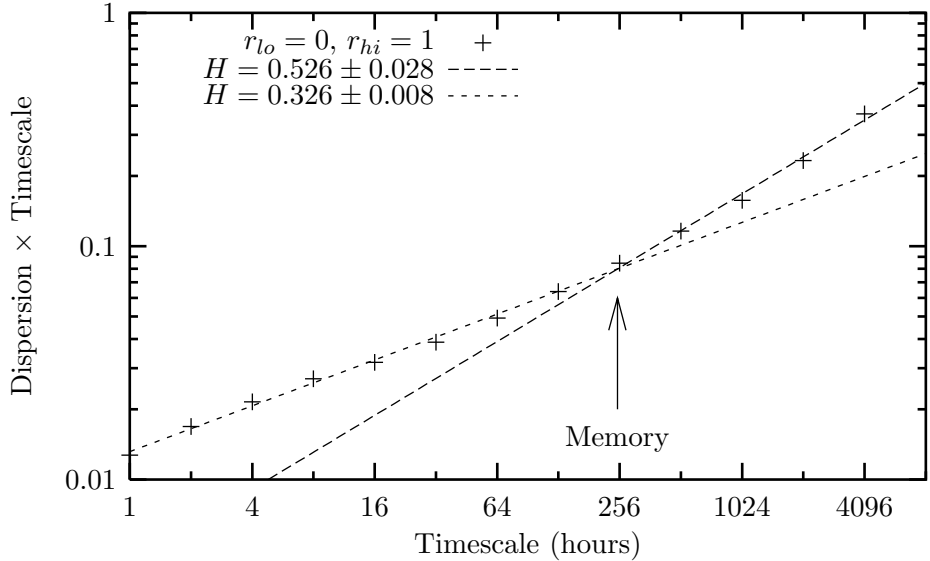}
	\caption{A reproduction of \fig{results2DsemHurstCrossover} except with regularly sampled returns at an ``hourly'' interval (instead of tickwise).  Short timescale anticorrelations crossing over to uncorrelated returns at long timescales are still observed so the effect is not an artifact of sampling tickwise.}
\label{fig:results2DsemHurstCrossover1Hr}
\end{figure}

The reader may be concerned that the timescale used for calculating the memory is not linear but {\em trading time}---the cumulative number of trades executed since the start of the experiment---and the crossover phenomenon may be an artifact of this sampling.  Evidence indicates that trading time is the more natural timescale \cite{clark73, palagyi99}, reducing biases introduced by regular sampling (see \ap{sampling}).  However, for completeness the data were also tested using regular sampling with largely the same results.  A sample plot is shown in \fig{results2DsemHurstCrossover1Hr} demonstrating crossover also occurs when returns are sampled at regular intervals.  The remaining discussion refers to tickwise sampling.

The crossover to $H\approx 1/2$ is not altogether surprising because on long enough timescales we expect the news process to be an important determination of the price movements, and the news is a simple, discretely-sampled Brownian motion with no memory ($H=1/2$).

\subsubsection{Yet another phase transition}

\begin{figure}
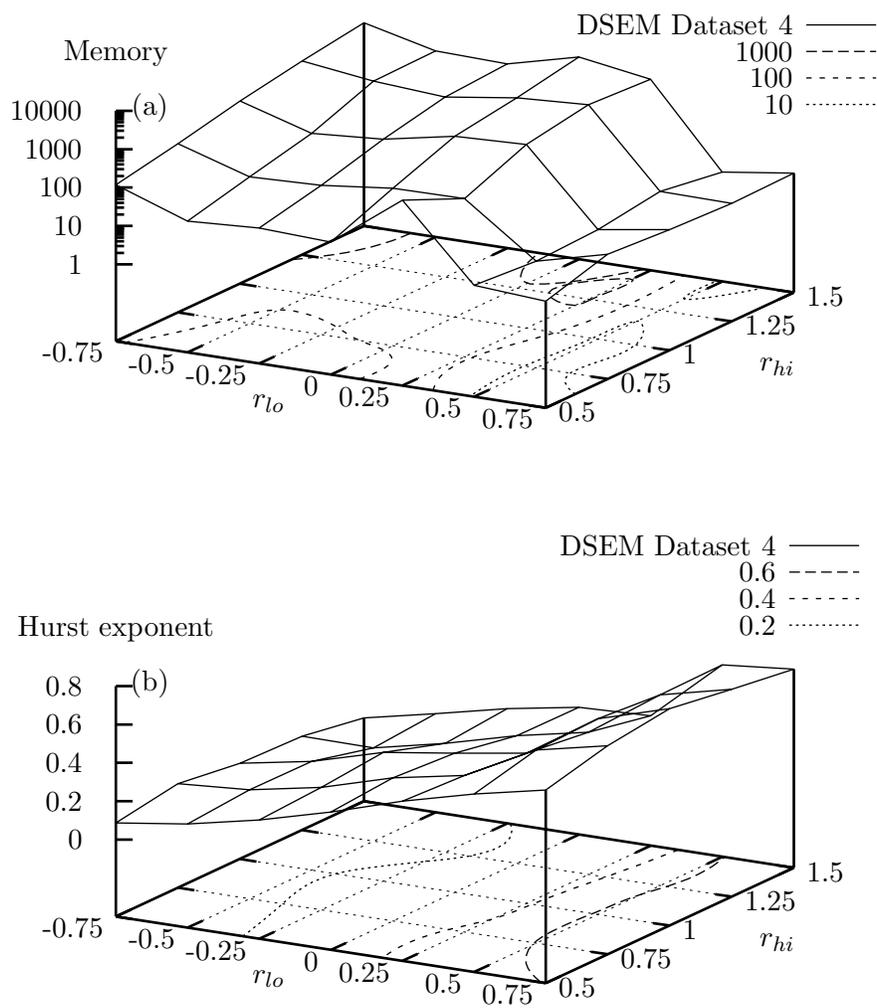
\centering
	\input{chResults2/dsemHurstMemory.tex} \\
	\input{chResults2/dsemHurstExponent.tex}
	\caption{Both the crossover point, or memory, (a) and Hurst exponent for short timescales (b) indicate that memory effects are minimized when $r_{lo}\geq 0.25$ in DSEM with a two-point price response distribution.  (The high values of the Hurst exponent for $r_{lo}>0.5$ (b) do not cause problems because the memory is very short in this region (a).)}
\label{fig:results2DsemHurstMemExp}
\end{figure}

The crossover point gives another estimate of the duration of correlations, or {\em memory} in the price series.  \fig{results2DsemHurstMemExp} shows that the memory depends strongly on the lower limit of the price response $r_{lo}$.  In fact, as this value crosses roughly $r_{lo}\approx 0.5$ a phase transition is apparent.  (While interesting, this transition will not be characterized further in this thesis, but may be analyzed in future work.)  For larger values of $r_{lo}$ the memory effects disappear very quickly, conforming to the efficient market hypothesis and empirical data.  Further, in this range $H$ is already quite close to one half, which also suggests the lack of a memory for any significant period.  (Personal experience suggests that the Hurst exponent has a typical error margin of $\pm 0.1$ so any value in $0.4\leq H\leq 0.6$ should be interpreted as potentially having no memory.)

Thus, the price returns in DSEM are observed to exhibit a realistic lack of (significant) memory when $r_{lo}$ exceeds roughly 0.25.  If DSEM is interpreted as representative of a real market, the question is naturally raised, ``Why do the agents choose this region of parameter space?''  The simplest (but unjustified and hardly satisfactory) explanation is simply that any other choice would provide arbitrage opportunities which could be taken advantage of by watching for trends in the price series.  So a rational agent would choose a nonzero price response in the expectation of these arbitrage opportunities, but ironically, in doing so, the memory (and opportunities) disappear!  In other words, expectations of information in the price series erase that very information.

\section{Volatility clustering}

In the last section we observed that the price series in DSEM always crossed over to a domain with no memory effects for long enough times.  This would seem to imply a lack of history-dependence in the time series.  However, it has been empirically observed that volatility (to be defined) has a {\em very} long memory.  This leads to the phenomenon of clustered volatility: high activity in the market is observed to cluster together, separated by spans of low activity.  

The simplest definition of volatility is the absolute value of the price return over some interval.  (This definition appears to be more prevalent than the square of the returns \cite{cizeau97, liu97, arneodo98, lux99}.)  Clustered volatility, by this definition, means there exists periods in which the price changes rapidly and dramatically, separated by other periods where few/small changes in the price occur.  Hence, there exist long-range temporal correlations in the absolute value of price returns.

As before, the Hurst exponent is a promising quantity to measure these correlations.  For the daily returns of the Dow Jones Industrial Average \cite{economagic99} shown in \fig{results2DjiacAutoCorr}, the Hurst exponent of the volatility is measured to be $H=0.852\pm 0.009$ demonstrating very strong positive correlations---high volatility tends to be followed by further high volatility and low by low.  Other studies measuring the Hurst exponent of the volatility from empirical data have concluded $H\approx 0.9$ \cite{cizeau97}, $0.63\leq H\leq 0.95$ \cite{liu97}, and $H\approx0.85$ \cite{lux99}.  Two other works I am aware of calculated the exponent of the autocorrelation function which decays as a power law with exponent $2H-2$ \cite{schepers92}, giving $H\approx 0.9$ \cite{arneodo98} and $H\approx 0.8$ \cite{cont97b}.  (The latter defined volatility as the squared return rather than its absolute value, but came to the same conclusions.)  These studies were performed on a variety of systems so, clearly, clustered volatility is universal.

\begin{figure}\centering
	\input{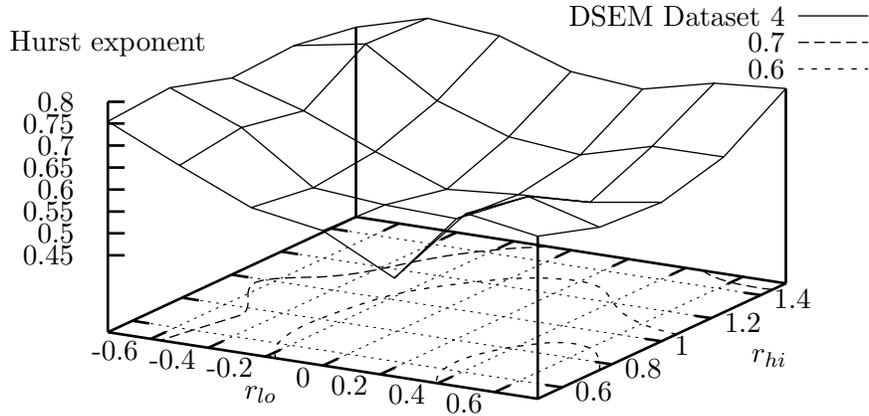}
	\caption{The Hurst exponent of the absolute returns, which measures the degree of clustered volatility, is strictly greater than one half for all parameter combinations in DSEM.  It is particularly high when the upper limit of the two-point distribution $r_{hi}$ is large or when the lower limit $r_{lo}$ is small.}
\label{fig:results2DsemVolHurst}
\end{figure}

Again, we seek to know whether DSEM also exhibits this property.  To test it, we continue with our analysis on a tickwise (number of trades executed) timescale and calculate the Hurst exponent for the absolute value of the returns.  The results are summed up in \fig{results2DsemVolHurst} which shows that the Hurst exponent measuring volatility clustering is always above one half over the whole parameter space, but significantly so when the upper limit $r_{hi}$ of the two-point price response distribution is large or the lower limit $r_{lo}$ is small.  

Overall, though, the Hurst exponents are somewhat smaller (the greatest value was $H=0.77$) than the empirical results ($H\approx 0.9$), suggesting that our search space should be expanded to larger values of $r_{hi}$.

\subsection{Shuffling}

As a check of the analysis the absolute return data for a particular run ($r_{lo}=0.75$, $r_{hi}=1.5$ with $H=0.72\pm 0.02$) were shuffled and the Hurst analysis of the shuffled data recalculated.  Shuffling destroys temporal correlations so the expected value of the exponent is one half.  In fact, the Dow Jones Industrial Average data yields $H=0.502\pm 0.010$ for the absolute returns when shuffled.  For the sample DSEM data the resultant exponent is $H=0.520\pm 0.006$.  Both are very close to one half, confirming that the high value for the unshuffled absolute-return data is due to temporal correlations (clustered volatility).

\section{Scaling and Clustered volatility}

In real markets all three properties of (1) scaling, (2) uncorrelated returns, and (3) clustered volatility are observed.  As we have seen, DSEM can replicate each of these features when the agents have a two-point price response ($r_{lo}$ and $r_{hi}$) and the values of these parameters are chosen appropriately.  Of particular interest is whether there is a region of parameter space in which all three of these phenomena are observed simultaneously.

In some of the previous plots contour lines were drawn to indicate separation into regions which did and did not exhibit the phenomenon of interest: \fig{results2DsemReturnTailsDt0} plotted the contour line $r_c=3$ to distinguish between parameter combinations which did ($r_c>3$) and did not exhibit scaling in the return distributions.  \fig{results2DsemHurstMemExp}(a) separates parameter combinations that do not have a long memory ($<100$) in the return series from those that do.  Finally, \fig{results2DsemVolHurst} measures, with the Hurst exponent, the clustered volatility where $H>0.6$ indicates the presence of clustered volatility while $H<0.6$ indicates its absence.

\begin{figure}\centering
	\input{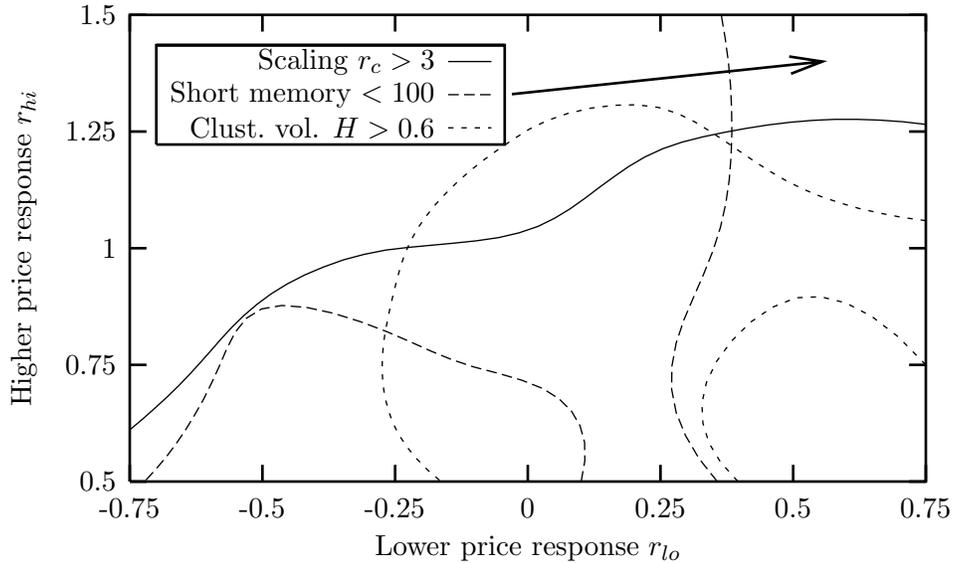}
	\caption{DSEM Dataset 4 ($N=100$ agents) is able to capture three important properties observed empirically when $r_{lo}>0.35$ and $r_{hi}>1.25$.  The curves are contours from previous plots: (1) characteristic return $r_c=3$ from \fig{results2DsemReturnTailsDt0} (solid line); (2) memory in return series $=100$ from \fig{results2DsemHurstMemExp}(a) (dashed line); and (3) Hurst exponent for the absolute returns $H=0.6$ from \fig{results2DsemVolHurst} (dotted line).}
\label{fig:results2DsemContours}
\end{figure}

These three contour curves are plotted together in \fig{results2DsemContours} showing that all three empirical properties are only observed in the upper right corner of the graph, when $r_{lo}>0.35$ and $r_{hi}>1.25$.  So DSEM is most realistic for these parameter combinations.  

One point to note regarding this region is that it spans the critical point at $r_p=r_1=1$ with the low end of the distribution below and the high end above.  It is also interesting that the value $r_{lo}=0$ (characterizing agents that do not use the return series as an indicator of performance) is not in the ``realistic'' region.  

\section{Wealth distribution}

Thus far we have explored only the temporal dynamics of the stock price.  But the distribution of wealth among agents may be of interest as well.  It is well known that incomes in many populations are distributed log-normally \cite{montroll83} (with a particular exception we will come to later).  So it is natural to ask how wealth is distributed in DSEM.  (Again, we neglect the analysis of CSEM.)

\subsection{Challenges}

Determining the wealth distribution in DSEM is problematic because long runs of many agents are required.  The durations must be long because the wealth distribution is initialized to a delta function (initially, all agents have the same amount of cash and stock) so a long transient is to be expected before any steady-state emerges.  

But this constraint must be balanced against the need for many agents.  Most of the simulations presented in this thesis consisted of $N=100$ interacting agents---a rather small number for any statistical description.  But computational limitations prevent serious investigation of larger systems because the number of operations grows as $N^2$.  So we must be content with the data we have collected so far.

\subsection{Log-normal distribution}

\begin{figure}\centering
	\input{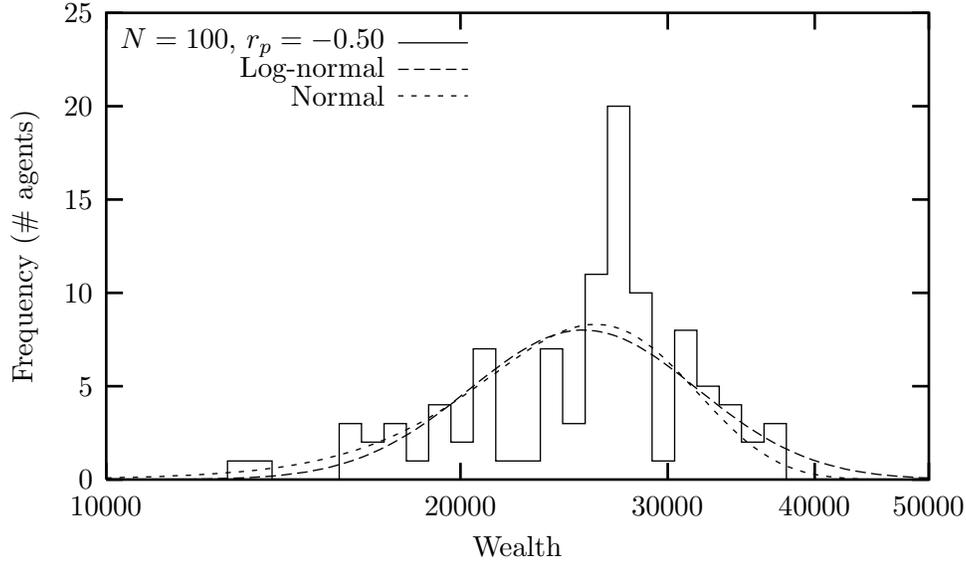}
	\caption{Sample distribution of agents' wealth from DSEM Dataset 3 ($N=100$, $r_p=-0.50$).  There is insufficient data to distinguish between a normal and a log-normal distribution.}
\label{fig:results2DsemWealthDist}
\end{figure}

A rigorous analysis of the distribution of wealth will not be attempted.  Instead, it is merely reported that a log-normal distribution was suitable in most cases for Datasets 1--3.  However, \fig{results2DsemWealthDist}---showing a typical distribution---highlights the difficulties in establishing the proper distribution: the small agent numbers combined with the narrow range of wealths observed allows one only to say that the distribution is unimodal, but nothing more.

Even so, simply knowing that the distribution is unimodal is satisfactory.  Any other result would be surprising because the agents in Datasets 1--3 are all of a similar character, varying (continuously) only in their news responses $r_n$ and frictions $f$, as shown in \tbl{resultsDsemDataset1}.

\subsection{Two-point price response}

Worth further investigation are the data from Dataset 4 (\tbl{results2DsemDataset4}) where the agents varied {\em discontinuously} in their price responses.  In this case the market consists of two distinct populations with fundamentally different behaviours, so one would not expect the wealth to be distributed unimodally.  A reasonable alternative is that each population has a log-normal wealth distribution producing a bimodal distribution over the whole market.

\begin{figure}\centering
	\input{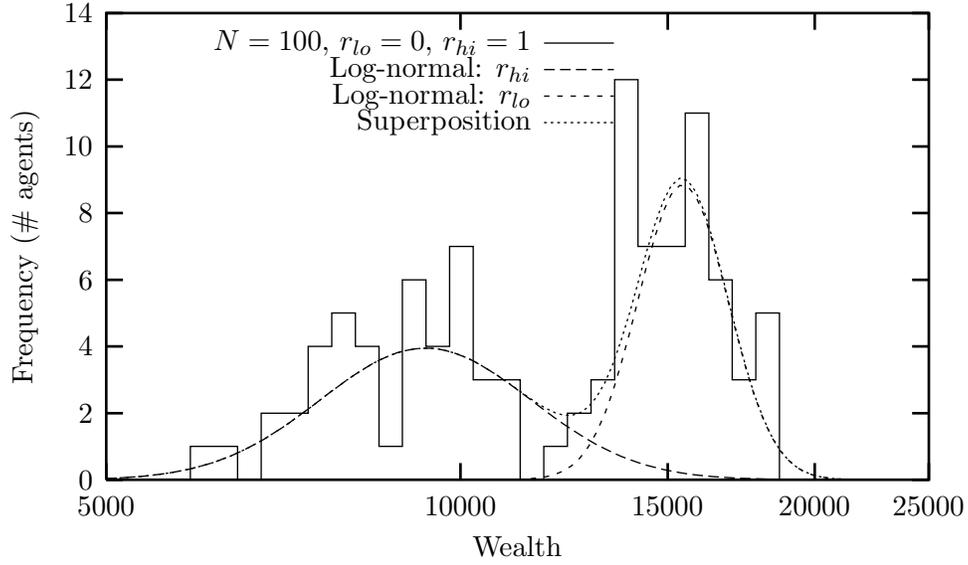}
	\caption{Sample distribution of agents' wealth from DSEM Dataset 4 ($N=100$, $r_{lo}=0$, $r_{hi}=1$).  The log-normal curves are calculated from each sub-population, revealing a strongly bimodal nature.}
\label{fig:results2DsemRp0and1WealthDist}
\end{figure}

A representative distribution is shown in \fig{results2DsemRp0and1WealthDist}, confirming the bimodal hypothesis.  Interestingly, the agents with $r_p=r_{lo}=0$ outperform (have more wealth) than their $r_p=r_{hi}=1$ counterparts.  

\begin{figure}\centering
	\input{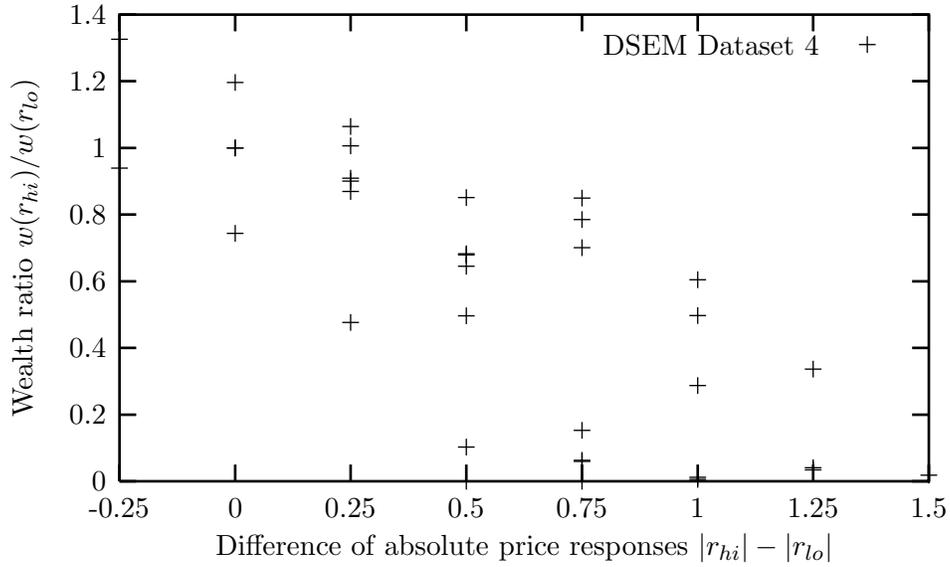}
	\caption{In DSEM with a two-point price response the wealth of each of the sub-populations $w(r_p)$ depends strongly on the magnitude of the price response $\abs{r_p}$.  The population with the smallest absolute price response ($r_{hi}$ to the left of zero and $r_{lo}$ to the right) consistently has more wealth as indicated by the ratio of wealth between the two sub-populations.}
\label{fig:results2DsemWealthRatio}
\end{figure}

To determine the generality of this result the average wealth for each sub-population was computed for all the simulations in Dataset 4. The hypothesis that the sub-population with the smallest absolute price response $\abs{r_p}$ would outperform the more reactive agents was tested by comparing the difference between the absolute values of the responses $\abs{r_{hi}}-\abs{r_{lo}}$ and the ratio of wealth held by each sub-population ($w(r_{hi})$ vs.\ $w(r_{lo})$).  If valid then we should find the ratio of the wealths obey $w(r_{hi})/w(r_{lo})>1$ when $\abs{r_{hi}}<\abs{r_{lo}}$ and vice versa.  \fig{results2DsemWealthRatio} demonstrates this hypothesis holds very well---with a linear correlation for the plotted data of --76\%---suggesting that the ``best'' strategy is to ignore the price fluctuations, $r_p=0$.

This raises an interesting question: if a zero price response is best, why would agents choose non-zero values?  We have seen in previous sections that realistic market phenomena such as scaling and clustered volatility only emerge in when the price responses are set far from zero.  If DSEM is meant to represent real investor behaviour, why do investors base their decisions on price fluctuations when the model indicates this is detrimental?  

This issue is currently being investigated by allowing the agents to ``learn'' from their past mistakes as will be discussed in \sect{concFuture}.  The purpose of this research is to determine if nonzero values of the price response parameter emerge spontaneously in the dynamics via the learning process.

\subsubsection{Pareto's law of income distribution}

In 1897 V.\ Pareto noticed that incomes tend to be distributed log-normally over the majority of the sample data excepting the tail of the distribution (the highest one percent of the incomes) which decay as a power law, an observation which holds in many countries to this day \cite{montroll83}.  CSEM and DSEM were deliberately constructed to keep a record of agents' wealths so that this claim could be tested.  However, testing for this property requires even larger system sizes than we have explored so far, since the highest percentile of a population of even $N=1000$ consists of only ten agents---inadequate for statistical analysis.

Larger systems are currently under investigation but insufficient data were available as of completion of this dissertation.

\section{Summary}

In this chapter a number of unusual qualities of empirical markets were explored in the context of the Centralized and Decentralized Stock Exchange Models (CSEM and DSEM, respectively).  CSEM was unable to reproduce even the first of these: scaling in the tail of the price return distribution.  DSEM was also unable to produce this effect until the restriction that all agents maintain the same price response parameter $r_p$ was relaxed and instead two values were allowed---$r_{lo}$ and $r_{hi}$---thereby splitting the population into two distinct ``types.''  Then, for particular values of the parameters, scaling was observed over a range of returns in excess of three standard deviations.

Two other empirical phenomena were also explored: the lack of correlations in the return series but the presence of long-range correlations in the volatility (defined as the absolute value of the return).  Having failed the first test CSEM was not tested but DSEM was able to capture both these properties, again in a suitable region of parameter space.  

All three of these properties were observed simultaneously in DSEM when $0.5\leq r_{lo}<1$ and $r_{hi}\geq 1.25$, spanning the critical point at $r_p=r_1$ discovered in \sect{resultsDsemPhaseTrans1}.  The significance of this result is unclear but some thoughts on the matter are discussed in \ch{conclusions}.  But first the results of some interesting experiments with some real stocks are discussed.
\chapter{Experiments with a hypothetical portfolio}

\label{ch:portfolio}

This chapter is somewhat of a departure from the rest of the thesis.  It describes some experimental results obtained purely to satiate my own curiosity.  As such, the experiments are less rigorous than they could be (in particular, the dataset is quite small) and the results should not be taken too seriously.  However, these experiments may yield valuable insight for the reader because they describe a real-world application of some of the theory discussed in prior chapters.

\section{Motivation}

The agents in \ch{dsem} trade using a {\em fixed investment strategy} (FIS) which states that they should keep a fixed fraction of their capital in stock and the remainder as cash, to minimize risk and maximize returns.  As was discussed, the theory underlying it has two important assumptions: (1) that there are no costs associated with trading, and (2) that moments of the return distribution higher than two (in particular, the kurtosis) are negligible on short timescales.  

I was curious how well FIS would work in a real market environment where these assumptions may not hold so I constructed a hypothetical portfolio to track real stocks.  Sandbox Entertainment (\url{http://www.sandbox.net/business/}) provides an online simulated stock market called ``PortfolioTRAC'' which gives users an imaginary bankroll of \$100,000 and allows them to invest it in stocks listed on the major American markets.  The simulation uses real trading prices and allows daily trades.  Although it requires a few other idealizations, it is quite thorough and supports such complexities as short positions, limit and stop orders, broker fees, and daily interest on cash.  Note that trades are only processed once per day (after closing) so limit and stop orders execute based on closing prices.

I began with \$100,000 on January 4, 1999 and decided to divide my assets uniformly among 9 stocks and one cash account.  \tbl{portInitial} lists my initial portfolio after commissions have been accounted for.  The FIS goal was to keep one tenth of my total capital in each stock.

\begin{table}
\begin{center} \begin{tabular}{l|l|r|r|r}
\hline \hline
Symbol& Company         & Price   & Shares& Value \\ 
\hline
AAPL  & Apple Comp Inc  &	\$41.25	& 244   & \$10,065.00 \\
AMD   & Adv Micro Device& \$28.00 & 345   & \$9,660.00 \\
AU    & Anglogold Ltd   & \$19.56 & 514   & \$10,055.13 \\
CHV   & Chevron Corp    & \$82.06 & 120   & \$9,847.50 \\
EK    & Eastman Kodak Co& \$71.19 & 141   & \$10,037.44 \\
IMNX  & Immunex Corp    & \$116.50& 82    & \$9,553.00 \\
MSFT  & Microsoft Corp  & \$141.00& 69    & \$9,729.00 \\
NSCP  & Netscape Comm   & \$63.25 & 157   & \$9,930.25 \\
RG    & Rogers Comm     & \$8.56  & 1139  & \$9,752.69 \\
      & Cash            &         &       & \$11,004.89 \\
      & Total           &         &       & \$99,634.89 \\ 
\hline \hline
\end{tabular}
\end{center}
\caption{Initial holdings of a hypothetical portfolio on January 4, 1999.}
\label{tbl:portInitial}
\end{table}

\section{Choice of companies}

My choice of companies was not completely random: I chose Apple Computers (AAPL), Advanced Micro Devices (AMD) and Netscape Communications (NSCP) because they were all ``underdogs'' in their respective industries and would have to be innovative and aggressive to survive.  Similarly, I chose Eastman Kodak (EK) because, although currently a large, stable company, I expected the emerging digital camera technology to threaten its dominance and I wanted to see how it fared.  I chose the cable company Rogers Communications (RG) because I was interested in the newly available cable modem technology which they were investing in.  Microsoft Corporation (MSFT) seemed a low-risk choice to balance my high-risk portfolio.  My focus to this point had been in the high technology sector so I determined to diversify: in the petroleum sector I chose Chevron Corporation (CHV) for its apparent low-risk and because it was the most recent gas station I had visited, and I chose Anglogold (AU) as a gold stock simply because of its catchy symbol.  I couldn't think of a last company I was interested in so I let my wife choose Immunex Corporation (IMNX) from the biomedical sector.  Although these choices were biased by my own interests it was hoped that they would prove sufficiently representative to test the performance of the fixed investment strategy.  (As will be seen, this portfolio correlated strongly with the Nasdaq composite index.)

\section{Friction}

The derivation of FIS in \ch{dsem} and other sources \cite{merton92, maslov98} neglected commissions; they considered a completely fluid portfolio, capable of adjusting instantly to infinitesimal price changes.  This market simulation was more realistic, with commissions which were handled as follows: in each trade a \$39.95 charge was levied for the first 1000 shares traded (bought or sold) and \$0.04 per share over 1000.  Obviously, it would be unprofitable to trade on every minuscule price fluctuation so a {\em friction} $f$ was introduced.  Orders are not placed until a stock's price $p$ exceeds a threshold as given by \eqs{dsemLimitPriceBuy}{dsemLimitPriceSell}.

\subsection{Minimum friction}

Under some particular conditions it is possible to estimate how large the friction needs to be for profitable trading in a commission-enabled market.  To calculate the necessary scale of the friction consider an imaginary scenario: we begin with a total capital of $w$ divided uniformly between a cash account and $N-1$ stocks, so the ideal investment fraction is $i=1/N$.  The scenario consists of 
\begin{enumerate}
	\item{a single stock's price moving to a trade limit (buy or sell),}
	\item{the stock being rebalanced (traded),}
	\item{returning to its original price, and}
	\item{being rebalanced again.}
\end{enumerate}  

In this scenario we assume all the other stocks are unchanged, each maintaining a value $iw$.  We are interested in what the minimum friction $f_{min}$ can be such that we don't lose any money given an absolute transaction cost $T$.

We begin with the fluctuating stock at its ideal price 
\begin{equation}
	p^* = \frac{iw}{s},
\end{equation}
where $s$ is the number of shares held of the stock.

So the limit prices are
\begin{equation}
	p_\pm = p^* (1+f)^{\pm 1},
\end{equation}
where $p_+$ is the sell limit and $p_-$ is the buy limit.

If the stock moves to one of the limits while all others remain constant, then our wealth (before trading) will become
\begin{equation}
	w_\pm = (1-i) w + s p_\pm
\end{equation}
and the quantity to be traded, from \eq{dsemOptimalHoldings}, will be
\begin{eqnarray}
	\Delta s_\pm & = & s^*(p_\pm) - s \\
	             & = & (1-i)s\left[ (1+f)^{\mp 1} - 1 \right]
\end{eqnarray}
maintaining the same notation (the upper symbol of $\pm$ and $\mp$ indicates an initial rise in price, and the lower indicates an initial drop).

The trade also changes our cash holdings by
\begin{eqnarray}
	\Delta c_\pm & = & -\Delta s_\pm p_\pm - T \\
	             & = & -\Delta s_\pm p^* (1+f)^{\pm 1} - T
\end{eqnarray}
where $T$ is the transaction cost (in dollars).

Now we assume the stock's price returns to its original value $p^*$ and we trade to recover our original portfolio $\Delta s'_\pm = -\Delta s_\pm$ (for simplicity), yielding another change in cash
\begin{equation}
	\Delta c_\pm' = \Delta s_\pm p^* - T
\end{equation}
so the net change is
\begin{equation}
	\Delta \hat{c} = \Delta s_\pm p^* \left[ 1 - (1+f)^{\pm 1} \right] - 2 T.
\end{equation}

Inserting the computation for $\Delta s_\pm$ gives a net change
\begin{eqnarray}
	\Delta \hat{c} & = & (1-i) i w  \left[ (1+f)^{\mp 1}- 1 \right] \left[ 1 - (1+f)^{\pm 1} \right] - 2 T \\
	               & = & (1-i) i w \left[ f + \frac{1}{1+f} - 1 \right] - 2 T,
\end{eqnarray}
regardless of whether the stock price rose then fell or fell and then recovered.

After some algebra we find the condition requiring a profit $\Delta \hat{c} > 0$ holds when $f>f_{min}$ where
\begin{equation}
\label{eq:portMinFriction}
	f_{min} = \frac{TN^2}{w(N-1)} \left[1 + \sqrt{1 + \frac{2w(N-1)}{TN^2}} \right].
\end{equation}
which simplifies to
\begin{equation}
	f_{min} \approx N\sqrt{ \frac{2T}{w(N-1)} }
\end{equation}
in the limit $w\gg T$.

This informal derivation is only meant to set a scale for the minimum friction, it is not meant to be rigorous.  A more detailed calculation may be possible by assuming each stock's price moves as geometric Brownian motion but the derivation would be cumbersome and the benefit dubious.

When I began trading with my hypothetical portfolio in the beginning of 1999 I arbitrarily chose $f=10\%$, a fortuitous choice, as it turns out, because anything less than $f_{min}= 9.90\%$ might have been a losing strategy.

Since \eq{portMinFriction} sets the break-even friction it is best to set the actual friction somewhat higher.  Once the minimum friction for my portfolio had been estimated (December 1999) I chose a dynamic value of $f = 2f_{min}$.

\section{FIS Experimental results}

In this section, the results of using the fixed investment strategy (with friction) on a hypothetical portfolio will be discussed.

\subsection{Events}

The experiment began on January 4, 1999 and ran until May 12, 2000 for a total of 343 trading days.  The portfolio was rebalanced faithfully, as needed almost every day (excepting a few rare and brief vacations).  Note that the simulation only executed trades after closing so intra-day trading was not supported and the trading price was always the stock's closing price.  Also note that the simulation did not constrain orders to be in round lots and most orders, in fact, were odd sizes.

\begin{table}\centering
\begin{tabular}{r|l}
	\hline \hline
	Date & Event \\ 
	\hline
	January 4, 1999 & Experiment started \\
	March 23, 1999 & Takeover: 1 NSCP $\rightarrow$ 0.9 AOL \\
	March 26, 1999 & Stock split: 2-for-1 IMNX \\
	March 29, 1999 & Stock split: 2-for-1 MSFT \\
	August 27, 1999 & Stock split: 2-for-1 IMNX \\
	November 11, 1999 & Stock split: 2-for-1 AOL \\
	December 10, 1999 & Tolerance changed from 10\% to $2f_{min}$ \\ 
	March 21, 2000 & Stock split: 3-for-1 IMNX \\
	April 14, 2000 & NASDAQ correction \\
	May 12, 2000 & Experiment ended \\
	\hline \hline
\end{tabular} 
\caption{Events relating to the hypothetical portfolio which occurred during the course of the experiment.}
\label{tbl:portEvents}
\end{table}

A list of important events occurring over the course of the experiment is shown in \tbl{portEvents}.  The majority of events consist of stock splits, a division of the shares owned by each shareholder of a company such the stake held by each is unchanged.  For example, a 2-for-1 stock split means each share is split into two, each worth half its original value.  Although theoretically a stock split should not affect an investor's capital in a company, stock splits are considered good news and often drive the stock's price up both before and after the split.  The main reason is that a split lowers the price of a stock and thereby makes it accessible to more potential investors---increasing demand.

Another interesting event which occurred during the experiment run was the takeover of Netscape Communications by America Online in March, 1999. AOL purchased Netscape for roughly \$4 billion and each share of Netscape stock was converted to 0.9 shares of AOL.  Takeovers tend to engender a great deal of speculation which can precipitate large fluctuations in the stock's value.

On the book-keeping side, the only change in methodology was the move from a constant friction of $f=10\%$ to a floating value of $f=2f_{min}$, as given by \eq{portMinFriction}, on December 10, 1999 (giving $f=15\%$ at the time).  The main consequence was a somewhat decreased trading frequency.

The most exciting event was the correction in the high-technology sector (which dominates my portfolio) in the week of April 14, 2000, evinced by an over-35\% drop in the Nasdaq Composite index from its all-time high only weeks earlier \cite{johansen00}.  This is a particularly fortunate occurrence because it tests the ability of the FIS to handle drawdowns.  Market-wide fluctuations of this magnitude are rare but an important consideration when devising a trading strategy.  This aspect of the experiment will be discussed below.

\subsection{Performance}

\begin{table}
\begin{center} \begin{tabular}{l|l|r|r|r}
\hline \hline
Symbol& Company         & Price   & Shares& Value \\ 
\hline
AAPL  & Apple Comp Inc  &	\$107.63	& 159   & \$17,112.38 \\
AMD   & Adv Micro Device& \$85.69 & 216   & \$18,508.50 \\
AOL		& America Online Inc& \$55.38 & 325 & 17,996.88 \\
AU    & Anglogold Ltd   & \$20.06 & 965   & \$19,360.31 \\
CHV   & Chevron Corp    & \$94.25 & 190   & \$17,907.50 \\
EK    & Eastman Kodak Co& \$56.13 & 310   & \$17,398.75 \\
IMNX  & Immunex Corp    & \$35.00 & 456    & \$15,960.00 \\
MSFT  & Microsoft Corp  & \$68.81 & 269    & \$18,510.56 \\
RG    & Rogers Comm     & \$25.56 & 732  & \$18,711.75 \\
      & Cash            &         &       & \$18,792.39 \\
      & Total           &         &       & \$180,259.02 \\ 
\hline \hline
\end{tabular}
\end{center}
\caption{Final holdings of a hypothetical portfolio on May 12, 2000.}
\label{tbl:portFinal}
\end{table}

The final state of the portfolio at the end of the experiment is shown in \tbl{portFinal}.  In this section the performance of the fixed investment strategy will be evaluated.  As a control, a simple ``Buy-and-Hold'' Strategy (BHS) with the initial portfolio shown in \tbl{portInitial} held fixed, will be contrasted with the fixed investment strategy (FIS).

\begin{figure}
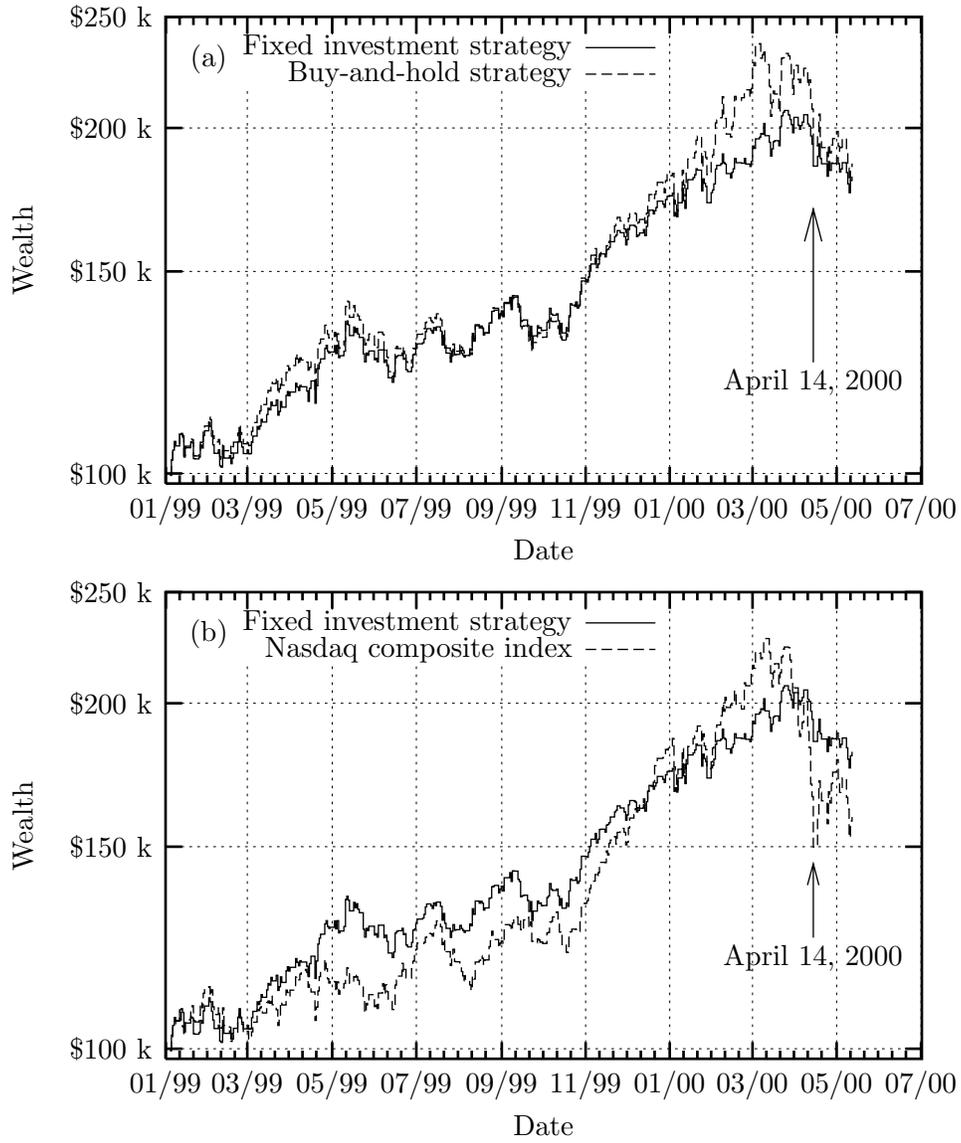
\centering
	\input{chPortfolio/portValueA.tex} \\
	\input{chPortfolio/portValueB.tex}
	\caption{Historical wealth using FIS versus (a) the Buy-and-Hold strategy and (b) the Nasdaq Composite Index over the same interval (rescaled to be equal at the start of the experiment).}
\label{fig:portValue}
\end{figure}

\fig{portValue}, which shows the evolution of total capital for both strategies, on the surface seems to indicate BHS outperforms FIS.  BHS reaches a high of \$237,000, a full 14\% higher than the maximum achieved with FIS.  Also, BHS maintained a higher capital on 292 of the 343 days (85\%) the market was open.  Evidently, FIS does not perform well in real-world applications.

However, a closer inspection suggests FIS should not be discarded too rashly.  For example, consider the market correction on and around the week of April 14, 2000.  The Nasdaq Composite peaked at 5,049 points on March 10 and fell to a low of 3,321 on April 14, a drop of 34\%.  The Buy-and-hold strategy fared somewhat better, dropping to \$180,000 for a drawdown of 25\%.  But the fixed investment strategy suffered the smallest decrease---down only 15\% to \$176,000, finishing with almost the same value as BHS.  (It should be noted that Maslov and Zhang \cite{maslov99} demonstrated that the FIS is the most aggressive possible strategy that keeps the {\em risk}---measured as the expected drawdown from the maximum---bounded.)

This suggests FIS is less susceptible to large fluctuations.  By rebalancing the portfolio, one moves capital out of stocks which may be overvalued and into safer companies which may be more resilient to perturbations.  In this sense, FIS reduces risk.

\begin{figure}\centering
	\input{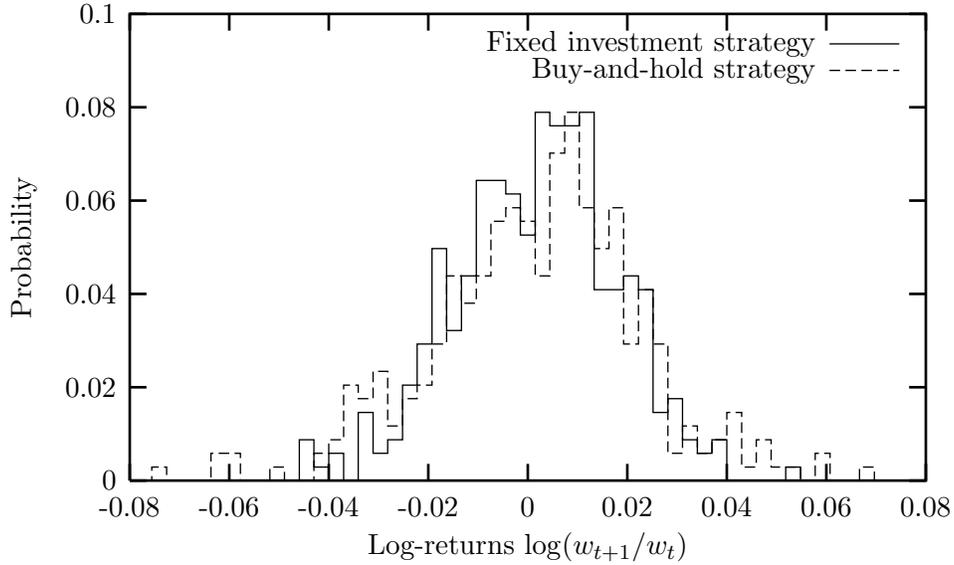}
	\caption{Histograms of log-returns of capital $r_{t+1} = \log(w_{t+1}/w_t)$ for both strategies.  Notice BHS exhibits more large fluctuations (fatter tails) than FIS.}
\label{fig:portHistogram}
\end{figure}

\begin{table}\centering
\begin{tabular}{r|r|r|r|r}
\hline \hline
Symbol& Mean & Std.\ Dev.\ & Skewness & Kurtosis \\ 
\hline
FIS   & 0.00178 & 0.016 & -0.18 & 0.12 \\
BHS   & 0.00183 & 0.021 & -0.23 & 0.82 \\
Nasdaq& 0.00119 & 0.022 & -0.58 & 1.65 \\
\hline
AAPL  & 0.00274 & 0.040 & 0.04 & 0.54 \\
AMD   & 0.00342 & 0.046 & 0.23 & 2.76 \\
AOL   & 0.00134 & 0.044 & 0.33 & 1.32 \\
AU    & 0.00042 & 0.029 & 0.60 & 3.93 \\
CHV   & 0.00050 & 0.020 & 0.39 & 0.95 \\
EK    & -0.00062& 0.019 & 0.25 & 5.40 \\
IMNX  & 0.00375 & 0.064 & 0.29 & 1.80 \\
MSFT  & -0.00010& 0.029 & -0.90& 5.18 \\
RG    & 0.00361 & 0.034 & 0.66 & 0.80 \\
\hline
Averages&0.00166& 0.032 & 0.08 & 2.11 \\ 
Significance& & & 0.21 & 0.53 \\
\hline \hline
\end{tabular}
\caption{First four moments of the distribution of log-returns for each stock, the two trading strategies under review and the Nasdaq Composite Index.  The skewness characterizes the asymmetry of the distribution and the kurtosis indicates the presence of outliers.  The average skewness is not found to be significant but the kurtosis is.}
\label{tbl:portStats}
\end{table}

This can be seen in \fig{portHistogram} which demonstrates that BHS is more prone to large fluctuations (both positive and negative).  To test whether these histograms are compatible with the Gaussian hypothesis \cite{osborne59} the first four moments of the distributions are calculated in \tbl{portStats}.  The moments of the daily returns of the Nasdaq Composite index and each of the stocks in the portfolio over the same interval are also shown.

The means for all the distributions are all small, negligible in comparison to the standard deviations (which had an average of 3\% daily).  The skewness, given by
\begin{equation}
	{\rm Skew}(\{r_i\}) = \frac{1}{N} \sum_i \left[ \frac{r_i - \bar{r}}{\sigma_r} \right]^3
\end{equation}
where $\sigma_r$ is the standard deviation of the returns, indicates the degree of asymmetry in the distribution.  The sign of the skewness indicates which tail of the distribution contains more outliers.  The skewness uncertainty for a (symmetric) Gaussian-distributed sampling of $N$ points is $\sqrt{15/N}$ \cite[Ch.\ 14]{press92}, thereby setting a scale for deciding if a particular skewness is significant or not.  Interestingly,  \tbl{portStats} indicates that individual stocks tend to be skewed positively but the portfolios and the Nasdaq index are negatively skewed.  This suggests that negative movements tend to be correlated between stocks but positive movements are not.

The (excess) kurtosis is a measure of the spread of the distribution.  A positive kurtosis indicates the presence of many outliers or ``fat tails''.  The kurtosis is defined as
\begin{equation}
	\kurt(\{r_i\}) = \left\{ \frac{1}{N} \sum_i \left[ \frac{r_i - \bar{r}}{\sigma_r} \right]^4 \right\} - 3
\end{equation}
where 3 is subtracted in order to fix the kurtosis at zero for a normal distribution.  The standard deviation of the kurtosis from a dataset of size $N$ sampled from a Gaussian is $\sqrt{96/N}$ so the Gaussian hypothesis is rejected if the kurtosis if found to be greater. 

The table shows that almost every calculated kurtosis is significant, indicating fat tails for these daily return distributions.  The only exception is the fixed investment strategy which only has a kurtosis of 0.12 (versus a significance level of 0.53).  This further confirms the hypothesis that FIS reduces risk: by rebalancing the portfolio the frequency of large fluctuations is reduced.  (Another favourable consequence is that the return distribution appears to converge to a Gaussian, as was assumed in the derivation of FIS.)

Nevertheless, it is undeniable that BHS outperformed FIS in the experiment.  But given the atypical trend seen in the portfolios, this conclusion may not be generalizable.  Both portfolios realized almost a 100\% growth over the first year, a gain which can hardly be expected to be repeated often (except, perhaps, during other speculative bubbles).  A more typical realization may have proven FIS superior.  By more typical is meant a smaller trend, relative to the scale of the fluctuations.  As can be seen in \fig{portValue} FIS performs best in the presence of fluctuations and slowly loses ground against BHS in the presence of an upwards trend.  

Even if this experiment doesn't demonstrate that FIS is optimal it still shows that it is a reasonable investment strategy, and a suitable choice for the agents in DSEM (\ch{dsem}).  The striking feature of FIS is that it reduces the kurtosis (risk of large events) and thereby misses large downturns (and upturns) in the market.

\section{Log-periodic precursors}

In this section the results of another experiment performed, using the same hypothetical portfolio, will be examined.  At issue is whether there exists a reliable method to forecast imminent crashes in the market.  First, some background theory is necessary.

\subsection{Scale invariance}

Scale invariance is a property of some systems such that a change of scale in a parameter $x' = \lambda x$ only has the effect of changing the scale of some observable $F'=F/\mu$, such that
\begin{equation}
\label{eq:portScaling}
	F(x) = \mu F(\lambda x).
\end{equation}
The above scaling relation has a power law solution $F(x)=Cx^z$ where
\begin{equation}
	z = -\frac{\log \mu}{\log \lambda}
\end{equation}
and $C$ is an arbitrary constant.

The important point to notice is that the scalings along both axes are related by $\mu = \lambda^{-z}$.  No matter how much the control parameter $x$ is scaled by (even infinitesimally, $\lambda\rightarrow 1$), it is always possible to rescale the observable so that it is invariant.  This is known as {\em continuous} scale invariance.

\subsection{Discrete scale invariance and complex exponents}

In contrast, {\em discrete} scale invariance only allows fixed-size rescalings of the parameter.  To see how this comes about we begin by substituting the solution $F(x)=Cx^z$ into the renormalization equation (\eq{portScaling}),
\begin{eqnarray}
	Cx^z           & = & \mu C \lambda^z x^z \\
	 \Rightarrow 1 & = & \mu \lambda^z.
\end{eqnarray}

Now notice that $1=e^{2\pi i n}$ for any integer $n$.  Applying this and taking the logarithm of both sides gives
\begin{equation}
	2\pi i n = \log \mu + z \log \lambda
\end{equation}
which has the solution
\begin{equation}
	z = -\frac{\log \mu}{\log \lambda} + i \frac{2\pi n}{\log \lambda}.
\end{equation}

For the scaling relation to hold $z$ must be a constant, which can only hold when $n=0$ (which allows $\lambda$ to take on any value, recovering continuous scale invariance) or when $\lambda$ is some fixed constant, the {\em preferred} scaling ratio.  Hence, the only invariant transformations are the discrete rescalings $x' = \lambda x$ with corresponding scalings in the observable $F' = F/\mu$ (for some fixed $\mu$).

\subsection{Log-periodic precursors}

So far this might all look like mathematical trickery to the reader but the theory does have testable consequences.  If we use the notation $z_n=\alpha + i \omega_n$ with
\begin{eqnarray}
\label{eq:portScalingAlpha}
	\alpha & = & -\frac{\log \mu}{\log \lambda} \\
	\omega_n & = & \frac{2\pi n}{\log \lambda}
\end{eqnarray}
then $F_n(x) = C_n x^\alpha x^{i \omega_n}$ is a solution to the scaling relation for each $n$ and the general solution is the linear combination over all integers $n$,
\begin{eqnarray}
	F(x) & = & x^\alpha \sum_n C_n x^{i \omega_n} \\
	     & = & x^\alpha \sum_n C_n \exp(i \omega_n \log x) \\
	     & = & x^\alpha \left[ C_0 + e^{i \omega \log x} \sum_{n\neq 0} C_n \exp(i \omega (n-1) \log x) \right]
\end{eqnarray}
where we have defined $\omega=2\pi/\log \lambda$, for convenience.

The final form of $F(x)$ indicates that the function has a periodic component with angular frequency $\omega$.  Expanding the periodic component as a Fourier series gives, to first order,
\begin{equation}
	F(x) \approx x^\alpha \left[ C_0 + C_1' \cos(\omega \log x + \phi) \right]
\end{equation}
where $\phi$ is an unknown phase constant.

This argument, a variation of those presented in Refs. \cite{sornette95, saleur96, sornette98b}, concludes that discrete scale invariance leads naturally to log-periodic (in $x$) corrections to the scaling function $F$.

\subsection{Critical points}

Near a critical point many properties of a system exhibit power law scaling relations as described above.  Therefore they are prime test-cases for the existence of complex exponents characterized by log-periodic precursors.  

Seismicity, studied in the context of critical phenomena, have been successfully modeled as self-organizing (with the build up of stress) to a critical point in time $t_c$ characterized by an earthquake, a sudden release of energy \cite{bak93, sornette95, groleau97}.  In the neighbourhood of the critical time (small $\abs{t_c-t}$) the stress exhibits classic power laws seen in critical phenomena.  One important goal in seismology is forecasting the time of occurrence $t_c$ of large earthquakes.  It has been argued that log-periodic fluctuations are present both before (foreshocks) and after (aftershocks) large events and that the precursors improve earthquake forecasts considerably \cite{sornette95b, johansen96, saleur96}.  

The premise is that the rate of change of the regional strain $\epsilon$ exhibits critical scaling near the critical point,
\begin{equation}
	\dot{\epsilon} = F(\abs{t_c-t})
\end{equation}
so that the strain (a measurable quantity) obeys
\widebox{\begin{equation}
\label{eq:portLogPeriodic}
	\epsilon = A + \abs{t_c-t}^{\alpha+1}\left[ B + C \cos(\omega \ln \abs{t_c-t} + \phi) \right]
\end{equation}}
in the vicinity of $t_c$.  The curve is fit to known data by tuning the seven model parameters ($A$, $B$, $C$, $t_c$, $\alpha$, $\omega$, and $\phi$) and the forecast of $t_c$ is read off from the best fit to the data.  This method has significantly improved precision over curve fits neglecting log-periodicity ($C=0$), validating the adoption of the three extra parameters.

\subsection{Application to financial time series}

The same group of researchers who developed the concept of log-periodic precursors in seismology have recently turned their attention to the stock market, arguing that market crashes should be predictable by the same methodology \cite{sornette96, johansen99, johansen00}.  

Johansen et al. \cite{johansen99} construct a theory for price fluctuations with the risk of crash such that price series obeys precisely the relationship given in \eq{portLogPeriodic}.  The basic argument is that stock prices enjoy exponential growth but with some risk of crash.  As time progresses the risk accumulates and the exponential growth rate increases to compensate for the risk (to remunerate rational investors for their risk).  At some point in time the risk diverges and a critical point emerges.

The fundamental component of the theory is that the instantaneous risk of crash (which they call the ``hazard rate'') is assumed to obey a scaling relation like \eq{portScaling} with a control parameter $t_c-t$.  Since it is related to the rate of change of the price, \eq{portLogPeriodic} arises.

In theory then, financial crashes should be predictable by curve fitting to the price series.  In practice, though, this is an extremely difficult task: while searching through the seven-dimensional parameter space for the optimum fit one often gets stuck in local optima, missing the global one.  This complaint has been raised against the theory \cite{laloux98} and is acknowledged by Johansen et al. \cite{johansen00}.

Another problem with the research is that the experiments are all performed on known crashes {\em after they have occurred}!  This introduces two problems: Firstly---with no disrespect intended---it may bias the results.  If one knows there was a crash at such-and-such a time it would be very difficult to be satisfied with a curve-fit which made no such prediction.  One would probably suspect the parameters were stuck in a local minimum and tweak them.  This is perfectly natural but without foreknowledge one might have accepted the results without prejudice.

Secondly, all the curve fits were performed around well-established crashes.  It would be as useful to test the theory during other periods when no crashes occur in order to test for ``false positives.''  If the theory predicts too many crashes when none actually occur it is of no use.

In order to avoid these pitfalls I conducted a ``blind'' experiment to test the ability of \eq{portLogPeriodic} to forecast crashes.  As I was already running my FIS experiment it was convenient to use it as my input data for the curve fit.  Instead of forecasting a crash in a single stock, then, I was attempting to forecast a crash in a portfolio of nine stocks (and one cash account).  But this is not seen as problematic since the other studies used composite market indices instead of individual stocks, as well \cite{sornette96, laloux98, johansen99, johansen00}.

\subsection{Experimental design}

The experiment consisted of collecting portfolio wealth data $w_t$ and fitting the curve given by \eq{portLogPeriodic} with $\epsilon = w$.  The experiment began on February 14, 2000 and ran through May 12, 2000 but the dataset used was the entire historical set from the FIS experiment (which began on January 4, 1999).  

The dataset consisted of sets of date-wealth pairs which were only collected on days when a trade was executed.  At the beginning of the experiment this consisted of 113 points which grew to 134 points by the close of the experiment.

The fitting over the seven model parameters was performed using Microsoft Excel's {\em Solver Add-in} which uses the Generalized Reduced Gradient (GRG2) nonlinear optimization technique \cite{lasdon78}.  The GRG2 method is suitable for problems involving up to 200 variables and 100 constraints.  The optimization condition was the minimization of the sum of the squared deviations ($\chi^2$ nonlinear least-squares fitting).

The fit was performed on a logarithmic price scale on the basis that it is the relative (fractional) fluctuations in capital which are fundamental, not the absolute variations.  Fitting on a linear scale, then, would significantly bias the curve to fit better at greater wealths at the expense of the fit at lesser wealths.  So the fit consisted of minimizing 
\begin{equation}
	\chi^2 = \sum_t \left[ \ln w(t) - \ln w_t \right]^2
\end{equation}
where $w_t$ is the actual wealth at time $t$ and $w(t)$ is the fitting function as given by \eq{portLogPeriodic}.

The {\em Solver} routine did not provide a measure of the quality of the fit or an estimate of the fitted parameters' uncertainties but an estimate of the quality is provided by the $\chi^2$ measure itself.  If the fit is of high quality then the data should be randomly distributed around the curve with a total $\chi^2$ variance proportional to $N-7$ \cite[Ch.\ 15]{press92}.  Hence, the ratio $(N-7)/\chi^2$ should be independent of the number of data points $N$ acquired.  A small value indicates a large $\chi^2$ variance and a poor fit, while a large value indicates a good fit.  Hence, the {\em quality} of the fit $Q$, defined as
\begin{equation}
	Q \equiv \frac{N-7}{\chi^2},
\end{equation}
is a dimensionless (strictly positive) quality which increases as the fit gets better.  Using $Q$ allows us to compare fits at different times $t$ with different amounts of data $N$.  Note that the parameter $Q$ is only useful so far as ordering the fits: if $Q_i>Q_j$ for fits $i$ and $j$ (possibly at different times) then $i$ is a better fit---more likely to explain the data and with more meaningful parameter values (in particular, the forecasted crash date $t_c$).

Every day of the experiment a new data point was recorded (if a trade had been executed) and then the curve was refit to the dataset generating a new forecast for the next market crash $t_c$.  The forecasted date of the crash, date the forecast had been generated and the quality of fit $Q$ were then recorded.  A new forecast was made everyday, even in the absence of new data, because the critical time $t_c$ was constrained to occur in the future.  

Nonlinear curve fitting is basically a parameter space exploration which depends crucially on the initial choice of parameters.  The initial parameter set, at the initiation of the experiment, was chosen by first trying to establish a good power law fit (with $C=0$) and then refitting over all seven parameters.  Subsequent fits all began with parameter values that were produced by the last fit with one important exception: the critical point $t_c$ was always initialized to be the current day.  The motivation was to avoid getting stuck in sub-optimal solutions at later times and miss an impending crash.  It was preferable to impose a bias towards imminent events.  It is still possible to converge to sub-optimal solutions but it was decided that false positives were preferable to false negatives.

It is important to stress that all the forecasts from this experiment were true predictions, tabulated as the experiment progressed for analysis later.  The calculations were not performed after-the-fact so the results are not biased by foreknowledge.

\subsection{Results}

\begin{figure}\centering
	\input{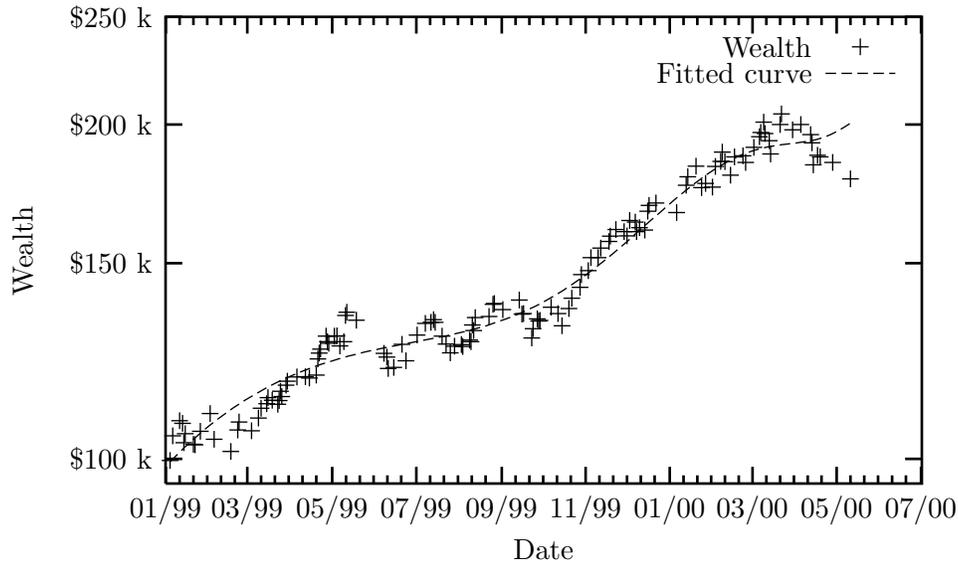}
	\caption{Sample fit of \eq{portLogPeriodic} to portfolio wealth on May 12, 2000.  The best fit parameters indicate a crash is anticipated on or around $t_c=$July 4, 2000.}
\label{fig:portLogPeriodicFit}
\end{figure}

Each day, a fit of \eq{portLogPeriodic} to the portfolio wealth data was performed giving a fitted curve similar to the one shown in \fig{portLogPeriodicFit} and the value of $t_c$ the fitting procedure converged upon was interpreted as a forecast of the time of the next crash.  

The experiment ran for 63 (week-)days and predicted a remarkable 30 crashes in that period.  Obviously, the theory predicts too many false positives.  However, it may still have some merit if the false positives have some correlation with returns.

\begin{figure}
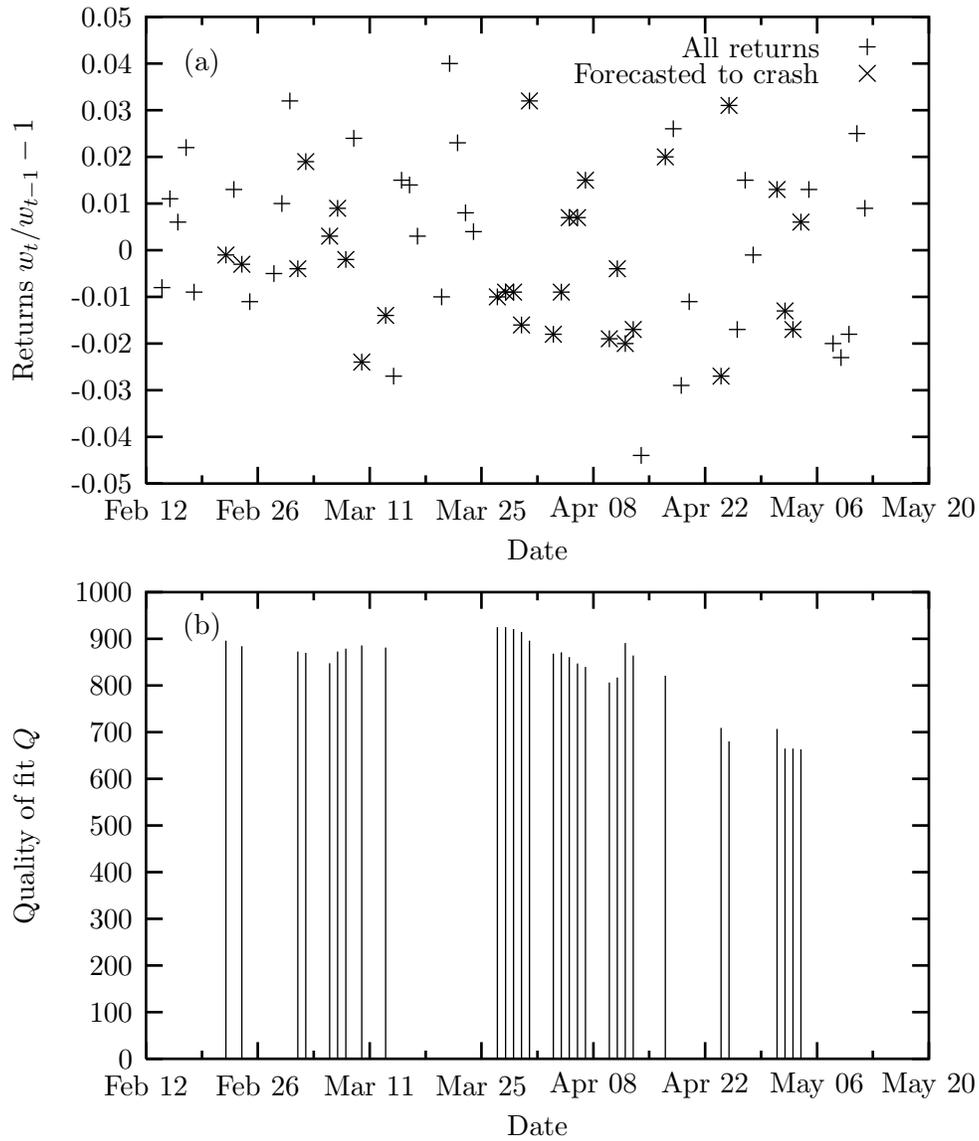
\centering
	\input{chPortfolio/crashReturns.tex} \\
	\input{chPortfolio/crashQuality.tex}
	\caption{Daily wealth returns ($w_t/w_{t-1}-1$) are shown along with the dates forecasted to crash in (a).  The qualities of the curve fits corresponding to the forecasted crashes, which suggest the reliability of the predictions, are shown in (b).}
\label{fig:portCrashReturnsQuality}
\end{figure}

The dates of forecasted crashes and their actual returns (fractional change of wealth) are plotted in \fig{portCrashReturnsQuality}.  Notice the increased numbers of forecasts for a crash in early April, in agreement with the observed decline on the 14th.  However, besides that, it is difficult to discern any pattern from the graph so some further statistical analysis is in order.

\begin{table}\centering
\begin{tabular}{r|r|r}
\hline \hline
Data set & Mean return & Std.\ Dev. \\ 
\hline
All data & $0.0\%$ & $1.8\%$ \\
Forecasted & $-0.2\%$ & $1.5\%$ \\
Not Forecasted & $0.2\%$ & $1.9\%$ \\
\hline \hline
\end{tabular}
\caption{Average values and standard deviations of the daily portfolio returns ($w_t/w_{t-1}-1$) for all data and separately for days a crash was forecasted and not forecasted.}
\label{tbl:portCrashStats}
\end{table}

We want to determine if there is a statistically significant signal in the returns on the days the market was forecasted to crash versus other days so the dataset is split into two: ``Forecasted'' and ``Not Forecasted.''  The mean returns and standard deviations were computed for both data sets (and the entire dataset) as shown in \tbl{portCrashStats}.  There does appear to be a small deviation between returns on forecasted days versus not-forecasted days but the deviation is insignificant when compared to each dataset's standard deviation.  

The likelihood that the two datasets come from the same underlying distribution can be calculated using the Kolmogorov-Smirnov (K-S) test which compares the cumulative probability distribution of the two samples \cite[Ch.\ 14]{press92}.  The calculation estimates a 24\% chance that the underlying distributions are the same, which is still statistically significant so the log-periodic precursor prediction method is not conclusive.

Recall that there was a (fortuitous) correction in the markets on and around April 14 as indicated by \fig{portValue}(b).  It would be interesting to know whether this forecasting method ``saw it coming.''  Interestingly, Johansen and Sornette submitted a paper to the LANL preprint archive on April 16 claiming to have predicted, as early as March 10, a major event between March 31 and May 2 \cite{johansen00}.

\begin{table}\centering
\begin{tabular}{r|r|r}
\hline \hline
Data set & Mean return & Std.\ Dev. \\ 
\hline
Forecasted & $-0.4\%$ & $1.4\%$ \\
Not Forecasted & $0.5\%$ & $1.9\%$ \\
\hline \hline
\end{tabular}
\caption{Same as \tbl{portCrashStats} except only including data up until the observed decline on April 14.}
\label{tbl:portCrashStatsApril14}
\end{table}

To test whether the crash around April 14 was predictable the data from \fig{portCrashReturnsQuality} are reused neglecting everything after April 14.  (Notice the quality of the fit declined markedly after April 14, suggesting the reliability of the later predictions is dubious.)  The average returns and their standard deviations for both ``Forecasted'' and ``Not forecasted'' dates is again shown in \tbl{portCrashStatsApril14}, with a somewhat more significant difference betweens the means (the standard deviations are almost unchanged).  Applying the K-S test now yields a much less significant 11.7\% likelihood that the datasets are samples from the same underlying distribution.

In conclusion, it appears that there may be some value in interpreting market crashes as critical phenomena with log-periodic precursors but the predictive advantage of doing so is limited.  The main difficulty lies in fitting seven nonlinear model parameters to a given dataset---often the fitting algorithm converges to a suboptimal solution, thereby forecasting an erroneous crash date $t_c$.

\subsection{Universality of scaling ratio}

In this section an open problem in the theory of log-periodic precursors will be presented.  

It has been observed that the scaling ratio $\lambda$ in \eq{portLogPeriodic} seems to be universal, almost always converging to a value near $\lambda\approx 2.5-3.0$ \cite{sornette95b, johansen96, sornette96, johansen99, johansen00}.  (Note this corresponds to a universal log-periodic frequency $\omega=2\pi/\ln\lambda \approx 5.5-7.0$.)  The emergence of a universal scaling ratio has come as a surprise to researchers \cite{johansen99, johansen00} since it describes some natural hierarchy within the specific system of interest and is not expected to be general.

Another peculiarity is that log-periodic fluctuations occur in some systems which do not have an obvious discrete scale invariance, such as the stock markets.  In the derivation of log-periodicity, discrete scale invariance was a fundamental ingredient, without which it did not emerge.  Why then might markets, which are not suspected to have any discrete scale invariant structures, exhibit log-periodicity?

It is my belief that these two idiosyncrasies are tied together: with the lack of a preferred scaling ratio a natural ratio is chosen, Euler's constant, $\lambda=e\approx 2.72$.  The log-periodic frequency is then $\omega=2\pi\approx 6.28$, in agreement with observation.  A mechanism that might produce this preferred scaling ratio is unknown and this issue is only discussed here to generate interest in the problem.  The discovery of a mechanism whereby $\omega$ is fixed would be a great boon to forecasting because this parameter is one of the most problematic for the optimization routine.  (Incidentally, fixing $\omega=2\pi$, the next crash (in the hypothetical portfolio) is forecasted to occur in the third week of October, 2000.)

\section{Summary}

In this chapter two experiments were performed with a hypothetical portfolio of nine stocks.  In the first experiment it was observed that the fixed investment strategy (FIS) performs sufficiently well to justify its application in the Decentralized Stock Exchange Model (DSEM).  Although it underperformed when compared to a trivial ``Buy-and-hold'' strategy, this is attributed to the strong upward trend in the portfolio over the course of the experiment.  In each case when the climb was interrupted the FIS managed to ``catch up to'' and surpass the Buy-and-hold strategy, only to lag behind again when the trend re-emerged.  The FIS also had the favourable property that it significantly reduced the kurtosis of the distribution of returns, essentially taming the largest fluctuations.  This may be relevant to derivative pricing theory \cite{black73} which assumes Gaussian-distributed increments with no excess kurtosis.

The second experiment tested a method for forecasting financial crashes.  The method relies on log-periodic oscillations in the price series which accelerate as the time of the crash approaches.  The data suggest that log-periodic precursors probably do exist but they offer little, if any, prediction advantage because the method requires solving an optimization problem involving seven nonlinear parameters.  Thus, the optimization procedure tends to get stuck in local, sub-optimal regions of the parameter landscape, frequently producing false-positive forecasts.  It would be interesting to discover whether stochastic optimization techniques, such as simulated annealing \cite{plischke94}, could provide better forecasts.

\chapter{Concluding remarks}

\label{ch:conclusions}

\section{Review}

Traditional economic theory interprets stock markets as equilibrium systems driven by exogenous events.  But this theory is incapable of explaining some peculiarities---such as the prevalence of large fluctuations---which are observed to be universal across all markets.  Instead, these phenomena are traditionally attributed to the exogenous driving factors.
The goal of this thesis was to discover whether these anomalies may arise directly from simple interactions between a large number of investors, and not depend on extraordinary external influences.

\subsection{Anomalous market properties}

Some of the peculiarities observed in the markets and not explainable by traditional economic theory follow: 

\subsubsection{Scaling}

Firstly, the distribution of returns (be they price returns for a particular stock or index returns) contain too many outliers to be adequately described by the default Gaussian distribution.  In \ch{results2} some alternatives were presented which properly capture the extra ``weight'' contained in the distribution tails.  Empirical analysis suggests the distribution is best described by a L\'evy distribution with exponent $\alpha\approx 1.40$ \cite{mantegna95} which is truncated for large returns by either a decaying exponential or a power law with an exponent near three.

\subsubsection{Clustered volatility}

Secondly, although the price series has no (significant) memory---supporting the hypothesis that markets are efficient, containing no arbitrage opportunities---the same cannot be said for market volatility.  Volatility, which describes the degree of excitation or uncertainty in the market and is quantified most simply by the absolute value of the price returns, exhibits extremely long temporal correlations.  High volatility tends to follow high and low follows low, resulting in clusters of activity.  This conflicts with traditional economic theory which states that fluctuations should be regular and uncorrelated.

To test the hypothesis that these properties may emerge spontaneously from the interactions of many simple investors, two market simulations, the Centralized and Decentralized Stock Exchange Models (CSEM and DSEM) were constructed in Chapters \ref{ch:csem} and \ref{ch:dsem}, respectively.

\subsection{Centralized stock exchange model}

CSEM was a traditional simulation, building on similar models developed over the last few years.  Its main features include centralized trading (all traders deal with a single {\em market maker}), synchronous updating and forecasting of returns.  Each forecast was deliberately nudged by a normally-distributed amount with standard deviation $\sigma_\epsilon$, the forecast error.  It was discovered that as the forecast error was reduced the system underwent a second-order (critical) phase transition near $\sigma_c\approx 0.08$, below which the price diverged (or would have if it wasn't artificially bounded).

When the distribution of the price returns was computed it was discovered that CSEM was only able to produce an overabundance of outliers (compared with the Gaussian) below the critical point, precisely in the regime where the price series is known to be unrealistic.  Above the critical point the distribution fit very well to a Gaussian.  Thus, CSEM is unable to capture the anomalous ``fat tails'' phenomenon observed empirically.  Since it failed this first test, it was not tested for any of the other properties mentioned above.  Instead, focus was shifted to the decentralized model.  (In retrospect, CSEM may have been abandoned too rashly.  By allowing multiple values of the control parameter, as in DSEM, more realistic dynamics may be realizable.  This hypothesis will be tested.)

\subsection{Decentralized stock exchange model}

DSEM arose from dissatisfaction with the structure of CSEM: synchronous, centralized trading was replaced with asynchronous, decentralized trading directly between market participants and the need for forecasting was eliminated with a simple {\em fixed investment} strategy in which agents trade in order to maintain a balance between stock and cash.  To drive the dynamics the ideal investment fraction was allowed to be affected by exogenous news events (modeled as a discrete Brownian process) and endogenously by price movements.

The dynamics were observed to have three phases of existence, depending on the strength of the agents' response to price movements: when the price response was in the region $r_1>r_p>r_2$ autocorrelations in the price series were relatively weak but as the price response passed the critical point $r_1=1$ very strong positive correlations emerged and the price diverged rapidly.  The third phase was found when the price response dropped below $r_2\approx -0.33$, revealing a first-order phase transition.  Below this point the price series was strongly anticorrelated.

When all agents were forced to share the same price response scaling in the price return distribution could not be induced except in the phases which exhibited unrealistic memory effects.  However, if the price response was sampled from a two-point distribution, scaling (with a realistic truncation for large returns) was found for a number of simulations, the best predictor for scaling being that the upper price response exceeded one, $r_{hi}>1$.  For those runs which did exhibit scaling the exponent was found to be $\alpha=1.64\pm 0.25$, which compares favourably with the best known empirical quantity $1.40\pm 0.05$ \cite{mantegna95}.

Having found that DSEM could capture this anomalous property of empirical markets it was also tested for memory effects, again using the two-point price response distribution.  It was found that the price series did not have a significant memory provided that the lower bound of the price response was in the region $0.5\leq r_{lo}<1$.  Similarly, volatility clustering was observed when the upper limit exceeded $r_{hi}>1.25$ or when the lower limit was below $r_{lo}<-0.5$.  

All three requirements were met when $0.5\leq r_{lo}<1$ and $r_{hi}>1.25$.  What this means for real markets will be discussed below.  But first we review the remainder of the thesis.

\subsection{Fixed investment strategy}

DSEM was constructed on the principle of the fixed investment strategy (FIS) which states that one should adjust one's portfolio in order to maintain a balance between the capital invested in a risky stock and the capital held in a safe(r) asset.  In \ch{portfolio} the results of an experiment intended to test the credibility of the FIS in a ``real-world'' situation (with trading costs, etc.) were reported.

It was discovered that the FIS actually underperformed when compared with a simple ``Buy-and-hold'' strategy, at least over this particular realization.  This is probably due to the strong trend observed in the portfolio over the course of the experiment, in which the capital nearly doubled.  The FIS is designed to take advantage of fluctuations in the price series and is sub-optimal in the presence of a long-term trend.  

However, the experiment did reveal an interesting (and possibly advantageous) feature of the FIS: it minimized the risk in the sense that it reduced the frequency of large events (both up and down) as measured by the excess kurtosis.  By applying the FIS large fluctuations were scaled down bringing them in line with the Gaussian distribution which is typically assumed.  Of course, it should be remembered that these conclusions are less than rigorous, being the result of a single brief experiment with a particular portfolio.

\subsection{Log-periodic precursors}

While the FIS experiment was running the hypothesis that market crashes are heralded by log-periodic precursors was also tested.  The theory derives from discrete scale invariance and suggests that, in some cases, systems approaching a critical event may exhibit accelerating oscillations in the power law describing the critical point.  

It has been suggested that detecting these oscillations may improve predictions of the critical event time and recent work in seismology is promising.  But the financial data from the FIS experiment indicate that, even if log-periodic precursors do exist, technical optimization difficulties prevent any accurate forecasts of large fluctuations therefrom.

\section{Conclusions to be drawn from this research}

The main point the reader should draw from this thesis is that it is possible to replicate realistic market dynamics with a many-agent model with simple driving forces.  DSEM was driven by a simple (discrete) Brownian motion without fat tails and having no memory, but through the interactions of the agents both fat tails and long memories (in the volatility) emerged.  Similarly, these properties may arise endogenously in real economic systems, and appeals to anomalous external events to explain them may be unwarranted.

Interestingly, the most realistic simulations were observed when the price response (control parameter) was centered around a critical point at $r_p=r_1=1$.  If DSEM is assumed to properly capture the essence of real markets the question is naturally raised: ``Why are the markets tuned to this region of parameter space?''  The fact that this region encompasses a critical point is suggestive of a concept called {\em self-organized criticality} (SOC) which claims that many dynamical systems spontaneously evolve towards a critical point \cite{bak87, bak88, bak93}.  The problem with this description is that it adds nothing to our knowledge: it does not tells us how or why the market self-organizes.  

In a simple economic model involving producers and consumers it was discovered that the system self-organizes to the critical state in order to maximize efficiency \cite[Ch.\ 11]{bak96}.  On one side of the critical point the supply outweighs demand and on the other the reverse is true.  In this example it is easy to see why the market would self-organize.  To test whether a similar process could drive DSEM to the critical state DSEM has been extended to allow the agents to adjust their {\em preferences} (news response and price response parameters) when their current choices are performing poorly.  This is discussed further below.

Another interesting consequence of the observation that the price response is centered around $r_p=1$ is that---if DSEM is at all meaningful---real investors do watch (and base decisions on) trends in stock prices.  In DSEM, to get realistic behaviour, even the least responsive agents had to have $r_{lo}\geq 0.5$ which can roughly be interpreted as the perceived autocorrelation between successive returns.  DSEM suggests that there do not exist any (pure) {\em fundamentalist} traders (who respond only to fundamental information about the company and are unconcerned with the stock's price movements) in real markets.  Unfortunately, while an interesting hypothesis, it is not clear how this assertion could be tested empirically.

\section{Relation to other work in the field}

Quite a few market models have been developed over the last few years.  In this section some of these models are contrasted with CSEM and DSEM.

We begin by comparing how the price is chosen in the models.  Recall that in CSEM the price is set by an auctioneer in order to balance supply and demand.  In DSEM, however, the price is simply the most recent trading price.  In most of the models reviewed the price was set by an external {\em market maker} as in CSEM \cite{palmer94, youssefmir94, levy95, caldarelli97, cont97, chen98, stauffer98, busshaus99, chang99, chowdhury99, eguiluz99, iori99, lux99, stauffer99} the only exceptions being reaction-diffusion models \cite{bak97, tang99} in which buyers and sellers diffuse in price space and a trade is executed when they meet.  DSEM provides a new mechanism for allowing the price to emerge directly from the agents' decisions.

Another major difference between the CSEM and DSEM is in how they are updated: in CSEM trades are executed synchronously, once per day while DSEM allows trading in real time, with agents choosing their own activation times.  On this front it appears that asynchronous updating is becoming more prevalent \cite{youssefmir94, chen98, chowdhury99, eguiluz99} with more of the older sources choosing parallel updating \cite{palmer94, levy95, arthur97, bak97, caldarelli97, busshaus99, iori99}.  This is fortunate because a mounting volume of evidence suggests that parallel updating may introduce spurious artifacts into simulation dynamics \cite{huberman93, bersini94, rajewsky97, rolf98, blok99}.

The preferred litmus test for each of these models is whether they can reproduce fat tails in the price return distribution and many of them can \cite{caldarelli97, cont97, stauffer98, busshaus99, chang99, chowdhury99, dhulst99, eguiluz99, stauffer99}.

The Cont-Bouchaud percolation model \cite{cont97} has received a great deal of attention lately \cite{stauffer98, chang99, chowdhury99, eguiluz99, stauffer99}.  It is characterized by a network of information which produces herding effects.  The advantage of the model is that analytic results exist \cite{cont97, dhulst99} which predict that the price return distribution should have a (truncated) power law distribution (with a scaling exponent $\alpha=3/2$).  It has also been demonstrated to exhibit clustered volatility \cite{chang99, stauffer99}.  DSEM provides an alternative explanation which does not require herding.  However, it would be interesting to know what the consequences of herding would be, which brings us to directions for future research.

\section{Avenues for further work}

\label{sect:concFuture}

I conclude this thesis with some thoughts on how DSEM may be extended to produce new insights and on further statistical properties which could be tested:

As discussed above, one of the most pressing issues is whether scaling and clustered volatility can emerge spontaneously without requiring tuning of the price response parameters.  This can be tested by allowing the agents to choose their preferences (response parameters) as they see fit.  To do so, a {\em meta-strategy} is required which controls when an agent adjusts its preferences and by how much.  An arbitrary but reasonable choice is to allow preference adjustments when the agent's performance is demonstrably poor: for instance, if the agent sells shares at a price below the average price it bought them for.  When this occurs the agent randomly shifts its preferences by some amount.  This has been recently coded into DSEM and research is ongoing.  

Another interesting direction to explore is the extension of DSEM to support multiple stocks.  This idea was inspired by Bak et al. \cite{bak97} in which they described adding a new stock as adding a new dimension in price space.  It is well known that the dimensionality is one of the few factors which can impact the character of a critical point \cite{plischke94} so it would be interesting to see how the critical point in DSEM would be affected.

On the surface CSEM and DSEM are quite different.  However, it should be possible to modify DSEM such that all trades are handled by a centralized control or market maker.  The agents could respond to orders called out by an auctioneer in similar manner to CSEM.  Discovering whether scaling and clustered volatility are robust to these changes would be very informative.

On the experimentation side, there are a number of statistical properties which could be tested for.  One of these is an asymmetry between up- and down-movements in the price series.  Roehner and Sornette \cite{roehner98} found that peaks tend to be sharp but troughs (lows) tend to be flat.  Since DSEM is symmetric in its response to up- and down-moves it would be surprising if this asymmetry could be replicated.

Another interesting property which is currently being tested (but did not make it into this thesis) is Pareto's law for the distribution of incomes which states that the richest segment of the population have incomes in excess of that predicted by the log-normal distribution (which fits the majority of the population).  This is thought to be an amplification effect whereby the richest individuals are able to leverage their wealth to increase their income faster than others \cite{montroll83}.  Data are being collected to test for this effect in DSEM.

Beyond that, the price series contains more information than just the distribution of returns.  For instance, the intra-trade interval and bid-offer spread are also interesting with testable distributions \cite{eliezer98}.

Finally, evidence is mounting that the distribution of empirical returns is truncated by an inverse cubic power law \cite{gopikrishnan98, gopikrishnan99, plerou99} rather than the exponential assumed in \sect{results2PriceFluct}.  It would be useful to determine which hypothesis DSEM obeys.  To do so, much larger datasets are required in order to determine the distribution of very large returns (since it is difficult to distinguish the two hypotheses on scales studied in this dissertation).  Alternatively, the moments of the distribution could be explored: if the exponential truncation holds then all moments should be finite but the inverse cubic implies that the $k$-th moment should diverge as the index $k$ increases to three.  Either way, it would be valuable to determine if the distribution of returns in DSEM is truncated by an inverse cubic as appears to be the case for empirical data.

In short, many exciting possibilities remain for future research into DSEM.


\appendix    

\chapter{Discounted least-squares curve fitting}

\label{ap:dls}

In this appendix the standard method of least-squares curve fitting is modified in order to make it more amenable to time series.  In particular the goal is to use time series data for forecasting by extrapolating from historical data.  As will be shown this method can require fewer computations and less storage.  Also, by discounting historical data extrapolated forecasts become more robust to outliers.

The reader should keep in mind that, despite the similarity of notation with standard least-squares curve fitting, the following is specifically meant to be applied to time series, where the relevance of past data are {\em discounted} as newer data arrive.

This appendix borrows heavily from Press et al.'s excellent discussion of generalized least-squares curve fitting \cite[Sect.\ 15.4]{press92} which is highly recommended.

\section{Least-squares curve fitting}

We use the index $i$ to label our data points where $i=0$ indicates the most recently acquired datum and $i=1,2,3,\dots$ indicate successively older data.  Each point consists of a triplet $(x,y,\sigma)$ where $x$ is the independent variable (eg. time), $y$ is the dependent variable, and $\sigma$ is the associated measurement error in $y$.

We wish to fit data to a model which is a linear combination of {\em any} $M$ specified functions of $x$.  The general form of this kind of model is
\begin{equation}
\label{eq:fitFunction}
	y(x) = \sum_{j=1}^M a_j X_j(x)
\end{equation}
where $X_1(x),\ldots,X_M(x)$ are arbitrary fixed functions of $x$, called the {\em basis functions}.  For example, a polynomial of degree $M-1$ could be represented by $X_j(x) = x^{j-1}$.  (Note that the functions $X_j(x)$ can be wildly nonlinear functions of $x$.  In this discussion ``linear'' refers only to the model's dependence on its {\em parameters} $a_j$.)  

A merit function is defined
\begin{equation}
\label{eq:chiSq}
	\chi^2 = \sum_{i=0}^N \left[ \frac{ y_i - \sum_j a_j X_j(x_i) }{\sigma_i} \right]^2.
\end{equation}
which sums the (scaled) squared deviations from the curve of all $N$ points.  The goal is to minimize $\chi^2$.

The derivative of $\chi^2$ with respect to all $M$ parameters $a_j$ will be zero at the minimum
\begin{equation}
\label{eq:chiSqMin}
	0=\sum_i \frac{1}{\sigma_i^2} \left[ y_i - \sum_j a_j X_j(x_i) \right] X_k(x_i), \; k=1,\ldots,M
\end{equation}
giving the best parameters $a_j$.

If we define the components of an $M\times M$ matrix $[\alpha]$ by
\begin{equation}
\label{eq:defAlpha}
	\alpha_{kj} = \sum_i \frac{ X_j(x_i) X_k(x_i) }{ \sigma_i^2 }
\end{equation}
and a vector $[\beta]$ of length $M$ by
\begin{equation}
\label{eq:defBeta}
	\beta_k = \sum_i \frac{ y_i X_k(x_i) }{ \sigma_i^2 }
\end{equation}
then \eq{chiSqMin} can be written as the single matrix equation
\begin{equation}
\label{eq:normalEq}
	[\alpha] \cdot {\bf a} = [\beta]
\end{equation}
where $\bf a$ is the vector form of the parameters $a_j$.  

Eqs.\ \ref{eq:chiSqMin} and \ref{eq:normalEq} are known as the {\em normal equations} of the least-squares problem and can be solved for the vector parameters $\bf a$ by {\em singular value decomposition} (SVD) which, although slower than other methods, is more robust and is not susceptible to round-off errors \cite[Ch.\ 2]{press92}.

\section{Discounting}

The discussion above applies to all linear least-squares curve fitting.  The variation proposed here is to discount the relevance of historical data as new data arrive.  This was motivated by time series where the fitting parameters may vary slowly.  

\begin{figure} 
	\psfrag{standard}[c][c]{standard}
	\psfrag{N}[c][c]{$N$}
	\psfrag{Data Windowing}[c][c]{Data Windowing}
	\psfrag{discounted}[c][c]{discounted}
	\psfrag{weight}[c][c]{weight}
	\psfrag{data}[c][c]{data}
	\psfrag{old}[c][c]{old}
	\psfrag{new}[c][c]{new}
	\psfrag{0}[c][c]{0}
	\psfrag{1}[c][c]{1}
\includegraphics[width=\columnwidth]{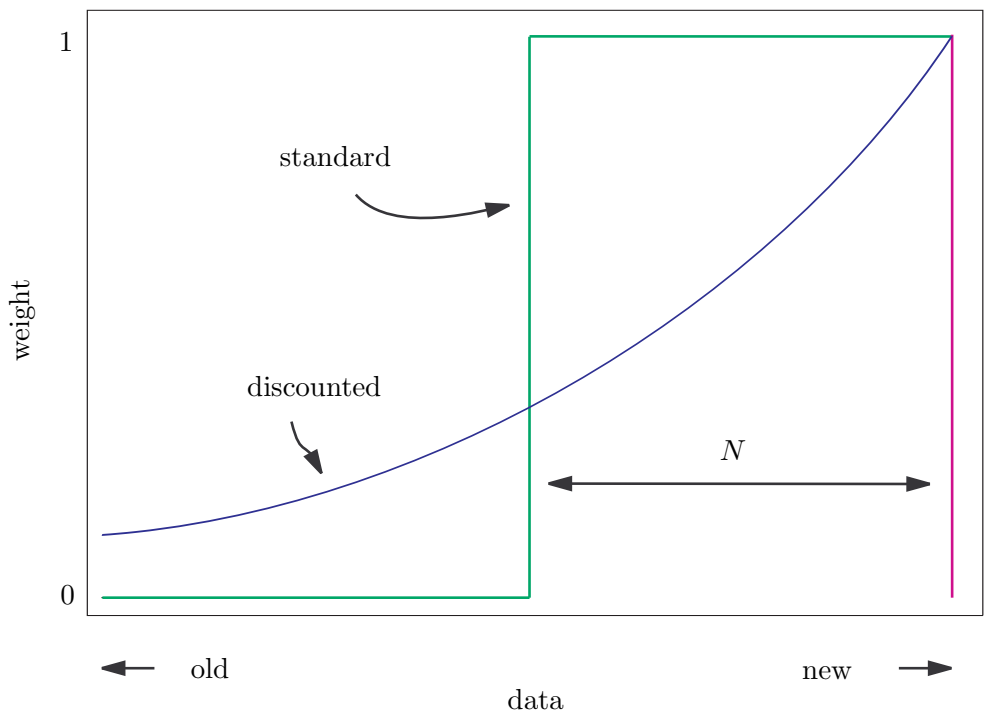}
	\caption{Comparison of weightings using standard and discounted windows.}
\label{fig:dataWindow}
\end{figure}

Fitting time series is typically handled with a moving window over the last $N$ data points.  Each of the last $N$ points is weighted equally and all prior data is discarded as shown in \fig{dataWindow}.  The discontinuous weighting function can introduce discontinuities in the fitting parameters $a_j$ as the data is updated, particularly when an {\em outlier} (a strongly atypical $y$-value) is suddenly discarded.

These discontinuities can be avoided by steadily discounting old data as new data arrive.  As will be shown, this method also has computational and resource advantages.

As before, we use the index $i$ to label our data points with larger $i$ indicating older data.  As a new datum arrives $(x_0,y_0,\sigma_0)$ we shift the indices of prior data and scale up the errors by some factor $0<\gamma<1$
\begin{equation}
	(x_{i+1}, y_{i+1}, \sigma_{i+1}) \leftarrow (x_i, y_i, \sigma_i/\gamma).
\end{equation}

If we define $\sigma_i^*$ as the original value of $\sigma_i$ then after applying $i$ of the above operations
\begin{equation}
	\sigma_i = \sigma_i^* / \gamma^i
\end{equation}
so, since $\gamma<1$, the historical deviations grow exponentially as new information is acquired.  Increasing the error effectively decreases the weight of a datum in the fitting procedure.  

Calculation of the covariance matrix and the uncertainties of the parameters proceeds as with standard least-squares fitting (see \cite[Ch.\ 15]{press92}, for instance) so I will just mention the main result, namely that the inverse of $[\alpha]$
\begin{equation}
\label{eq:defC}
	{\bf C} = [\alpha]^{-1}
\end{equation}
gives the covariances of the fitting parameters
\begin{equation}
	\cov{a_j}{a_k} = C_{jk}
\end{equation}
and the variance of a single parameter is, of course,
\begin{equation}
	\var{a_j} = C_{jj}.
\end{equation}

\section{Storage and updating}

So far we have made no mention of $N$, the number of data points to be fit.  From \fig{dataWindow} it appears we need to store the entire history to apply this technique.  But notice that as we acquire a new datum $(x_0,y_0,\sigma_0)$, from Eqs.\ \ref{eq:defAlpha} and \ref{eq:defBeta}, the matrix $[\alpha]$ and vector $[\beta]$ update as
\begin{equation}
\label{eq:updateAlpha}
	\alpha_{kj} \leftarrow \frac{ X_j(x_0) X_k(x_0) }{\sigma_0^2} + \gamma^2 \alpha_{kj}
\end{equation}
and
\begin{equation}
\label{eq:updateBeta}
	\beta_j \leftarrow \frac{ X_j(x_0) y_0 }{\sigma_0^2} + \gamma^2 \beta_j
\end{equation}
so it appears we need not store any data points, but should just store $[\alpha]$ and $[\beta]$ and update them as new data are accumulated.

A useful measure we have neglected to calculate so far is $\chi^2$, the chi-square statistic itself.  In (partial) matrix notation \eq{chiSq} can be written
\begin{eqnarray}
	\chi^2 & = & \sum_i \frac{y_i^2}{\sigma_i^2} + {\bf a}^T\cdot [\alpha]\cdot {\bf a} - {\bf a}^T\cdot [\beta] - [\beta]^T\cdot {\bf a} \\
	       & = & \sum_i \frac{y_i^2}{\sigma_i^2} + {\bf a}^T\cdot ([\alpha]\cdot {\bf a} - [\beta]) - [\beta]^T\cdot {\bf a} \\
	       & = & \sum_i \frac{y_i^2}{\sigma_i^2} - [\beta]^T\cdot {\bf a} \\
\end{eqnarray}
which appears to still depend on the data history in the first term.  Let us define this term as a new variable $\delta$,
\begin{equation}
	\delta \equiv \sum_i \frac{y_i^2}{\sigma_i^2}.
\end{equation}
Then, similarly to Eqs.\ \ref{eq:updateAlpha} and \ref{eq:updateBeta}, $\delta$ can be updated as more information is accumulated
\begin{equation}
	\delta \leftarrow \frac{y_0^2}{\sigma_0^2} + \gamma^2 \delta
\end{equation}
without requiring the entire data history.

Finally, it may be useful to record the number of points accumulated.  But because each point loses relevance as it gets ``older'' we should likewise discount this measure, giving an effective memory
\begin{equation}
\label{eq:updateN}
	N^* \leftarrow 1 + \gamma^2 N^*
\end{equation}
(not to be confused with the number of parameters $M$.)

\begin{figure}\centering
	\input{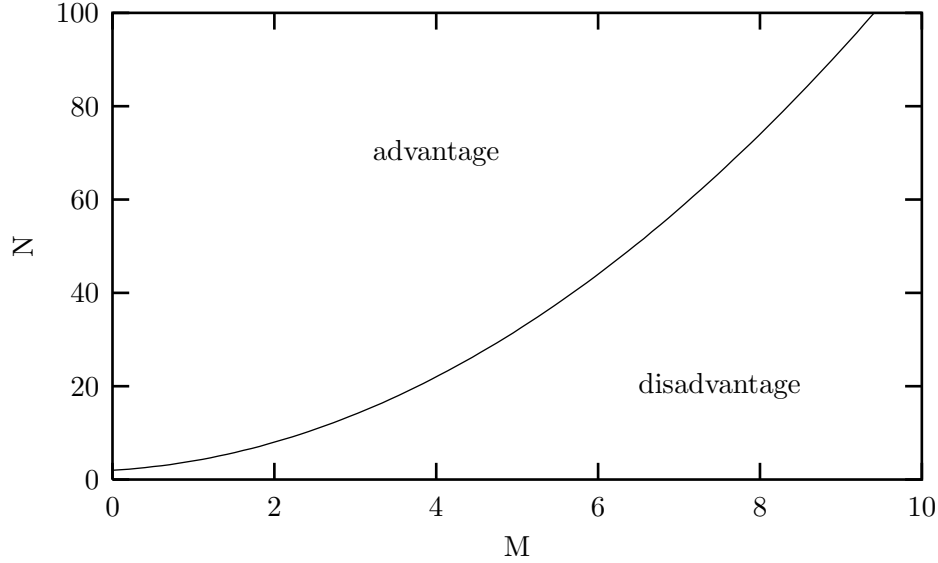}
	\caption{Discounted least-squares fitting has a computational storage advantage over moving windows of $N$ data points when $N>M^2+M+2$ where $M$ is the number of parameters to be fitted.}
\label{fig:storage}
\end{figure}

So, to store all relevant historical information we need only remember $[\alpha]$, $[\beta]$, $\delta$, and $N^*$ for a total of $M^2+M+2$ numbers, regardless of how many data points have been acquired.  \fig{storage} shows that for many practical problems discounted least-squares fitting requires less storage than the standard moving window.  Although it has not been tested, I expect a similar condition to hold for processing time.

As the reader can justify, all of these values should be initialized (prior to any data) with null values: $[\alpha]=\bf 0$, $[\beta]=\bf 0$, $\delta=0$, and $N^*=0$.

\section{Memory}

For traditional least-squares fitting it is well known that if the measurement errors of $y_i$ are distributed normally then the method is a {\em maximum likelihood estimation} and the expectation value of \eq{chiSq} evaluates to
\begin{equation}
	\expect{\chi^2} = N-M
\end{equation}
because each term $(y_i-y(x_i))/\sigma_i$ should be distributed normally with mean zero and variance one and there are $N-M$ degrees of freedom to sum the variances over.

Similarly with discounting, assuming $(y_i-y(x_i))/\sigma_i^*$ has variance one (notice this is the unscaled error), 
\begin{eqnarray}
	\expect{\chi^2} & = & \sum_{i=0}^N \gamma^{2i} \expect{ \left[ \frac{y_i - y(x_i)}{\sigma_i^*} \right]^2 } - M \\
	                & = & \sum_i \gamma^{2i} - M \\
\label{eq:expectChiSq}
	                & = & N^* - M
\end{eqnarray}
from \eq{updateN}.

Notice that as the amount of data collected grows
\begin{equation}
	N_{max}^* \equiv \lim_{N\rightarrow \infty} N^* = \frac{1}{1-\gamma^2}
\end{equation}
which relates the discounting factor $\gamma$ to the effective memory $N^*$.  Conversely, it is more natural to set $\gamma$ such that it produces the desired memory via
\begin{equation}
	\gamma(N_{max}^*) = \sqrt{1 - \frac{1}{N_{max}^*}}.
\end{equation}

\section{Unknown measurement errors}

On occasion measurement uncertainties are unknown and least-squares fitting can be used to recover an estimate of these uncertainties.  Be forewarned that this technique assumes normally distributed (around the curve) $y$ data with identical variances.  If this is not the case, the results become meaningless.  It also precludes the use of a ``goodness-of-fit'' estimator (such as the incomplete gamma function, see \cite[Sect.\ 6.2]{press92} because it {\em assumes} a good fit.

We begin by assuming $\sigma_i^*=1$ for all data points and proceeding with our calculations of $\bf a$ and $\chi^2$.  If all (unknown) variances are equal $\sigma^*\equiv \sigma_i^*$ then \eq{expectChiSq} actually becomes
\begin{equation}
	\expect{\chi^2} = (N^*-M)\sigma^{*\, 2}
\end{equation}
so the actual data variance is best estimated by
\begin{equation}
\label{eq:calcSigma}
	\sigma^{*\, 2} = \frac{\chi^2}{N^*-M}.
\end{equation}

We can update our parameter error estimates by recognizing that, from Eqs.\ \ref{eq:defAlpha} and \ref{eq:defC}, the covariance matrix is proportional to the variance in the data, so
\begin{equation}
\label{eq:updateC}
	C_{jk} \leftarrow \sigma^{*\, 2} C_{jk}.
\end{equation}

\section{Forecasting}

Forecasting via curve fitting is a dangerous proposition because it requires extrapolating into a region beyond the scope of the data, where different rules may apply and, hence, different parameter values.  Nevertheless, it is often used simply for its convenience.  We assume the latest parameter estimations apply at the forecasted point $x$ and simply use \eq{fitFunction} to predict \begin{equation}
	y_f = y(x) = \sum_j a_j X_j(x).
\end{equation}

The uncertainty in the prediction can be estimated from the covariance matrix.  Recall, the definition of variance is
\begin{equation}
	\var{z} \equiv \expect{ \left( z - \expect{z} \right)^2 }
\end{equation}
and the covariance between two variables is defined as
\begin{equation}
	\cov{z_1}{z_2} \equiv \expect{ \left( z_1 - \expect{z_1} \right)\left( z_2 - \expect{z_2} \right) }
\end{equation}
so \eq{fitFunction} has variance
\begin{eqnarray}
	\var{y(x)} & = & \var{ \sum_j a_j X_j(x) } \\
	           & = & \expect{ \left( \sum_j \left( a_j - \expect{a_j} \right) X_j(x) \right)^2 } \\
	           & = & \sum_{jk} X_j(x) \expect{ \left( a_j - \expect{a_j} \right)\left( a_k - \expect{a_k} \right) } X_k(x) \\
	           & = & \sum_{jk} X_j(x) C_{jk} X_k(x)
\end{eqnarray}
where $\bf C$ is the covariance matrix with possible updating, in the absence of measurement errors, according to \eq{updateC}.

The above gives the uncertainty in $y(x)$ but in the derivation it was assumed that the observed $y$-values were distributed normally around the curve where $y(x)$ represents the mean of the distribution.  Similarly for the prediction, $y(x)$ is the prediction of the mean with its own uncertainty---on top of which there is the measurement uncertainty of data around the mean $\sigma_{\rm meas}$.  These two uncertainties are mutually independent so the variances of the two simply add to give the cumulative variance of the prediction
\begin{eqnarray}
	\var{y_f} & = & \var{y(x)} + \sigma_{\rm meas}^2 \\
	          & = & \sum_{jk} X_j(x) C_{jk} X_k(x) + \sigma_{\rm meas}^2.
\end{eqnarray}

\subsection{Unknown measurement errors}

If the measurement errors are not known in advance, but are calculated from \eq{calcSigma} then the above formula should be rewritten
\begin{equation}
	\var{y_f} = \sigma^{*\, 2} \left( \sum_{jk} X_j(x) C_{jk} X_k(x) + 1 \right)
\end{equation}
where $C_{jk}$ in this equation, are the covariances {\em without} rescaling.



\section{Summary}

Discounted least-squares curve fitting differs from the traditional linear least-squares method in that the uncertainties of older data are artificially amplified as new data are acquired, effectively discounting the relevance of older data.  Discounting provides a very efficient method of storing the entire data series in only $M^2+M+2$ values, where $M$ is the number of parameters to be fit, regardless of the length of the series.  Discounting also smooths the fit, reducing the effects of outliers.

It has been demonstrated how discounted least-squares can be used for forecasting.  Whether it is valid depends very much on the time series in question, and its consistency.  If the fitting parameters vary on time scales of the same order or smaller than the memory $N^*$ of the fit then the forecasts will not be reliable.  (Of course, a suitable model of the time series is necessary as well.)

I have found no evidence of discounting being applied to curve fitting before; the only similar procedure I have found is ``exponential smoothing'', a technique which uses damping coefficients to smooth forecasts.  However, being such a simple premise I am confident this technique has already been discovered, I just don't know where to look.
\chapter{Sampling discrete processes}

\label{ap:sampling}

Frequently computer simulations generate synthetic Brownian motion via a simple random walk at discrete intervals.  Sampling of such a process to get the distribution of increments $p(x)$ can be problematic because of introduced artifacts which bias the statistics.

One often-used statistic is the (excess) kurtosis, defined as
\begin{equation}
	\kurt [x] = \frac{\mu_4}{\mu_2^2} - 3
\end{equation}
where $\mu_k$ is the $k$'th (centered) moment of the distribution
\begin{equation}
	\mu_k = \expect{ \left[ x - \expect{x} \right]^k }.
\end{equation}
The kurtosis is useful because it quantifies the ``weight'' of the distribution tail (far from the mean).  For the Gaussian the excess kurtosis is zero (because $\mu_4=3\mu_2^2$) compared to which a negative kurtosis indicates less weight in the tails and a positive indicates more.

Difficulties arise, however, if the Brownian motion is generated by a discrete process as will be demonstrated in two examples below.  Unless great care is taken, the kurtosis may be artificially inflated by regular sampling.

\section{Simple random walk}

\begin{figure}\centering
	\input{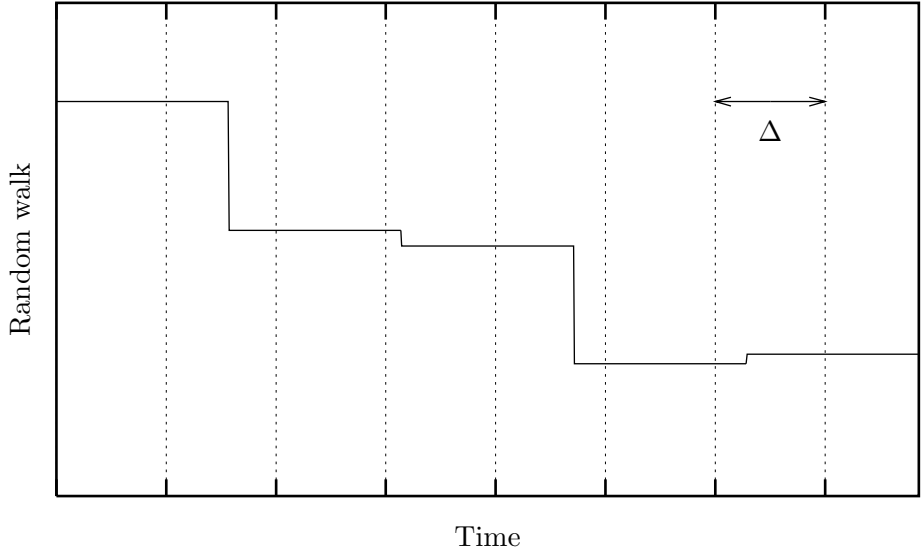}
	\caption{When a random walk is generated at some regular interval and sampled at another, $\Delta$, the number of jumps between samples will vary.}
\label{fig:samplingSampling}
\end{figure}

Consider a discrete Brownian process with normally-distributed (zero mean, unit variance) jumps at regular intervals of $\tau=1$ (without loss of generality).  If this process is sampled at regular intervals of $\Delta\neq \tau$, as demonstrated in \fig{samplingSampling}, some intervals will have more ``jumps'' than others so the distribution of increments will not be Gaussian.

To be precise, let $\Delta\equiv n+r$ where $n\equiv \floor{\Delta}$ is the largest integer not greater than $\Delta$ (the {\em floor} of $\Delta$) and $0\leq r < 1$ is the remainder.  Then each interval will span at least $n$ jumps, spanning $n+1$ with the probability $r$.  Since each jump $x$ is normally distributed $N(x;0,1)$ with zero mean and unit variance, $j$ jumps are also normally distributed with zero mean and variance $j$, denoted by $N(x;0,j)$.  The distribution of increments of the random walk, sampled at intervals of $\Delta=n+r$ is then given by
\begin{equation}
	RW(x;0,\Delta)=(1-r)N(x;0,n) + r N(x;0,n+1).
\end{equation}

Calculating the first four moments of the increment distribution is very straight-forward since
\begin{equation}
	\mu_k[RW(x;0,\Delta)] = (1-r) \mu_k[N(x;0,n)] + r \mu_k[N(x;0,n+1)]
\end{equation}
and the normal distribution has moments $\mu_1=0$, $\mu_2[N(x;0,j)]=j$, $\mu_3=0$, and $\mu_4=3\mu_2^2$.  Therefore, the moments of $RW$ are
\begin{eqnarray}
	\mu_1 & = & 0 \\
	\mu_2 & = & (1-r)n + r(n+1) = n+r = \Delta \\
	\mu_3 & = & 0 \\
	\mu_4 & = & 3(1-r) n^2 + 3r (n+1)^2 = 3(\Delta^2 + r(1-r)).
\end{eqnarray}
Notice that the variance of the distribution is simply $\Delta$, exactly the same as for {\em continuous} Brownian motion sampled at intervals of $\Delta$.  

\begin{figure}\centering
	\input{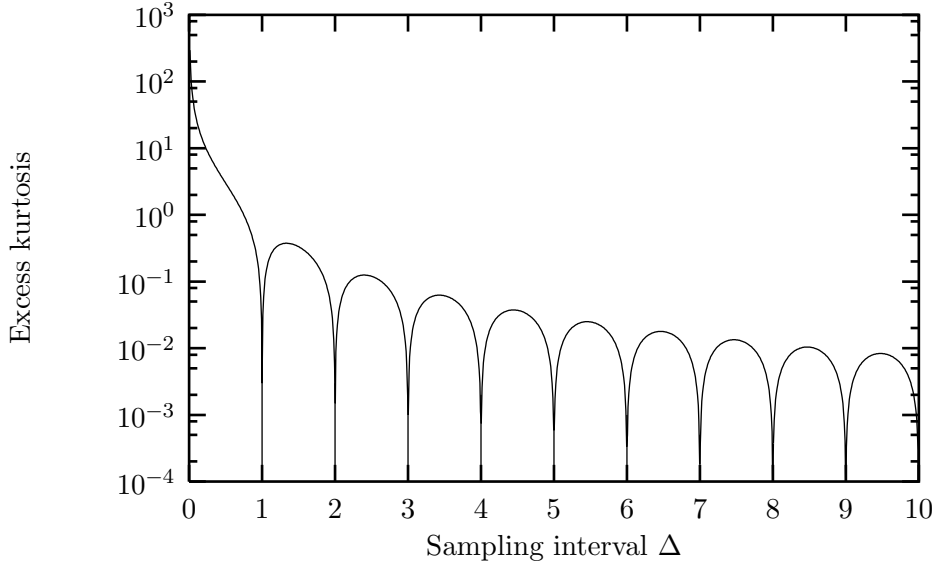}
	\caption{The kurtosis is only zero at integer values of the sampling interval $\Delta$ and diverges as the sampling interval approaches zero.}
\label{fig:samplingKurtRW}
\end{figure}

In fact, all three of the lowest moments are identical to Brownian motion, lulling us into a false sense of security.  However, the fourth moment differs and the excess kurtosis, which is zero for Brownian motion, is now
\begin{equation}
	\kurt[RW(x;0,\Delta)] = 3 \frac{r(1-r)}{\Delta^2}
\end{equation}
which, on the surface, would seem to indicate the distribution has fat tails.  The kurtosis is only zero at integer values of $\Delta$ ($r=0$) and is a maximum for any $n$ when $r=n/(1+2n)$ as shown in \fig{samplingKurtRW}.

In particular, the kurtosis diverges as the sampling rate accelerates
\begin{equation}
	\kurt[RW] \rightarrow \frac{3}{\Delta} \mbox{ as } \Delta\rightarrow 0,
\end{equation}
a result of the Dirac delta function $N(x;0,0)$ dominating the distribution, scaling the variance down faster than the fourth moment.

Even though all the evidence presented suggests that the distribution of increments in the random walk truly does have fat tails when sampled at non-integer intervals $\Delta$, it is actually just an artifact of sampling.  

Since we are getting an overlap of two Gaussian distributions, with variances $n$ and $n+1$, the center of the distribution is dominated by the smaller variance contribution and the tails are dominated by the larger variance.  Hence, the second moment of the random walk is scaled down by the smaller variance but the fourth moment is scaled up by the larger.  The net effect is the illusion of fat tails in the distribution.  

\begin{figure}\centering
	\input{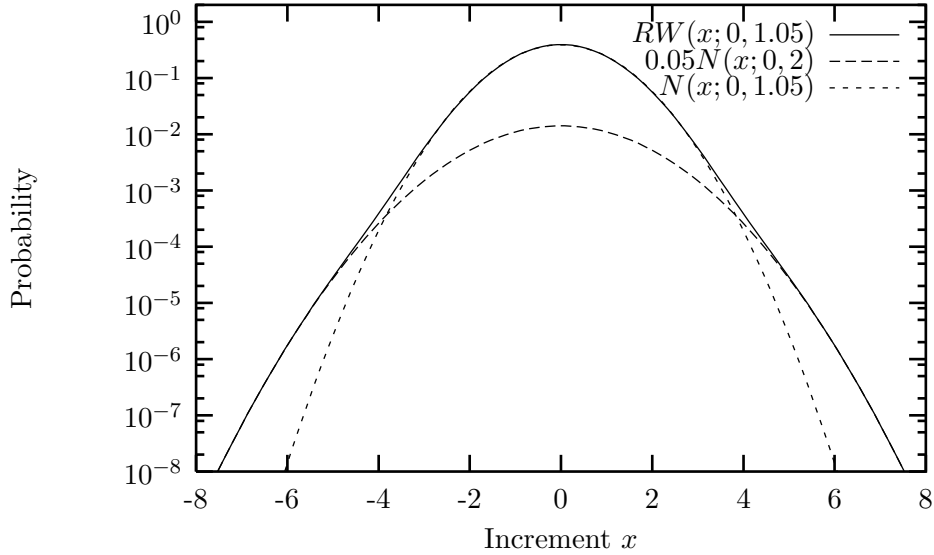}
	\caption{The distribution of increments for the random walk appears to have fatter tails than a normal distribution with the same variance when sampled at intervals of $\Delta=1.05$.  However, the tails still drop off as $e^{-x^2}$.}
\label{fig:samplingDistRW}
\end{figure}

However, the tails of the distribution still fall off as $e^{-x^2}$, as demonstrated in \fig{samplingDistRW}, so the term ``fat tails'' is misleading, usually being reserved for simple exponential or power law tails.  

Notice that the center of the distribution behaves as a normal with variance $\Delta$ and the tail also behaves as a normal, with variance $n+1$, but weighted by $r$.  The crossover between the two regimes, after some algebra, is found to be
\begin{equation}
	x_c = \sqrt{ \frac{ \left[ \ln(n+1) - 2 \ln(r\sqrt{\Delta}) \right] \Delta (n+1) }{1-r} }.
\end{equation}
This indicates the scale of increments, $x\approx \pm x_c$, for which the distribution will appear most strongly non-Gaussian.

Next we consider a process generated at Poisson intervals rather than regular.

\section{Poisson Brownian motion}

In this section we again consider a discrete Brownian motion but, in this case, the intervals between the jumps are Poisson-distributed instead of regular.  The Poisson distribution gives the probability of $j$ events within a time interval $t$ given an average event rate $\tau\equiv 1$ (without loss of generality),
\begin{equation}
	P(j,t)= e^{-t} \frac{t^j}{j!}.
\end{equation}
Given normally-distributed jump sizes the distribution of $j$ jumps is $N(x;0,j)$ so the distribution of increments of the Poisson Gaussian process at intervals of $\Delta$ is
\begin{equation}
	PG(x;0,\Delta)=\sum_{j=0}^\infty P(j,\Delta) N(x;0,j).
\end{equation}

The analytic solution for the distribution of increments is challenging but the moments of the distribution are relatively easy to compute,
\begin{eqnarray}
	\mu_k[PG(x;0,\Delta)] & = & \int dx \sum_{j=0}^\infty P(j,\Delta) N(x;0,j)\, x^k \\
	          & = & \sum_{j=0}^\infty P(j,\Delta) \int dx\, N(x;0,j)\, x^k \\
	          & = & \sum_{j=0}^\infty P(j,\Delta)\, \mu_k[N(x;0,j)],
\end{eqnarray}
depending directly on the moments of the normal distribution (which were presented in the last section).

From the identity $e^x\equiv \sum_j x^j/j!$, the first four moments of the Poisson Gaussian are
\begin{eqnarray}
	\mu_1 & = & 0 \\
	\mu_2 & = & \Delta \\
	\mu_3 & = & 0 \\
	\mu_4 & = & 3 \Delta (\Delta+1).
\end{eqnarray}

Again, the first three moments are unchanged from the normal distribution but the kurtosis becomes
\begin{equation}
	\kurt[PG(x;0,\Delta)] = \frac{3}{\Delta}
\end{equation}
for all $\Delta$.  (This form was also observed for the random walk in the limit $\Delta \rightarrow 0$.)  So, again the kurtosis diverges as the sampling interval drops towards zero.

\begin{figure}\centering
	\input{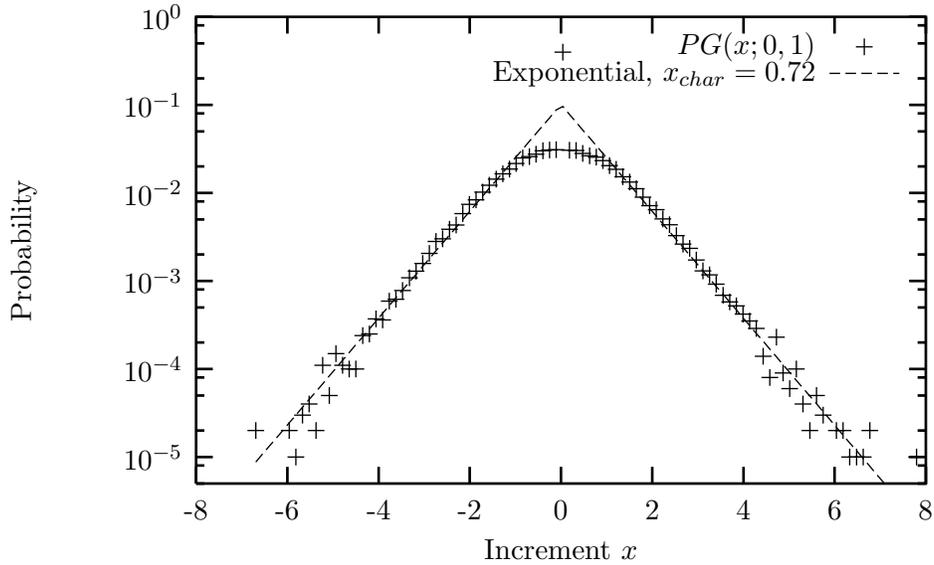}
	\caption{Discrete Brownian motion with Poisson-distributed jump intervals has tails which fall off exponentially (with a decay constant of 0.72), instead of as $e^{-x^2}$, when sampled at regular intervals ($\Delta=1$).}
\label{fig:samplingDistPG}
\end{figure}

In the case of the random walk we found the excess kurtosis arose from the overlap of two Gaussians but the tails still fell off as $e^{-x^2}$.  However, for Poisson Brownian motion the distribution tails are much heavier.  A synthetic dataset generated from a Poisson Brownian motion sampled at intervals of $\Delta=1$ (\fig{samplingDistPG}) shows that the tails fall off only exponentially,  

\section{Sampling}

Evidently, by generating a synthetic Brownian motion at non-uniform intervals, the illusion of fat tails can be achieved by simply sampling the process regularly.  However, the underlying process is still generated by Gaussian-distributed jumps and, over long timescales, still looks like Brownian motion.

The easiest way to avoid these artifacts is to not sample the process in ``real time'' but in ``event time.''  That is, take a single sample after each event.  Then, the underlying jump process will be revealed without any complications from zero or multiple events per sample.

Unfortunately, in some cases the available data do not allow for the determination of individual events.  In this case, a very high frequency sampling is recommended and all intervals with zero increment should be discarded.  High frequency sampling minimizes the likelihood that multiple jumps could occur in any one interval but increases the likelihood of zero increments.  By discarding these null events only the intervals with a single increment remain.  (This also discards actual jump events of size zero but this should have a minimal bias on the statistics since a jump size of identically zero has a negligible probability measure.)

\chapter{Long-range memory: The Hurst exponent}

\label{ap:hurst}

Brownian motion (in one dimension) is a random walk on the line where the step length is given by a mean zero Gaussian (normal) probability distribution.  Since each of the steps are independent the cumulative position $X$ is known to obey
\begin{eqnarray}
  \expect{X(t)-X(0)} & = & 0 \\
  \expect{\left[X(t)-X(0)\right]^2}^{1/2} & \propto & \abs{t}^{1/2}
\end{eqnarray}
so the standard deviation from the origin grows as $t^{1/2}$.  Mandelbrot and Van Ness \cite{mandelbrot68} introduced {\em fractional} Brownian motion (fBm) as a generalization to processes which grow at different rates $t^H$ 
\begin{equation}
  \expect{\left[X_H(t)-X_H(0)\right]^2}^{1/2} \propto \abs{t}^H
\end{equation}
where $0<H<1$ is called the Hurst exponent.

\begin{figure}
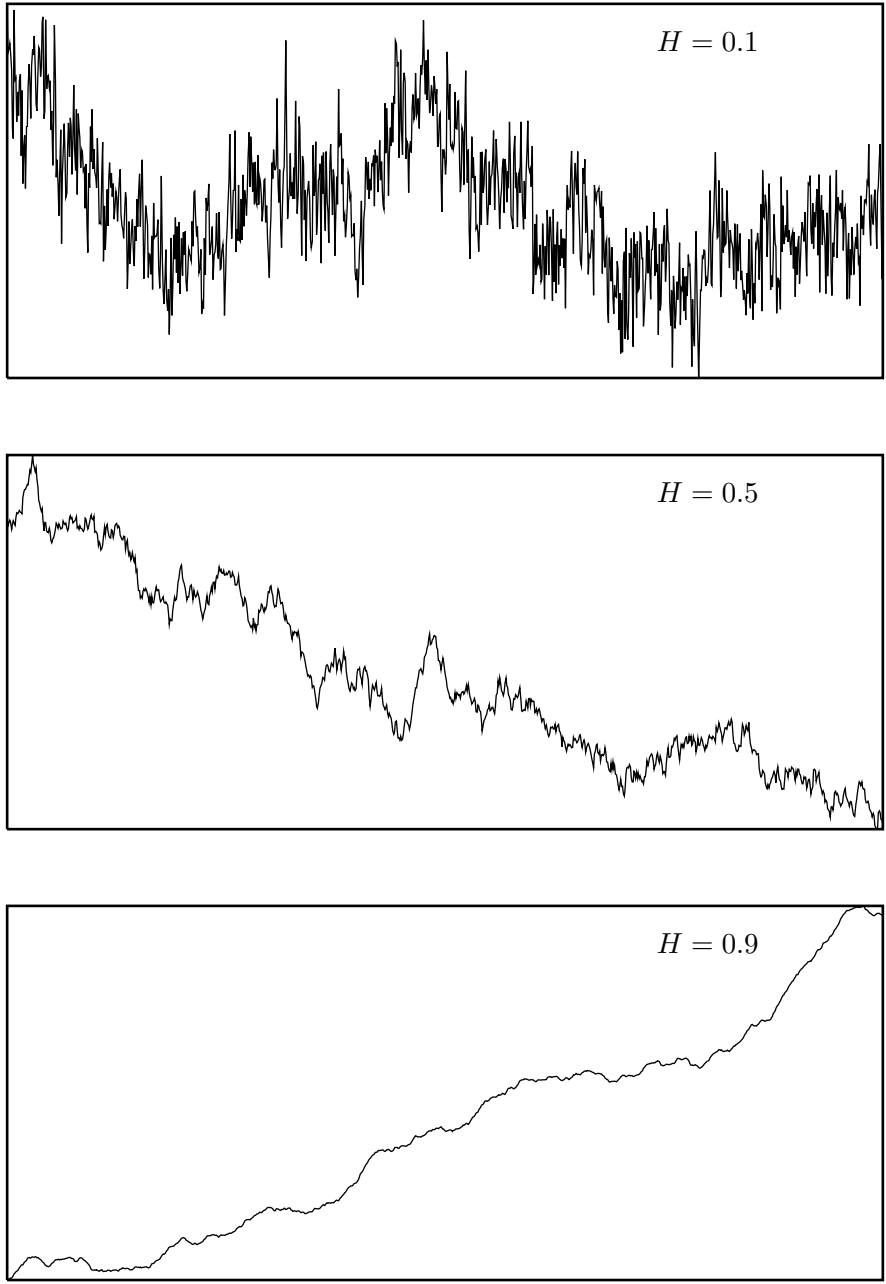
\centering
	\input{apHurst/H0.1.tex}\\
	\vspace{0.25in}
	\input{apHurst/H0.5.tex}\\
	\vspace{0.25in}
	\input{apHurst/H0.9.tex}
  \caption{Sample fractional Brownian motion time series with different Hurst exponents: antipersistent $H=0.1$ (top) has negative long-range correlations, uncorrelated $H=0.5$ (center) is standard Brownian motion, and persistent $H=0.9$ (bottom) has positive long-range correlations.}
\label{fig:hurstSamples}
\end{figure}

Successive increments $\xi_H$ of a fractional Brownian motion are called fractional Gaussian noise (fGn)
\be
  \xi_H(t) = X_H(t+\delta) - X_H(t)
\ee
where $\delta$ can always be rescaled to one (to be discussed).  The autocorrelation function (which measures the covariance of a data series with itself at some lag $\tau$) is formally defined as
\be
  C(\tau) \equiv \frac{
    \expect{ \left[ \xi_H(t) - \expect{\xi_H(t)} \right] 
      \left[ \xi_H(t-\tau) - \expect{\xi_H(t-\tau)} \right]
    }
  }{
    \left\{
      \expect{ \left[ \xi_H(t) - \expect{\xi_H(t)} \right]^2 }
      \expect{ \left[ \xi_H(t-\tau) - \expect{\xi_H(t-\tau)} \right]^2 }
    \right\}^\half
   }
\label{eq:hurstAutoCorr}.
\ee
For an fGn process the definition is \cite{mandelbrot68,mandelbrot71}
\be
  C(\tau)  =  \half \left( \abs{\tau+1}^{2H} - 2 \abs{\tau}^{2H} + \abs{\tau-1}^{2H} \right)
\ee
which is obviously zero for $H=1/2$ (except for $\tau=0$ where the autocorrelation is always one) while for general $H\neq 1/2$ and large $\tau$
\bea
\label{eq:hurstAutoCorrLimit}
  \lim_{\tau\rightarrow\infty} C(\tau)
    & \propto & \tau^{2H} \left[ (1+\tau^{-1})^{2H} - 2 + (1-\tau^{-1})^{2H} \right] \\
    & \propto & \tau^{2H} \left[ (1+2H\tau^{-1} + H(2H-1)\tau^{-2}) - 2 + (\cdots) \right] \\
    & \propto & \tau^{2H} \left[ \tau^{-2} \right] \\
    & \propto & \tau^{2H-2} \\
\eea
so correlations decay slowly and the resulting fractional Brownian motion exhibits long memory effects.  Correlations are positive for $H>1/2$ ({\em persistence}) and negative for $H<1/2$ ({\em antipersistence}) as shown in \fig{hurstSamples}.  (Note that fBm is not the only framework for generating long range memory effects: for instance, fractional ARIMA(0,$d$,0) processes also exhibit scaling with an exponent $H = d + 1/2$ \cite{taqqu95}.)

As for standard Brownian motion, all fBm series are self-affine \cite{rambaldi94, taqqu95}
\be
  X_H(at) \stackrel{d}{=} a^H X_H(t)
\label{eq:hurstAffine}
\ee
meaning that the series appears statistically identical under rescaling the time axis by some factor $a$ and the displacement $X_H$ by $a^H$.  Hence, fBm lacks any characteristic time scale and when generating or sampling an fBm series, an arbitrary step length of one unit may be used without loss of generality \cite{yin96}.  Self-affine signals can be described by a {\em fractal} dimension $D$ which is related to the Hurst exponent through $D=2-H$ for fBm \cite{bassingthwaighte94,raymond99}.  (The fractal dimension $D$ can be loosely interpreted as the ``number of dimensions'' the signal fills up.  For example, notice that in \fig{hurstSamples} the $H=0.1$ signal ``fills in'' significantly more space than $H=0.9$ and, consequently, has a higher fractal dimension.)

\begin{figure}
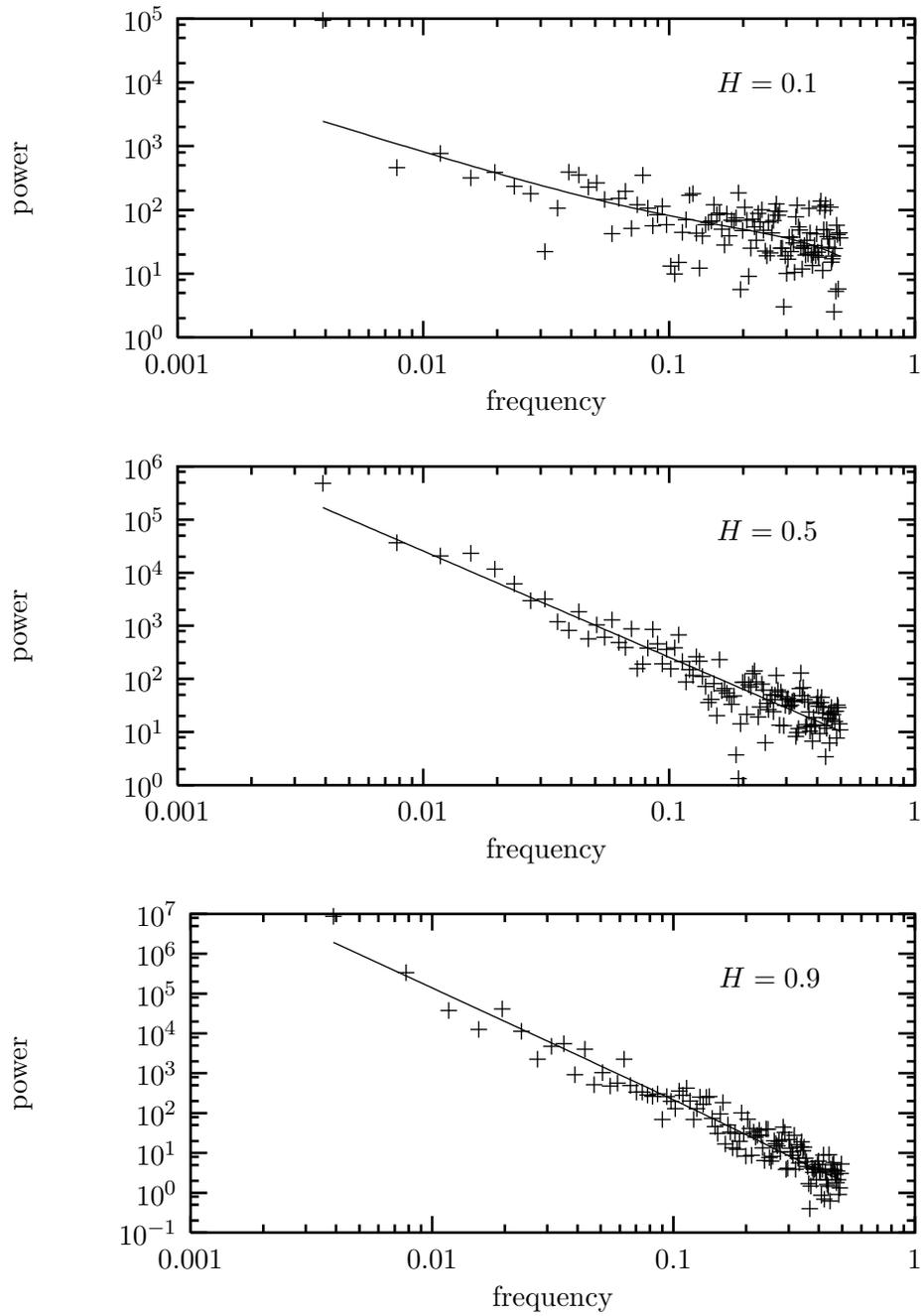
\centering
	\input{apHurst/psd0.1.tex}\\
	\vspace{0.25in}
	\input{apHurst/psd0.5.tex}\\
	\vspace{0.25in}
	\input{apHurst/psd0.9.tex}
  \caption{Power spectral densities for the fractional Brownian motion time series shown in \fig{hurstSamples}.  The points are from finite samples of 1000 points each and the line represents the theoretical spectrum.  For low frequencies the power spectrum is well approximated by a power law $1/f^{2H+1}$.}
\label{fig:hurstPsd}
\end{figure}

The power spectrum (defined as the amplitude-squared contributions from the frequencies $\pm f$, $S(f) \equiv \abs{F_H(f)}^2 + \abs{F_H(-f)}^2$ where $F_H$ is the Fourier transform of $X_H$ \cite{press92}) of fBm also demonstrates scaling behaviour.  The exact spectrum is difficult to compute but for low frequencies it can be approximated by a power law $S(f) \sim 1/f^{2H+1}$ \cite{taqqu95} (see \fig{hurstPsd}) which corresponds to long-term spatial correlations.  {\em Flicker} or $1/f^\alpha$ noise with $\alpha \approx 1$ is ubiquitous in nature (see Ref.\ \cite{bak88} and references therein) and some of it may be attributable to long-memory fBm processes \cite{pilgram98}.  Note that from the definition of the Fourier transform the derivative of fBm, fractional Gaussian noise, also has a low frequency power law spectrum but with exponent reduced by 2, i.e.. $1/f^{2H-1}$.

Fractional Brownian motion has been criticized because it lacks a physical interpretation and because the process has an unrealistic {\em infinite} memory \cite{mandelbrot83, mesa93}.  However, it suits our purposes here because it is a mathematically elegant extension of standard Brownian motion which introduces long-range memory effects and can be characterized by a single parameter $H$.  Hence, it is an ideal experimental control for testing procedures of measuring the Hurst coefficient in real data sets.

\section{Synthesis}

Before we can test various methods of estimating the Hurst exponent, we need some {\em control} data sets with known exponents.  This data must be synthesized from the first principles of fractional Brownian motion as defined above.  The computational difficulty is that for $H\neq 1/2$ fBm has an infinite dependence on its history so approximations are required.  A number of generators have been proposed \cite{mandelbrot71, rambaldi94, yin96, carmona98} but most are slow and/or inaccurate.  One of the most common techniques, Successive Random Addition (SRA) \cite{voss85} is very fast but its correlation function does not match that of fBm.

Another technique, the Spectral Synthesis Method (SSM) \cite{schepers92}, uses the scaling behaviour $1/f^{2H+1}$ of the power spectrum to generate synthetic data in frequency space and then inverse Fourier transform the data to recover the desired time series.  Although simple and fast---the Fast Fourier Transform (FFT) algorithm only requires on the order of $N \log N$ operations---it fails because the power law in the frequency domain only applies for low frequencies, as mentioned above.

I prefer the generator by Vern Paxson \cite{paxson97} because it is quick and accurate.  It also uses a Fourier transform but it uses an accurate approximation to the fBm correlation function to generate a proper power spectrum.  

The basic algorithm for generating a data set of $N$ points with Hurst exponent $H$ follows:

\begin{enumerate}
  \item{Find the smallest integer $N_8$ which is a power of 2 and is not smaller than $8 N$.}
  \item{Generate a discrete power spectrum for $f_i=i/N_8,\, i=1,\ldots, N_8/2$ using Paxson's equations \cite[Eqs. (4-6)]{paxson97} given here for convenience (see Ref.\ \cite{sinai76} for derivation):
    \be
      S(f) = {\cal A}(f,H)\left[ \abs{2\pi f}^{-2H-1} + \tilde{\cal B}_3''(f,H) \right]
    \ee
    where
    \be
      {\cal A}(f,H) = 2\sin(\pi H) \Gamma(2H+1)\left(1-\cos(2\pi f)\right)
    \ee
    and
    \begin{eqnarray}
      \tilde{\cal B}_3''(f,H) & = & \left[ 1.0002 - 0.000842 f \right] \tilde{\cal B}_3'(f,H) \\
      \tilde{\cal B}_3'(f,H) & = & \tilde{\cal B}_3(f,H) - 2^{-7.65 H -7.4}
    \end{eqnarray}
    are improved approximations of 
    \be
      \tilde{\cal B}_3(f,H) \approx a_1^d + b_1^d + a_2^d + b_2^d + a_3^d + b_3^d + \frac{a_3^{d'} +b_3^{d'} + a_4^{d'} + b_4^{d'}}{8\pi H}
    \ee
    where
    \bea
      d & = & -2H-1 \\
      d' & = & -2H \\
      a_k & = & 2\pi (k+f) \\
      b_k & = & 2\pi (k-f).
    \eea
  }
  \item{Choose a zero-amplitude null component of the power spectrum $S(0) = 0$ to detrend the fGn increments in real space (zero mean).}
  \item{Multiply each component of the power spectrum by a Poisson distributed uniform deviate $\eta_f$ with mean $\expect{\eta}=1$
    \be
      S(f) \leftarrow \eta_f S(f).
    \ee
    This simulates the noise associated with a real data series, for which uncertainties in the power spectrum are multiplicative \cite[p.\ 552]{press92}.
  }
  \item{Construct the complex Fourier space representation of the series $f_i$, $i=-N_8/2,\ldots,+N_8/2$ from the power spectrum using random phases $0 \leq \theta_f < 2\pi$
    \bea
      F_H(f) = \sqrt{S(\abs{f})}\, e^{i\theta_f}.
    \eea
    Randomizing the phases does not disturb the power spectrum and ensures the finite-sample correlation function converges to the proper theoretical form in the limit $N\rightarrow \infty$ \cite{yin96}.
  }
  \item{Compute the inverse Fourier transform $\xi_H(t_i), i=1\ldots N_8$ and discard the imaginary components to get a fractional Gaussian noise series
    \be
      \xi_H(t) = \Re \left({\cal F}^{-1}\left[ F_H(f) \right] \right).
    \ee
  }
  \item{Pick a random subset of length $N$ of the series and discard the remainder.  This minimizes wrap-around effects from the Fast Fourier Transform \cite{press92,bassingthwaighte95,caccia97,pilgram98} and gives the illusion of a non-stationary series (to simulate real data, for which the stationarity may be difficult to decide).  Note that Paxson does not consider subsampling in his original algorithm.}
  \item{Finally, to convert to a fractional Brownian motion, simply integrate
    \be
      X_H(t) = X_H(t-1) + \xi_H(t).
    \ee
  }
\end{enumerate}

Paxson's method is accurate \cite{paxson97}, computationally simple, and fast (most of the computation is in the Fourier transform so it still only requires on the order of $N \log N$ operations).

\section{Analysis}

One's first instinct to check for long-range correlations in a data set may be to simply test how quickly the autocorrelation function (\eq{hurstAutoCorr}) decays with large lags.  This proves to be a poor choice however, because antipersistent data is difficult to distinguish from uncorrelated, the correlations can be mistaken for noise fluctuations around zero.

A method to reliably estimate the Hurst coefficient from a time series would be a useful method of testing for and quantifying long-range correlations.  The oldest and still most common method is due to Hurst \cite{hurst51} who noticed that the range $R$ of the depth (or cumulated influx) of water behind a dam over a span of time $\tau$ was related to the standard deviation $S$ of the influx over the same period through
\be
	R/S \propto \tau^H
\ee
where $H$ should be $1/2$ for random, uncorrelated processes \cite{feder88}.  Hurst's method, Rescaled Range or $R/S$ analysis, was to sample non-overlapping subsets of length $\tau$ from a time series and calculate the average $R/S$ statistic.  Repeating over a wide range of $\tau$-values and recording the data on double-logarithmic graph the Hurst exponent should emerge as the slope of a straight line, $\log(R/S) = H\log \tau + C$.

Unfortunately, despite its extensive usage \cite{mandelbrot83, lo91, chen97, liebovitch97, gammel98, oliver98} Hurst's rescaled range analysis has been shown to be a poor estimator of $H$ \cite{feder88, mesa93, bassingthwaighte94, caccia97} with a consistent bias towards $H=0.7$ and requiring a large data set for convergence.

Another common technique for testing for correlations is shuffling the order of the data and comparing the statistics of the original data with the shuffled.  Shuffling destroys the correlations in the data but care must be taken to detrend the data as well.  Persistent data series are characterized by large low-frequency components which make the data series appear non-stationary (notice, for example, the trend in the $H=0.9$ series in \fig{hurstSamples}).  Shuffling without first removing this trend would not destroy these low-frequency correlations which extend throughout the entire dataset.  Shuffling is a valuable way of testing for correlations but, in itself, doesn't specify any statistical techniques for distinguishing the original from the shuffled series, and we have already seen that the correlation function and rescaled range analysis are inadequate.

A number of alternatives have been proposed including autocorrelation analysis \cite{schepers92}, Fourier analysis \cite{schepers92, pilgram98}, and maximum likelihood estimators.  The advantage of the latter is that they are not graphical techniques but numerical---they simply return the best estimate of the Hurst exponent directly.  Unfortunately, they require (at least) an assumption about the form of the long-range dependence (such as fBm or fractional ARIMA) and perform poorly if the assumption is incorrect \cite{taqqu97}.

Each of the above methods suffers from biases and slow convergence (a large dataset is required to reduce the bias).  However, two methods have been consistently better, requiring smaller datasets and exhibiting less bias \cite{bassingthwaighte95, cannon97, raymond99}: dispersional analysis and scaled-window variance analysis.  Both of these methods are graphical, producing a power law relationship from which the exponent can be read off as the slope of the line when using double-logarithmic axes.

\subsection{Dispersional analysis}

\label{sect:hurstDispersion}

Dispersional analysis, also known as the Aggregated Variance method \cite{taqqu95}, averages the differenced fGn series over bins of width $\tau$ and calculates the variance of the averaged dataset.  Given a fGn series $\xi_H(i), \, i=1,\ldots,N$ a particularly simple but effective version of the algorithm follows:

\begin{enumerate}
	\item{Set the bin size to $\tau=1$.}
	\item{Calculate the standard deviation of the $N$ data points and record the point $(\tau,\tau\cdot\sigma_\tau))$. \label{enum:hurstDispStart}}
	\item{Average neighbouring data points and store in the original dataset
		\be
			\xi_H(i) \leftarrow \half \left[ \xi_H(2i-1) + \xi_H(2i) \right]
		\ee
		and rescale $N$ and $\tau$ appropriately
		\bea
			N & \leftarrow & N/2 \\
			\tau & \leftarrow & 2 \tau.
		\eea
	}
	\item{As long as more than four data points remain ($N>4$) return to Step \ref{enum:hurstDispStart}.  (The reader may prefer to require more than four bins to reduce noise.)}
	\item{Perform a linear regression on the log-log graph
		\be
			\log\left(\tau\cdot\sigma_\tau\right) = H \log \tau + C;
		\ee
		the calculated slope is the best estimate of $H$.}

\end{enumerate}

Recording $\tau\cdot\sigma_\tau$ in Step \ref{enum:hurstDispStart} instead of just the standard deviation $\sigma_\tau$ is not standard but it simplifies the regression because the Hurst exponent can be simply be read off the graph instead of $H-1$.

\fig{hurstMethods} shows that Dispersional analysis performs significantly better than rescaled range analysis.

\subsection{Scaled Window Variance analysis}

The other method well-received method, Scaled Window Variance analysis (SWV), also known as Detrended Fluctuation Analysis \cite{peng94} or Residuals of Regression \cite{taqqu95}, applies to the cumulated fBm series instead.  Given a fBm series $X_H(i), \, i=1,\ldots,N$ my own variation of the algorithm follows:

\begin{enumerate}
	\item{Split the series into $M\equiv \floor{N/\tau}$ (where $\floor{x}$ is the floor operator---returning the greatest integer not greater than $x$) evenly-spaced bins of size $\tau=16$ (SWV is inaccurate for smaller $\tau$ \cite{raymond99})
		\be
			X_H^{(k)}(j) = X_H\left( (k-1)\kappa + j \right), \, j=1\ldots\tau
		\ee
		where
		\be
			\kappa = \floor{ \frac{N-\tau}{M-1} }
		\ee
		This allows the option of setting a minimum and maximum on the number of bins $M_{min} \leq M \leq M_{max}$.  Setting $M_{min}$ larger than $N/\tau$ will necessarily result in overlapping bins but this effect has been tested \cite{caccia97} and the benefit of the larger sample size outweighs the influence of cross-correlations introduced.  
	}
	\item{For each bin $k,\, k=1\ldots M$, detrend the local series by subtracting off
		\be
			\overline{X}_H^{(k)}(j) = mj+b
		\ee
		Three options for calculating the trendline have been tested \cite{cannon97}:
		\begin{enumerate}
			\item{No detrending: $m=0, b=0$.  This is only recommended for $N < 2^9$ data points.}
			\item{Bridge detrending.  Form a line between the first and last point in the bin:
				\bea
					m & = & \frac{1}{\tau-1}\left[ X_H^{(k)}(\tau) - X_H^{(k)}(1) \right] \\
					b & = & X_H^{(k)}(1) - m
				\eea
				Recommended for $N > 2^{12}$ data points.
			}
			\item{Linear detrending.  Perform a linear least-squares regression over the entire bin to calculate $m$ and $b$.  Recommended for intermediate $N$.
			}
		\end{enumerate}
	}
	\item{Calculate the residuals after subtracting off the trend line
		\be
			\widehat{X}_H^{(k)}(j) = X_H^{(k)}(j) - \overline{X}_H^{(k)}(j)
		\ee
	}
	\item{Calculate the standard deviation of the residuals in each bin $\sigma_\tau^{(k)}$ and compute the average and standard deviation of these samples
		\bea
			\sigma_\tau & \equiv & \expect{\sigma_\tau^{(k)}} \\
			\Delta_\tau & \equiv & \sqrt{\expect{\left( \sigma_\tau^{(k)} - \sigma_\tau\right)^2} }
		\eea
	}
	\item{If the average standard deviation $\sigma_\tau$ is non-zero convert it to a log-scale $\sigma_{\log\tau} \equiv \log \sigma_\tau$ and plot $\sigma_{\log\tau} \pm \Delta_{\log\tau}$ versus $\log \tau$.  The uncertainty on the log scale can be approximated by 
		\be
			\Delta_{\log\tau} \approx \frac{ \Delta_\tau }{ \sigma_\tau } \log e
		\ee
	}
	\item{Double the bin size $\tau \leftarrow 2 \tau$ and repeat while $N > 2\tau$.}
	\item{Perform a linear regression on the log-log graph $\log \sigma_\tau = H \log \tau + C$; the calculated slope is the best estimate of $H$.}
\end{enumerate}

\begin{figure}
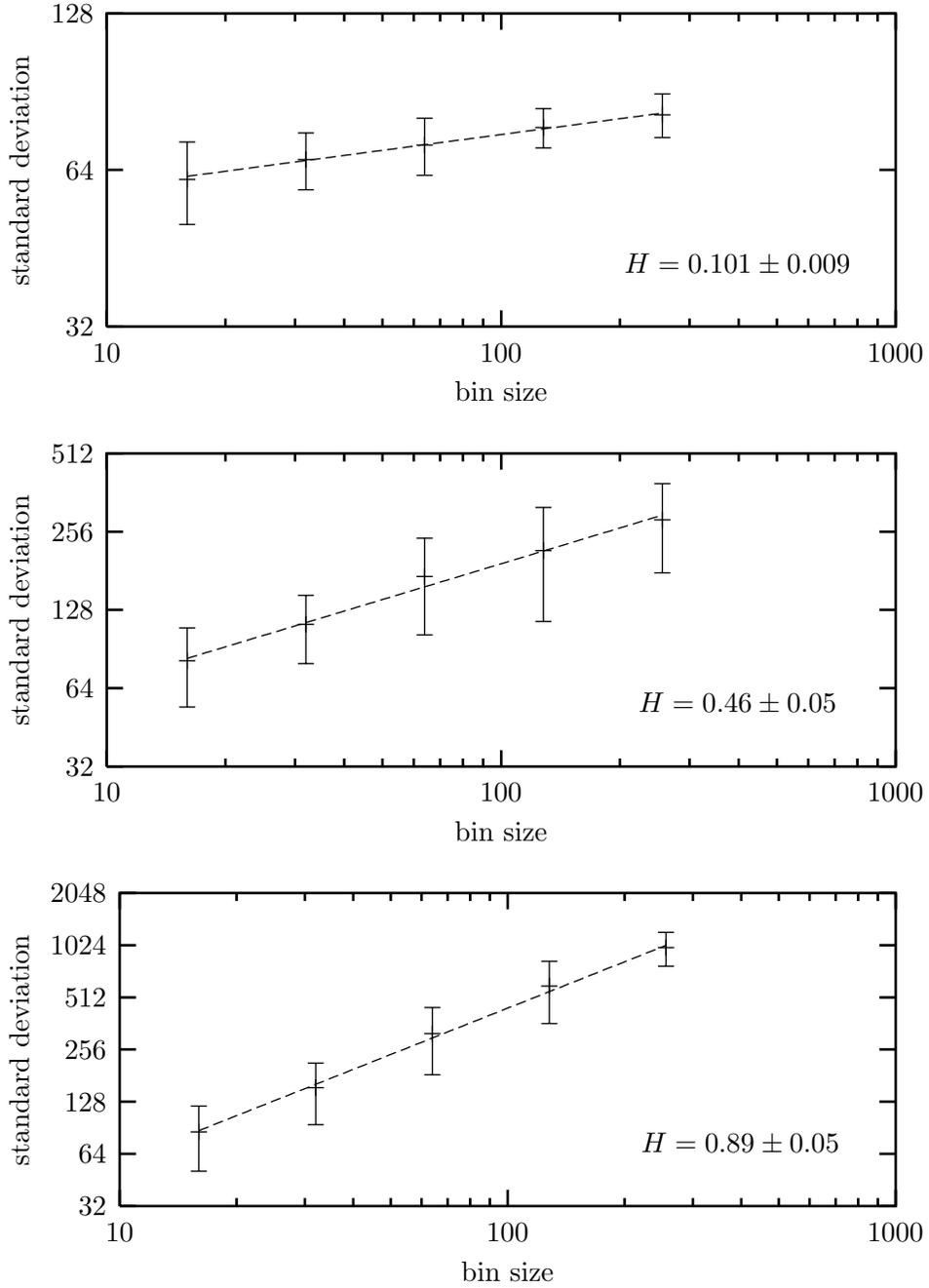
\centering
	\input{apHurst/swv0.1.tex}\\
	\vspace{0.25in}
	\input{apHurst/swv0.5.tex}\\
	\vspace{0.25in}
	\input{apHurst/swv0.9.tex}
  \caption{Scaled window variance analyses for the fractional Brownian motion time series shown in \fig{hurstSamples} (exact $H$=0.1, 0.5, and 0.9, respectively).  The estimated values of $H$ shown represent the best fit slopes of the lines.  The analysis used $M_{min}=4$ (see the text).}
\label{fig:hurstSwv}
\end{figure}

\begin{figure}\centering
	\input{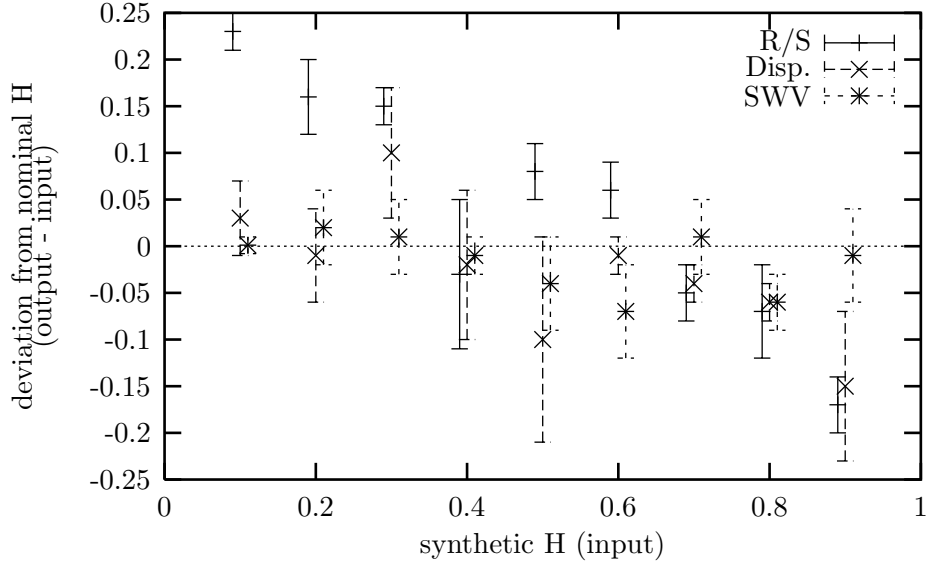}
  \caption{Comparison of Hurst estimators using synthetic datasets of 1000 points each.  The scaled-window variance method (SWV, $\ast$) performs significantly better than rescaled range analysis ($R/S$, $+$) and marginally better than dispersional analysis (Disp., $\times$).  (The points are offset slightly to improve readability.)}
\label{fig:hurstMethods}
\end{figure}

Sample fits using the SWV method (with at least $M_{min}=4$ bins) are shown in \fig{hurstSwv} using the same data as before.  Notice that these data sets are rather small (1000 data points each) but even so, the accuracy is remarkably good.  When compared with Hurst's rescaled range analysis (see \fig{hurstMethods}) it becomes clear that the SWV method is superior (also edging out the Dispersional method).

Another good feature of the SWV method is that, like all graphical techniques, it clearly reveals multifractal behaviour.  At some critical scale, the fractal dimension may crossover to a new value.  This is characterized in graphical techniques by a discontinuity in the slope of the log-log graph.  In particular, transitions to $H=0.5$ are often observed for large $\tau$, indicating a transition from correlated behaviour over short time scales to uncorrelated on long time scales.  The memory duration can simply be read off the graph as the transition point $\tau$.

\subsection{L\'evy Flight}

\label{sect:hurstLevy}

Despite its advantages, SWV fails in one respect: it is unable to distinguish between long-range correlations and uncorrelated L\'evy flight.  L\'evy flight is similar to (traditional) Brownian motion in that it is a cumulated series of independent, identically-distributed ({\em iid}) increments, but in this case, the increments are L\'evy distributed instead of normally distributed.

Normal or Gaussian distributions are well known to obey the following {\em stability} property: if $x_1$ and $x_2$ are both Gaussian-distributed random variables then their sum
\be
	x \equiv x_1 + x_2
\ee
is also Gaussian-distributed.  Paul L\'evy discovered a general class of distributions which have the stability property \cite{mandelbrot83, montroll83, montroll87}.  L\'evy distributions generally have no closed analytical form but can be defined in terms of their characteristic function $f(k)$ (the Fourier-space representation of the probability distribution) \cite{weron96, montroll87}
\be
	\ln f(k;\alpha,\beta) = \left\{ \begin{array}{ll}
		-\abs{k}^\alpha \left( 1 - i \beta\; \tan\frac{\pi\alpha}{2}\; \mbox{sign}(k) \right)
			& \alpha\not=1 \\
		-\abs{k} \left( 1 + \frac{2}{\pi} i \beta\; \ln\abs{k}\; \mbox{sign}(k) \right)
			&  \alpha=1
	\end{array} \right.
\ee
where $0<\alpha\leq 2$ is a characteristic exponent and $-1\leq \beta \leq 1$ is the skewness.  Special cases of the L\'evy distribution include the Gaussian ($\alpha=2$, $\beta=0$) and Cauchy ($\alpha=1$, $\beta=0$) distributions.

The stability property says that the sum of a large number of {\em iid} L\'evy random variables will also be a L\'evy random variable with the same $\alpha$ and $\beta$ in apparent violation of the Central Limit Theorem.  The paradox is resolved by recognizing that L\'evy distributions with $\alpha<2$ have power-law tails far from the origin
\be
	p(x) \sim \frac{1}{\abs{x}^{\alpha+1}}\, \mbox{ as }\, \abs{x}\rightarrow \infty
\ee
so the variance $\expect{x^2}$ is infinite for $\alpha<2$ whereas the Central Limit Theorem assumes a finite variance.

Another interesting property of L\'evy distributions is that the cumulative L\'evy flight $X_\alpha$ is self-affine, scaling as
\be
  X_\alpha(at) \stackrel{d}{=} a^{1/\alpha} X_\alpha(t)
\ee
which parallels the scaling relation Eq.\ \ref{eq:hurstAffine} for fractional Brownian motion.  A consequence of this is that Hurst coefficient estimators which depend on this scaling property may erroneously predict positive long-term correlations with
\be
	H = 1/\alpha
\ee
when applied to uncorrelated L\'evy flights with $1<\alpha<2$.

\begin{figure}\centering
	\input{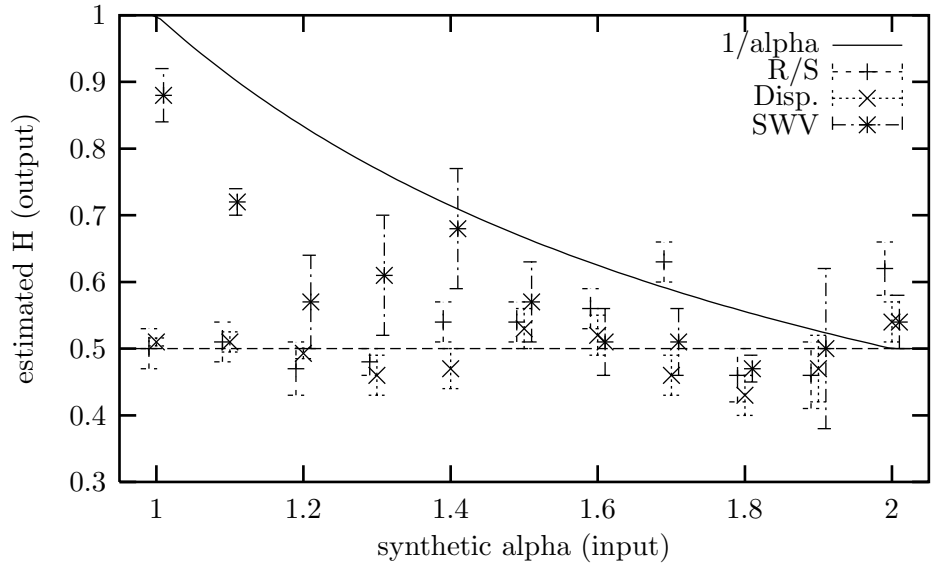}
  \caption{Comparison of Hurst estimators on uncorrelated L\'evy flight with characteristic exponent $\alpha$ using synthetic datasets of 1000 points each.  Rescaled range ($R/S$, $+$) and dispersional analysis (Disp., $\times$) perform well but scaled window variance analysis (SWV,$\ast$) performs poorly, especially for small $\alpha$, tending towards the $1/\alpha$ curve.  (The points are offset slightly to improve readability.)}
\label{fig:hurstLevyHurst}
\end{figure}

To test for this effect, data sets of 1000 symmetric ($\beta=0$) L\'evy distributed random variables with were synthetically generated (using a simple and elegant algorithm explained in Ref.\ \cite{weron96}) and cumulated to produce a one-dimensional L\'evy flight.  The synthetic data was then analyzed using the rescaled range, dispersional, and scaled-window variance techniques.  The results shown in \fig{hurstLevyHurst} indicate that SWV is sensitive to L\'evy noise whereas $R/S$ and Dispersion are not.  (Note also that the Fourier spectrum of L\'evy flight still approximates a power law $1/f^2$ with exponent 2 (indicating no correlations).)

\section{Conclusions}

In summary, to test for long-range correlations in a data set Dispersional analysis is recommended.  If more precision is required (especially for $H$ near 1) and the increments are Gaussian-distributed, a scaled window variance analysis should be performed.

\begin{figure}\centering
    \psfrag{fractional Brownian motion}[c][c]{\Large fractional Brownian}
    \psfrag{Levy motion}[c][c]{\Large L\'evy}
    \psfrag{Gaussian}{\Large Brownian}
    \psfrag{H}{\Large $H$}
    \psfrag{alpha}[c][c]{\Large $\alpha$}
    \includegraphics[width=0.6\columnwidth]{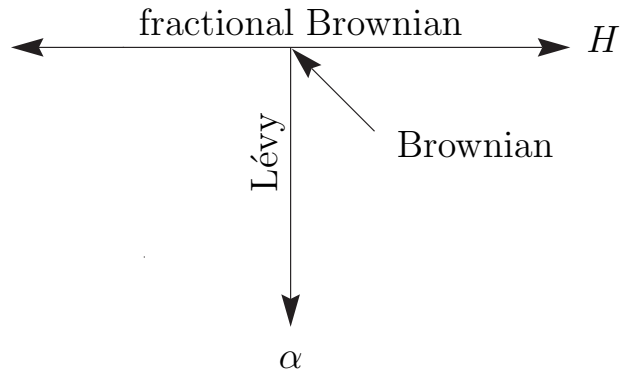} \\
  \caption{Schematic representation of relation between fractional Brownian motion and L\'evy flight.  Traditional Brownian motion sits at the intersection ($H=1/2$, $\alpha=2$).  The natural extension into the two-space is fractional L\'evy motion which has correlated, non-Gaussian increments.}
\label{fig:hurstFbmLevySpace}
\end{figure}

In this discussion we have explored tests for long-range correlations in fractional Brownian motion (correlated with Gaussian increments) and L\'evy flight (uncorrelated with non-Gaussian increments).  These two extensions to Brownian motion are not exclusive.  \fig{hurstFbmLevySpace} shows how that they are complementary notions and the two ideas can be combined to produce fractional L\'evy motion (fLm) with correlated, non-Gaussian increments.  There is very little literature on the subject but it may be a useful model for some natural phenomena \cite{kogon96}.  I am unaware of any efficient algorithm to synthesize fLm but it would begin with the correlation function in \eq{hurstAutoCorr}, which still applies.

\end{document}